\RequirePackage{fix-cm}
\documentclass[superscriptaddress,secnumarabic,amssymb,amsmath,nobibnotes,aps,prd,showkeys,showpacs,nofootinbib,twocolumn,a4paper,epjc3]{svjour3}
\smartqed  
\RequirePackage{graphicx}
\usepackage{graphicx}
\usepackage{placeins}
\usepackage{float}

\RequirePackage{latexsym}
\RequirePackage[numbers,sort&compress]{natbib}
\RequirePackage[colorlinks,citecolor=blue,urlcolor=blue,linkcolor=blue]{hyperref}
\journalname{Eur. Phys. J. C}
\usepackage{graphicx}
\usepackage{dcolumn}
\usepackage{bm}
\usepackage[
scale=0.7, marginratio={1:1, 2:3}, ignoreall,
text={7in,10in},centering,
margin=1.5in,
total={6.5in,8.75in}, top=1.2in, left=0.9in, includefoot,
height=10in,a5paper,hmargin={3cm,0.8in},
]{geometry}
\RequirePackage{graphicx}
\RequirePackage{mathptmx}      
\usepackage{adjustbox}
\RequirePackage{latexsym}
\RequirePackage[colorlinks,citecolor=blue,urlcolor=blue,linkcolor=blue]{hyperref}
\usepackage[english]{babel}
\usepackage{amssymb,amsmath,amsfonts,graphicx, graphics, setspace}
\usepackage{bigints}
\usepackage{anysize}
\usepackage{float}
\usepackage{fancyhdr}
\usepackage{subfigure}
\usepackage{pdflscape}
\usepackage{epsfig}
\usepackage{epsf}
\usepackage{epstopdf}
\usepackage{hyperref} 
\newtheorem{thm}{Theorem}

\newtheorem{lem}[thm]{Lemma}
\newtheorem{rem}{Remark}

\usepackage{amsmath}
\usepackage{amsfonts}
\usepackage{amssymb}
\usepackage{subfigure}
\usepackage{epstopdf}
\usepackage{epsf}
\usepackage{url} 
\usepackage{pdflscape} 
\usepackage{amsmath}
\usepackage{amssymb} 
\usepackage{epsfig} 
\usepackage{amsfonts}
\usepackage{graphicx}
\usepackage{times,fancyhdr}
\usepackage{enumitem}
\setlist[enumerate,2]{label=\roman*)}
\newcommand{\sfrac}[2]{{\textstyle{#1\over#2}}}
\def\case#1/#2{\textstyle\frac{#1}{#2}}

\newcommand{\be}{\begin{equation}}
\newcommand{\ee}{\end{equation}}
\newcommand{\ben}{\begin{eqnarray}}
\newcommand{\een}{\end{eqnarray}}
\def\udot{\dot{u}}
\usepackage{amsmath}
\usepackage{amsfonts}
\usepackage{amssymb}
\usepackage{widetext}
\usepackage{epstopdf}
\usepackage{epsf}
\def\ex{e_1{}^1}
\def\ey{e_2{}^2}
\def\ez{e_3{}^3}
\def\R{{}^3\!R}
\def\S{{}^3\!S_+}
\def\e{\mathbf{e}}
\def\sp{\sigma_+}
\def\udot{\dot{u}}
\def\ex{e_1{}^1}
\def\ey{e_2{}^2}
\def\ez{e_3{}^3}
\def\R{{}^3\!R}
\def\S{{}^3\!S}
\def\y{\vartheta}
\def\z{\zeta}

\begin{document}
\title{Averaging generalized scalar field cosmologies III: Kantowski--Sachs and closed Friedmann--Lemaître--Robertson--Walker models}
\author{Genly Leon \thanksref{e1,addr1} \and  Esteban Gonz\'alez \thanksref{e3,addr2} \and Samuel Lepe \thanksref{e4,addr3}  \and Claudio Michea \thanksref{e5,addr1} \and  Alfredo D. Millano \thanksref{e6,addr1}}
\thankstext{e1}{genly.leon@ucn.cl}
\thankstext{e3}{gonzalez.estebanb@gmail.com}
\thankstext{e4}{samuel.lepe@pucv.cl}
\thankstext{e5}{claudio.ramirez@ce.ucn.cl}
\thankstext{e6}{alfredo.millano@alumnos.ucn.cl}
\institute{Departamento  de  Matem\'aticas,  Universidad  Cat\'olica  del  Norte, Avda. Angamos  0610,  Casilla  1280  Antofagasta,  Chile\label{addr1} \and Direcci\'on de Investigaci\'on y Postgrado, Universidad de Aconcagua, Santiago, Chile. \label{addr2}
\and Instituto de F\'isica, Facultad de Ciencias, Pontificia Universidad Cat\'olica de Valpara\'iso, 
Av. Brasil 2950, Valpara\'iso, Chile \label{addr3}
}
\date{\today}

\maketitle

\begin{abstract}
Scalar field cosmologies with a generalized harmonic potential and matter with energy density $\rho_m$, pressure $p_m$, and  barotropic equation of state (EoS)  $p_m=(\gamma-1)\rho_m, \; \gamma\in[0,2]$  in Kantowski-Sachs (KS) and  closed Friedmann--Lemaître--Robertson--Walker (FLRW) metrics are investigated. We use methods from non--linear dynamical systems theory and  averaging  theory considering  a time--dependent perturbation function $D$.  We define a regular dynamical system over a compact phase space, obtaining global results. That is, for KS metric the global late--time  attractors  of  full and time--averaged systems are  two anisotropic contracting solutions,  which are non--flat locally rotationally symmetric (LRS) Kasner and  Taub (flat LRS Kasner) for $0\leq \gamma \leq 2$, and  flat FLRW matter--dominated universe if  $0\leq \gamma \leq \frac{2}{3}$. For closed FLRW metric late--time  attractors of full and averaged systems are  a flat matter--dominated FLRW universe for  $0\leq \gamma \leq \frac{2}{3}$ as in KS and Einstein-de Sitter solution for $0\leq\gamma<1$. Therefore, time--averaged system determines future asymptotics of  full system. Also, oscillations entering the system through  Klein-Gordon (KG) equation can be controlled and smoothed out when $D$ goes monotonically to zero, and incidentally  for the whole $D$-range  for KS and for closed FLRW  (if $0\leq \gamma< 1$) too. However, for $\gamma\geq 1$ closed FLRW  solutions of the full system depart from the solutions of the averaged system as $D$ is large. Our results are supported by numerical simulations. 
\end{abstract}


\maketitle

\section{Introduction}

Scalar fields have played important roles in the physical description of the universe \cite{Foster:1998sk,Miritzis:2003ym,Dania&Yunelsy,Leon:2008de,Giambo:2009byn,Giambo:2008ck,Leon:2010ai,Leon:2014rra,Fadragas:2014mra,vandenHoogen:1999qq,Copeland:1997et,Tzanni:2014eja,Giambo:2019ymx,Cid:2017wtf,Alho:2014fha,Guth:1980zm,Guth:1980zm2,Linde:1983gd,Linde:1986fd,Linde:2002ws,Guth:2007ng}, particularly, in the inflationary scenario. For example, chaotic inflation is a model of cosmic inflation in which  the potential term takes the form of the harmonic potential $V(\phi)= \frac{m_\phi^2 \phi^2}{2}$ \cite{Linde:1983gd,Linde:1986fd,Linde:2002ws,Guth:2007ng}. 
Scalar field models can be examined by means of qualitative techniques of dynamical systems \cite{Coddington55,Hale69,AP,wiggins,perko,160,Hirsch,165,LaSalle,aulbach,TWE,coleybook,Coley:94,Coley:1999uh,bassemah,LeBlanc:1994qm,Heinzle:2009zb}, which allow a stability analysis of the solutions. Complementary, asymptotic methods and averaging theory  \cite{dumortier,fenichel,Fusco,Berglund,holmes,Kevorkian1,Verhulst} are helpful to obtain relevant information about the solutions space of scalar field cosmologies: (i) in the vacuum and (ii) in the presence of matter \cite{Leon:2021lct,Leon:2021rcx}. In this process  one idea is  to construct a time--averaged version of the original system. By solving this version the oscillations of the original system are smoothed out \cite{Fajman:2020yjb}. This can be achieved  for Bianchi I, flat FLRW, Bianchi III, and negatively curved FLRW metrics  where  the Hubble parameter  $H$   plays the role of a time--dependent perturbation function which controls the magnitude of the error between the solutions of  full and  time--averaged problems as $H \rightarrow 0$ \cite{Leon:2021lct,Leon:2021rcx}. 

\noindent The conformal algebra of Bianchi III and Bianchi V spacetimes, which admits a proper conformal Killing vector, was studied in \cite{Mitsopoulos:2019afs}. In \cite{Paliathanasis:2016rho} the method of Lie symmetries was applied for Wheeler-De Witt equation in Bianchi Class A cosmologies for minimally coupled scalar field gravity and Hybrid Gravity in General Relativity (GR). Several invariant solutions were determined and classified
according to the form of the scalar field potential by means of these symmetries.

\noindent Based on references 
\cite{Leon:2020pfy,Leon:2019iwj,Leon:2020ovw,Leon:2020pvt} we started the   ``Averaging generalized scalar field cosmologies'' program \cite{Leon:2021lct}. The idea is to use asymptotic methods and averaging theory   to obtain relevant information about the solutions space of scalar field cosmologies in the presence of matter with energy density $\rho_m$ and pressure $p_m$ with  a barotropic  EoS  $p_m=(\gamma-1)\rho_m$ (with barotropic index $\gamma\in[0,2]$) minimally coupled to a scalar field with generalized harmonic potential \eqref{pot}.  This research program has three steps according to three cases of study:  (I)  Bianchi III and open FLRW model \cite{Leon:2021lct}, (II)  Bianchi I and flat  FLRW model \cite{Leon:2021rcx}, and (III) KS and closed FLRW. 

\noindent In reference \cite{Leon:2021lct} Bianchi III metrics were studied and written conveniently as 
\begin{align}
\label{metricLRSBIII}
   &  ds^2= - dt^2 + A^2(t) dr^2 
 + B^2 (t) \mathbf{g}_{H^2},
\end{align}
where $\mathbf{g}_{H^2}=   d \vartheta^2 +  \sinh^2 (\vartheta)d \zeta^2$ denotes the 2-metric of negative constant curvature on hyperbolic 2-space and the lapse function $N$ was set one.  Moreover, functions $A(t)$ and $B(t)$ are interpreted as scale factors. There we calculate the characteristic scale factor $\ell$ to obtain  
\begin{equation}
    \ell = \ell_0 (A B^2)^{1/3}.
\end{equation}
The formal definition of $\ell$ is given in eq.  \eqref{characteristic-length}. For FLRW, it is the scale factor of the universe. By convention we set $t=0$ as the present time.

In reference \cite{Leon:2021lct}  late--time attractors of original and time--averaged systems for LRS Bianchi III were found to be the following: 
\begin{enumerate}
    \item  A solution with the asymptotic metric
 \begin{align}
\label{scaling2}
   &  ds^2= - dt^2 + \ell_{0}^2 \left(\frac{3 \gamma  H_{0} t}{2}+1\right)^{\frac{4}{3
   \gamma }} \left (dr^2 + \mathbf{g}_{H^2}\right),
\end{align} for  $0\leq \gamma \leq \frac{2}{3}$,    
where $\ell_0$ is the current value of the characteristic scale factor $\ell$ and  $H_0$ is the current value of the Hubble factor. It represents a matter--dominated flat FLRW universe. 

\item A solution with the asymptotic metric 
\begin{align}
\label{scaling2b}
   &  ds^2= - dt^2 + \ell_{0}^2 \left(\frac{3 \gamma  H_{0} t}{2}+1\right)^{\frac{4}{3
   \gamma }} \left( dr^2 +  \mathbf{g}_{H^2}\right),
\end{align}    
for  $\frac{2}{3}<\gamma <1$, which represents a  matter-curvature scaling solution. 

\item A solution with asymptotic metric
\begin{align}
   &  ds^2= - dt^2 + c_1^{-2} dr^2 +  \frac{(3
   H_{0} t+2)^2}{4 c_1} \mathbf{g}_{H^2}, \label{metricD}
\end{align}
for $1\leq \gamma\leq 2$, where $c_1$ is a constant. It  corresponds to the Bianchi III form of flat spacetime (\cite{WE} p 193, eq.  (9.7)).
\end{enumerate}

In reference \cite{Leon:2021lct} the  open FLRW model was studied, whose metric is given by 
\begin{align}
\label{mOpenFLRW}
   &  ds^2= - dt^2 + A^2(t) dr^2 
 + A^2(t) \sinh^2(r) d\Omega^2,
\end{align}
where $d\Omega^2= d\y^2 + \sin^2 \y\, d\z^2$ and $A(t)$ is the scale factor of the isotropic and homogeneous universe. The late--time attractors are the following:
\begin{enumerate}
    \item A solution with asymptotic metric     
    \begin{align}
\label{scaling}
   &  ds^2= - dt^2 + \ell_{0}^2 \left(\frac{3 \gamma  H_{0} t}{2}+1\right)^{\frac{4}{3
   \gamma }} \left (dr^2 +\sinh^2 (r) d\Omega^2 \right), 
\end{align}
for $0\leq \gamma \leq \frac{2}{3}$, corresponding to a flat matter--dominated FLRW universe. 
\item A solution with asymptotic metric
\begin{align}
   &  ds^2= - dt^2 + \ell_{0}^2 \left(H_0 t+1\right)^{2} \left( dr^2+
\sinh^2 (r) d\Omega^2 \right), \label{Milne}
\end{align}
for $\frac{2}{3}<\gamma <2$, corresponding to a curvature dominated Milne  solution (\cite{Milne}; \cite{WE} Sect. 9.1.6, eq.  (9.8), \cite{Misner:1974qy,Carroll:2004st,Mukhanov:2005sc}). 
In all metrics  the matter--dominated flat  FLRW universe represents quintessence fluid if $0< \gamma < \frac{2}{3}$. 
\end{enumerate}
\noindent The chosen barotropic equation of state can mimic one of several fluids of interest in early--time and late--time cosmology. Typical values are   $\gamma =2$ corresponding to stiff matter, $\gamma =\frac{4}{3}$ corresponding to radiation, $\gamma =1$ corresponding to cold dark matter (CDM), $\gamma=\frac{2}{3}$ corresponding to Dirac-Milne universe, and $\gamma=0$ corresponding to cosmological constant (CC). According to our stability analysis,  the  ranges $0\leq \gamma <2, \; 0\leq \gamma \leq \frac{2}{3}, \; 0<\gamma <1$, and $1<\gamma <2$ are found. 
Special cases $\gamma=1$ and  $\gamma =2$, corresponding to bifurcations parameters where the stability changes, are treated separately. It is important to mention that stiff matter is a component present in a very early evolution, and had a role before the last scattering epoch. The last scattering epoch is an important cornerstone in the cosmological history since
after that, the cosmic microwave background (CMB) photons freely travelled  through the universe providing a photographic picture of the
universe at that epoch. Also, radiation is an early relevant cosmic component, although even today we have (tiny) traces of it. Moreover, CDM is an important component in current cosmology.
 The range $%
0 <\gamma <\frac{2}{3}$ corresponds to a quintessence and $ \gamma = 0 $ represents the CC (``omnipresent" in cosmic evolution). In the current state of cosmology  the dark components satisfy  $0\leq \gamma <\frac{2}{3}$ (dark energy) and $\gamma=1$ (CDM). The Dirac-Milne universe is characterized by $ \gamma = \frac{2}{3}$. 

For FLRW  metric,  the characteristic length scale $\ell$ coincides with the scale factor of the universe. Thus, Friedmann's usual scheme leads to
\begin{eqnarray*}
3H^{2} &=&\rho_m -3\frac{k}{\ell^{2}}, \\
\dot{\rho}_m+3H\gamma \rho_m  &=&0\implies \rho_m \left( z\right) =\rho_m
\left( 0\right) \left( 1+z\right) ^{3\gamma },
\end{eqnarray*}%
where $z$ is the redshift and we use $k$ to indicate if the model is a closed ($k=1$), flat ($k=0$), or open ($k=-1$). If  $\gamma =\frac{2}{3}$ and using  $1+z=\ell_{0}/\ell$, we have from the above equations
\begin{eqnarray*}
\rho_m \left( z\right)  &=&\rho_m \left( 0\right) \left( 1+z\right) ^{2} \\
&\implies &H\left( z\right) =\sqrt{\frac{1}{3}\left( \rho_m \left(
0\right) -\frac{3k}{\ell_{0}^{2}}\right) }\left( 1+z\right) .
\end{eqnarray*}%
Hence, if $k=0$,
\[
H\left( z\right) =\sqrt{\frac{\rho_m \left( 0\right) }{3}}\left( 1+z\right) .
\]%
For vacuum $\rho_m=0$ and  $k=-1$ (open case)
\begin{equation}
\label{DiracMilne}
H\left( z\right) =\frac{1}{\ell_{0}}\left( 1+z\right).     
\end{equation}
\noindent
Thus, we obtain a Dirac-Milne universe. The behavior $ H \left (
z \right) \approx \left (1 + z \right) $ is also satisfied in presence of matter ($\rho_m\neq 0$) and for $k = 0$.  More generically, for  $k=-1$  and a fluid that dilutes over time for which $ \gamma> \frac{2}{3} $, then, 
\begin{eqnarray*}
H\left( z\right)  &=&\sqrt{\frac{1}{3}\left( \rho_m \left( 0\right) \left(
1+z\right) ^{3\left( \gamma -\frac{2}{3}\right) }+\frac{3}{\ell_{0}^{2}}\right) }\left(1+z\right).
\end{eqnarray*}
For $ \gamma> \frac{2}{3}$, the dominant term  as $z\rightarrow -1$ is given by \eqref{DiracMilne}. 
Namely, the asymptotic evolution is towards a Dirac-Milne type evolution. On the contrary, for $\gamma<\frac{2}{3}$ the universe becomes matter--dominated.

\noindent Following our research program,  in reference \cite{Leon:2021rcx} case (II) was studied. Late--time attractors of the original and time--averaged systems for Bianchi I and flat FLRW are the following. 
\begin{enumerate}
    \item For $0\leq \gamma <1$  the late--time  attractor is  a matter--dominated FLRW universe  mimicking de Sitter, \newline quintessence, or zero acceleration solutions  with asymptotic metric 
     \begin{align}
   &  ds^2= - dt^2 + \ell_{0}^2 \left(\frac{3 \gamma  H_{0} t}{2}+1\right)^{\frac{4}{3
   \gamma }} dr^2 \nonumber \\
   & + \ell_{0}^2 \left(\frac{3 \gamma  H_{0} t}{2}+1\right)^{\frac{4}{3
   \gamma }}  \left[ d \vartheta^2 +  \vartheta^2 d \zeta^2\right]. \label{eq60}
\end{align}   

\item For $1<\gamma\leq 2$ the late--time  attractor is an equilibrium solution with asymptotic metric
\begin{align}
   &  ds^2= - dt^2 + c_1^{-2} {t^{\frac{4}{3}}} dr^2  +  {c_2^{-1}}{t^{\frac{4}{3}}} \left[ d \vartheta^2 +  \vartheta^2 d \zeta^2\right], \label{eq61}
\end{align}   
 \noindent where $c_1$ and  $c_2$ are constants. This solution can be associated with Einstein--de Sitter solution (\cite{WE}, Sec 9.1.1 (1)) with $\gamma= 1$).
\end{enumerate}
This paper, which is the third of the series, is devoted to the case (III) KS and positively curved FLRW metrics.  We will prove that  the quantity 
\begin{equation}
D=\sqrt{H^2 + \frac{\R}{6}}, \label{GEN-D}
\end{equation} where  $\R$ is the 3-Ricci curvature of spatial surfaces (if the congruence $\mathbf{u}$ is irrotational),  plays the role of a time--dependent perturbation function which controls the magnitude of the error between the solutions of  full and  time--averaged problems. The analysis of the system  is therefore reduced to study the corresponding time--averaged equations as the time--dependent perturbation function $D$ goes monotonically to zero  for a finite time interval.  The region where the perturbation parameter $D$ changes its monotony from monotonic decreasing to monotonic increasing is analyzed by a discrete symmetry and  by defining the variable $T=D/(1+D)$ that maps $[0, \infty)$ to a finite interval $[0,1)$. Consequently, the  limit $D\rightarrow +\infty$ corresponds to $T=1$ and the limit   $D\rightarrow 0$ corresponds to $T=0$.

The paper is organized as follows. In section \ref{motivation} we motivate our choice of potential and the topic of averaging in the context of differential equations. In section \ref{Sect2} we introduce the model under study. In section  \ref{SECT:II} we apply  averaging methods to analyze periodic solutions of a scalar field  with self-interacting potentials within the class of generalized harmonic potentials \cite{Leon:2019iwj}. In section  \ref{SECT3.5} KS model is studied  by using $D$--normalization, rather than  Hubble--normalization, because Hubble factor is not monotonic for closed universes. FLRW models with $k=+1$ (positive curvature) are studied in section \ref{SECT:IIIB}. 
In section  \ref{SECT:III} we study the resulting time--averaged systems for KS and positively curved FLRW models. In particular, in section \ref{KS}  KS model is studied.  The FLRW model  with $k=+1$ is studied in section \ref{FLRWclosed}. In section \ref{SECT:5} a regular dynamical system defined on a compact phase space  is constructed. This allows to find  global results  for KS and closed FLRW models. Finally, in section \ref{Conclusions} our main results are discussed.  
In \ref{gLKSFZ11}  the proof of the main theorem is given. In  \ref{numerics}  numerical evidence supporting the results of section \ref{SECT:II} is presented.

\section{Motivation}
\label{motivation}

\subsection{The generalized harmonic potential}

In this research we investigate a scalar field $\phi$ with  generalized harmonic potential
\begin{small}

\begin{equation}\label{pot}
 V(\phi)= \mu ^3 \left[b f \left(1-\cos \left(\frac{\phi }{f}\right)\right)+\frac{\phi ^2}{\mu}\right], \;    b> 0,
\end{equation}
\end{small} where $\mu ^3 b f\ll 1$ is interpreted as a perturbation parameter. When $b=0$, the parameter $\mu$ is related to the mass of the standard harmonic potential by  $\mu^2=\frac{m_\phi^2}{2}$. The potential satisfies  $V(\phi) = \mu ^2 \phi ^2+\mathcal{O}\left(1\right) \; \text{as} \; \phi\rightarrow \pm \infty$. Near the global minimum $\phi=0$, the  potential  takes the form $V(\phi) = \phi ^2 \left(\frac{b \mu ^3}{2 f}+\mu ^2\right)-\frac{\phi ^4 \left(b \mu
   ^3\right)}{24 f^3}+O\left(\phi ^5\right)$. Neglecting quartic terms, we have the corrected ``mass term'' $m_\phi^2=\left(\frac{b \mu ^3}{f}+ 2\mu ^2\right)$. 
   \noindent
Then, potential \eqref{pot}
can be re-expressed as 
\begin{small}
\begin{equation}
\label{pot_v2}
    V(\phi)=\mu ^2 \phi ^2 + f^2 \left(\omega ^2-2 \mu ^2\right) \left(1-\cos \left(\frac{\phi
   }{f}\right)\right),
\end{equation} 
\end{small}
by introducing a new parameter $\omega$ through  equation $b \mu ^3+2 f \mu ^2-f \omega ^2=0$. 
\noindent
Near the global minimum $\phi=0$, we have from \eqref{pot_v2}
that $V(\phi) = \frac{\omega ^2 \phi ^2}{2}+\mathcal{O}\left(\phi ^4\right) \; \text{as} \; \phi\rightarrow 0$. 
That is, $\omega^2$ can be related to the mass of the scalar field near its global minimum. The applicability of this re-parametrization will be discussed at the end of section \ref{section2-3}.  

Potential \eqref{pot_v2} has the following generic features:
\begin{enumerate} 
    \item$V$ is a real-valued smooth function  $V\in C^{\infty} (\mathbb{R})$  with  $\lim_{\phi \rightarrow \pm \infty} V(\phi)=+\infty$. 
        \item $V$ is an even function  $V(\phi)=V(-\phi)$.
    \item  $V(\phi)$ has always a local minimum at $\phi=0$;  $V(0)=0, V'(0)=0, V''(0)= \omega^2> 0$, what makes it suitable to describe oscillatory behavior in cosmology. 
    \item There is a finite number of values $\phi_c \neq 0$ satisfying $2 \mu ^2 \phi_c +f \left(\omega ^2-2 \mu ^2\right) \sin \left(\frac{\phi_c
   }{f}\right)=0$, which are local maxima or local minima depending on whether  $V''(\phi_c):= 2 \mu ^2+\left(\omega ^2- 2 \mu ^2\right) \cos \left(\frac{\phi_c }{f}\right)<0$ or $V''(\phi_c)>0$. For $\left|\phi_c\right| >\frac{f(\omega^2-2 \mu^2)}{2 \mu^2}= \phi_*$  this set is empty. 
    \item There exist 
    $V_{\max}= \max_{\phi\in [-\phi_*,\phi_*]} V(\phi)$   and $V_{\min}= \min_{\phi\in [-\phi_*,\phi_*]} V(\phi)=0$. The function $V$ has no upper bound  but it has a lower bound equal to zero.
  \end{enumerate}
\noindent
Potentials \eqref{pot} or \eqref{pot_v2} are related but not equal to the monodromy potential of  \cite{Sharma:2018vnv} used in the context of loop-quantum gravity, which is a particular case of the general monodromy potential  \cite{McAllister:2014mpa}. 
The potential studied in references \cite{Sharma:2018vnv,McAllister:2014mpa} for $p=2$, i.e.,
$V(\phi)= \mu^3 \left[\frac{\phi^2}{\mu} + b f \cos\left(\frac{\phi}{f}\right)\right]$, $b\neq 0$, is not good to describe the late--time FLRW universe driven by a scalar field as shown by references  \cite{Leon:2019iwj,Leon:2020ovw,Leon:2020pvt}  because it has two symmetric local negative minima  which are related to  Anti-de Sitter solutions. 
Setting  $\mu=\frac{\sqrt{2}}{2}$ and $b \mu=2$ in eq.  \eqref{pot} we recover the potential  
\begin{equation}
\label{EQ:23}
    V(\phi)= \frac{\phi^2}{2}+ f\left[1- \cos\left(\frac{\phi}{f}\right)\right], 
\end{equation}
that  was studied by \cite{Leon:2019iwj,Leon:2020ovw}.  Setting 
$\mu=\frac{\sqrt{2}}{2}$,  $\omega=\sqrt{2}$, we have 
\begin{equation}
\label{pot28}
 V(\phi)= \frac{\phi ^2}{2} + f^2 \left[1-\cos \left(\frac{\phi }{f}\right)\right]. 
\end{equation}
Potentials \eqref{EQ:23} and \eqref{pot28} provide non-negative local minima  which can be related to a late--time accelerated universe. Generalized harmonic potentials \eqref{pot_v2},  \eqref{EQ:23} and  \eqref{pot28}
belong to the class of potentials studied in \cite{Rendall:2006cq}.  Meanwhile, the potential  
$V(\phi)=\mu^4 \left[1- \cos \left(\frac{\phi}{f}\right)\right]$, where $\mu$ is a parameter, is relevant for axion models \cite{DAmico:2016jbm}. In  \cite{Balakin:2020coe} axionic dark matter with  modified periodic potential   $V(\phi, \Phi_*)= \frac{m_A^2 {\Phi_*}^2}{2 \pi^2}\left[1- \cos \left(\frac{2 \pi \phi}{\Phi_*}\right)\right]$, where $\Phi_*$ is a parameter describing the basic state
of the axion field,  has been studied in the framework of the axionic extension of Einstein-aether theory. This periodic potential has minima at $\phi =n \Phi_*$, where $n \in \mathbb{Z}$, whereas maxima
are found when $n \rightarrow m +
\frac{1}{2}$. Near the minimum $\phi =n \Phi_* + \psi$  when  $|\psi|$ is small, $V \rightarrow \frac{m_A^2 \psi^2}{2}$, where $m_A$ is the axion rest mass.

\noindent In reference \cite{Chakraborty:2021vcr}  an axion model given
by two canonical scalar fields $\phi_1$ and $\phi_2$ coupled via the potential
\begin{small}
\begin{align}
& V\left(\phi_1,\phi_2\right)=\mu_1^4\left[1-\cos\left(\frac{\phi_1}{f_1}\right)\right]   +\mu_2^4\left[1-\cos\left(\frac{\phi_2}{f_2}\right)\right] \nonumber \\
& +\mu_3^4\left[1-\cos\left(\frac{\phi_1}{f_1}-n\frac{\phi_2}{f_2}\right)\right],    
\end{align}
\end{small}
was investigated by combining standard dynamical systems tools and averaging techniques to investigate oscillations in Einstein-KG equations. As in references \cite{Leon:2021lct,Leon:2021rcx} methods from the theory of averaging nonlinear dynamical
systems allow to prove that time--dependent systems
and their corresponding time--averaged versions have
the same future asymptotic behavior. Thus, oscillations entering a nonlinear system through  KG equation can be controlled and smoothed out as the Hubble factor $H$  tends monotonically to zero. 

\subsection{Simple example of averaging problem}
One approximation scheme which can be used to approximately solve the ordinary differential equation  $\dot{\mathbf{x}}= \mathbf{f}(\mathbf{x}, t,\varepsilon)$ with $\varepsilon\geq 0$ and $\mathbf{f}$ periodic in $t$ is to solve the unperturbed problem $\dot{\mathbf{x}}= \mathbf{f}(\mathbf{x}, t,0)$ by setting $\varepsilon=0$ at first and then, with the use of the approximated unperturbed solution, to formulate variational equations in standard form which can be averaged. For example, consider the usual equation of a damped harmonic oscillator 
 \begin{equation}
 \label{harm-osc}
 \ddot \phi + \omega^2 \phi = \varepsilon (-2 \dot \phi),
 \end{equation}
 with given $\phi(0)$  and $\dot\phi(0)$, where $\omega^2$ is the undamped angular frequency of the oscillator, and the parameter $\epsilon=\zeta \omega$, with $\zeta$ the damping ratio, is considered as a small parameter. The unperturbed problem 
\begin{equation}
 \ddot \phi +\omega^2 \phi = 0   
\end{equation} 
admits the solution 
 \begin{equation}
\dot\phi(t)= r_0 \omega \cos (\omega t-\Phi_0), \;  \phi(t)= r_0 \sin (\omega t-\Phi_0),     
 \end{equation}
 where $r_0$ and $\Phi_0$ are constants depending on  initial conditions. 
 Using the variation of constants  we propose the solution for the perturbed problem \eqref{harm-osc} as 
 \begin{small}
 \begin{equation}
 \label{amplitud-phase}
 \dot{\phi}(t)= r(t) \omega \cos (\omega t-\Phi(t)), \;  \phi(t)  = r(t) \sin (\omega t-\Phi(t)),
 \end{equation}
 \end{small}
 such that
 \begin{small}
 \begin{equation}
 \label{eqAA25}
 r=\frac{\sqrt{\dot{\phi}^2(t)+\omega ^2 \phi^2(t)}}{\omega }, \;  \Phi =\omega t-\tan ^{-1}\left(\frac{\omega \phi
   (t)}{\dot \phi(t)}\right). 
 \end{equation}
 \end{small}
 This procedure is called the amplitude-phase transformation  in chapter 11 of \cite{Verhulst}. 
  
 \noindent
 Then,  eq. \eqref{harm-osc} 
 becomes  
 \begin{equation}
 \label{eq4}
 \dot r= -2 r \varepsilon  \cos ^2( \omega t-\Phi), \;  \dot\Phi = - \varepsilon \sin (2 (\omega t-\Phi )).
 \end{equation}
From eq. \eqref{eq4} it follows that $r$ and $\Phi$ are  slowly varying functions of time,
 and the system takes the form $\dot y= \varepsilon f(y)$.  The idea is to consider only nonzero average of the right--hand--side  keeping $r$ and $\Phi$
 fixed  and leaving out the terms with average zero and ignoring the slow--varying dependence of $r$ and $\Phi$ on $t$ through the averaging process \begin{equation}
\label{timeavrg}
      \bar{\mathbf{f}}(\cdot):=\frac{1}{L} \int_{0}^L \mathbf{f}(\cdot, t) dt, \quad L=\frac{2 \pi}{\omega}.  
\end{equation}
Replacing $r$ and $\Phi$ by their averaged approximations $\bar{r}$ and $\bar{\Phi}$ we obtain the system
 \begin{align}
 \label{eq6}
 & \dot {\bar{r}} = - \varepsilon \bar{r}, 
\quad  \dot{\bar{\Phi}} = 0. 
 \end{align}
   Solving eq. \eqref{eq6} with $\bar{r}(0)=r_0$ and $\bar{\Phi}(0)= \Phi_0$, we obtain  $\bar{\phi}= r_0 e^{-\varepsilon t} \sin (\omega t-\Phi_0)$ which is an accurate approximation of the exact solution
   \begin{align*}
     & \phi(t)=  -r_0  e^{-t \varepsilon }  \sin (\Phi_0) \cos
   \left(t \sqrt{\omega ^2-\varepsilon ^2}\right) \nonumber \\
   & -\frac{r_0 e^{-t
   \varepsilon } \sin \left(t \sqrt{\omega ^2-\varepsilon ^2}\right) (\varepsilon 
   \sin (\Phi_0)-\omega  \cos (\Phi_0))}{\sqrt{\omega
   ^2-\varepsilon ^2}},
   \end{align*} due to 
\begin{align*}
\bar{\phi}(t) - \phi(t)  =   \frac{r_0 \varepsilon  e^{-t \varepsilon } \sin (\Phi_0) \sin (t  \omega )}{\omega }+\mathcal{O}\left(\varepsilon^2  e^{-t \varepsilon }\right),
\end{align*}
as $\varepsilon \rightarrow 0^+$.

\subsection{General class of systems with a time--dependent perturbation function}
\label{section2-3}
Let us consider for example the Einstein--KG system
\begin{align}
\label{KGharmonic}
   & \ddot \phi + \omega^2 \phi = -3 H \dot \phi, \\
   &\dot{H}= -\frac{1}{2}\dot\phi^2. \label{Friedmann}
\end{align} The similarity between \eqref{harm-osc} and \eqref{KGharmonic}  suggests to treat the latter as a perturbed harmonic
oscillator as well  and to apply averaging in an analogous way. However, 
in contrast to $\varepsilon$, $H$ is time--dependent and it is governed by evolution equation \eqref{Friedmann}. Then, a surprising feature of such approach is the possibility of exploiting the fact that $H$ is strictly decreasing and goes to zero by promoting  Hubble parameter $H$  to a time--dependent perturbation function in \eqref{KGharmonic} controlling the magnitude of the error between solutions of the  full and time--averaged problems. 
Hence, with strictly decreasing $H$ the error should decrease as well. 
Therefore, it is possible to obtain  information about the large-time behavior of the more complicated full system via an analysis of the simpler averaged system of equations by means of dynamical systems techniques. This result is based on the monotony of $H$ and its sign invariance. 

\noindent
With this in mind, in \cite{Fajman:2021cli} the long-term behavior of solutions of a general class of spatially homogeneous cosmologies, when $H$ is positive strictly decreasing in $t$ and $\lim_{t\rightarrow \infty}H(t)=0$, was studied. 
However, this analysis is not valid  when the Hubble parameter is not a monotonic function as in the case of this study.

\section{Spatially homogeneous and anisotropic spacetimes}
\label{Sect2}
The spatially homogeneous but anisotropic spacetimes are known as either  Bianchi or KS cosmologies.  
In Bianchi models
the spacetime manifold is foliated along  the time axis with three dimensional
homogeneous hypersurfaces. On the other hand, the isometry group of KS spacetime is $\mathbb {R} \times SO(3)$ and it does not act simply transitively on spacetime, nor  does it possess a subgroup with simple transitive action. Hence, this model is spatially homogeneous but it does not belong to the Bianchi classification.  KS model approaches a closed FLRW model \cite{KS1,KS2,KS3,KS4} when it isotropizes. In  GR the Hubble parameter $H$ is always monotonic for Bianchi I and Bianchi III. For Bianchi I the anisotropy decays on time for
$H>0$. Therefore, isotropization occurs \cite{nns1}.  
Generically, in KS as well as for closed FLRW, the Hubble parameter is non monotonic and  anisotropies would increase rather than
vanish as $H$ changes the sign.  We refer the reader to \cite{Byland:1998gx,Fadragas:2013ina,Collins:1977fg,Weber:1984xh,Gron:1986ua,LorenzPetzold:1985jm,Barrow:1996gx,Clancy:1998ka,Rendall:1998rb,Carr:1999qr,Solomons:2001ef,Calogero:2009mi,Leon:2010pu,Leon:2013bra,Alvarenga:2015jaa,Zubair:2016ccy,Latta:2016jix,Camci:2016yed,Jamal:2017cut,VanDenHoogen:2018anx,Barrow:2018zav,Fajman:2019wut,Leon:2020cge,deCesare:2020swb,Coley:2008qd,WE}  and references therein  for applications of KS models, spatially homogeneous, and LRS metrics. 
The typical behavior of KS metric for perfect fluids, Vlasov matter, etc., is that the generic solutions are past and future asymptotic to the non--flat LRS Kasner vacuum solution, which have a big--bang
(or big--crunch). Moreover, there exist non--generic solutions which are past (future) asymptotic to the anisotropic
Bianchi I matter solution and others to the flat Friedman
matter solution. The qualitative properties of positive-curvature models and the KS models with a barotropic fluid and a non--interacting scalar field with exponential potential $V(\phi)=V_0 e^{\lambda\phi}$, being $\lambda$ a constant, were examined, e.g., in \cite{coleybook}. The main results are the following. For positively curved FLRW models  and for $\lambda^2>2$  all the solutions start from and recollapse to a singularity, and they are not generically inflationary. For $\lambda^2<2$ the universe can either  recollapse or expand forever. The KS model exhibits similar global properties of the closed FLRW models. In particular, for $\lambda^2>2$, all initially expanding solutions reach  a maximum expansion and after that recollapse. These solutions are not inflationary nor does they isotropize.  For $\lambda^2<2$ the  models generically recollapse or expand forever to power-law inflationary flat FLRW solution. Intermediate behavior of KS as compared with closed FLRW is rather different.

\subsection{General relativistic $1+3$ orthonormal frame formalism}

In this section we follow references \cite{WE,vanElst:1996dr} where the $1+3$ orthonormal frame formalism was presented. 

\noindent A cosmological model $(\mathcal{M}, \mathbf{g}, \mathbf{u})$  (representing the universe at a particular averaged scale) is defined by specifying the spacetime geometry through a Lorentzian metric $\mathbf{g}$ defined on the manifold $\mathcal{M}$  and a family of fundamental observers  whose congruence of worldlines is represented by the four velocity field  $ \mathbf{u}$, which will usually be the matter four velocity. 

\noindent
The following index conventions for tensors are used. Covariant spacetime indices are denoted by letters
from the second half of the Greek alphabet ($\mu, \nu, \rho \ldots = 0 \ldots 3$)  with spatial coordinate indices symbolized
by letters from the second half of the latin alphabet  ($i, j, k, \ldots = 1 \ldots 3$). Orthonormal frame spacetime
indices are denoted by letters from the first half of the latin alphabet  ($a, b, c \ldots = 0 \ldots 3$)  with spatial frame indices chosen from the first half of the Greek alphabet ($\alpha, \beta, \gamma, \ldots =1 \ldots 3$). 
The ``symmetrization'' of two indices is indicated using parenthesis, while the ``antisymmetrization'' of two indices is indicated by square brackets  and they are defined respectively by
$v_{(a b)}=\frac{1}{2} \left(v_{a b} + v_{b a}\right),\quad 
v_{[a b]}=\frac{1}{2} \left(v_{a b} - v_{b a}\right),$
and  $\delta_b^a$ is the Kronecker's delta  which is equal to $1$ if $a=b$ or equal to zero if $a \neq b$.  A system of units in which $8\pi G=c=\hslash=1$ is used. 

\noindent The following symbols are used. $R$ is the scalar curvature of the spacetime, $g_{\mu
\nu}$ are the metric components, $g$ is the determinant of the metric,  $\phi$ is the scalar field, and $V(\phi)$ is the scalar field potential defined by \eqref{pot}.   A semicolon ``$;$'' as well as $\nabla_a$ indicates covariant derivatives.  

\noindent It is common to describe cosmological models in terms of a basis of vector fields $\{\e_a\}$  and a dual basis of 1-forms $\{{\bm{\omega}}^a\}$, $a=0,1,2,3$. Any vector $\mathbf{X}$ can be written as $\mathbf{X}= X^a \e_a$ in terms of this basis.  The components of the metric tensor $\mathbf{g}$ relative to this basis are given by \begin{equation}
   g_{a b}= \mathbf{g}(\e_a, \e_b). 
\end{equation}
The line element can be symbolically written as  
\begin{equation}
  d s^2=  g_{a b} {\bm{\omega}}^a {\bm{\omega}}^b.   
\end{equation}
In any coordinate chart  there is a natural basis, namely $\{\e_a\}$ can be chosen to be a coordinate basis $\{\partial/\partial x^i\}$  with the dual basis being the coordinate 1-forms $\{d x^i\}$, where  $d x^i(\partial/\partial x^j)= \delta^{i}_{ j}$. 
The general basis vector fields $\e_a$ and  1-forms ${\bm{\omega}}^a$ can be written in terms of a coordinate basis as follows 
\begin{equation}
    \e_a = e_a^i (x^j) \frac{\partial}{\partial x^i}, \quad {\bm{\omega}}^a= \omega^a_{i}(x^j) dx^i.
\end{equation}
Thus, any vector field $\mathbf{X}$ can be interpreted as a differential operator which acts on scalars  as
\begin{equation}
 \mathbf{X} f = X^i \frac{\partial f}{\partial x ^i}. 
\end{equation}
In particular, 
\begin{equation}
\e_a f= e_{a}^i \frac{\partial f}{\partial x^i}.    
\end{equation}
In terms of a coordinate basis  the components of the metric tensor $\mathbf{g}$ relative to this basis are given by \begin{equation}
   g_{i j}= \mathbf{g}\left( \frac{\partial f}{\partial x^i},  \frac{\partial f}{\partial x^j}\right). 
\end{equation}
The line element can be  symbolically written  as   
\begin{equation}
  d s^2=  g_{i j} d x^i dx^j.   
\end{equation} 
\noindent Another special type of basis is the orthonormal frame, in which the four vector fields $\e_a$ are mutually orthogonal and of unit length  with $\e_0$ timelike. The vector fields, thus, satisfy $ \mathbf{g}(\e_a, \e_b)= \eta_{a b}$,  where $\eta$ is the Minkowski metric  $\eta_{a b}= \text{diag}(-1,1,1,1)$. 

\noindent Given any basis  of vector fields $\e_a $, the commutators $[\e_a, \e_b]$ are vector fields and hence they can be written as a linear combination of the basis vectors $[\e_a, \e_b] = \gamma_{a b}^c \e_c$. The coefficients $\gamma_{a b}^c(x^i)$ are called the commutation functions (which are 24 functions). For a coordinate basis  $\{\partial/\partial {x^i}\}$ the commutation functions are all zero. If we use an orthonormal frame, the 24 commutation functions are the basic variables  and the gauge freedom is an arbitrary Lorentz transformation. 

\noindent Applying the Jacobi identity for vector fields to $\gamma_{a b}^c$ we obtain a set of 16 identities (Eqs. 1.18 in \cite{WE})  
\begin{equation}
\label{frameid}
 \e_{[c} \gamma^d_{a b ]} - \gamma^d_{e [c} \gamma^{e} _{a b]}=0,
\end{equation}  
where we denote by $\e_a(f)$ the action of the vector field $\e_a$ on a scalar $f$ as the differential operator 
$\e_a f= e_{a}^i \frac{\partial f}{\partial x^i}$ using coordinate basis $\{\partial/\partial x^i\}$.
 The identities \eqref{frameid}, in conjunction with Einstein field equations rewritten as \footnote{Where $\Lambda$ is the cosmological constant. See eq.  (23) of reference \cite{vanElst:1996dr}.}
 \begin{equation}
     R_{a b}=T_{a b}- \tfrac12 T_c^c g_{a b}+\Lambda g_{a b},
 \end{equation}
 and conservation equations 
 \begin{equation}
     \nabla_b T_a^b=0, 
 \end{equation}
give first-order evolution equations for some of the commutation functions and for some stress-energy tensor components   and also provide a set of constraints involving only spatial derivatives,  which is referred as the orthonormal frame formalism (Sect. 1.4, \cite{WE} pages 31-35  and section 2, \cite{vanElst:1996dr} pages 2675 - 2682). For the moment we set $\Lambda=0$. 

\noindent It is well-known that a unit timelike vector field $\mathbf{u}(x^i)$, $u^a u_{a}=-1$, determines a projection tensor
\begin{equation}
h_{a b}:= g_{a b}+ u_a u_b,    
\end{equation} which at each point projects into the 3-space  orthogonal to $\mathbf{u}$. It follows that 
\begin{equation}
 h_{a}^c h_{c}^b = h_{a}^b, \quad h_{a b} u^{b}=0, \quad h_{a}^a=3.    
\end{equation} 
Therefore, we can define two derivatives,  one along the vector $u^a$  defined by
$\dot{T^{a..b}_{c..d}}=u^{e}\nabla_{e}{T^{a..b}_{c..d}}$, 
 and a projected derivative defined as $
D_{e} T^{a..b}_{c..d}= h^{a}_{f} h^{p}_{c}...h^{b}_{g} h^{q}_{d} h^{r}_{e}\nabla_{r} T^{f..g}_{p..q}$. 
\newline
When $\mathbf{u}= \partial/\partial_t$  and $f$ is an scalar, $\dot f= f_{;b} u^{b}= \frac{\partial  f}{\partial t}$ reduces to the usual time derivative of a function.

\noindent The covariant derivative of a unit timelike vector field $\mathbf{u}(x^i)$, $u^a u_{a}=-1$ can be decomposed in its irreducible parts as follows (see \cite{WE} page 18, \cite{kramer} page 70) 
\begin{equation}
    u_{a; b}= -\udot_{a} u_b +\omega_{a b} +\sigma_{a b} + \theta h_{a b}/3, \label{119}
\end{equation}
where  $\sigma_{a b}$ is symmetric and trace-free, $\omega_{a b}$ is antisymmetric  and  $\sigma_{a b} u^{b}=0= \omega_{a b}u^{b}$. 
\noindent
Physically, a timelike unit time vector is generally chosen as the four velocity of the fluid and the quantities $ \udot_ {a}, \theta, \omega_ {ab}, \omega^a, \sigma_ {ab} $  are called the acceleration vector, rate of expansion scalar, vorticity tensor, vorticity vector, and the rate of shear tensor, respectively. The magnitude of the shear tensor is 
\begin{equation}
    \sigma^2 = \sfrac12 \sigma_{a b} \sigma^{a b}. \label{magnitudesigma}
\end{equation}
\noindent It follows that
    \begin{align} 
    & \theta:= u^a_{;a}, \nonumber\\
    & \udot_{a}:= u_{a; b}u^{b}, \quad \udot_a u^a=0, \nonumber\\
    & \omega_{a b}:= u_{[a; b]}+ \udot_{[a} u_{b]},   \nonumber\\
    & \omega^{a} := \sfrac12 \eta^{a b c d} u _b \omega_{c d}, \nonumber\\
    & \sigma_{a b} := u_{(a; b)} + \udot_{(a} u_{b)}-\theta h_{a b}/3,  \label{decomp}
\end{align}
where  $\eta^{a b c d}$ is the totally antisymmetric permutation tensor such that $\eta_{0123}= \sqrt{-g}, \eta^{0123}= -1/\sqrt{-g}$, where $g$ denotes the determinant of the spacetime metric tensor $\mathbf{g}$ with Lorentzian signature. 

\noindent In Cosmology  it is useful to define a representative length along the worldlines of $\mathbf{u}$  describing the volume expansion (contraction) behavior of the congruence completely  by the equation  
\begin{equation}\label{characteristic-length}
\frac{\dot{\ell}}{\ell} = \frac{u^\mu \nabla_\mu \ell}{\ell}:= H,
\end{equation}
where $H$ is the Hubble parameter defined by 
\begin{equation}
    H:= \frac{1}{3} \theta.
\end{equation}
\noindent In the orthonormal frame formalism, the components of the connection are simplified to 
\begin{equation}
    \Gamma_{a b c}= \sfrac12 \left(\gamma^{d}_{c b} \eta_{a d}+ \gamma^{d}_{a c} \eta _{b d} -\gamma^d_{b a} \eta_{c d}\right).
\end{equation}
The spatial frame vectors are denoted by $\{\e_{\alpha}\}$, where the indices  chosen from the first half of the Greek alphabet run from $1$ to $3$ and $\epsilon_{\alpha \beta \gamma}$ denotes the alternating symbol ($\epsilon_{123}=+1$). Notice that greek indices are raised and lowered with the spatial metric tensor $g_{\alpha \beta}= \delta_{\alpha \beta}$. 

\noindent The commutation functions are decomposed into algebraically simple quantities, some of which have a direct physical or geometrical interpretation. Firstly, the commutation function $\gamma^c_{a b}$ with one zero index  can be expressed as functions of geometrical quantities \eqref{decomp} of the timelike congruence $\mathbf{u}=\e_0$ and the quantity 
\begin{equation}
    \Omega^{\alpha}= \sfrac12 \epsilon^{\alpha \mu \nu } e_\mu^{i} e_{\nu i; j} u^j,
\end{equation}
which is the local angular velocity of the spatial frame $\{\e_{\alpha}\}$ with respect to a Fermi-propagated spatial frame.

\noindent Eq.  (1.61) of \cite{WE} gives
\begin{align}
    & \gamma^{\alpha}_{0 \beta}= -\sigma^{\alpha}_{\beta} - H \delta^{\alpha}_{\beta} - \epsilon^{\alpha}_{\beta \mu} (\omega^{\mu} + \Omega^{\mu}), \nonumber\\
    & \gamma^{0}_{0 \alpha}= \udot_\alpha, \quad \gamma^{0}_{\alpha \beta}= -2 \epsilon_{\alpha \beta}^{\mu} \omega_\mu,
\end{align}
where  $\sigma_{\alpha \beta}$, $\omega_\alpha$, and $\udot_\alpha$ are  spatial components of  $\sigma_{a b}$, $\omega_a$ and $\udot_a$.
Secondly, spatial components $\gamma^{\mu}_{\alpha \beta}$ are decomposed  into a 2-index symmetric object $n_{\alpha \beta}$ and a 1-index object $a_{\alpha}$ as follows  
\begin{equation}
\label{EQ_42}
    \gamma^{\mu}_{\alpha \beta}= \epsilon_{\alpha \beta \nu} n^{\mu \nu} + a_{\alpha} \delta_{\beta}^{\mu} -  a_{\beta} \delta_{\alpha}^{\mu}.
\end{equation}
In order to incorporate the variety of matter sources in Einstein field equations 
 the standard decomposition of the stress-energy tensor $T_{a b}$ with respect to the timelike vector $\mathbf{u}$ is used,  
\begin{equation}
    T_{a b}= \mu u_a u_b + 2 q_{(a} u_{b)} + p h_{a b} + \pi_{a b},
\end{equation}
where $\mu$ is the total energy density, $q_{a}$ is the energy current density, $p$ is the isotropic pressure and  $\pi_{a b}$ is the anisotropic pressure tensor. It follows that
\begin{equation}
    q_a u^a=0, \quad \pi_{a b} u^b=0, \quad \pi_a^a=0, \quad \pi_{a b}= \pi_{b a}.
\end{equation}
If the congruence $\mathbf{u}$ is irrotational ($\omega_{a b}=0$), the curvature of the 3-spaces orthogonal to the congruence can be expressed as  
\begin{equation}
 \R_{a b c  d} = h_{a}^p h_{b}^q h_{c}^r h_{d}^s R_{p q r s}   -\Theta_{a c} \Theta_{b d} + \Theta_{a d} \Theta_{b c},
\end{equation}  
where $\Theta_{a c} $ is the rate of expansion tensor  given by 
\begin{equation}
    \Theta_{a b}= \sigma_{a b} +\sfrac13 \theta h_{a b}, 
\end{equation}
and 
$R_{a  b c d}$ is the Riemann tensor defined  by 
\begin{equation}
 R^{a}_{b c d}  = \e_c (\Gamma^{a}_{b d}) - \e_d (\Gamma^{a}_{b c}) +\Gamma^{a}_{f c} \Gamma^{f}_{b d} - \Gamma^{a}_{f d} \Gamma^{f}_{b c} -\Gamma^{a}_{b f} \gamma^{f}_{c  d}.
\end{equation}
\noindent 
The trace-free spatial Ricci tensor is defined by 
\begin{equation}
    \S_{a b}= \R_{a b} -\tfrac13 \;\R h_{a b},
\end{equation}
where  
\begin{equation}
    \R_{a b}= h^{p q} \; \R _{a p b q}, \; \R = h^{p q} \; \R_{p q}. 
\end{equation}
Also, the Weyl conformal curvature tensor can be expressed as
\begin{equation}
    C^{a b}_{c d}= R^{a b}_{c d} - 2 \delta^{[ a}_{[c} R^{b]}_{d]} + \sfrac13 R \delta^{[ a}_{[c} \delta^{c]}_{d]}.
\end{equation}
It is useful to define the electric part $E_{a b}$ and magnetic part $H_{a b}$ of the Weyl conformal curvature tensor relative  to $\mathbf{u}$ according to 
\begin{equation}
    E_{a c}= C_{a b c d} u^b u^d, \quad H_{a c}= \sfrac12 \eta_{a b}^{s t} C_{s t c d} u^{b} u^{d}.
\end{equation}
These tensors are symmetric, trace-free, and satisfy $E_{a b} u^b =0= H_{a b} u^b$. 

\noindent When performing the $1+3$ decomposition  
for an irrotational congruence $\mathbf{u}_0$ the components of Weyl tensor are reduced to  
\begin{align}
   E_{\alpha\beta} &  = H \sigma_{\alpha \beta} -\left( \sigma_{\alpha}^{\mu} \sigma_{\mu \beta} -\sfrac23 \sigma^2 \delta_{ \alpha \beta}\right) + \S_{\alpha \beta} -\sfrac12 \pi_{\alpha \beta},\\
    H_{\alpha \beta}& = (\e_\mu -a_\mu) \sigma_{\nu (\alpha} \epsilon_{\beta)}^{\mu \nu} -3 \sigma^{\mu}_{(\alpha} n_{\beta) \mu} \nonumber \\
    & + n_{\mu \nu } \sigma^{\mu \nu}\delta_{\alpha \beta} + \sfrac12 u_{\mu}^{\mu} \sigma_{\alpha \beta},
\end{align}
where the magnitude of the shear tensor $\sigma^2$ is defined by \eqref{magnitudesigma}. 

\noindent The equations of the orthonormal frame formalism are the following.  The Einstein field equations (1.65), (1.66), (1.67), and (1.68) in \cite{WE}, where $\S_{\alpha \beta}$ and $\R$ are given by (1.69) and (1.70) in \cite{WE}, respectively, the Jacobi identities (1.71), (1.72), (1.73), (1.74),  and  (1.75) in \cite{WE} and the contracted Bianchi identities (1.76) and (1.77) in \cite{WE}. This formalism was revisited in \cite{vanElst:1996dr} where the authors discuss their applications in detail.  Bianchi identities for  Weyl curvature tensor are presented in  \cite{vanElst:1996dr} in a fully expanded form, as they are given a central role in the extended formalism. By specializing the general $1 + 3$ dynamical equations it was illustrated
how a number of interesting problems can be obtained. In particular, the simplest choices of
spatial frames for spatially homogeneous cosmological models, locally rotationally symmetric
spacetime geometries, cosmological models with an Abelian isometry group $G_2$, and ``silent'' dust cosmological models were discussed.

\subsubsection{Specialization for spherically symmetric models}
In the “resource” paper \cite{Coley:2008qd}  the $1+3$ orthonormal frame formalism (as developed in  \cite{WE,vanElst:1996dr}) was specialized to write down the evolution equations for spherically symmetric models as a well-posed system of first order partial differential equations (PDEs) in two variables. This  “resource” paper reviews a number of well-known results properly cited in \cite{Coley:2008qd} and they are simply gathered together because of  their functionality  and in this context it serves to define all of the quantities. Therefore, we refer researchers interested in the formalism to reference \cite{Coley:2008qd} and references therein, and present essential equations now.   

\noindent The metric for the spherically symmetric models is given by
\begin{small}
\begin{align}
\label{sphsymm}
   &  ds^2= - N^2(t,r) dt^2 + \left[{e_1}^1(t,r)\right]^{-2} dr^2 + \left[{e_2}^2(t,r)\right]^{-2} d\Omega^2,\nonumber \\
   & d\Omega^2= d\y^2 + \sin^2 \y\, d\z^2,
\end{align}
\end{small}
where $N$, $\ex$, and $\ey$ are functions of $t$
and $r$  and $N$ is the lapse function. The   Killing vector fields (KVF) in spherically symmetric spacetime  are given by \cite{kramer} $\partial_\z,\quad \cos \z \ \partial_\y - \sin \z \cot \y \ \partial_\z,  \sin \z \ \partial_\y + \cos \z \cot \y \partial_\z$. 
Frame vectors in coordinate form are 
\[ \e_0 = N^{-1} \partial_t
    ,\quad
    \e_1 = \ex \partial_r
        ,\quad
        \e_2 = \ey \partial_\y
        ,\quad
        \e_3 = \ez \partial_\z,\] 
where $\ez = \ey / \sin \y$. 
\noindent We see that  frame vectors $\e_2$ and $\e_3$
tangent to the spheres are not group-invariant because commutators
$[\e_2, \partial_\z]$ and $[\e_3, \partial_\z]$
are zero  but not with the other two Killing vectors.
Frame vectors $\e_0$ and $\e_1$
orthogonal to the spheres are group-invariant and the
correspondingly commutator reads  
\begin{equation}
[\e_0,\e_1]=\udot \e_0-(H-2\sp)\e_1.
\end{equation}     
We use the symbol $\e_a$ (in reference \cite{WE} $\pmb{\partial}_a$ is used ) to denote the action of the vector field $\e_a$ on a scalar $f$ as  a differential operator.

\noindent Geometric  objects of the $1+3$ formalism (kinematic variables, spatial commutations functions)  as well as the matter components  are deduced as follows. 

\noindent First, we define the four velocity vector field by $\mathbf{u}=\e_0$ representing the congruence of worldlines. The representative length $\ell(t,r)$ along worldlines of $\mathbf{u}=\e_0$ describing the volume expansion (contraction) behavior of the congruence is reduced to \cite{vanElst:1996dr} 
\begin{align}\label{defH}
\frac{\e_0 \ell (t,r)}{\ell (t,r)}= H(t,r), 
\end{align}
where the Hubble parameter  $H(t,r)$ is brought to  
\begin{align}\label{defH}
H(t,r):=     -\frac{1}{3} \e_0 \ln\left[ {e_1}^1(t,r) ( {e_2}^2(t,r))^2\right]. 
\end{align}
Furthermore,  we have the following restrictions on the kinematic variables (rate of shear tensor, vorticity tensor, acceleration vector) 
\begin{align*}
  & \sigma_{\alpha\beta} = \text{diag}(-2\sp,\sp,\sp), \; \omega_{\alpha\beta} =0, \;  \udot_\alpha =(\udot_1,0,0).
\end{align*}
The anisotropic parameter $\sigma_{+}(t,r)$ in $\sigma_{\alpha \beta}$ is defined by
\begin{align}
\label{defsigma}
\sigma_+ := \frac{1}{3} \e_0 \ln\left[ {e_1}^1(t,r) ( {e_2}^2(t,r))^{-1}\right].
\end{align} 
\noindent The acceleration vector is calculated as $\udot_a= u_{a; b}u^{b}$ obtaining only one non-zero component in terms of the spatial derivatives of the lapse function given by $\udot_1 = \e_1 \ln N.$ 

\noindent We have restrictions on  spatial commutation functions (1-index objects and 2-index symmetric objects in eq.  \eqref{EQ_42}) 
\begin{equation*}
    a_\alpha = (a_1, a_2, 0),\quad
    n_{\alpha\beta} = \left( \begin{array}{ccc}
            0 & 0 & n_{13}  \\
            0 & 0 & 0   \\
            n_{13} & 0 & 0 \end{array} \right),
\end{equation*}
where%
\begin{equation*}
    a_1 = \e_1 \ln\ey,\quad
    a_2 = n_{13} = - \frac12 \ey \cot \y.
\end{equation*}
The dependence of $a_2$ and $n_{13}$ on $\y$ is due to the fact
that the chosen orthonormal frame is not group-invariant. However, this is not a concern  since   dependence on $\y$ will be hidden.

\noindent On the matter components we have restrictions as follows, 
\begin{equation*}
    q_\alpha = (q_1,0,0),\quad
    \pi_{\alpha\beta} = \text{diag}(-2\pi_+,\pi_+,\pi_+).
\end{equation*}
The frame rotation $\Omega_{\alpha\beta}$  is  zero.

\noindent Furthermore, $n_{13}$ only appears in equations together with
$\e_2 n_{13}$ in the form of the Gaussian curvature of the spheres
\begin{equation}
    {}^2\!K := 2(\e_2 - 2 n_{13}) n_{13}, 
\end{equation}
which simplifies to 
\begin{align}
\label{calcK}
& {}^2\!K= - \ey \partial_\y  \left( \ey \cot \y \right) - (\ey)^2 \cot^2 \y \nonumber \\
& = (\ey)^2  \left(-\partial_\y   \cot \y   - \cot^2 \y \right) =  (\ey)^2. 
\end{align}
Thus, the dependence on $\y$ is hidden in the equations. In the following   we will
use ${}^2\!K$ as a dynamical variable instead of  $\ey$.

\noindent Spatial curvatures also simplify to 
\begin{equation*}
      {}^3\!S_{\alpha\beta} = \text{diag}(-2\; \S,\S,\S), 
\end{equation*}
with $\R$ and $\S$ given by 
\begin{align}
 &       \R = 4 \e_1 a_1 - 6 a_1^2 + 2 {}^2\!K, \label{3Ricci}
\\
&        \S_+= - \tfrac13 \e_1 a_1 + \tfrac13 {}^2\!K.
\end{align}
Weyl curvature components simplify to 
\begin{equation*}
    E_{\alpha\beta} = \text{diag}(-2 E_+,E_+,E_+),\quad
    H_{\alpha\beta} = 0, 
\end{equation*}
with $E_+$ given by
\begin{equation*}
      E_+ = H\sp + \sp^2 + \S_+- \tfrac12 \pi_+. 
\end{equation*}
To simplify notation  we will write ${}^2\!K,\ \udot_1,\ a_1$ as $K,\ \udot,\ a$.

\noindent
To summarize,  essential variables are
\begin{equation*}
    N, \ex,\ K,\ H,\ \sp,\ a,\ \mu,\ q_1,\ p,\ \pi_+,  
\end{equation*}
and   auxiliary variables are
\begin{equation*}
    \R,\ \S,\ \udot.
\end{equation*}
So far, there are no evolution equations for $N$, $p$ and $\pi_+$. They need to be specified by a temporal gauge (for $N$) and by a fluid model (for $p$ and $\pi_+$). Recall that  the total energy density and total isotropic pressure of the matter fields are $\mu$ and $p$, respectively. 

\noindent Evolution equations  including  a non-negative CC  $\Lambda$ are 
\begin{subequations}
\label{gen-eqs}
\begin{align}
    \e_0 \ex &= (-H+2\sp) \ex, \label{55a}
\\
    \e_0 K &= -2(H+\sp)K, \label{55b}
\\
    \e_0 H &= - H^2 - 2 \sp^2 + \tfrac13 (\e_1 + \udot - 2 a)\udot
      \nonumber\\
      & - \tfrac16(\mu+3p) + \tfrac13 \Lambda,  \label{55c}
\\
    \e_0 \sp &= -3H \sp - \tfrac13(\e_1 + \udot + a)\udot
        - \S_++ \pi_+, \label{evol:sigma}
\\
    \e_0 a &= (-H+2\sp) a - (\e_1 + \udot)(H+\sp),  \label{55e}
\\
    \e_0 \mu &= -3H(\mu+p) - (\e_1+2\udot-2a)q_1 - 6\sp\pi_+,  \label{55f}
\\
    \e_0 q_1 &= (-4H+2\sp)q_1 - \e_1 p -(\mu+p)\udot
          \nonumber \\
          & + 2(\e_1+\udot-3a)\pi_+.  \label{55g}
\end{align}
\end{subequations}
Eqs. \eqref{55a} and \eqref{55b} come from the definition of $H$ in \eqref{defH} and the definition of $\sigma_+$ in \eqref{defsigma}, respectively. Eq.  \eqref{55c} is the realization of eq.  (1.65) in \cite{WE} for spherically symmetric metrics and by including a CC.  The same occurs for eq.  \eqref{evol:sigma} which is the realization of eq.  (1.66) in \cite{WE} in our case of study. Eq. \eqref{55e} is the realization of the Jacobi identity (1.72)  in \cite{WE} in our case. Eq. \eqref{55f} is the realization of the contracted Bianchi identity (1.76) in \cite{WE} and eq.  \eqref{55g} is the realization of the contracted Bianchi identity (1.77) in \cite{WE} for spherically symmetric metrics and by including a CC. 

\noindent Constraint equations are:
\begin{subequations}
\begin{align}
    0 &= 3H^2 + \tfrac12 \R - 3 \sp^2 - \mu - \Lambda, \label{gauss-eq}
\\
    0 &= -2 \e_1(H+\sp) + 6 a \sp + q_1, \label{Codazzi}
\\
    0 &= (\e_1 - 2 a) K,  \label{56c}
\end{align}
\end{subequations}
where  spatial curvatures (given by eqs. (1.69) and (1.70) in \cite{WE}) are reduced to
\begin{align*}
    \R  = 4 \e_1 a - 6 a^2 + 2 K, \; 
    \S_+ = - \tfrac13 \e_1 a + \tfrac13 K.
\end{align*}
\noindent
Eq. \eqref{gauss-eq} comes from Einstein equation (1.67) in \cite{WE}.
\noindent
Eq. \eqref{Codazzi} is Einstein equation  (1.68) in \cite{WE} for $\alpha=1$. 
\noindent
Eq.  \eqref{56c} is the definition $a = \e_1 \ln\ey$ combined with the identity $K=(\ey)^2$. 
\subsection{Special cases with extra Killing vectors}

Spherically symmetric models with more than three KVF are either
spatially homogeneous or static. Let us  discuss the spatially
homogeneous cosmological models.
Spatially homogeneous spherically symmetric models consist of two
disjoint sets of models,  KS models and  
FLRW models.
Static and self-similar spherically symmetric models have been studied in
\cite{Coley:2008qd,Nilsson:2000zg,Carr:1998at,Carr:1999qr,Carr:1999rv,Goliath:1998mx}.

\subsubsection{The Kantowski-Sachs models}

Spatially homogeneous spherically symmetric models (that have four
Killing vectors with the fourth being $\partial_r$) are the so-called
KS models \cite{kramer}.
The metric (\ref{sphsymm}) simplifies to
\begin{align}
        & ds^2 = - N^2(t) dt^2 + (\ex(t))^{-2} dr^2
                + (\ey(t))^{-2} d\Omega^2, \nonumber \\
                & d\Omega^2 =(d\y^2 + \sin^2 \y\, d\z^2); \label{metric}
\end{align}
i.e., $N$, $\ex$, and $\ey$ are now independent of $r$.

\noindent Spatial derivative terms of type $\e_1(\cdot )$ in eqs. \eqref{gen-eqs}- \eqref{gauss-eq} vanish and, as
result, $a=0=\udot$. Since $\udot=0$, the temporal gauge is
synchronous and we can set $N$ to any positive function of $t$.

\noindent
Spatial curvatures are given by
\begin{align*}
        \R  = 2 K, \; 
        \S_+ = \tfrac13 K.
\end{align*}
\noindent Constraint \eqref{Codazzi} restricts the source by
$ q_1 = 0$. Meanwhile, functions 
$p$, and $\pi_+$  are still unspecified.

\noindent Evolution equations \eqref{gen-eqs} for KS models with unspecified
source reduce to
\begin{subequations}
\label{KS-gen}
\begin{align}
        \e_0 \ex &= (-H+2\sp) \ex ,
\\
        \e_0 K &= -2(H+\sp)K ,
\\
        \e_0 H &= - H^2 - 2 \sp^2
                - \tfrac16(\mu+3p) + \tfrac13 \Lambda ,
\\
        \e_0 \sp &= -3H \sp - \tfrac13 K + \pi_+ ,
\\
        \e_0 \mu &= -3H(\mu+p) - 6\sp\pi_+.
\end{align}
\end{subequations}
\noindent The remaining constraint equation \eqref{gauss-eq} reduces to 
\be
        0 = 3H^2 + K - 3 \sp^2 - \mu - \Lambda, 
        \label{Gauss}
\ee
where we have substituted in eq. \eqref{gauss-eq} the relation $\tfrac12 \R = K$ valid for KS metric.

 \paragraph{Kantowski-Sachs models for perfect fluid and homogeneous scalar field.}
\label{SECT3.5a}
In equations \eqref{KS-gen} and in the restriction \eqref{Gauss} we can replace the expressions $\pi_+=0, \Lambda=0, \mu= \frac{1}{2}{\dot{\phi}}^2 + V(\phi)+ \rho_m, p= \frac{1}{2}{\dot{\phi}}^2 + V(\phi)+ (\gamma-1)\rho_m$. Assuming that the energy--momentum of the scalar field and  matter are separately conserved and setting $N\equiv 1$, we obtain the following equations for KS metric for perfect fluid and homogeneous scalar field 
\begin{align}
    & \ddot{\phi}= -3 H \dot{\phi} - V'(\phi), \\
    & \dot{\rho}_m= -3 \gamma H \rho_m, \\
    & \dot{K} = -2 ({\sigma_+} +H) K, 
\\
    & \dot{H}= -H^2 -2 {\sigma_+}^2 -\frac{1}{6} (3 \gamma -2) \rho_m -\frac{1}{3} {\dot{\phi}^2} +\frac{1}{3} V(\phi), 
\\
    & \dot{{\sigma_+}}= -3 H {\sigma_+} -\frac{K}{3},
\\
    & 3 H^2 + K = 3 {\sigma_+}^2 +\rho_m+\frac{1}{2}{\dot{\phi}}^2 +V(\phi). \label{GaussKS}
\end{align}
Again, in eq. \eqref{GaussKS} we have substituted the relation $\tfrac12 \R = K$ valid for KS metric. 
As commented before, when $\mathbf{u}= \partial/\partial_t$ the dot derivative of a scalar  $f$,  given by  $\dot f= f_{;b} u^{b}= \frac{\partial  f}{\partial t}$, denotes  the usual time derivative.   

\subsubsection{The FLRW models}

Spatially homogeneous spherically symmetric models, that are not
KS, are  FLRW models (with or without
$\Lambda$). The source must be a comoving perfect fluid (or vacuum).

\noindent The metric has the form
\begin{align}
\label{metricFLRW}
       &  ds^2 = - N^2(t) dt^2 + \ell^2(t) dr^2
               \nonumber \\
               & + \ell^2(t) f^2(r) (d\y^2 + \sin^2 \y\, d\z^2),
\end{align}
with
\be
\label{fx_FLRW}
    f(r) = \sin r,\ r,\ \sinh r,
\ee
for closed, flat, and open FLRW models, respectively. $\ell(t)$ is the scale factor of the universe.
The frame coefficients are given by $\ex = \ell^{-1}(t)$ and $\ey =
\ell^{-1}(t) f^{-1}(r)$. Then, $\sp = \frac13\e_0 \ln(\ex/\ey)$
vanishes.
$N=N(t)$ implies that $\udot=0$; i.e.,
the temporal gauge is synchronous  and we can set $N$ to any positive
function of $t$.
The Hubble scalar $H = \e_0 \ln \ell(t)$ is also a function of
$t$. For the spatial curvatures, $\S$ vanishes because eq. (\ref{fx_FLRW})
implies $\e_1 a = K$, which is consistent with
the fact that the frame vector $\e_1$ is not group-invariant. 

\noindent The evolution equation \eqref{evol:sigma} for $\sp$ and the constraint \eqref{Codazzi} imply
that $\pi_+ =0= q_1$, i.e., the source is a comoving perfect fluid  with
unspecified pressure $p$. Also, note that $\mu$ and $p$ only depend on $t$ and $p$ is not
specified yet.

\noindent From eq.  \eqref{calcK} the Gaussian curvature of the two spheres is  
$K = \ell^{-2} f^{-2}$. 
Meanwhile, from eq.  \eqref{3Ricci}  we obtain 
\begin{align*}
      &\R = 4 \e_1 a  - 6 a ^2 + 2 K \nonumber \\
      & = 4 \ex \partial_r (\ex \partial_r \ln \ey) - 6 (\ex \partial_r \ln \ey)^2 + 2 K \nonumber\\
      & = 4 (\ex)^2 \partial^2_{r,r} \ln \ey - 6  (\ex)^2 (\partial_r \ln \ey)^2 + 2 K
      \nonumber\\
      & = 2  \ell^{-2}  \left(2 \partial^2_{r,r} \ln (\ell^{-1} f^{-1})  - 3   (\partial_r \ln (\ell^{-1} f^{-1}))^2  +f^{-2}\right) \nonumber \\
       & =  2  \ell^{-2}  \left(-2 \partial^2_{r,r} \ln f  - 3   (\partial_r \ln f)^2  +f^{-2}\right).
\end{align*}
On the other hand,
\begin{align*}
    & \partial^2_{r,r} f= -k f, \\ 
    & (\partial_r f)^2= 1- k f^2, \\
    & (\partial_r \ln f)^2= f^{-2} (\partial_r f)^2= f^{-2} - k, \\
    & \partial^2_{r,r} \ln f = - f^{-2} (\partial_r \ln f)^2 + f^{-1} \partial^2_{r,r} f= - f^{-2},
    \end{align*}
for closed ($k=1$), flat ($k=0$), and open FLRW ($k=-1$), respectively. We used $k$ to indicate the choice of $f$ in eq. \eqref{fx_FLRW}. 
Then, 
 \begin{align}
 \label{equ87}
&\R  = 2  \ell^{-2}  \left(2 f^{-2}  - 3  f^{-2} + 3 k  +f^{-2}\right)\nonumber \\
&= \frac{6k}{\ell^2},\quad
    k = 1,0,-1.   
\end{align}

\noindent The evolution equations simplify to 
\begin{subequations}
\label{FLRW-gen}
\begin{align}
    \e_0 \ell &= H \ell,
\\
    \e_0 H    &= - H^2 - \tfrac16(\mu+3p) + \tfrac13 \Lambda,
\\
    \e_0 \mu  &= -3H(\mu+p).
\end{align}
\end{subequations}
The constraint becomes
\be
    0 = 3H^2 + \frac{3k}{\ell^2} - \mu - \Lambda,\quad k=1,0,-1, \label{FLRWGauss}
\ee
where we have substituted $\frac{\R}{2}=\frac{3k}{\ell^2}$ (valid for FLRW). 
\noindent
The vacuum ($\mu=0$) cases are: de Sitter model ($\Lambda>0$, $k=0$),
the model with $\Lambda>0$, $k=1$,
the model with $\Lambda>0$, $k=-1$,
Milne model ($\Lambda=0$, $k=-1$)
and Minkowski spacetime ($\Lambda=0$, $k=0$) which is also static.

\begin{enumerate}
\item The vacuum model with $\Lambda>0$, $k=1$ is past asymptotic to a
model with negative $H$, and it is future asymptotic to   de Sitter model
with positive $H$.  Indeed, for $\Lambda> 0$ we have  $3H ^ 2 = -1 / a ^ 2 + \Lambda$. So, the scale factor has to satisfy $a \geq a_{\text{crit}}=\sqrt{3 /\Lambda}$. Additionally, the scale factor $a$ is not monotonic because  $H$ changes its sign. The orbits generated at a negative value  of $H$ with $a > a_{\text{crit}}$ are such that the scale factor $a$ decreases monotonically until reaching the value $a_{\text{crit}}= \sqrt {3 /\Lambda}$ where $H$ becomes zero. Then, $H$ becomes positive because it is continuous and again the scale factor $a$ starts growing from the smallest critical value $a_{\text{crit}}$ to infinite at an exponential rate (in a de Sitter phase).

\item The vacuum model with $\Lambda>0$, $k=-1$  and positive $H$  is past asymptotic
to  Milne model  and  future asymptotic to  de Sitter model with
positive $H$.
\end{enumerate}

\paragraph{FLRW models for perfect fluid and homogeneous scalar field.}

Replacing in equations \eqref{FLRW-gen} and  \eqref{FLRWGauss}  expressions $\pi_+=0,   \Lambda=0,   \mu= \frac{1}{2}{\dot{\phi}}^2 + V(\phi)+ \rho_m,   p= \frac{1}{2}{\dot{\phi}}^2 + V(\phi)+ (\gamma-1)\rho_m$ and assuming that the energy--momentum of the scalar field and  matter are separately conserved, we obtain  for the metric \eqref{metricFLRW} the evolution/constraint equations  
\begin{subequations}
	\label{Non_minProb1FLRW0-1+1}
	\begin{align}
	&\ddot{\phi}= -3 H \dot{\phi}-V'(\phi),
\\
	&\dot{\rho}_m=- 3\gamma H\rho_m,
	\\
	&\dot{\ell} = \ell H, 
	\\
	& \dot{H}=-\frac{1}{2}\left(\gamma \rho_m+{\dot \phi}^2\right)+\frac{k}{\ell^2},
	\\
	& 3H^2=\rho_m+\frac{1}{2}{\dot{\phi}}^2+V(\phi)-\frac{3 k}{\ell^2},  \label{GaussFLRW}
	\end{align}
with  $k = 1,0,-1$ 
for closed, flat, and open FLRW, respectively.
\end{subequations}

\section{Averaging scalar field cosmologies}
\label{SECT:II}
For KS and positively curved FLRW metrics the Hubble scalar is not monotonic. This means that $H$ cannot be used as a time--dependent perturbation function as in references \cite{Leon:2021lct,Leon:2021rcx}. However, the function \eqref{GEN-D}   is monotonic in a finite time-interval before changing monotony. The region where the perturbation parameter changes its monotony can be analyzed by a discrete symmetry and  by introducing the variable $T=D/(1+D)$ that brings infinite to a finite interval.

\noindent
For KS the normalization factor \eqref{GEN-D}  becomes 
\begin{align}
D=\sqrt{H^2 + \frac{K}{3}}, \label{D-KS} 
\end{align}
where $K$ denotes  Gaussian spatial curvature of the 2-spheres. 
Using the calculation \eqref{equ87} we obtain for closed FLRW the  normalization factor
\begin{align}
D= \sqrt{H^2 + \frac{1}{\ell^2}}. \label{varsclosedFLRW}
\end{align}

 \subsection{Kantowski-Sachs}
\label{SECT3.5}
We define normalized variables 
\begin{small}
\begin{align}
& Q=\frac{H}{D}, \;  \Sigma=\frac{{\sigma_+}}{D},  \; \Phi= t \omega -\tan^{-1}\left(\frac{\omega \phi}{\dot{\phi}}\right),  \nonumber \\
& \Omega:= \frac{\omega r}{\sqrt{6 } D}=\sqrt{\frac{\omega^2 \phi^2+{\dot\phi}^2}{6 D^2}},    \label{varsKS}
\end{align}
\end{small}
\noindent where $D$, defined by \eqref{D-KS}, is the dominant quantity in eq. \eqref{GaussKS}. The function $Q$ is the Hubble factor normalized with $D$.  Such a solution is classified as contracting if ${Q}<0$, since $H$ and $Q$ have the same sign due to $D>0$. 
 
\noindent The function $\Sigma$ is a  measure of the anisotropies of the model normalized with $D$. When $\Sigma\rightarrow 0$, the solution isotropizes.  $\Phi$ is the ``phase'' in the amplitude-phase transformation \eqref{amplitud-phase} and $r$ is defined in  \eqref{eqAA25}. 
 
\noindent The function $\Omega^2$ is the total energy of the harmonic oscillator (at the  minimum) normalized with $3D^2$. Indeed, $\omega^2 r^2/2$ represents the total energy density of the pure harmonic oscillator with potential $\frac{\omega^2}{2}\phi^2$. This is the energy of the oscillator when oscillations are smoothed out and the scalar field approaches its minimum value.  
 
\noindent Recall that the potential \eqref{pot_v2} satisfies $V(\phi) = \frac{\omega ^2 \phi ^2}{2}+\mathcal{O}\left(\phi ^4\right) \; \text{as} \; \phi\rightarrow 0$.  
 
 Then, we obtain  
\begingroup\makeatletter\def\f@size{8}\check@mathfonts
\begin{subequations}
\label{unperturbed1KS}
\begin{align}
  & \dot{\Omega}=  -\frac{b \mu ^3}{\sqrt{6}   D}  \cos (t \omega -\Phi ) \sin  \left(\frac{\sqrt{6}   D \sin (t \omega -\Phi ) \Omega }{f \omega }\right)  \nonumber \\
  & -\frac{b f \gamma  Q   \Omega  \mu ^3}{  D} \sin ^2  \left(\frac{\sqrt{\frac{3}{2}}   D \sin (t \omega -\Phi
   ) \Omega }{f \omega }\right) \nonumber\\
   & +\frac{\left(\omega ^2-2 \mu ^2\right) \Omega }{2 \omega } \sin (2 t \omega -2 \Phi )  \nonumber \\
   &  +3 D  Q   \Omega  \cos ^2(t \omega -\Phi )   \left(\left(\gamma 
   \left(\frac{\mu ^2}{\omega ^2}-\frac{1}{2}\right)+1\right) \Omega ^2-1\right)  \nonumber \\
   & +\frac{1}{2}   D \Omega  \Big(-2 \Sigma  Q  ^2 +2 \Sigma   +3 \left(-(\gamma -2) \Sigma ^2-\frac{2 \gamma  \mu ^2 \Omega ^2}{\omega ^2}+\gamma \right)
   Q  \Big),   
\\
     & \dot{\Sigma}=-\frac{b
   f \gamma  Q    \Sigma  \mu ^3}{  D} \sin ^2  \left(\frac{\sqrt{\frac{3}{2}}   D \sin (t \omega -\Phi ) \Omega }{f \omega }\right)  \nonumber \\
   & -\frac{3}{2}   D (\gamma -2)Q   \Sigma  \Omega ^2  \cos ^2(t \omega -\Phi ) \nonumber \\
   & +  D
   \Bigg(-\frac{3 \gamma  \mu ^2 Q    \Sigma  \Omega ^2}{\omega ^2} \sin ^2(t \omega -\Phi ) \nonumber \\
   & -\frac{1}{2} \left(2 Q  ^2+3 (\gamma -2) \Sigma  Q  -2\right) \left(\Sigma ^2-1\right)\Bigg),
\\
& \dot{Q}=-\frac{b f \gamma  \left(Q  ^2-1\right)  \mu ^3}{  D} \sin ^2  \left(\frac{\sqrt{\frac{3}{2}}   D \sin (t \omega -\Phi ) \Omega }{f \omega }\right) \nonumber \\
& -\frac{3}{2}   D (\gamma -2)
   \left(Q  ^2-1\right) \Omega ^2  \cos ^2(t \omega -\Phi )\nonumber \\
   & +\frac{1}{4}   D \left(1-Q  ^2\right) \Bigg(6 (\gamma -2) \Sigma ^2+4 Q   \Sigma -6 \gamma +4  \nonumber \\
   & +\frac{12 \gamma  \mu ^2  \Omega ^2}{\omega ^2} \sin ^2(t \omega -\Phi )\Bigg),
\\
   & \dot{\Phi}= -\frac{b  \mu ^3}{\sqrt{6}   D \Omega } \sin (t \omega -\Phi ) \sin
    \left(\frac{\sqrt{6}   D \sin (t \omega -\Phi ) \Omega }{f \omega }\right) \nonumber \\
    & +\frac{\left(\omega ^2-2 \mu ^2\right)}{\omega } \sin ^2(t \omega -\Phi ),
\\
   & \dot{D}= b f  \mu ^3 \gamma  Q   \sin ^2  \left(\frac{\sqrt{\frac{3}{2}}   D \sin (t \omega -\Phi ) \Omega }{f \omega }\right) \nonumber \\
   & +\frac{3}{2}   D^2 (\gamma -2)  Q   \Omega ^2 \cos ^2(t \omega -\Phi )\nonumber \\
   & +\frac{1}{4}
     D^2 \Bigg(4 \Sigma  Q  ^2 -4 \Sigma + 6 (\gamma -2) \Sigma ^2 Q \nonumber\\
     & +6  \gamma  \left(\frac{2 \mu ^2 \sin ^2(t \omega -\Phi ) \Omega ^2}{\omega ^2}-1\right) Q  \Bigg),
\end{align}
\end{subequations}
\endgroup
and the deceleration parameter is 
\begingroup\makeatletter\def\f@size{7}\check@mathfonts
\begin{align}
  & q=  -\frac{b f \gamma  \mu ^3}{  D^2 Q  ^2}  \sin
   ^2 \left(\frac{\sqrt{\frac{3}{2}}   D \sin (t \omega -\Phi ) \Omega }{f \omega }\right)   -\frac{3 (\gamma -2) \cos ^2(t \omega -\Phi ) \Omega ^2}{2 Q  ^2}  \nonumber \\
   & +\frac{3 (2-\gamma) \Sigma
   ^2-\frac{6 \gamma  \mu ^2 \sin ^2(t \omega -\Phi ) \Omega ^2}{\omega ^2}+3 \gamma -2}{2 Q  ^2}.
\end{align}
\endgroup
System  \eqref{unperturbed1KS} is invariant for the simultaneous change  \newline $(t, \Sigma, Q, \Phi) \mapsto (-t, -\Sigma,  -Q, -\Phi)$. 
Setting the constant $b \mu ^3+2 f \mu ^2-f \omega ^2=0\implies f=\frac{b \mu ^3}{\omega ^2-2 \mu ^2}$,
the fractional energy density of matter $\Omega_m:= \frac{\rho_m}{3 H^2}= \frac{\rho_m}{3 Q^2 D^2}$ is parametrized by the equation
\begingroup\makeatletter\def\f@size{8}\check@mathfonts
\begin{align}
  &Q^2 \Omega_m= 1  -{\Sigma^2}  -\Omega ^2  +\Omega ^2 \left(1-\frac{2 \mu ^2}{\omega
   ^2}\right) \sin ^2(\Phi -t \omega ) \nonumber \\
  & + \frac{2 b^2 \mu ^6 }{3 D^2 \left(2 \mu ^2-\omega ^2\right)} \sin ^2 \left(\frac{\sqrt{\frac{3}{2}} D \Omega  \left(2 \mu ^2-\omega ^2\right) \sin (\Phi -t \omega )}{b \mu ^3 \omega }\right)\nonumber \\
  & = 1 -\Sigma^2-\Omega ^2 + \mathcal{O}(H^2).
\end{align} 
\endgroup
Assuming $\omega ^2>2 \mu ^2$ and  setting  $f=\frac{b \mu ^3}{\omega ^2-2 \mu ^2}>0$,  we obtain 
\begingroup\makeatletter\def\f@size{8}\check@mathfonts
\begin{align}
    \dot{\mathbf{x}}=  &\mathbf{f}(\mathbf{x}, t) D + \mathcal{O}(D^2), \;  \mathbf{x}= \left(\Omega, \Sigma, Q,  \Phi \right)^T, \label{EQ:96a}\\
    \dot{D}= & -\frac{1}{2} D^2 \Bigg(2 \Sigma (1-Q^2 +3 Q \Sigma) + 3 \gamma Q (1-\Sigma^2-\Omega^2)\nonumber \\
    & + 6 Q \Omega ^2 \cos^2 (\Phi -t \omega ) \Bigg)+ \mathcal{O}(D^3), \label{EQ:96b}
\end{align}
\endgroup
where
\begingroup\makeatletter\def\f@size{6.5}\check@mathfonts
\begin{align}
\label{EQ:97}
 & \mathbf{f}(\mathbf{x}, t) \nonumber \\
 & = 
   \left(\begin{array}{c}
\frac{1}{2}  \Omega  \Big(3 Q \left(\Omega ^2-1\right) (-\gamma +2\cos^2 (\Phi -t \omega ))  -\Sigma  \left(2 Q^2+3 (\gamma -2) Q \Sigma -2\right)\Big) \\\\
\frac{1}{2}  \Big(\left(\Sigma ^2-1\right) \left(-2 Q^2-3 (\gamma -2) Q \Sigma +2\right)   +3 Q \Sigma  \Omega ^2 (-\gamma +2\cos^2 (\Phi -t \omega ))\Big) \\\\
\frac{1}{2}  \left(Q^2-1\right) \Big(-3 (\gamma -2) \Sigma ^2+3
   \gamma -2 Q \Sigma   +3 \Omega ^2 (-\gamma +2\cos^2 (\Phi -t \omega ))-2\Big)\\\\
-\frac{3}{2}  Q \sin (2 (t \omega -\Phi))
      \end{array}
   \right).
\end{align}
\endgroup
Replacing $\dot{\mathbf{x}}= \mathbf{f}(t, \mathbf{x}) D$ with $\mathbf{x}= \left(\Omega, \Sigma, Q,  \Phi \right)^T$ and $\mathbf{f}(t, \mathbf{x})$ as in eq. \eqref{EQ:97}
by $\dot{\mathbf{y}}= H  \bar{f}(\mathbf{y})$ with $\mathbf{y}=\left(\bar{\Omega}, \bar{Q}, \bar{\Sigma}, \bar{\Phi} \right)^T$ with the time averaging \eqref{timeavrg} we obtain the time--averaged system 
\begingroup\makeatletter\def\f@size{7.5}\check@mathfonts
\begin{align}
& \dot{\bar{\Omega}}= \frac{1}{2} D \bar{\Omega}  \left(\bar{Q} \left(-3 \gamma  \left(\bar{\Sigma}^2+\bar{\Omega} ^2-1\right)  -2 \bar{Q}\; \bar{\Sigma} +6 \bar{\Sigma}^2+3 \bar{\Omega}^2-3\right)+2 \bar{\Sigma}\right), \label{eq65a}
\\
& \dot{\bar{\Sigma}}=\frac{1}{2} D \left(\left(\bar{\Sigma}^2-1\right) \left(-2 Q^2-3 (\gamma -2) \bar{Q} \; \bar{\Sigma} +2\right) -3 (\gamma -1) \bar{Q} \; \bar{\Sigma} \; \bar{\Omega} ^2\right), \label{eq65b}
\\
& \dot{\bar{Q}}= -\frac{1}{2} D \left(\bar{Q}^2-1\right) \left(3 \gamma  \left(\bar{\Sigma}^2+\bar{\Omega}
   ^2-1\right)  +2 \bar{\Sigma} (Q-3 \bar{\Sigma})-3 \bar{\Omega}^2+2\right), \label{eq65c}
\\
& \dot{\bar{\Phi}}=0, \label{eq65d} 
\\
& \dot{D}=  -\frac{1}{2} D^2 \Big(2 \bar{\Sigma} (1-\bar{Q}^2 +3 \bar{Q} \; \bar{\Sigma})  + 3 \bar{Q} \; \bar{\Omega}^2  + 3 \gamma \bar{Q} (1-\bar{\Sigma}^2-\bar{\Omega}^2)\Big). \label{eq65e}
\end{align}
\endgroup
Proceeding in an analogous way as in references \cite{Alho:2015cza,Alho:2019pku} we implement a local nonlinear transformation    
\begin{small}
\begin{align}
&\mathbf{x}_0:=\left(\Omega_{0}, \Sigma_{0}, Q_{0}, \Phi_{0}\right)^T  \mapsto \mathbf{x}:=\left(\Omega, \Sigma, Q, \Phi\right)^T, \nonumber \\
& \mathbf{x}=\psi(\mathbf{x}_0):=\mathbf{x}_0 + D \mathbf{g}(D, \mathbf{x}_0,t), \label{AppKSquasilinear211}
\\
& \mathbf{g}(D, \mathbf{x}_0,t)= \left(\begin{array}{c}
    g_1(D , \Omega_{0}, \Sigma_{0}, Q_{0}, \Phi_{0}, t)\\
    g_2(D , \Omega_{0}, \Sigma_{0}, Q_{0}, \Phi_{0}, t)\\
    g_3(D , \Omega_{0}, \Sigma_{0}, Q_{0}, \Phi_{0}, t)\\
    g_4(D , \Omega_{0}, \Sigma_{0}, Q_{0}, \Phi_{0}, t)\\
 \end{array}\right),   \label{eqT55}
\end{align}
\end{small}
\noindent 
where $D$ is the normalization factor and its evolution equation is given by \eqref{eq65e}.
Taking time derivative in both sides of \eqref{AppKSquasilinear211} with respect to $t$ we obtain 
\begin{small}
\begin{align}
    & \dot{\mathbf{x}}= \dot{\mathbf{x}_0}+ \dot{D} \mathbf{g}(D, \mathbf{x}_0,t)  \nonumber \\
    & + D \Bigg(\frac{\partial }{\partial t} \mathbf{g}(D, \mathbf{x}_0,t) + \dot{D} \frac{\partial }{\partial D} \mathbf{g}(D, \mathbf{x}_0,t)    + \mathbb{D}_{\mathbf{x}_0} \mathbf{g}(D, \mathbf{x}_0,t) \cdot \dot{\mathbf{x}_0}\Bigg), \label{EQT56}
    \end{align}
    \end{small}
    where 
   $\mathbb{D}_{\mathbf{x}_0} \mathbf{g}(D, \mathbf{x}_0,t)$ 
is the  Jacobian matrix of $\mathbf{g}(D, \mathbf{x}_0,t)$ with respect to the vector  $\mathbf{x}_0$.  The function $\mathbf{g}(D, \mathbf{x}_0,t)$ is conveniently chosen. 
\newline By substituting eq. \eqref{EQ:96a}, which can be written as
\begin{equation}
   \dot{\mathbf{x}}= D \mathbf{f}(\mathbf{x}_0 + D \mathbf{g}(D, \mathbf{x}_0,t),t) + \mathcal{O}(D^2),
\end{equation}
along with eqs.
\eqref{eq65e} and \eqref{AppKSquasilinear211} in eq. \eqref{EQT56} we obtain 
\begin{small}
\begin{align}
       & \Bigg(\mathbf{I}_4 + D \mathbb{D}_{\mathbf{x}_0} \mathbf{g}(D, \mathbf{x}_0,t)\Bigg) \cdot \dot{\mathbf{x}_0}= D \mathbf{f}(\mathbf{x}_0 + D \mathbf{g}(D, \mathbf{x}_0,t),t) \nonumber \\
       & -D \frac{\partial }{\partial t} \mathbf{g}(D, \mathbf{x}_0,t) -\dot{D} \mathbf{g}(D, \mathbf{x}_0,t) -D \dot{D} \frac{\partial }{\partial D} \mathbf{g}(D, \mathbf{x}_0,t)  + \mathcal{O}(D^2), 
\end{align}
\end{small}
where 
$\mathbf{I}_4$ is the $4\times 4$ identity matrix.

\noindent               
Then, we obtain 
  \begin{small}  
  \begin{align}
  \dot{\mathbf{x}_0} =  &\Bigg(\mathbf{I}_4 + D  \mathbb{D}_{\mathbf{x}_0} \mathbf{g}(D, \mathbf{x}_0,t)\Bigg)^{-1} \nonumber \\
 & \cdot \Bigg(D  \mathbf{f}(\mathbf{x}_0 + D \mathbf{g}(D, \mathbf{x}_0,t),t)-D \frac{\partial }{\partial t} \mathbf{g}(D, \mathbf{x}_0,t) \nonumber \\
 & -\dot{D} \mathbf{g}(D, \mathbf{x}_0,t) -D \dot{D} \frac{\partial }{\partial D} \mathbf{g}(D, \mathbf{x}_0,t) + \mathcal{O}(D^2)\Bigg).  
\end{align}
\end{small}
Using eq.   \eqref{eq65e}, we have $ \dot{D}= \mathcal{O}(D^2)$. Hence,
\begin{small}
\begin{align}
    \dot{\mathbf{x}_0} =  &\underbrace{\Bigg(\mathbf{I}_4 - D \mathbb{D}_{\mathbf{x}_0} \mathbf{g}(0, \mathbf{x}_0,t) +  \mathcal{O}(D^2)\Bigg)}_{4\times 4 \: \text{matrix}} \nonumber\\
    & \cdot \underbrace{\Bigg(D \mathbf{f}(\mathbf{x}_0, t)-D \frac{\partial }{\partial t} \mathbf{g}(0, \mathbf{x}_0,t) +   \mathcal{O}(D^2) \Bigg)}_{4\times 1 \; \text{vector}} \nonumber\\
  & = \underbrace{D \mathbf{f}(\mathbf{x}_0, t)-D \frac{\partial }{\partial t} \mathbf{g}(0, \mathbf{x}_0,t) +   \mathcal{O}(D^2)}_{4\times 1 \; \text{vector}}.\label{eqT59}
    \end{align} 
    \end{small}
The strategy is to use eq.   \eqref{eqT59} for choosing conveniently $\frac{\partial }{\partial t} \mathbf{g}(0, \mathbf{x}_0,t)$ to prove that 
\begin{align}
 & \dot{\Delta\mathbf{x}_0}= -D G(\mathbf{x}_0, \bar{\mathbf{x}}) +   \mathcal{O}(D^2), \label{EqY60}
  \end{align}
where $\bar{\mathbf{x}}=(\bar{\Omega},  \bar{\Sigma}, \bar{\Phi})^T$ and  $\Delta\mathbf{x}_0=\mathbf{x}_0 - \bar{\mathbf{x}}$. The function $G(\mathbf{x}_0, \bar{\mathbf{x}})$ is unknown at this stage. 
By construction we neglect the dependence of $\partial g_i/ \partial t$ and $g_i$ on $D$, i.e., we assume $\mathbf{g}=\mathbf{g}(\mathbf{x}_0,t)$ because the dependence of $D$ is dropped out along with higher order terms in  eq.   \eqref{eqT59}. 
\newline 
Next, we solve a partial differential equation  for $\mathbf{g}(\mathbf{x}_0,t)$ given by:  
\begin{align}
     & \frac{\partial }{\partial t} \mathbf{g}(\mathbf{x}_0,t) = \mathbf{f}(\mathbf{x}_0, t) - \bar{\mathbf{f}}(\bar{\mathbf{x}}) + G(\mathbf{x}_0, \bar{\mathbf{x}}), \label{eqT60}
\end{align}
\noindent  where we have considered $\mathbf{x}_0$ and $t$ as independent variables. 
The right hand side of eq. \eqref{eqT60} is almost periodic with period $L=\frac{2\pi}{\omega}$ for large times. Then, implementing the average process \eqref{timeavrg} on right hand side of eq. \eqref{eqT60}, where slow-varying dependence of quantities $\Omega_{0}, \Sigma_{0}, Q_0,  \Phi_0$ and  $\bar{\Omega},  \bar{\Sigma}, \bar{Q}, \bar{\Phi}$  on $t$ is ignored through  averaging process, we obtain \begin{align}
    & \frac{1}{L}\int_0^{L} \Bigg[\mathbf{f}(\mathbf{x}_0, s) - \bar{\mathbf{f}}(\bar{\mathbf{x}}) +G(\mathbf{x}_0, \bar{\mathbf{x}}) \Bigg] ds  \nonumber \\
    & = \bar{\mathbf{f}}( {\mathbf{x}}_0)-\bar{\mathbf{f}}(\bar{\mathbf{x}} )+G(\mathbf{x}_0, \bar{\mathbf{x}}). \label{newaverage}
\end{align}
Defining 
\begin{equation}
  G(\mathbf{x}_0, \bar{\mathbf{x}}):=  -\left(\bar{\mathbf{f}}( {\mathbf{x}}_0)-\bar{\mathbf{f}}(\bar{\mathbf{x}})\right),
\end{equation} the average \eqref{newaverage} is zero so that $\mathbf{g}(\mathbf{x}_0,t)$ is bounded.
\newline 
Finally, eq.   \eqref{EqY60} transforms to 
\begin{align}
 & \dot{\Delta\mathbf{x}_0}= D \left(\bar{\mathbf{f}}( {\mathbf{x}}_0)-\bar{\mathbf{f}}(\bar{\mathbf{x}})\right) +   \mathcal{O}(D^2),  \label{EqY602}
  \end{align}
and eq.   \eqref{eqT60} 
is simplified to 
\begin{align}
     & \frac{\partial }{\partial t} \mathbf{g}(\mathbf{x}_0,t) = \mathbf{f}(\mathbf{x}_0, t) - \bar{\mathbf{f}}( \mathbf{x}_0). \label{eqT602}
\end{align}

\begin{thm}
\label{KSLFZ11} Let be defined the functions  
$\bar{\Omega}, \bar{\Sigma}, \bar{Q}, \bar{\Phi}$,  and $D$ satisfying  time--averaged equations \eqref{eq65a}, \eqref{eq65b}, \eqref{eq65c}, \eqref{eq65d}, and \eqref{eq65e}.  Then, there exist continuously differentiable functions $g_1, g_2, g_3$, and $g_4$   such that   $\Omega, \Sigma, Q$, and $\Phi$  are locally given by eq. \eqref{AppKSquasilinear211},  where $\Omega_{0}, \Sigma_{0}, Q_{0}, \Phi_0$ are order zero approximations of  $\bar{\Omega}, \bar{\Sigma}, \bar{Q}, \bar{\Phi}$ as $D\rightarrow 0$. Then,  functions $\Omega_{0}, \Sigma_{0}, Q_{0}, \Phi_0$ and  $\bar{\Omega},  \bar{\Sigma}, \bar{Q}, \bar{\Phi}$  have the same limit on a time scale $t D =\mathcal{O}(1)$.
Setting $\Sigma=\Sigma_0=0$ analogous results are derived  for positively curved FLRW model. 
\end{thm}
\textbf{Proof.} The proof is given in \ref{gLKSFZ11}.
\noindent
Theorem \ref{KSLFZ11} implies that $\Omega , \Sigma, Q$,  and $\Phi$  evolve at first order in $D$ according to the time--averaged equations
\eqref{eq65a}, \eqref{eq65b}, \eqref{eq65c},  \eqref{eq65d}, and \eqref{eq65e} on a time scale $t D =\mathcal{O}(1)$.

According to eq. \eqref{EQ:96b} (eq. \eqref{eq65e}, respectively), we have that $D$ is monotonic decreasing when  $0< {Q}<1, {\Sigma}^2+ {\Omega}^2<1, {\Sigma}(1- {Q}^2 +3  {Q}  {\Sigma})>0$ ($0<\bar{Q}<1, \bar{\Sigma}^2+\bar{\Omega}^2<1, \bar{\Sigma}(1-\bar{Q}^2 +3 \bar{Q} \bar{\Sigma})>0$, respectively). Unfortunately, since these regions of full system or averaged system phase space are not invariant for the flow, the monotonicity of $D$ is not guaranteed for all $t$. 
\begin{rem}
\label{rem1}
The  initial region $0<Q<1, \Sigma^2+\Omega^2<1, {\Sigma}(1-Q^2 +3 Q \Sigma)>0$  is not invariant for the full system \eqref{unperturbed1KS} and for the averaged equations \eqref{eq65a}, \eqref{eq65b}, \eqref{eq65c}, \eqref{eq65d}, and \eqref{eq65e}.  Hence, although  $D(t)$ remains close to zero for $t< t^*$, where $t^*$ satisfies $\dot{D}(t^*)=0$, once the orbit crosses the  initial region, $D$ changes its monotony   and it will strictly increase without bound for $t>t^*$. Hence, Theorem \ref{KSLFZ11}  is valid on a time scale $t D =\mathcal{O}(1)$.
\end{rem}

\subsection{FLRW metric with positive curvature}
\label{SECT:IIIB}

In this section we will study the model with FLRW metric with positive curvature: \begin{align}
\label{closedmetricFLRW}
& ds^2 = - dt^2 + \ell^2(t) \Big[ dr^2+  \sin^2 r (d\y^2 + \sin^2 \y\, d\z^2)\Big].
\end{align}
For FLRW metric with positive curvature the field equations are obtained by setting $k=+1$ in eqs. \eqref{Non_minProb1FLRW0-1+1}. 

\noindent Using similar variables as in eqs. \eqref{varsKS} with $\Sigma=0$ and replacing the  normalization factor $D=\sqrt{H^2 + \frac{\R}{6}}$,  where  $\R$ denotes the 3-Ricci curvature of  spatial surfaces calculated in \eqref{equ87} by eq.   \eqref{varsclosedFLRW},  we obtain the system 
\begingroup\makeatletter\def\f@size{8}\check@mathfonts
\begin{subequations}
 \label{unperturbed1FLRWClosed}
\begin{align}
    & \dot \Omega= -\frac{b \gamma  f \mu ^3 Q  \Omega}{D} \sin ^2  \left(\frac{\sqrt{\frac{3}{2}} D \Omega  \sin (t \omega -\Phi)}{f \omega }\right) \nonumber \\
   &  -\frac{b \mu ^3}{\sqrt{6} D}  \cos (t \omega -\Phi ) \sin  \left(\frac{\sqrt{6} D \Omega 
   \sin (t \omega -\Phi)}{f \omega }\right)\nonumber \\
   & +\frac{3}{2} D Q  \Omega  \cos ^2(t \omega -\Phi )  \left(\Omega  ^2 \left(\gamma  \left(\frac{2 \mu ^2}{\omega ^2}-1\right)+2\right)-2\right) \nonumber \\
   & +\frac{3}{2}
   \gamma  D Q  \Omega  \left(1-\frac{2 \mu ^2 \Omega^2}{\omega ^2}\right) +\frac{\left(\omega ^2-2 \mu ^2\right) \Omega \sin (2 t \omega -2 \Phi)}{2 \omega },
\\
   & \dot{Q}=\frac{b \gamma  f \mu ^3 \left(1- Q^2\right)}{D} \sin
   ^2  \left(\frac{\sqrt{\frac{3}{2}} D \Omega \sin (t \omega -\Phi)}{f \omega }\right) \nonumber \\
   & -\frac{1}{4} D \left(1-Q^2\right) \left(6 \gamma -\frac{12 \gamma  \mu ^2 \Omega^2 \sin ^2(t \omega -\Phi)}{\omega
   ^2}-4\right) \nonumber \\
   & + \frac{3}{2} (\gamma -2) D \left(1-Q^2\right) \Omega^2 \cos ^2(t \omega -\Phi),
\end{align}
\begin{align}
   & \dot\Phi= 
  -\frac{b \mu ^3 \sin (t \omega -\Phi)}{\sqrt{6} D \Omega} \sin  \left(\frac{\sqrt{6} D \Omega \sin (t \omega -\Phi)}{f \omega
   }\right) \nonumber \\
   & -3 D Q \sin (t \omega -\Phi) \cos (t \omega -\Phi)  +\frac{\left(\omega ^2-2 \mu ^2\right) \sin ^2(t \omega -\Phi)}{\omega }, 
\\
   & \dot D=b \gamma  f \mu ^3 Q  \sin
   ^2 \left(\frac{\sqrt{\frac{3}{2}} D \Omega  \sin (t \omega -\Phi)}{f \omega }\right) \nonumber \\
   & +\frac{3}{2} \gamma  D^2 Q \left(\frac{2 \mu ^2 \Omega^2 \sin ^2(t \omega -\Phi)}{\omega ^2}-1\right) \nonumber \\
   & +\frac{3}{2} (\gamma -2)
  D^2 Q \Omega^2 \cos ^2(t \omega -\Phi), 
\end{align}
\end{subequations}
\endgroup
and the deceleration parameter is 
\begingroup\makeatletter\def\f@size{7.5}\check@mathfonts
\begin{align}
  & q= \frac{3 \gamma -2}{2 Q^2}-\frac{b \gamma  f \mu ^3}{D^2 Q^2}  \sin ^2  \left(\frac{\sqrt{\frac{3}{2}} D \Omega \sin (t \omega -\Phi)}{f \omega }\right)   \nonumber \\
  & -\Omega^2  \left(\frac{3 \gamma  \mu ^2 \sin ^2(t \omega -\Phi)}{\omega ^2
   Q^2}+\frac{3 (\gamma -2) \cos ^2(t \omega -\Phi)}{2 Q^2}\right).
\end{align}
\endgroup
The fractional energy density of matter $\Omega_m:= \frac{\rho_m}{3 H^2}= \frac{\rho_m}{3 Q^2 D^2}$ is parametrized by the equation
  \begin{small}
\begin{align}
  &Q^2 \Omega_m=1-\frac{2 b f \mu ^3}{3 D^2}  \sin^2   \left(\frac{\sqrt{6} D \Omega \sin (t \omega -\Phi)}{2 f \omega }\right) \nonumber\\
  &-\frac{2 \mu ^2 \Omega^2 }{\omega ^2}\sin ^2(t \omega -\Phi)-\Omega^2 \cos ^2(t
   \omega -\Phi).
\end{align}
\end{small}
Setting $f=\frac{b \mu ^3}{\omega ^2-2 \mu ^2}>0$, we obtain the series expansion near $D=0$  
\begin{small}
\begin{align}
& \dot{\mathbf{x}}= \mathbf{f}(\mathbf{x},t) D+\mathcal{O}\left(D^2\right), \;   \mathbf{x}= \left(\Omega, Q, \Phi \right)^T, \nonumber\\
 & \dot{D}= -\frac{3}{2} {Q} \left(\gamma (1- \Omega^2) +2\cos^2 (t \omega - \Phi
    ) \Omega  ^2 \right) D^2 +\mathcal{O}\left(D^3\right), \label{perturbed1FLRWClosed}
\end{align}
\end{small}
 where the vector function is defined as
\begingroup\makeatletter\def\f@size{7.5}\check@mathfonts
\begin{align}
   & \mathbf{f}(t, \mathbf{x})  = 
   \left(\begin{array}{c}
  -\frac{3}{2}  {Q} \Omega   \left(1-\Omega  ^2\right) \left(2
   \cos ^2(t \omega -\Phi  )-\gamma \right) \\\\
   \frac{1}{2}  \left(1-{Q}^2\right) \Big(2-3 \gamma  -3 \Omega
    ^2 \left(2 \cos ^2(t \omega -\Phi  )-\gamma
   \right)\Big)\\\\
   -\frac{3}{2}  {Q} \sin (2 t \omega -2 \Phi )
      \end{array}
   \right).
\end{align}
\endgroup
Systems  \eqref{unperturbed1FLRWClosed} and \eqref{perturbed1FLRWClosed} are invariant for the simultaneous change $(t,  Q, \Phi) \mapsto (-t,  -Q, -\Phi)$. 
Replacing $\dot{\mathbf{x}}= \mathbf{f}(t, \mathbf{x}) D$ with $\mathbf{x}= \left(\Omega, Q, \Phi \right)^T$, where $ \mathbf{f}(t, \mathbf{x})$ is given by eq. \eqref{perturbed1FLRWClosed}, 
by $\dot{\mathbf{y}}=D  \bar{\mathbf{f}}(\mathbf{y})$ with $\mathbf{y}=\left(\bar{\Omega}, \bar{Q}, \bar{\Phi} \right)^T$  with the time averaging \eqref{timeavrg}, we obtain for $\gamma \neq 1$ the following time--averaged system  
\begin{align}
    &\dot{\bar{\Omega}}=\frac{3}{2} (\gamma -1) D  \; \bar{Q} \bar{\Omega}  \left(1-{\bar{\Omega}}
   ^2\right), \label{eq72}\\
   &\dot{\bar{Q}}=-\frac{1}{2} D  \left(1-{\bar{Q}}^2\right) \left(3 \gamma 
   \left(1-{\bar{\Omega}} ^2\right)+3 {\bar{\Omega}}^2-2\right), \label{eq73}\\
   &\dot{\bar{\Phi}}=0, \label{eq74}\\
   & \dot{{D}}=  -\frac{3}{2} \bar{Q} \left(\gamma (1- \bar{\Omega}^2) + \bar{\Omega}^2 \right) D^2.
  \end{align}
\section{Qualitative analysis of time--averaged systems}
\label{SECT:III}
According to Theorem   \ref{KSLFZ11}, in KS metrics and positively curved FLRW models the function \eqref{D-KS} plays the role of a time--dependent perturbation function controlling the magnitude of error between  solutions of the full and time--averaged problems with the same initial conditions as $D\rightarrow \infty$. Thus,  oscillations are viewed as perturbations as far as $D$ is bounded. In the time--averaged system  Raychaudhuri equation decouples. Therefore, the analysis of the system is  reduced to study the corresponding time--averaged equations.

\subsection{Kantowski-Sachs}
\label{KS}
With time variable $\eta$ defined by $\frac{d f}{d \eta}= \frac{1}{D}\frac{d f}{d t}$  the time--averaged system   \eqref{eq65a}, \eqref{eq65b}, \eqref{eq65c},  \eqref{eq65d} transforms to  
\begingroup\makeatletter\def\f@size{7}\check@mathfonts
\begin{subequations}
\label{avrgsystKS}
\begin{align}
&\frac{d \bar{\Omega}}{d \eta}=\frac{1}{2}   \bar{\Omega}  \Bigg(\bar{Q} \Big(3 \gamma  \left(1-\bar{\Sigma}^2-\bar{\Omega} ^2\right)  -2 \bar{Q}\; \bar{\Sigma} +6 \bar{\Sigma}^2+3 \bar{\Omega}^2-3\Big)+2 \bar{\Sigma}\Bigg), \label{KSguidingC1} 
\\
& \frac{d \bar{\Sigma}}{d \eta}=\frac{1}{2}   \Big(\left(\bar{\Sigma}^2-1\right) \left(-2 \bar{Q}^2-3 (\gamma -2) \bar{Q} \; \bar{\Sigma} +2\right)  -3 (\gamma -1) \bar{Q} \; \bar{\Sigma} \; \bar{\Omega} ^2\Big), \label{KSguidingC2} 
\\
& \frac{d \bar{Q}}{d \eta}=-\frac{1}{2}   \left(\bar{Q}^2-1\right) \Big(3 \gamma  \left(\bar{\Sigma}^2+\bar{\Omega}
  ^2-1\right)  +2 \bar{\Sigma} (Q-3 \bar{\Sigma})-3 \bar{\Omega}^2+2\Big), \label{KSguidingC3}
\\
&  \frac{d{{\bar{\Phi}}}}{d {\eta}}=0,   
\\
&  \frac{d D}{d {\eta}}=  -\frac{1}{2} D \Big(2 \bar{\Sigma} (1-\bar{Q}^2 +3 \bar{Q} \; \bar{\Sigma})  + 3 \bar{Q} \; \bar{\Omega}^2  + 3 \gamma \bar{Q} (1-\bar{\Sigma}^2-\bar{\Omega}^2)\Big). 
\end{align}
\end{subequations}
\endgroup
where we have defined $\bar{\Omega}_m$  as
\begin{align}
\bar{Q}^2 \bar{\Omega}_m:= 1-\bar{\Sigma}^2- \bar{\Omega}^2, \label{eqOmega69}
\end{align}
and it was interpreted as the time--averaged values of $\Omega_m:=  \frac{\rho_m}{3 H^2}$. 
Then, the phase space is 
\begin{align}
    \left\{(\bar{\Omega}, \bar{\Sigma}, \bar{Q})\in \mathbb{R}^3: -1\leq \bar{Q} \leq 1, \bar{\Sigma}^2+ \bar{\Omega}^2 \leq 1\right\}.
\end{align}
Furthermore, we have the auxiliary equations
\begin{align}
\label{aux-eqs}
        & \dot{{e_1}^1}=  -D (Q-2 \Sigma){e_1}^1 \; \text{and}\; \dot{K}= -2 D (Q +\Sigma)K.
\end{align}
Evaluating the averaged values $Q=\bar{Q}, \Sigma=\bar{\Sigma}$ at eqs. \eqref{aux-eqs} and integrating the resulting equations, approximated solutions for ${e_1}^1$ and $K$  as functions of $t$ are obtained. 

\noindent Recall that a set of non-isolated singular points is said to be normally hyperbolic if the only eigenvalues with zero real parts are those whose corresponding eigenvectors are tangent to the set.  
Since by definition any point on a set of non-isolated singular points will have at least
one eigenvalue which is zero, all points in the set are non-hyperbolic. However, a set which is
normally hyperbolic can be completely classified according to its stability by considering the signs of eigenvalues in the remaining directions (i.e., for a curve  in the
remaining $n-1$ directions) (see \cite{aulbach}, pp. 36).
\newline

In the special case $\gamma=1$ there exist two lines of equilibrium points which are normally  hyperbolic 
\begin{enumerate}
    \item $K_{-}: (\bar{\Omega}, \bar{\Sigma}, \bar{Q})=(\bar{\Omega}_c, 0, -1)$ with eigenvalues \newline $\{\frac{3}{2},-1,0\}$, it is a saddle. 
    
    \item $K_{+}: (\bar{\Omega}, \bar{\Sigma}, \bar{Q})=(\bar{\Omega}_c, 0, 1)$ with eigenvalues \newline $\{-\frac{3}{2},1,0\}$, it is a saddle.
\end{enumerate}

Likewise, in the special case $\gamma=2$ there exist two lines of equilibrium points which are normally  hyperbolic 
\begin{enumerate}
    \item $L_{-}: (\bar{\Omega}, \bar{\Sigma}, \bar{Q})=(0, \bar{\Sigma}_c, -1)$ with eigenvalues\newline $\{-\frac{3}{2},0,-2(2+\bar{\Sigma}_c)\}$, it is stable for $\bar{\Sigma}_c\geq-2$ and a saddle for $\bar{\Sigma}_c<-2$ .
    \item $L_{+}: (\bar{\Omega}, \bar{\Sigma}, \bar{Q})=(0, \bar{\Sigma}_c, 1)$ with eigenvalues\newline $\{\frac{3}{2},0,-2(-2+\bar{\Sigma}_c)\}$, it is unstable for $\bar{\Sigma}_c\leq2$ and a saddle for $\bar{\Sigma}_c>2$. 
   
\end{enumerate}
Recall, the subindex $\pm$ indicates whether they correspond to contracting (``$-$'') or expanding  (``$+$'') solutions.   A solution is classified as expanding if ${Q}>0$ since $H$ and $Q$ have the same sign due to $D>0$.

The equilibrium points of the guiding system   \eqref{KSguidingC1}, \eqref{KSguidingC2}, \eqref{KSguidingC3}  are 
   \begin{enumerate}      
       \item $P_1: (\bar{\Omega},\bar{\Sigma},\bar{Q})=(0,-1,-1)$ with eigenvalues \newline $\left\{-2,-\frac{3}{2},3
   (\gamma -2)\right\}.$ 
   \begin{enumerate}
       \item It is a sink for $0\leq \gamma <2.$       
        \item It is nonhyperbolic for $\gamma=2$ (contained in $L_{-}$).
   \end{enumerate}
For $P_1$ we obtain  
   \begin{align*}
    & \dot{D}=3 D^2,\;  \dot{H}=-3 D^2, \; \dot{{\sigma_+}}= -3 D^2,\nonumber \\
    & \dot{{e_1}^1}=  -D{e_1}^1, \; \dot{K}= 4 D K.
   \end{align*}
Imposing initial conditions 
\begin{align}
\label{ini}
       &K(0)=1/c_1, \; e_1^1(0)=c_1, \nonumber \\
       & H(0)=H_0, \; D(0)=D_0, \; \sigma (0)= \sigma_0,
\end{align}
where $t=0, \tau(0)=0$ is the current time and $D_0^2=H_0^2+\frac{1}{3 c_1}$,  we obtain by integration 
   \begin{align*}
       & D(t)=\frac{{D_0}}{1-3 {D_0} t},\\
       & H(t)=\frac{{D_0}}{3 {D_0} t-1}+{D_0}+{H_0},\\
       & \sigma_+(t)=\frac{{D_0}}{3 {D_0} t-1}+{D_0}+ \sigma_{0},\\
       & e_1^1(t)^{-2}=\frac{1}{{c_1}^2 (3 {D_0} t-1)^{\frac{2}{3}}}, \;  K(t)^{-1}=c_1 (3 {D_0} t-1)^{\frac{4}{3}}.
   \end{align*}
 The asymptotic metric at  $P_1$ is given by 
      \begin{align}
   &  ds^2= - dt^2 + \frac{1}{{c_1}^2 (3 {D_0} t-1)^{\frac{2}{3}}} dr^2 \nonumber \\
   &   + c_1 (3 {D_0} t-1)^{\frac{4}{3}}  (d\y^2 + \sin^2 \y\, d\z^2). \label{metricKS-P1}
\end{align} 
   
   This point represents a non--flat LRS Kasner ($p_1=-\frac{1}{3}, p_2= p_3= \frac{2}{3}$)  contracting solution (\cite{WE} Sect. 6.2.2 and Sect. 9.1.1 (2)). This solution is singular at finite time  $t_0=\frac{1}{3D_0}$  and is valid for $t>t_0$. 
   
     \item $P_2: (\bar{\Omega},\bar{\Sigma},\bar{Q})=(0,1,1)$ with eigenvalues \newline $\left\{2,\frac{3}{2},6-3 \gamma
   \right\}.$ 
   \begin{enumerate}
       \item It is a source for $0\leq \gamma <2.$       
        \item It is nonhyperbolic for $\gamma=2$ (contained in $L_{+}$).
   \end{enumerate}
   Evaluating at $P_2$ we obtain 
   \begin{align*}
    & \dot{D}=-3 D^2,\;  \dot{H}=-3 D^2, \;  \dot{{\sigma_+}}= -3 D^2,\nonumber \\
    & \dot{e_1^1}=D e_1^1, \;  \dot{K}= 4 D K. 
   \end{align*}
Imposing initial conditions \eqref{ini}, we obtain by integration 
   \begin{align*}
       & D(t)=\frac{{D_0}}{3 {D_0} t+1},\\
       & H(t)={D_0} \left(\frac{1}{3 {D_0} t+1}-1\right)+{H_0},\\
       & \sigma_+(t)={\sigma_0}-\frac{3 {D_0}^2 t}{3 {D_0} t+1},\\
       & e_1^1(t)^{-2}=\frac{1}{{c_1}^2 (3 {D_0} t+1)^{\frac{2}{3}}},\; K(t)^{-1}=c_1 (3 {D_0}
   t+1)^{\frac{4}{3}}.
   \end{align*}
The asymptotic metric at  $P_3$ can be written as 
       \begin{align}
   &  ds^2= - dt^2 + \frac{(1-3 {D_0} t)^2}{{c_1}^2} dr^2 \nonumber \\
   &   + c_1  (d\y^2 + \sin^2 \y\, d\z^2). \label{KS-metric-P3}
\end{align}
    This point represents a non--flat LRS Kasner ($p_1=-\frac{1}{3}, p_2= p_3= \frac{2}{3}$)  expanding solution (\cite{WE} Sect. 6.2.2 and Sect. 9.1.1 (2)). It is valid for all $t$. 
    
     \item $P_3:  (\bar{\Omega},\bar{\Sigma},\bar{Q})=(0,1,-1)$ with eigenvalues \newline $\left\{-6,-\frac{3}{2},3
   (\gamma -2)\right\}.$ 
   \begin{enumerate}
          \item It is a sink for $0\leq \gamma <2.$       
        \item It is nonhyperbolic for $\gamma=2 $ (contained in $L_{-}$).
   \end{enumerate}
      Evaluating at $P_3$ we obtain
      \begin{align*}
    & \dot{D}=3 D^2,\;  \dot{H}=-3 D^2, \nonumber \\
    & \dot{{\sigma_+}}= 3 D^2, \;  \dot{e_1^1}=3 D e_1^1, \;  \dot{K}=0. 
   \end{align*}
Imposing initial conditions  
   \eqref{ini}, we obtain by integration 
   \begin{align*}
       & D(t)=\frac{{D_0}}{1-3 {D_0} t},\\
       & H(t)=\frac{{D_0}}{3 {D_0} t-1}+{D_0}+{H_0},\\
       & \sigma_+(t)=\frac{3 {D_0}^2 t}{1-3 {D_0} t}+{\sigma_0},\\
       & e_1^1(t)^{-2}=\frac{(1-3 {D_0} t)^2}{{c_1}^2},\; K(t)^{-1}=c_1.
   \end{align*}
This point represents a Taub (flat LRS Kasner) contracting solution ($p_1=1, p_2= 0, p_3= 0$)
\cite{WE} (Sect 6.2.2 and p 193, eq.    (9.6)).  This solution is singular at finite time  $t_0$ and is valid  for $t>t_0$.

          \item $P_4: (\bar{\Omega},\bar{\Sigma},\bar{Q})=(0,-1,1)$ with eigenvalues \newline $\left\{6,\frac{3}{2},6-3 \gamma
   \right\}.$ 
   \begin{enumerate}
       \item It is a source for $0\leq \gamma <2.$
       \item It is nonhyperbolic for $\gamma=2 $ (contained in $L_{+}$).
   \end{enumerate}
   Evaluating at $P_4$ we obtain 
      \begin{align*}
    & \dot{D}=-3 D^2,\;  \dot{H}=-3 D^2, \nonumber \\
    & \dot{{\sigma_+}}= 3 D^2, \; \dot{e_1^1}=-3 D e_1^1, \;  \dot{K}=0. 
   \end{align*}
Imposing initial conditions  
   \eqref{ini}, we obtain by integration  
   \begin{align*}
       & D(t)=\frac{{D_0}}{3 {D_0} t+1},\\
       & H(t)={D_0} \left(\frac{1}{3 {D_0} t+1}-1\right)+{H_0},\\
       & \sigma_+(t)=\frac{3 {D_0}^2 t}{3 {D_0} t+1}+{\sigma_0},\\
       & e_1^1(t)^{-2}=\frac{(3 {D_0} t+1)^2}{{c_1}^2}, \; K(t)^{-1}=c_1.
   \end{align*}
This point represents a Taub (flat LRS Kasner) expanding solution ($p_1=1, p_2= 0, p_3= 0$)
\cite{WE} (Sect 6.2.2 and p 193, eq.    (9.6)).
    \item $P_5: (\bar{\Omega},\bar{\Sigma},\bar{Q})=(0,0,-1)$ with eigenvalues \newline $\left\{3-\frac{3 \gamma
   }{2},-\frac{3}{2} (\gamma
   -1),2-3 \gamma \right\}.$ 
   \begin{enumerate}
       \item It is a source for $0\leq \gamma <\frac{2}{3}.$
       \item It is a saddle for $\frac{2}{3}<\gamma <1$ or $1<\gamma <2.$
       \item It is nonhyperbolic for $\gamma =\frac{2}{3}$ or $\gamma =1$ (contained in $K_{-}$) or $\gamma
   =2$ (contained in $L_{-}$).
   \end{enumerate}
         Evaluating at $P_5$ we obtain 
      \begin{align*}
    & \dot{D}=\frac{3}{2} \gamma   D^2,\;  \dot{H}=-\frac{3}{2} \gamma D^2, \; \dot{{\sigma_+}}= 0,\nonumber \\
    &\dot{e_1^1}= D e_1^1, \;  \dot{K}=2 D K. 
   \end{align*}
Imposing initial conditions  
   \eqref{ini} with $\sigma_0=0$, we obtain by integration  
   \begin{align*}
       & D(t)=\frac{2 {D_0}}{2-3 \gamma  {D_0} t},\\
       & H(t)=\frac{2 {D_0}}{3 \gamma  {D_0} t-2}+{D_0}+{H_0},\\
       & \sigma_+(t)=0,\\
       & e_1^1(t)^{-2}=\frac{\left(1-\frac{3 \gamma  {D_0} t}{2}\right)^{\frac{4}{3 \gamma
   }}}{{c_1}^2}, \;  K(t)^{-1}=c_1 \left(1-\frac{3 \gamma  D_0 t}{2}\right)^{\frac{4}{3 \gamma }}.
   \end{align*}
The corresponding solution is a  flat matter--dominated FLRW contracting solution with $\bar{\Omega}_m=1$. This solution is singular at finite time  $t_1=\frac{2}{3 \gamma D_0}$ and is valid  for $t>t_1$.
   
       \item $P_6: (\bar{\Omega},\bar{\Sigma},\bar{Q})=(0,0,1)$ with eigenvalues \newline $\left\{\frac{3 (\gamma
   -2)}{2},\frac{3 (\gamma
   -1)}{2},3 \gamma -2\right\}.$  
   \begin{enumerate}
       \item It is a sink for $0\leq \gamma <\frac{2}{3}.$
       \item It is a saddle for $\frac{2}{3}<\gamma <1$ or $1<\gamma <2.$
       \item It is nonhyperbolic for 
$\gamma =\frac{2}{3}$ or $\gamma =1$ (contained in $K_{+}$) or $\gamma =2 $   (contained in $L_{+}$).
    \end{enumerate}
Evaluating at $P_6$ we obtain 
      \begin{align*}
    & \dot{D}=-\frac{3}{2} \gamma   D^2,\;  \dot{H}=-\frac{3}{2} \gamma D^2, \nonumber \\
    & \dot{{\sigma_+}}= 0, \; \dot{e_1^1}= -D e_1^1, \;  \dot{K}=-2 D K. 
   \end{align*}      
Imposing initial conditions  
   \eqref{ini} with $\sigma_0=0$, we obtain by integration 
   \begin{align*}
       & D(t)=\frac{2 {D_0}}{3 \gamma  {D_0} t+2},\\
       & H(t)={D_0} \left(\frac{2}{3 \gamma  {D_0} t+2}-1\right)+{H_0},\\
       & \sigma_+(t)={0},\\
       & e_1^1(t)^{-2}=\frac{\left(\frac{3 \gamma  {D_0} t}{2}+1\right)^{\frac{4}{3 \gamma
   }}}{{c_1}^2},\;  K(t)^{-1}=c_1 \left(\frac{3 \gamma  D_0 t}{2}+1\right)^{\frac{4}{3 \gamma }}.
   \end{align*}
The asymptotic metric at $P_6$ is given by 
      \begin{align}
   &  ds^2= - dt^2 + \frac{\left(\frac{3 \gamma  {D_0} t}{2}+1\right)^{\frac{4}{3 \gamma
   }}}{{c_1}^2} dr^2 \nonumber \\
   & + c_1 \left(\frac{3 \gamma  D_0 t}{2}+1\right)^{\frac{4}{3 \gamma }}  (d\y^2 + \sin^2 \y\, d\z^2). \label{KS-metric-P6}
\end{align}  
   
     The corresponding solution is a  flat matter--dominated FLRW universe
   with $\bar{\Omega}_m=1$.
   $F_0$ represents a  quintessence fluid for $0<\gamma<\frac{2}{3}$ or a zero-acceleration model for $\gamma=\frac{2}{3}$. 
In the limit $\gamma=0$ we have $\ell(t)\propto  \left(1+\frac{3 \gamma  {D_0} t}{2}\right)^{\frac{2}{3
   \gamma }}\rightarrow \ell_0  e^{D_0 t}$, where $D_0=H_0$, i.e., there is a de Sitter solution.  
        \item $P_7: (\bar{\Omega},\bar{\Sigma},\bar{Q})=(1,0,-1)$ with eigenvalues \newline $\left\{-\frac{3}{2},1,3-3
   \gamma \right\}.$ This point is always a saddle because it has a negative and a positive eigenvalue. For $\gamma =1$ it is a nonhyperbolic saddle (contained in $K_{-}$). 
   Evaluating at $P_7$ we obtain
      \begin{align*}
    & \dot{D}= 3 \cos ^2(t \omega -\Phi) D^2+\mathcal{O}\left(D^3\right), \nonumber 
\\
   & \dot{H}= -3 \cos ^2(t \omega -\Phi) D^2+\mathcal{O}\left(D^3\right), \nonumber 
\\
   & \dot{{\sigma_+}}= 0, \; \dot{e_1^1}= D e_1^1, \;  \dot{K}=2 D K. 
   \end{align*}
In average  $\Phi$ is constant. Then, setting for simplicity $\Phi=0$ and  integrating the first two eqs. we obtain  
\begin{align*}
      & D(t)\approx -\frac{4 \omega }{4 \alpha \omega +6 t \omega +3 \sin (2 t \omega )}, \\
      & H(t)\approx  \frac{4 \omega }{4 \alpha \omega +6 t \omega +3 \sin (2 t \omega )},
     \end{align*}
 for large $t$.     Then, 
    \begin{equation}
        D(t)\approx -\frac{2}{3 t}, \; H(t)\approx \frac{2}{3 t},
    \end{equation}
    as $t\rightarrow \infty$. 
    Finally,
    \begin{equation}
        e_1^1(t)= \frac{c_1}{t^{\frac{2}{3}}}, \; K(t)= \frac{c_2}{t^{\frac{4}{3}}}, \; \sigma=0.     \end{equation}
The line element \eqref{metric} becomes
\begin{align}
   &  ds^2= - dt^2 + c_1^{-2} {t^{\frac{4}{3}}} dr^2 \nonumber \\
   &  +  {c_2^{-1}}{t^{\frac{4}{3}}} (d\y^2 + \sin^2 \y\, d\z^2).
\end{align}  
 Hence, the equilibrium point  can be associated with   Einstein-de Sitter solution (\cite{WE}, Sec 9.1.1 (1)) with $\gamma= 1$. It is a contracting solution.

       \item $P_8: (\bar{\Omega},\bar{\Sigma},\bar{Q})=(1,0,1)$ with eigenvalues \newline $\left\{\frac{3}{2},-1,3 (\gamma
   -1)\right\}.$ This point is always a saddle because it has a negative and a positive eigenvalue. For $\gamma =1$ it is a nonhyperbolic saddle (contained in $K_+$).
   
Evaluating at $P_8$ we obtain 
      \begin{align*}
    & \dot{D}=-3 \cos ^2(t \omega -\Phi) D^2+\mathcal{O}\left(D^3\right), \nonumber \\
\\
   & \dot{H}= -3 \cos ^2(t \omega -\Phi) D^2+\mathcal{O}\left(D^3\right), \nonumber
\\
   & \dot{{\sigma_+}}= 0, \; \dot{e_1^1}= -D e_1^1, \;  \dot{K}=-2 D K.
   \end{align*}
In average $\Phi$ is constant. Then, setting for simplicity $\Phi=0$ and  integrating the first two eqs. we obtain  \begin{align*}
      & D(t)\approx \frac{4 \omega }{-4 \alpha  \omega +6 t \omega +3 \sin (2 t \omega )}, \\
      & H(t)\approx  \frac{4 \omega }{-4 \alpha  \omega +6 t \omega +3 \sin (2 t \omega )}. 
     \end{align*}
    Then, 
    \begin{equation}
        D(t)\approx \frac{2}{3 t}, \; H(t)\approx \frac{2}{3 t},
    \end{equation}
    as $t\rightarrow \infty$. 
   Finally,
    \begin{equation}
        e_1^1(t)= \frac{c_1}{t^{\frac{2}{3}}}, \; K(t)= \frac{c_2}{t^{\frac{4}{3}}}, \;\sigma (t)=0.
    \end{equation}
The line element \eqref{metric} becomes
\begin{align}
   &  ds^2= - dt^2 + c_1^{-2} {t^{\frac{4}{3}}} dr^2 \nonumber \\
   &  +  {c_2^{-1}}{t^{\frac{4}{3}}} (d\y^2 + \sin^2 \y\, d\z^2).
\end{align}  
 Hence,  the equilibrium point  can be associated with Einstein-de Sitter solution (\cite{WE}, Sec 9.1.1 (1)) with $\gamma= 1$. It is an expanding solution.  
   
        \item $P_{9}: (\bar{\Omega},\bar{\Sigma},\bar{Q})=\left(0,\frac{2-3 \gamma
   }{\sqrt{(4-3 \gamma
   )^2}},-\frac{2}{\sqrt{(4- 3
   \gamma)^2}}\right)$ with eigenvalues \newline $\Big\{\frac{1}{3 \gamma
   -4}+1,\frac{3 \left(\gamma
   -\sqrt{2-\gamma }
   \sqrt{\gamma  (24 \gamma
   -41)+18}-2\right)}{6 \gamma
   -8}$, \newline $\frac{3 \left(\gamma
   +\sqrt{2-\gamma }
   \sqrt{\gamma  (24 \gamma
   -41)+18}-2\right)}{6 \gamma
   -8}\Big\}.$ It exists for $0\leq \gamma \leq \frac{2}{3}$ or
   $\gamma =2$.
   \begin{enumerate}
       \item It is a saddle for $0\leq \gamma <\frac{2}{3}.$
       \item It is nonhyperbolic for $\gamma =\frac{2}{3}$ or $\gamma =2.$
   \end{enumerate}
   
  Evaluating at $P_{9}$ for $0\leq \gamma <\frac{2}{3}$ we obtain  the following: 
\begin{align*}
   & \dot{D}= -\frac{3 \gamma D^2}{3 \gamma -4}, \; \dot{H}=
   -\frac{6 \gamma  D^2}{(3 \gamma -4)^2}, \nonumber \\
   & \dot{\sigma_{+}}= -\frac{3
   \gamma  (3 \gamma -2) D^2}{(3 \gamma -4)^2}, \nonumber \\
   & \dot{e_1^1}= \frac{6 (\gamma -1)D
  e_1^1}{3 \gamma -4},  \;  \dot{K}= -\frac{6 \gamma  D K}{3 \gamma -4}.
\end{align*}  
Imposing initial conditions  
   \eqref{ini}, we obtain by integration  
 \begin{align*}
     & D(t)= \frac{(3 \gamma -4) {D_0}}{3 \gamma  ({D_0} t+1)-4}, \\
     & H(t)=2 {D_0} \left(\frac{1}{4-3 \gamma }+\frac{1}{3 \gamma +3 \gamma  {D_0} t-4}\right)+{H_0},\\
     & \sigma_{+}(t)= \frac{3 (2-3 \gamma ) \gamma  {D_0}^2 t}{(3 \gamma -4) (3
   \gamma +3 \gamma  {D_0} t-4)}+{\sigma_0},\\
   & e_{1}^1(t)^{-2}=\frac{(3 \gamma -4)^{4-\frac{4}{\gamma }} (3 \gamma  ({D_0} t+1)-4)^{\frac{4}{\gamma }-4}}{c_1^2},\\
   & K(t)^{-1}=\frac{c_1 (4-3 \gamma  ({D_0} t+1))^2}{(4-3 \gamma
   )^2}.
 \end{align*}
The line element \eqref{metric} becomes
\begin{align}
   &  ds^2= - dt^2 + \frac{(3 \gamma -4)^{4-\frac{4}{\gamma }} (3 \gamma  ({D_0} t+1)-4)^{\frac{4}{\gamma }-4}}{c_1^2} dr^2 \nonumber \\
   &  +\frac{c_1 (4-3 \gamma  ({D_0} t+1))^2}{(4-3 \gamma
   )^2} \left[ d \vartheta^2 +  \sin^2(\vartheta) d \zeta\right].
\end{align}  
  This solution is singular at finite time  $t_2= \frac{1}{D_0}\left(\frac{4}{3 \gamma}-1\right)$ and is valid  for $t>t_2$.
  
  For $\gamma=2$ we have 
  \begin{align*}
   & \dot{D}= 3 D^2, \; \dot{H}= -3 D^2, \;  \dot{\sigma_{+}}= -6
   D^2, \nonumber \\
   & \dot{e_1^1}= -3 D e_1^1,\;  \dot{K}=6 D K.   
   \end{align*}
Imposing initial conditions  
   \eqref{ini}, we obtain by integration  
 \begin{align*}
     & D(t)=\frac{{D_0}}{1-3 {D_0} t}, \\
     & H(t)= \frac{{D_0}}{3 {D_0} t-1}+{D_0}+{H_0},\\
     & \sigma_{+}(t)=\frac{6 {D_0}^2 t}{3 {D_0} t-1}+{\sigma_0},\\
   & e_{1}^1(t)^{-2}=\frac{1}{(c_1-3 c_1 {D_0} t)^2}, \; K(t)^{-1}=c_1 (1-3 {D_0}
   t)^2.
 \end{align*}
  This solution is singular at finite time  $t_3= \frac{1}{3 D_0}$ and is valid  for $t>t_3$.
  
      \item $P_{10}: (\bar{\Omega},\bar{\Sigma},\bar{Q})=\left(0,\frac{3 \gamma
   -2}{\sqrt{(4-3 \gamma
   )^2}},\frac{2}{\sqrt{(4-3
   \gamma)^2}}\right)$ with eigenvalues \newline $\Big\{\frac{3 (\gamma -1)}{4-3
   \gamma },\frac{3
   \left(\gamma +\sqrt{2-\gamma
   } \sqrt{\gamma  (24 \gamma
   -41)+18}-2\right)}{8-6
   \gamma }$,\newline $\frac{-3 \gamma +3
   \sqrt{2-\gamma }
   \sqrt{\gamma  (24 \gamma
   -41)+18}+6}{6 \gamma
   -8}\Big\}.$  It exists for $0\leq \gamma \leq \frac{2}{3}$ or
   $\gamma =2$. 
   \begin{enumerate}
       \item It is a saddle for $0\leq \gamma <\frac{2}{3}.$
       \item It is nonhyperbolic for $\gamma =\frac{2}{3}$ or $\gamma =2.$
   \end{enumerate}
   
     Evaluating at $P_{10}$ for $0\leq \gamma <\frac{2}{3}$ we obtain  the following: 
\begin{align*}
   & \dot{D}= \frac{3 \gamma D^2}{3 \gamma -4}, \; \dot{H}=
   -\frac{6 \gamma  D^2}{(3 \gamma -4)^2}, \nonumber \\
   & \dot{\sigma_{+}}= -\frac{3
   \gamma  (3 \gamma -2) D^2}{(3 \gamma -4)^2}, \nonumber \\
   & \dot{e_1^1}=-\frac{6 (\gamma -1) D e_1^1}{3 \gamma -4}, \;  \dot{K}=\frac{6 \gamma  D K}{3 \gamma -4}.
\end{align*}  
   Imposing initial conditions  
   \eqref{ini}, we obtain by integration 
 \begin{align*}
     & D(t)= \frac{(4-3 \gamma ) {D_0}}{3 \gamma  ({D_0} t-1)+4}, \\
     & H(t)= 2 {D_0} \left(\frac{1}{3 \gamma -4}+\frac{1}{-3 \gamma +3 \gamma  {D_0} t+4}\right)+H_0,\\
     & \sigma_{+}(t)=\frac{3 \gamma  (3 \gamma -2) {D_0}^2 t}{(3 \gamma -4)
   (-3 \gamma +3 \gamma  {D_0} t+4)}+\sigma_0,\\
   & e_{1}^1(t)^{-2}=\frac{\left(\frac{-3 \gamma +3 \gamma  {D_0} t+4}{4-3 \gamma }\right)^{\frac{4}{\gamma }-4}}{c_1^2},\\
   & K(t)^{-1}= \frac{c_1 (3 \gamma  ({D_0} t-1)+4)^2}{(4-3 \gamma
   )^2}. 
 \end{align*}
  
The line element \eqref{metric} becomes
\begin{align}
   &  ds^2= - dt^2 + \frac{\left(\frac{-3 \gamma +3 \gamma  {D_0} t+4}{4-3 \gamma }\right)^{\frac{4}{\gamma }-4}}{c_1^2} dr^2 \nonumber \\
   &  + \frac{c_1 (3 \gamma  ({D_0} t-1)+4)^2}{(4-3 \gamma
   )^2} \left[ d \vartheta^2 + \sin^2(\vartheta) d \zeta\right].
\end{align}  

  For $\gamma=2$ we have  
  \begin{align*}
   & \dot{D}=- 3 D^2, \; \dot{H}= -3 D^2, \;  \dot{\sigma_{+}}= -6
   D^2, \nonumber \\
   & \dot{e_1^1}= 3 D e_1^1,\;  \dot{K}=-6 D K. 
  \end{align*}
      Imposing initial conditions  
   \eqref{ini}, we obtain by integration 
 \begin{align*}
     & D(t)=\frac{{D_0}}{3 {D_0} t+1}  , \\
     & H(t)= {D_0} \left(\frac{1}{3 {D_0} t+1}-1\right)+H_0,\\
     & \sigma_{+}(t)=\sigma_0-\frac{6 {D_0}^2 t}{3 {D_0} t+1}  ,\\
   & e_{1}^1(t)^{-2}=\frac{1}{(3 c_1 {D_0} t+c_1)^2} ,\\
   & K(t)^{-1}= c_1 (3 {D_0} t+1)^2.
 \end{align*}
   \end{enumerate}
      To study the dynamics at the invariant  boundary $\bar{\Sigma}^2+ \bar{\Omega}^2=1$ which corresponds to vacuum solutions, we introduce cylindrical coordinates 
\begin{small}
\begin{equation}
    \bar{\Sigma}= \cos \theta, \; \bar{\Omega}= \sin \theta,  \; \theta \in[0,\pi], \; \bar{Q}\in[-1,1]. 
\end{equation}
   \end{small}
The dynamics on the invariant surface is given by 
\begin{small}
\begin{subequations}
\label{dspolar}
\begin{align}
  & \theta'= \frac{1}{2} \sin (\theta ) \left(3 \bar{Q} \cos (\theta )-2 \bar{Q}^2+2\right),\\
  & \bar{Q}'= \frac{1}{4} \left(\bar{Q}^2-1\right) \left(-4 \bar{Q} \cos (\theta )+3 \cos (2 \theta )+5\right).
\end{align}
\end{subequations}
\end{small}

\begin{table}[t!]
\caption{\label{Tabtheta-Q} Equilibrium points of system \eqref{dspolar}. The eigenvalues are obtained evaluating the linearization matrix of \eqref{dspolar} at each point.}
\footnotesize\setlength{\tabcolsep}{9pt}
    \begin{tabular}{lccccc}\hline
Label  & \multicolumn{1}{c}{$\theta$} & \multicolumn{1}{c}{$\bar{Q}$} & \multicolumn{1}{c}{Eigenvalues} &  \multicolumn{1}{c}{Behavior}   \\\hline
        $P_1$ & $\pi$ & $-1$ & $\left\{-2,-\frac{3}{2}\right\}$ & sink  \\
        $P_2$ & $0$ & $1$ & $\left\{2,\frac{3}{2}\right\}$ &  source \\
        $P_3$ & $0$ & $-1$ & $\left\{-6,-\frac{3}{2}\right\}$ & sink \\ 
        $P_4$ &$\pi$ & $1$ & $\left\{6,\frac{3}{2}\right\}$ & source \\ 
        $P_7$ & $\frac{\pi}{2}$ & $-1$ & $\left\{\frac{3}{2},-1\right\}$ & saddle \\
        $P_8$ & $\frac{\pi}{2}$ & $1$ & $\left\{-\frac{3}{2},1\right\}$ & saddle \\
        \hline
    \end{tabular}
\end{table}
\noindent
In Table \ref{Tabtheta-Q}, the equilibrium points of system \eqref{dspolar} are presented. The eigenvalues are obtained evaluating the linearization matrix of \eqref{dspolar} at each point.

\begin{figure}[t]
    \centering
    \includegraphics[width=0.45\textwidth]{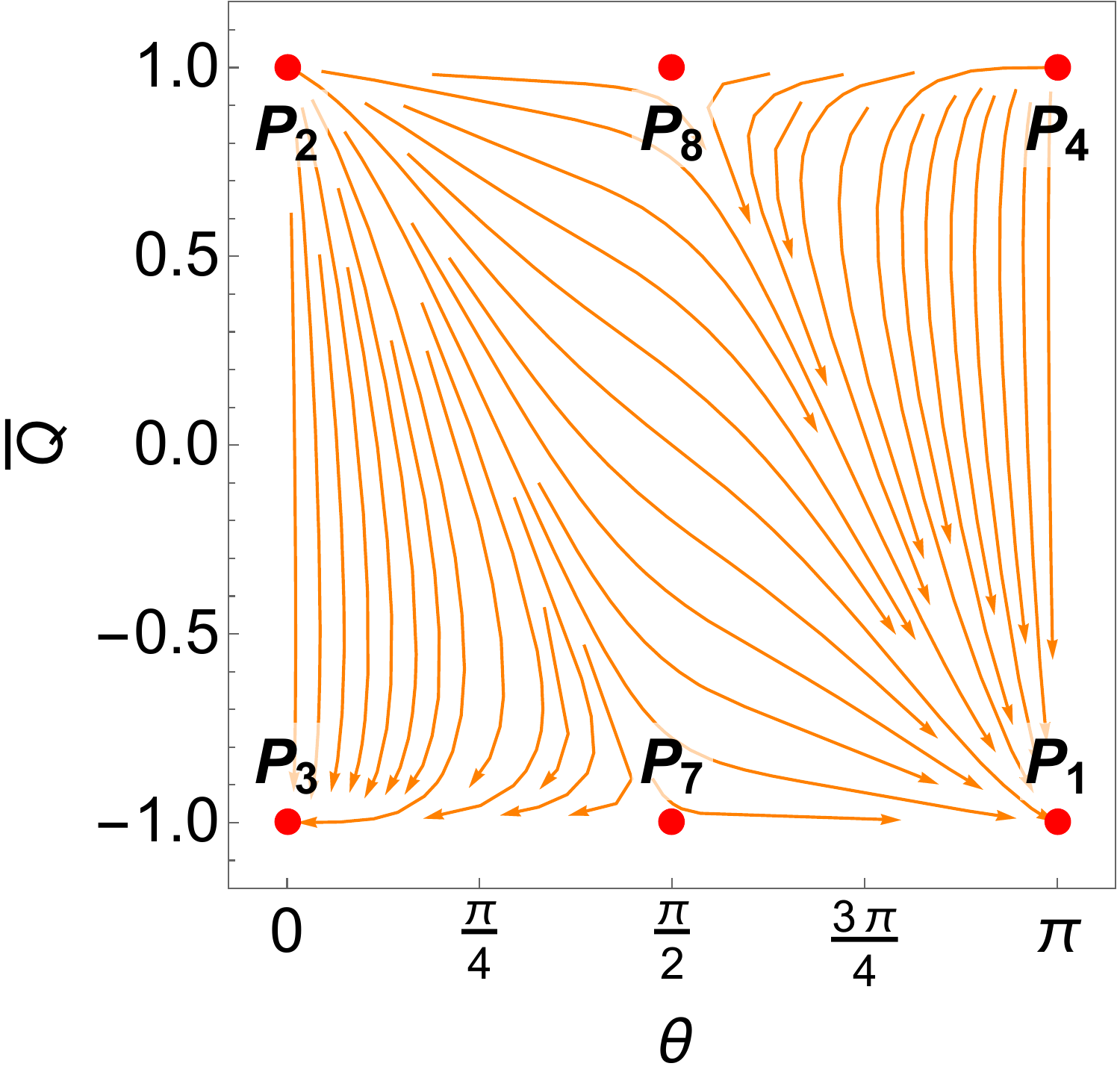}
    \caption{Orbits in the invariant set $(\theta , \bar{Q})$, with dynamics given by \eqref{dspolar}.}
    \label{KSphaseplotpolar}
\end{figure}

\noindent
In Figure \ref{KSphaseplotpolar} orbits in the invariant set $(\theta , \bar{Q})$  with dynamics given by \eqref{dspolar} are presented. There  it is illustrated that $P_7$ and $P_8$ are saddle points. 
\noindent
   \begin{figure*}[t]
    \centering
    \subfigure[\label{KSphaseplot3DCC} Orbits in the  phase space $(\bar{\Sigma}, \bar{Q}, \bar{\Omega})$. ]{\includegraphics[width=0.55\textwidth]{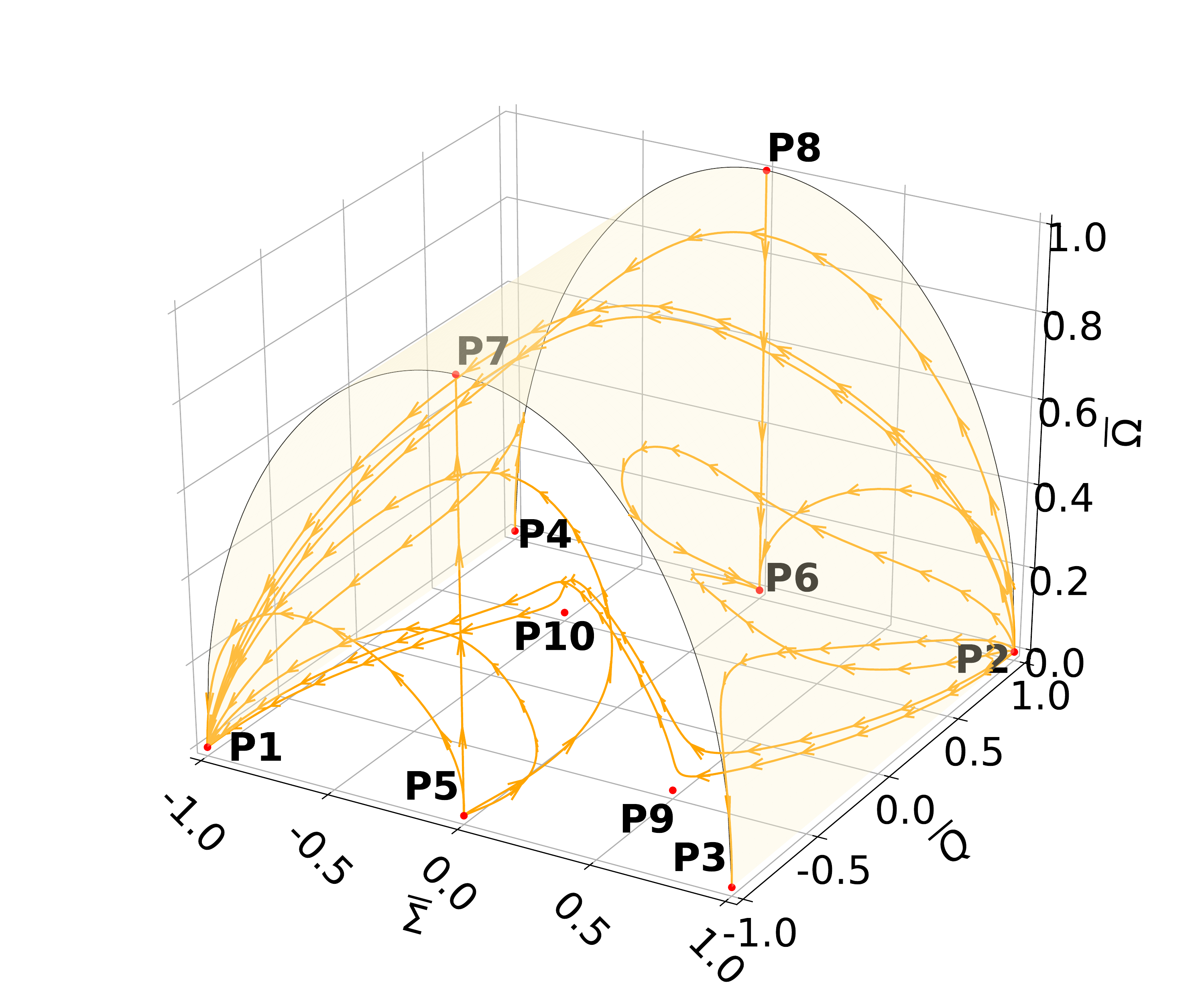}}
      \subfigure[\label{KSphaseplot3DCC1} Orbits in the invariant set $\bar{\Omega}=0$.]{\includegraphics[width=0.4\textwidth]{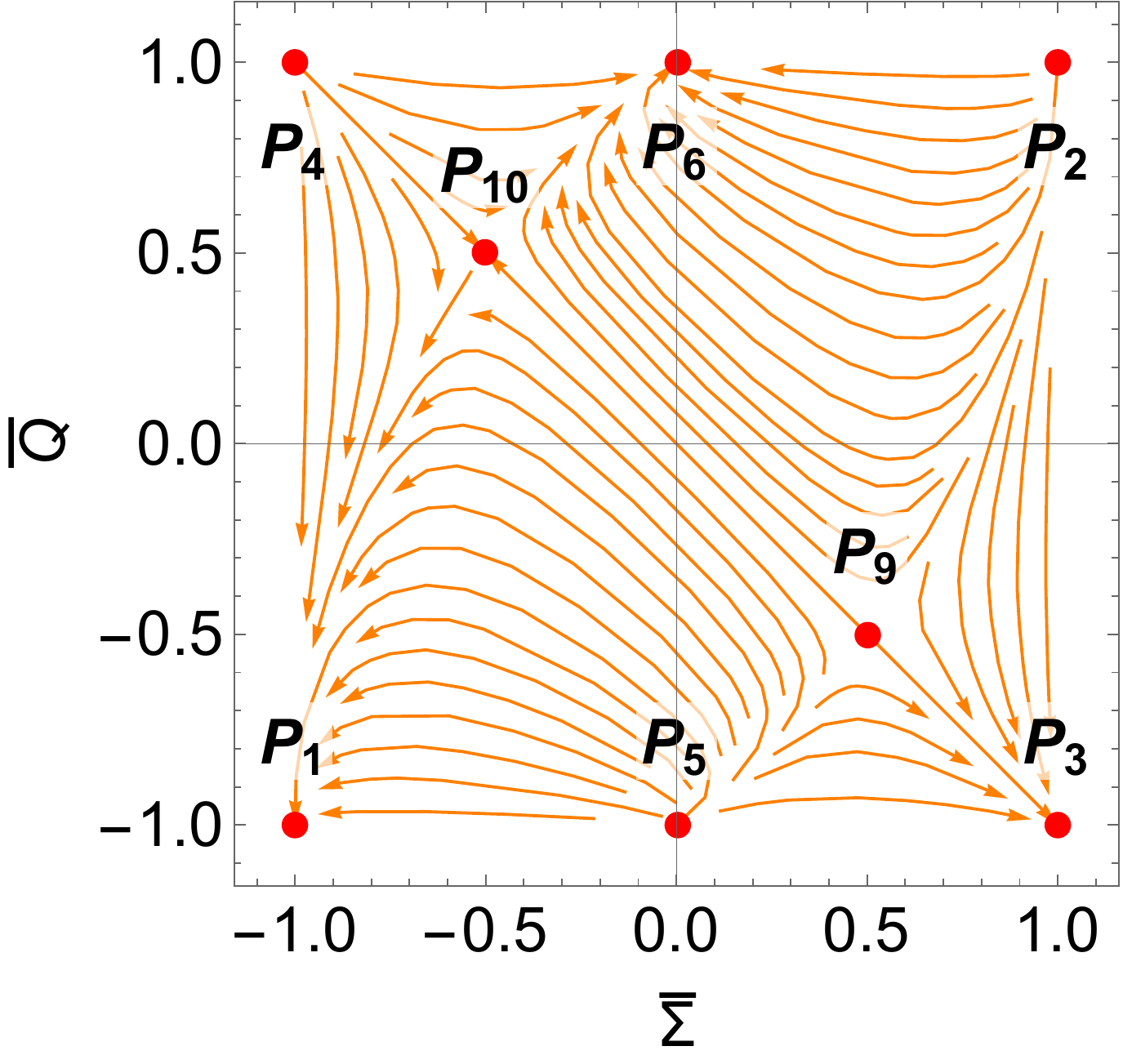}}
       \caption{Phase space of the guiding system  \eqref{KSguidingC1}, \eqref{KSguidingC2}, \eqref{KSguidingC3} corresponding to KS for  $\gamma=0$.}
    \label{Kantowski-Sachsphaseplot3DCC}
\end{figure*}
   
 \begin{figure*}[t]
    \centering
     \subfigure[\label{KSphaseplot3DBif} Orbits in the  phase space $(\bar{\Sigma}, \bar{Q}, \bar{\Omega})$.]{\includegraphics[width=0.55\textwidth]{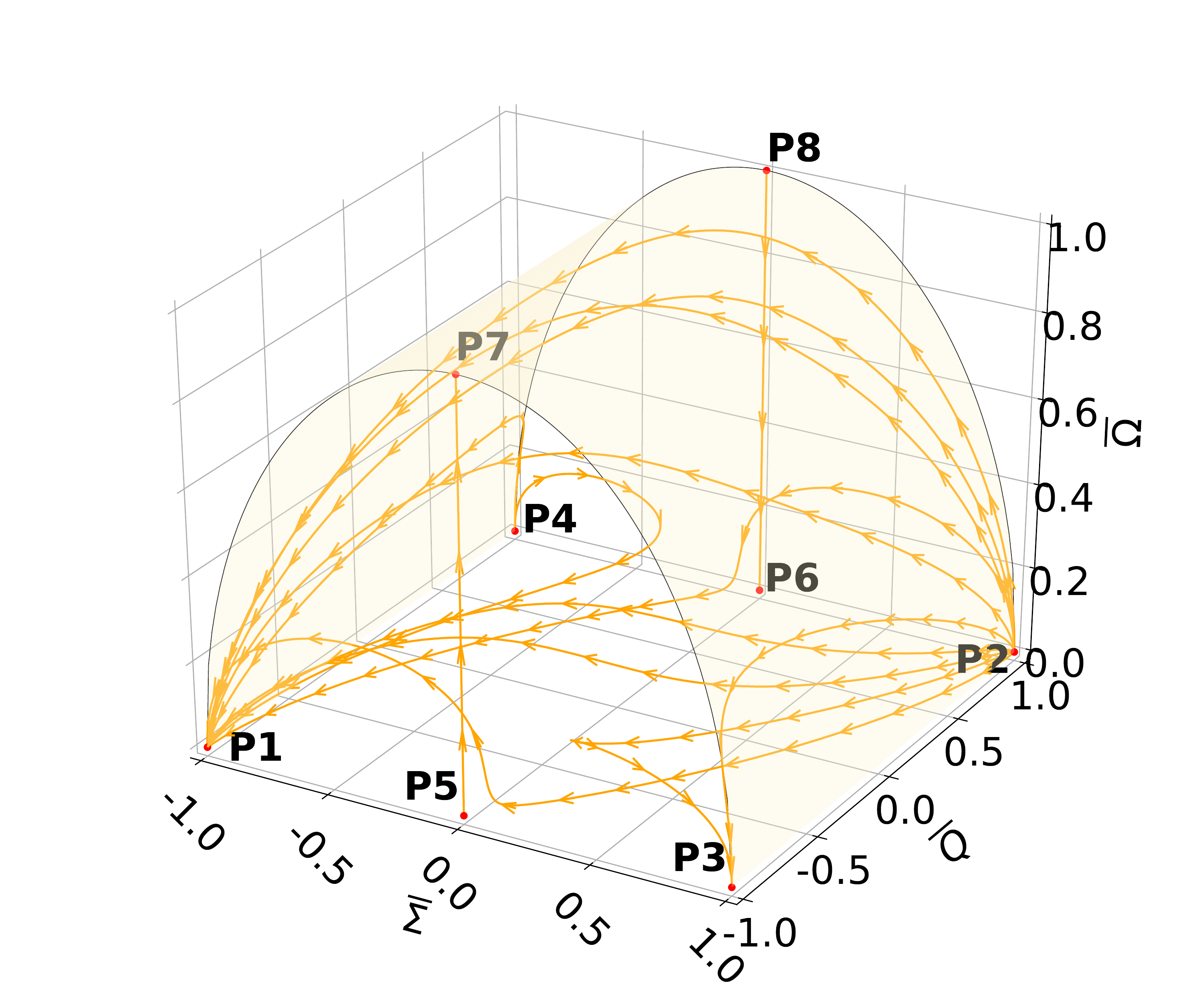}}
      \subfigure[\label{KSphaseplot3DBif1} Orbits in the invariant set $\bar{\Omega}=0$.]{\includegraphics[width=0.4\textwidth]{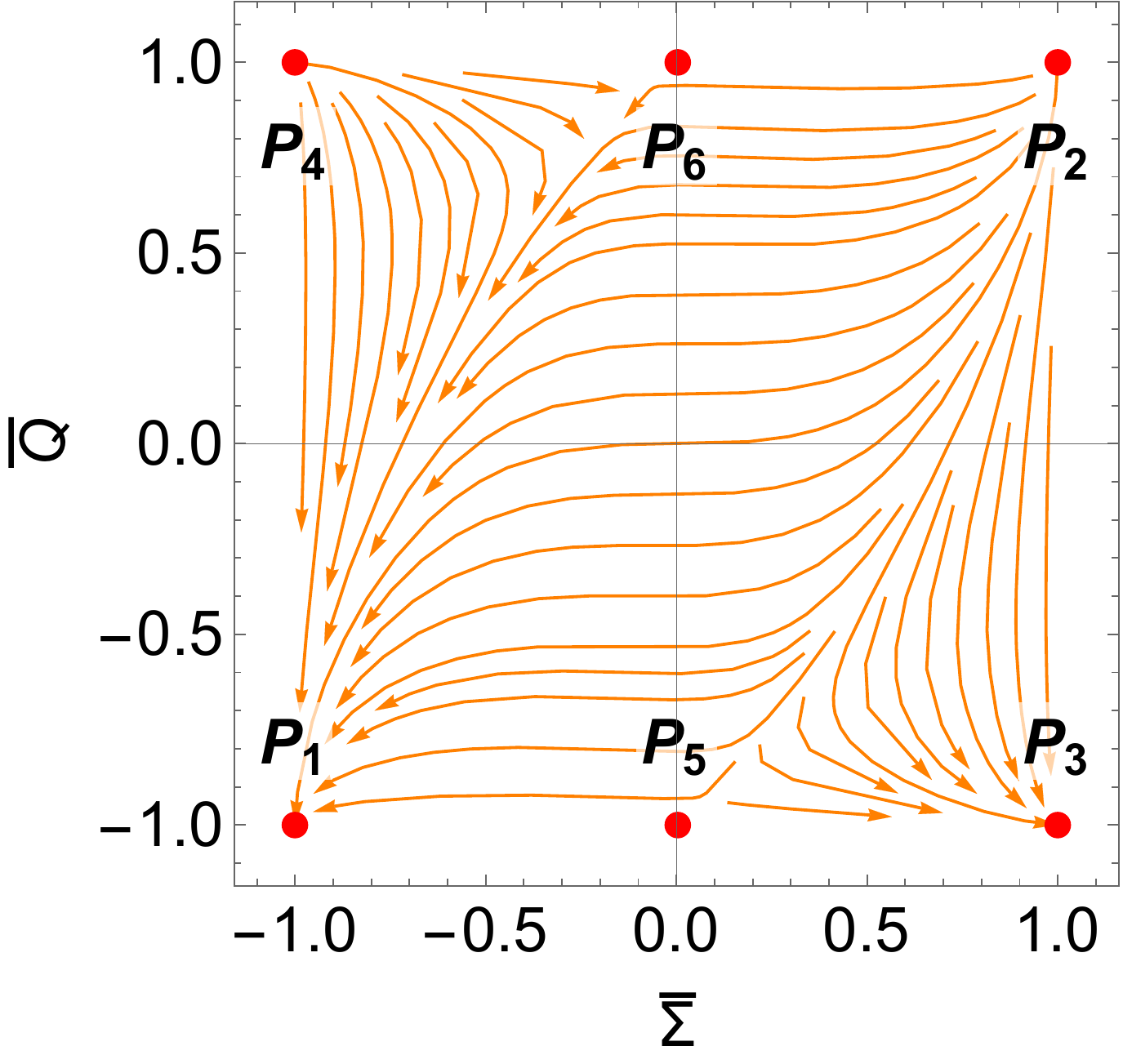}}
     \caption{Phase space of the guiding system  \eqref{KSguidingC1}, \eqref{KSguidingC2}, \eqref{KSguidingC3} corresponding to KS for  $\gamma=\frac{2}{3}$.}
    \label{Kantowski-Sachsphaseplot3DBiff}
\end{figure*}

\begin{figure*}[t]
    \centering
    \subfigure[\label{KSphaseplot3DDust} Orbits in the phase space $(\bar{\Sigma}, \bar{Q}, \bar{\Omega})$.]{\includegraphics[width=0.55\textwidth]{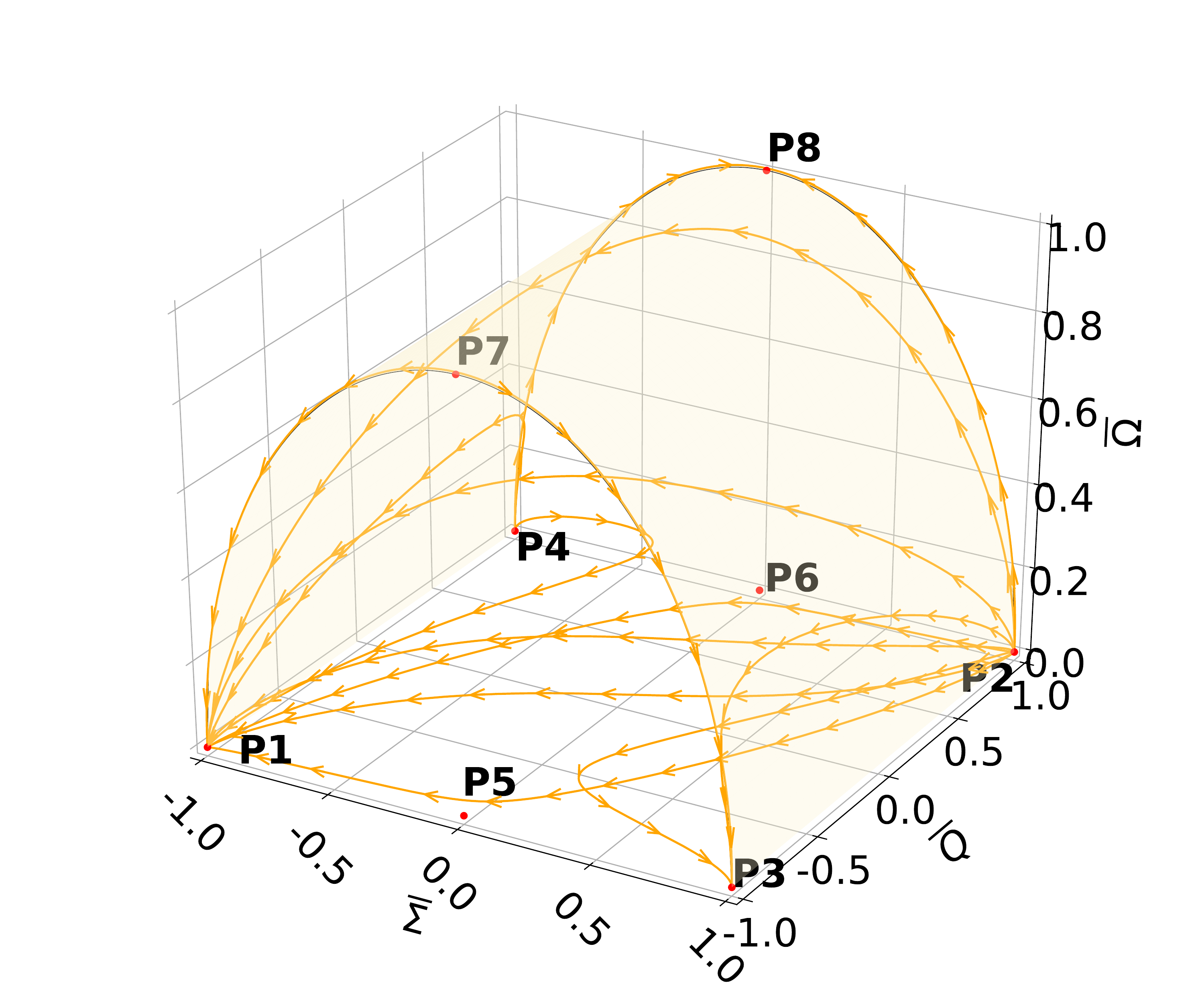}}
   \subfigure[\label{KSphaseplot3DDust1} Orbits in the invariant set $\bar{\Omega}=0$.]{\includegraphics[width=0.4\textwidth]{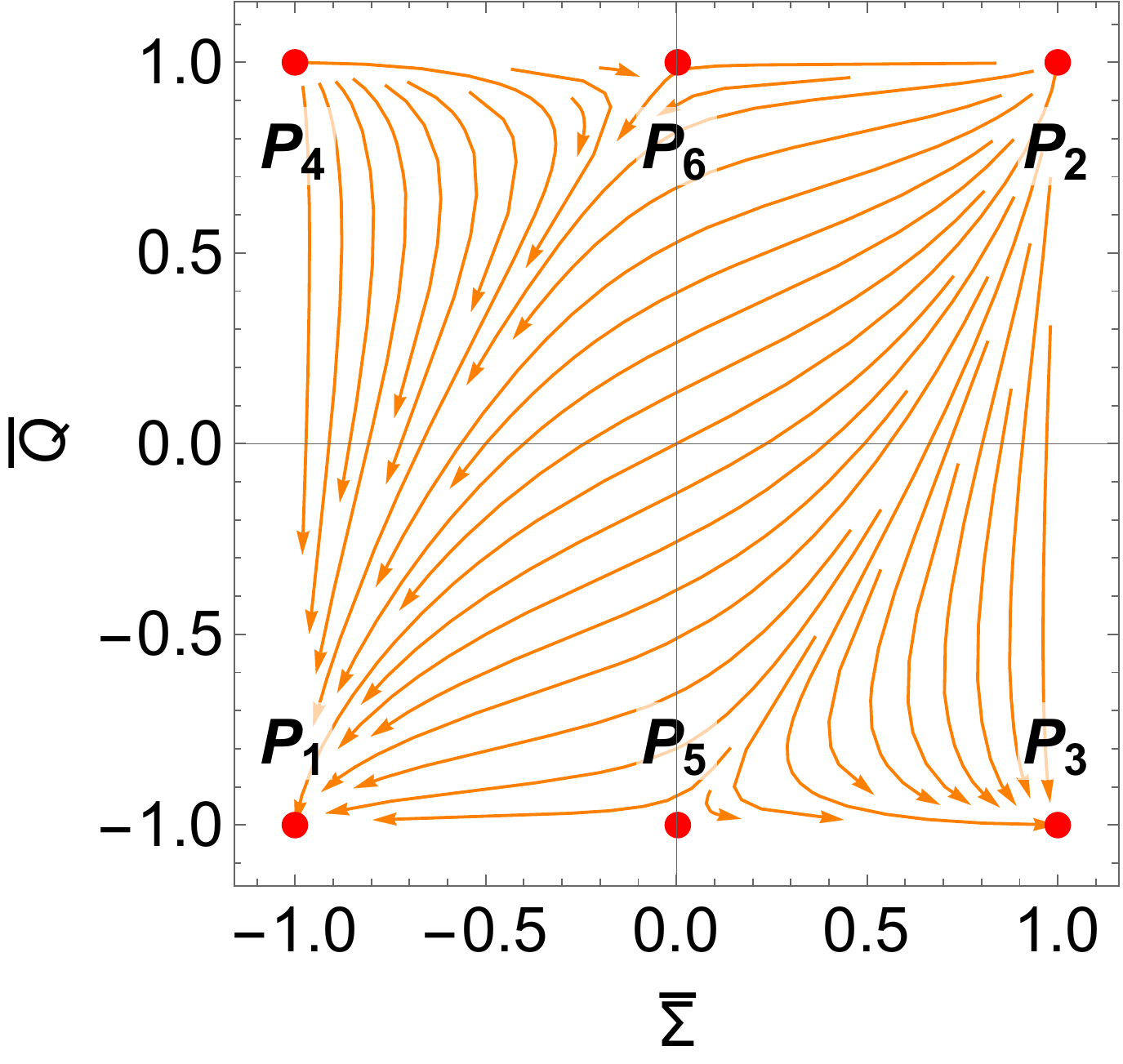}}
        \caption{Phase space of the guiding system  \eqref{KSguidingC1}, \eqref{KSguidingC2}, \eqref{KSguidingC3} corresponding to KS for $\gamma=1$.  \label{Kantowski-Sachsphaseplot3DDust}}
   \end{figure*}
   
   \begin{figure*}[t]
    \centering
        \subfigure[\label{KSphaseplot3DRad} Orbits in the phase space $(\bar{\Sigma}, \bar{Q}, \bar{\Omega})$.]{\includegraphics[width=0.55\textwidth]{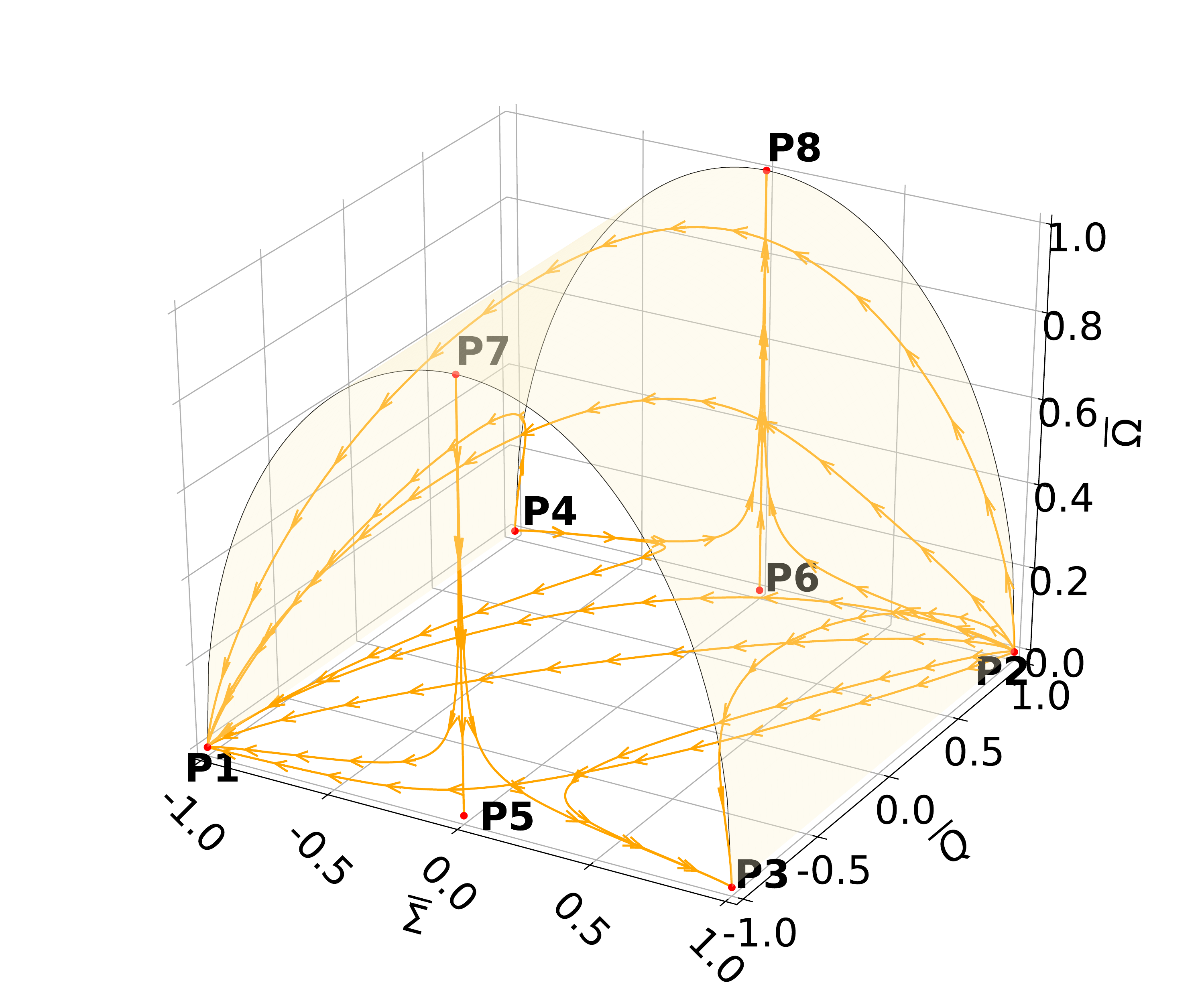}}
    \subfigure[\label{KSphaseplot3DRad1} Orbits in the invariant set $\bar{\Omega}=0$.]{  
    \includegraphics[width=0.4\textwidth]{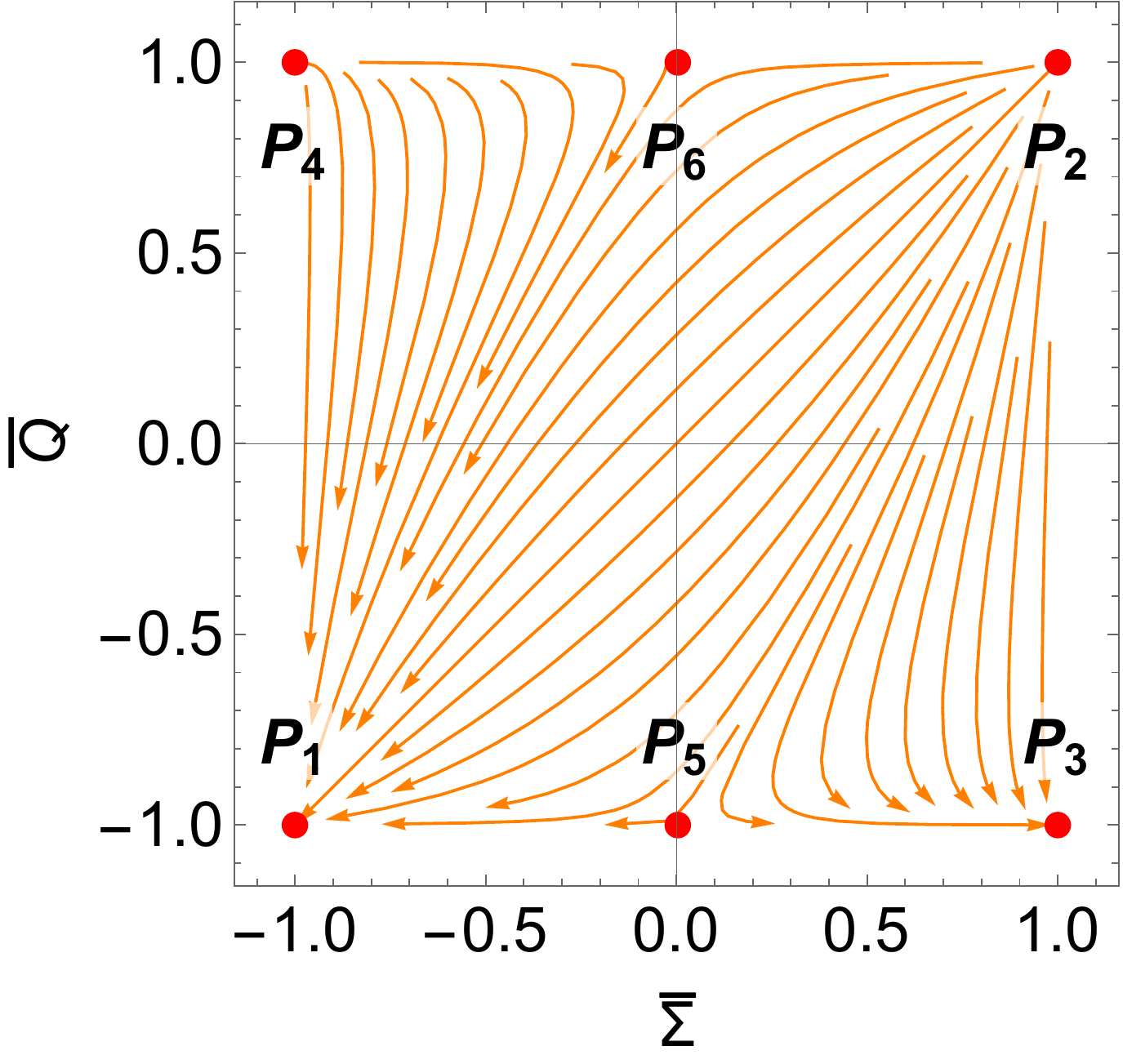}}
           \caption{Phase space of the guiding system  \eqref{KSguidingC1}, \eqref{KSguidingC2}, \eqref{KSguidingC3} corresponding to KS for $\gamma=\frac{4}{3}$. \label{Kantowski-Sachsphaseplot3DRad}}
   \end{figure*}
   
    \begin{figure*}[t]
    \centering
     \subfigure[\label{KSphaseplot3DStiff}  Orbits in the phase space $(\bar{\Sigma}, \bar{Q}, \bar{\Omega})$. ]{\includegraphics[width=0.55\textwidth]{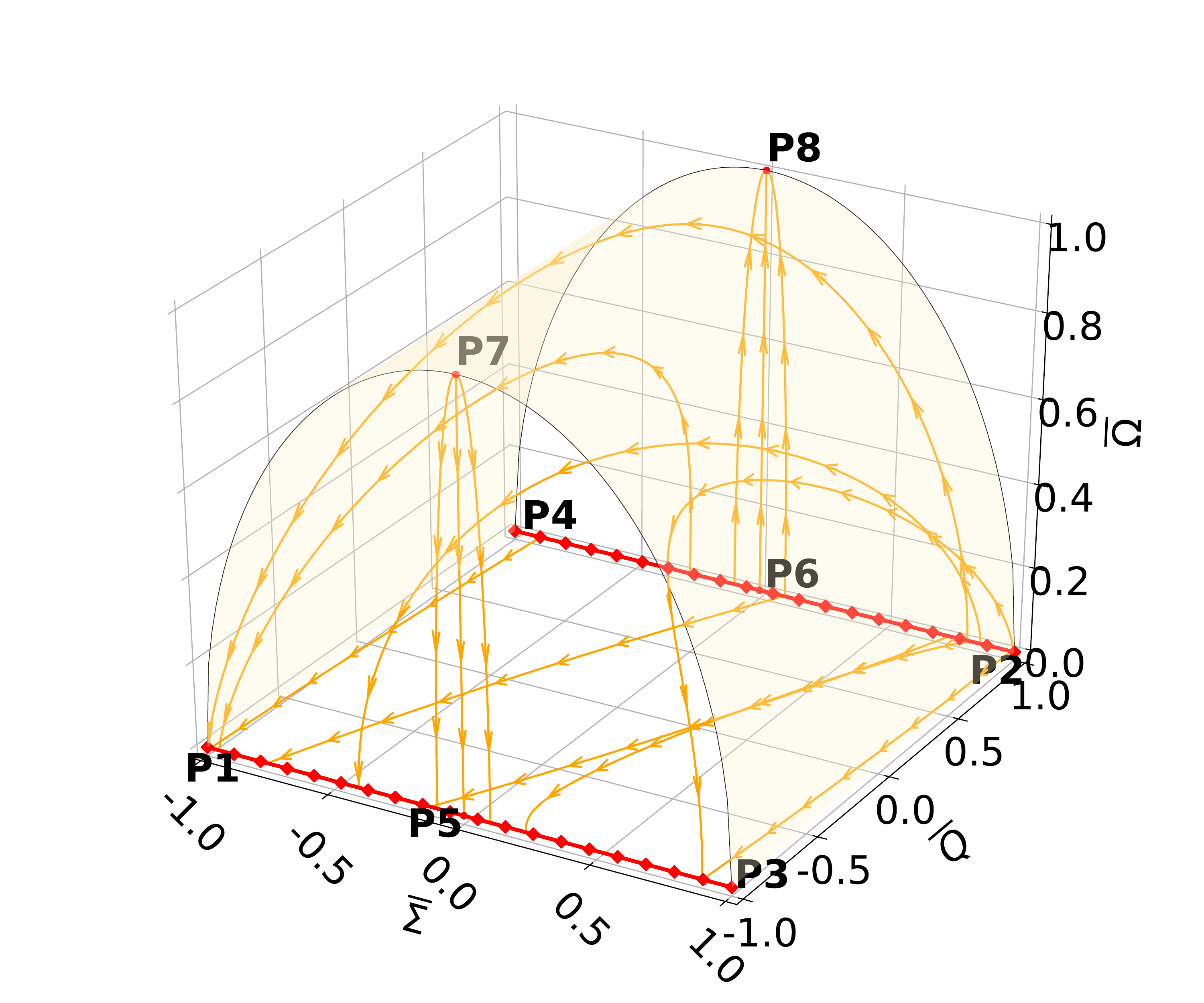}}
     \subfigure[\label{KSphaseplot3DStiff1}  Orbits in the invariant set $\bar{\Omega}=0$.]{\includegraphics[width=0.4\textwidth]{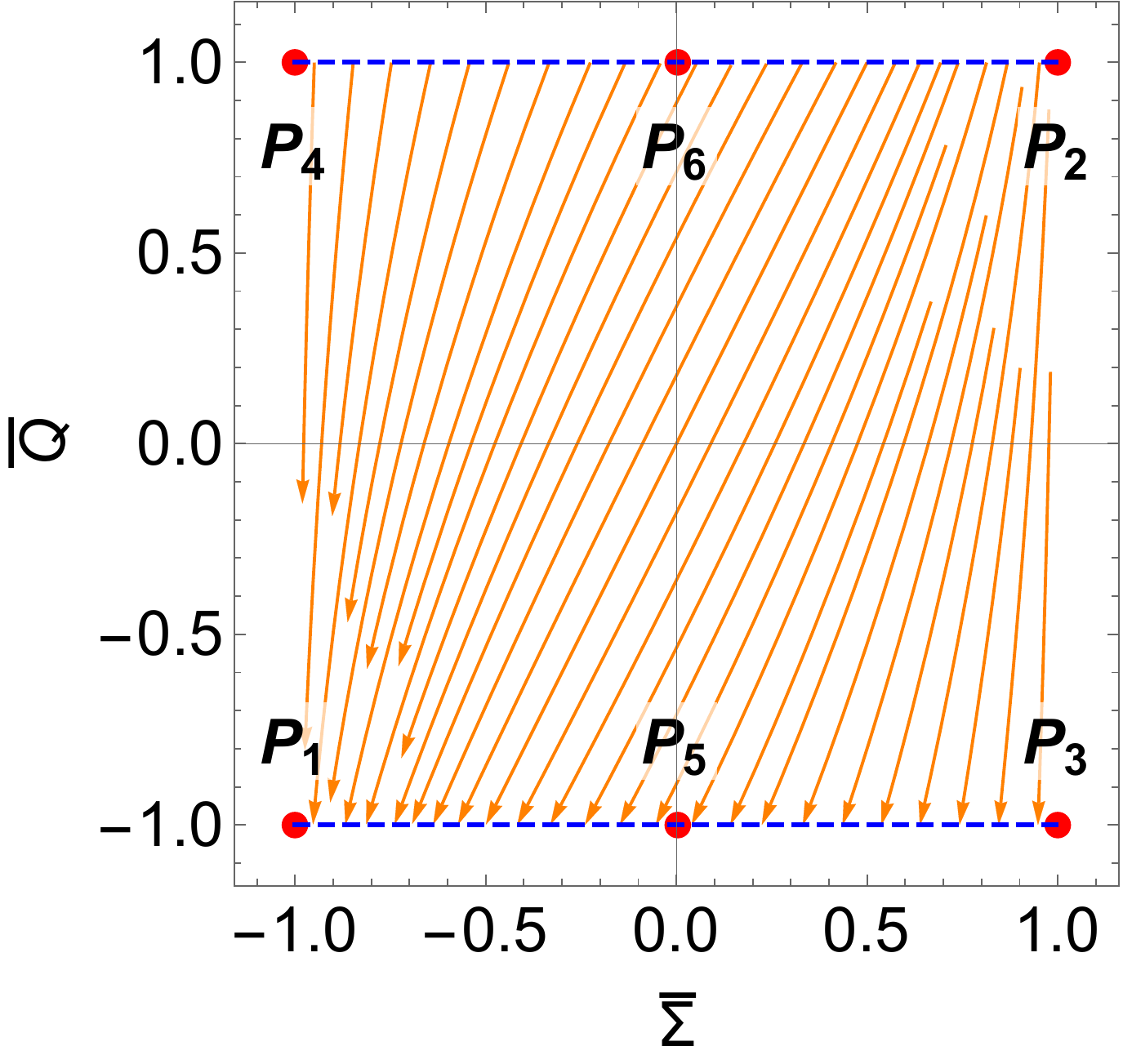}}
     \caption{\label{KSphaseplotStiff} Phase space of the guiding system  \eqref{KSguidingC1}, \eqref{KSguidingC2}, \eqref{KSguidingC3} corresponding to KS for barotropic index $\gamma=2$.}
   \end{figure*}
\noindent
In Figure \ref{Kantowski-Sachsphaseplot3DCC}  some orbits in the phase space of the guiding system \eqref{KSguidingC1}, \eqref{KSguidingC2}, \eqref{KSguidingC3} for $\gamma=0$ corresponding to the CC  are presented. In Figure \ref{KSphaseplot3DCC} orbits in the  phase space $(\bar{\Sigma}, \bar{Q}, \bar{\Omega})$  are displayed. In Figure \ref{KSphaseplot3DCC1} orbits in the invariant set $\bar{\Omega}=0$ are shown. Points $P_2$, $P_4$, and $P_5$ are early-time attractors. Points $P_1, P_3$, and $P_6$ are late--time attractors. Points $P_7$, $P_8$, $P_9$, and $P_{10}$ are saddle points. 

\noindent In Figure \ref{Kantowski-Sachsphaseplot3DBiff}  some orbits of the phase space of the guiding system  \eqref{KSguidingC3} for  $\gamma=\frac{2}{3}$ are presented.  In Figure \ref{KSphaseplot3DBif} some orbits in the  phase space $(\bar{\Sigma}, \bar{Q}, \bar{\Omega})$ are displayed.
 In Figure \ref{KSphaseplot3DBif1} some orbits in the invariant set $\bar{\Omega}=0$ are shown.
For the value of $\gamma=\frac{2}{3}$,  $P_5$ coincides with $P_9$  and $P_6$ coincides with $P_{10}$. They are nonhyperbolic.  Points $P_2$ and $P_4$ are early-time attractors. Points  $P_1$ and $P_3$ are late--time attractors. Points  $P_5$, $P_6$, $P_7$, and $P_8$ are saddle points. 

\noindent In Figure \ref{Kantowski-Sachsphaseplot3DDust} some orbits in the phase space of the guiding system  \eqref{KSguidingC1}, \eqref{KSguidingC2}, \eqref{KSguidingC3} for $\gamma=1$ which corresponds to dust are displayed.  In Figure \ref{KSphaseplot3DDust} some orbits in the phase space $(\bar{\Sigma}, \bar{Q}, \bar{\Omega})$ are presented. In Figure \ref{KSphaseplot3DDust1} some orbits in the invariant set $\bar{\Omega}=0$ are presented. Points $P_{2}$ and $P_{4}$ are early-time attractors. Points $P_{1}$ and $P_{3}$ are late--time attractors. Points $P_{5}$, $P_{6}$, $P_{7}$, and $P_{8}$ are saddle points.

\noindent
In Figure \ref{Kantowski-Sachsphaseplot3DRad} some orbits in the phase space of  guiding system  \eqref{KSguidingC1}, \eqref{KSguidingC2}, \eqref{KSguidingC3} for $\gamma=\frac{4}{3}$ corresponding to radiation are presented. In Figure \ref{KSphaseplot3DRad} some orbits in the phase space $(\bar{\Sigma}, \bar{Q}, \bar{\Omega})$ are displayed.
In Figure \ref{KSphaseplot3DRad1} some orbits in the invariant set $\bar{\Omega}=0$ are shown. In both diagrams \ref{Kantowski-Sachsphaseplot3DRad} and
\ref{KSphaseplotStiff} points $P_2$ and $P_4$ are early-time attractors.  Points $P_1$ and $P_3$ are late--time attractors. Points $P_5, P_6, P_7$,  and $P_8$ are saddle points. Points $P_9$ and $P_{10}$ do not exist.

\noindent In Figure \ref{KSphaseplotStiff} some orbits in the phase space of  guiding system  \eqref{KSguidingC1}, \eqref{KSguidingC2}, \eqref{KSguidingC3} for $\gamma=2$ which corresponds to stiff matter are presented. In Figure \ref{KSphaseplot3DStiff}  some orbits in the phase space $(\bar{\Sigma}, \bar{Q}, \bar{\Omega})$ are displayed. 
In Figure \ref{KSphaseplot3DStiff1}  some orbits in the invariant set $\bar{\Omega}=0$ are shown. 
The line connecting points $P_2, P_4$, and $P_6$ (denoted by a dashed blue line) is invariant, and it is the early-time attractor. The line connecting points $P_1, P_3$, and $P_5$ (denoted by a dashed blue line) is invariant  and it is the late--time attractor. Points $P_7$ and $P_8$ are saddle points. Points $P_9$ and $P_{10}$ do not exist.  
\subsubsection{late--time behavior in the reduce phase space}
Now, we study the dynamics in the reduced phase space $\mathbf{x}=(\Omega, \Sigma, Q)$ where  the effect of $D$ in the dynamics was neglected. The results from the linear stability analysis combined with Theorem \ref{KSLFZ11} lead to the following local results. 
\begin{thm}
\label{thm11}
Late--time  attractors of full system \eqref{unperturbed1KS} and time--averaged system  \eqref{avrgsystKS} as $D\rightarrow 0$ for Kantowski-Sachs line element are 
\begin{enumerate}
 \item[(i)]   The anisotropic solution $P_1$  with $\bar{\Omega}_m=0$ and line element \eqref{metricKS-P1}
   if  $0\leq \gamma < 2$. 
This point represents a non--flat LRS Kasner ($p_1=-\frac{1}{3}, p_2= p_3= \frac{2}{3}$)  contracting solution with $H<0$ (\cite{WE} Sect. 6.2.2 and Sect. 9.1.1 (2)). This solution is singular at finite time  $t_0=\frac{1}{3D_0}$ and is valid  for $t>t_0$. 

 \item[(ii)]   The anisotropic solution $P_3$  with $\bar{\Omega}_m=0$ and line element \eqref{KS-metric-P3}
   if  $0\leq \gamma < 2$, representing a contracting solution with $H<0$.
This point represents a Taub (flat LRS Kasner) contracting solution ($p_1=1, p_2= 0, p_3= 0$)
\cite{WE} (Sect 6.2.2 and p 193, eq.    (9.6)).

    \item[(iii)]  The flat matter--dominated FLRW universe $P_6$    with metric \eqref{KS-metric-P6}
  if  $0< \gamma < \frac{2}{3}$. $P_6$ represents a  quintessence fluid or a zero-acceleration (Dirac-Milne) model for $\gamma=\frac{2}{3}$. 
In the limit $\gamma=0$ we have \newline$\ell(t) \propto \left(\frac{3 \gamma  D_{0} t}{2}+1\right)^{\frac{2}{3
   \gamma }}\rightarrow  e^{D_0 t}$, where $D_0=H_0$, i.e., the de Sitter solution. 
 \end{enumerate}  
\end{thm}
 For global results when $D\in[0, \infty)$ see section \ref{SECT:5}. 
\subsection{FLRW metric with positive curvature}
\label{FLRWclosed}
In FLRW metric with positive curvature time--averaged system  \eqref{eq72}, \eqref{eq73}, and \eqref{eq74} transforms to 
\begin{small}
\begin{subequations}
\label{avrgsystFLRWClosed}
\begin{align}
   &\frac{d\bar{\Omega}}{d \eta}=\frac{3}{2} (\gamma -1) \bar{Q} \bar{\Omega}  \left(1-{\bar{\Omega}}
   ^2\right), \label{eq76}\\
   &\frac{d{\bar{Q}}}{d \eta}=-\frac{1}{2} \left(1-{\bar{Q}}^2\right) \left(3 \gamma 
   \left(1-{\bar{\Omega}} ^2\right)+3 {\bar{\Omega}}^2-2\right),  \label{eq77}
\\
& \frac{d{\bar{\Phi}}}{d\eta}=0,   \\
& \frac{d{D}}{d\eta}=  -\frac{3}{2} \bar{Q} \left(\gamma (1- \bar{\Omega}^2) + \bar{\Omega}^2 \right) D.
\end{align}
\end{subequations}
\end{small}
where $D$  is defined by eq.   \eqref{varsclosedFLRW}.

The equation for $\bar{\Omega}$ is simplified by setting $\bar{\Sigma}=0$ in eq. \eqref{eqOmega69}  and we define $\bar{\Omega}_k$  as $\bar{Q}^2 \bar{\Omega}_k:= 1-\bar{Q}^2$,
where $\bar{\Omega}_k$ is interpreted as the time--averaged values of $\Omega_k:= \frac{1}{a^2 H^2}$. 
Then, the phase space is 
\begin{align}
    \left\{(\bar{\Omega},  \bar{Q})\in \mathbb{R}^2: -1\leq \bar{Q} \leq 1, 0\leq  \bar{\Omega} \leq 1\right\}.
\end{align}
In some special cases we relax the condition $\bar{\Omega}\leq 1$ and  consider $\bar{\Omega}>1$. 
\begin{figure*}[t!]
    \centering
    \subfigure[]{\includegraphics[width=0.4\textwidth]{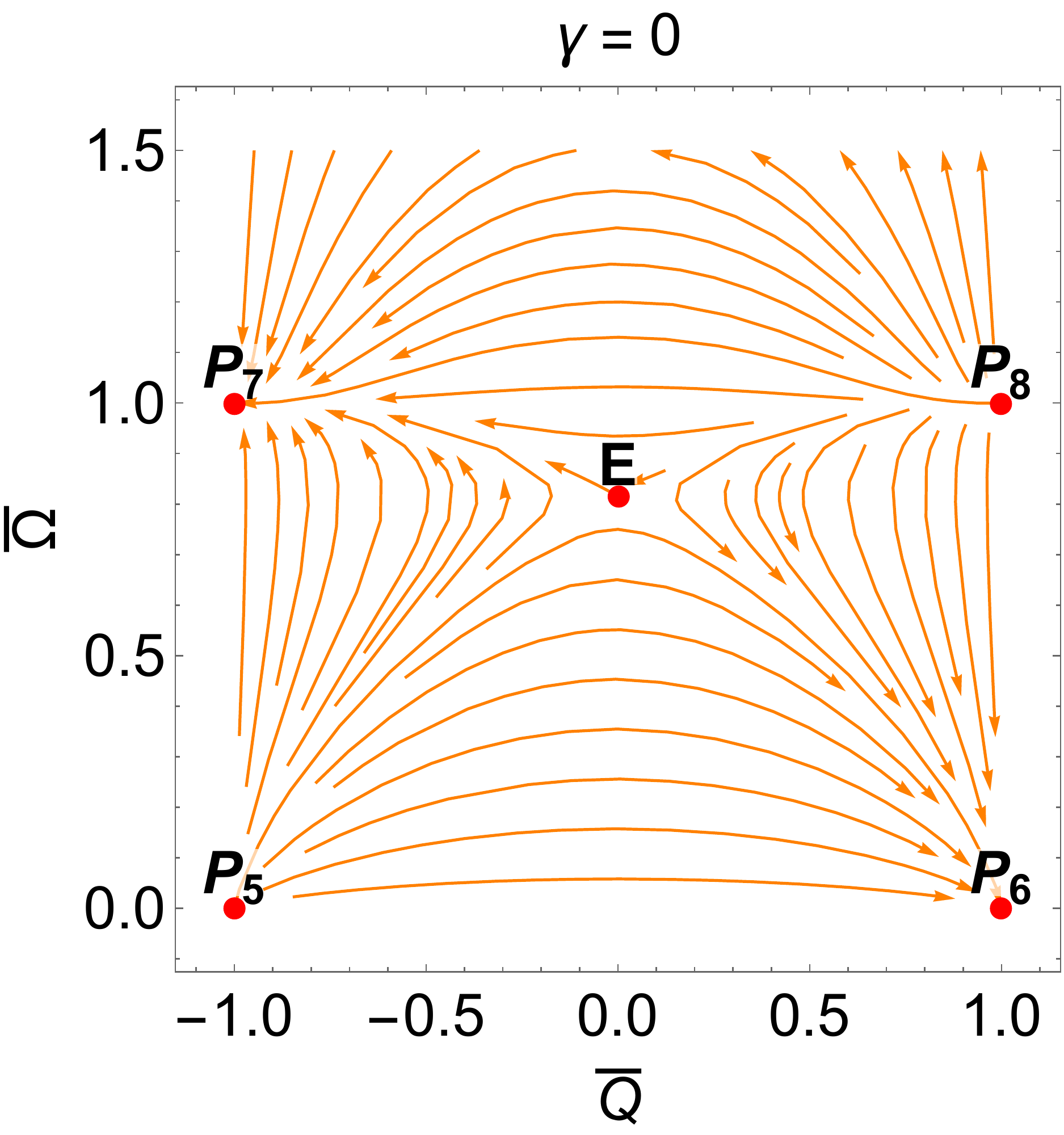}}
    \subfigure[]{\includegraphics[width=0.4\textwidth]{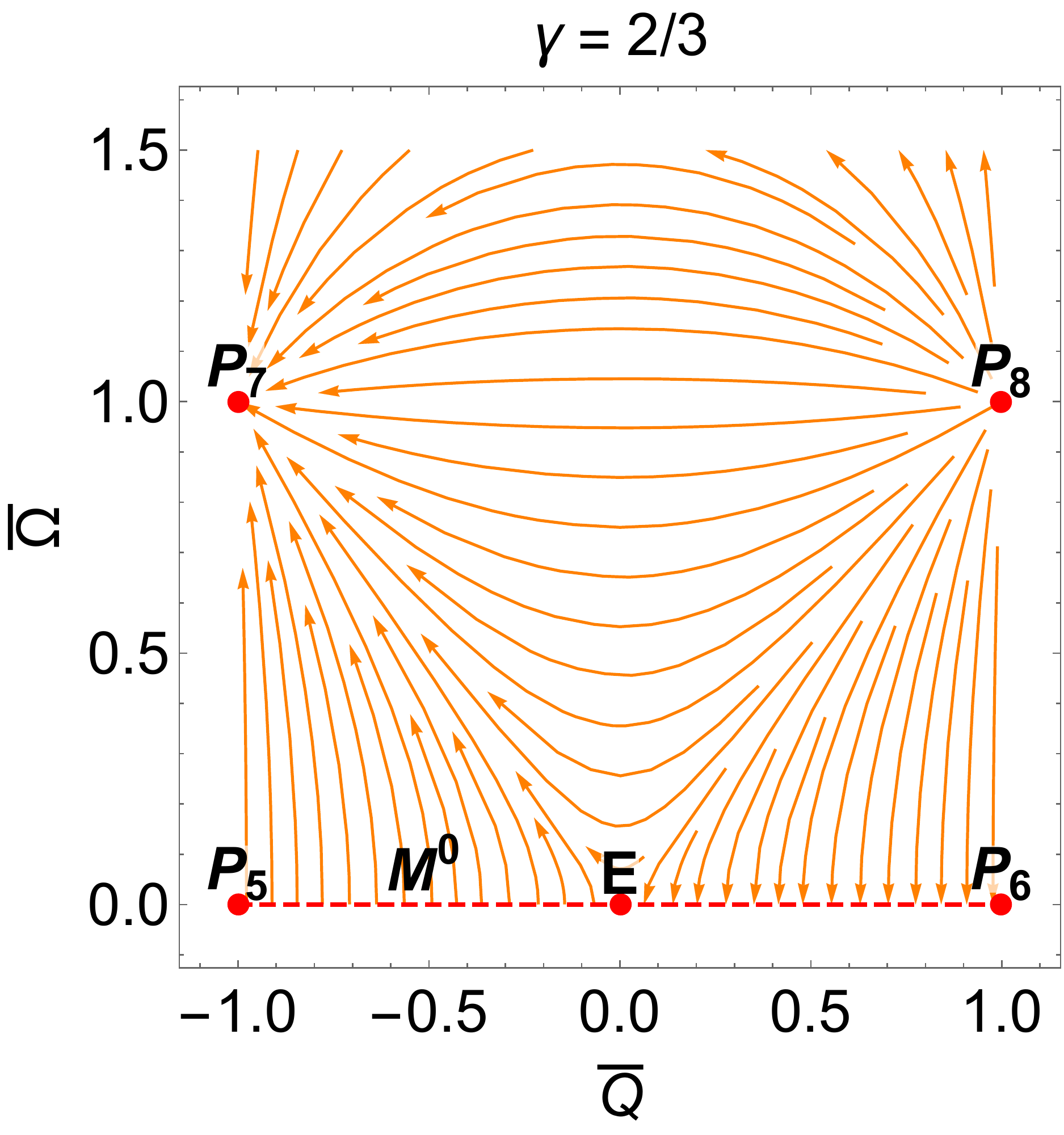}} 
    \subfigure[]{\includegraphics[width=0.4\textwidth]{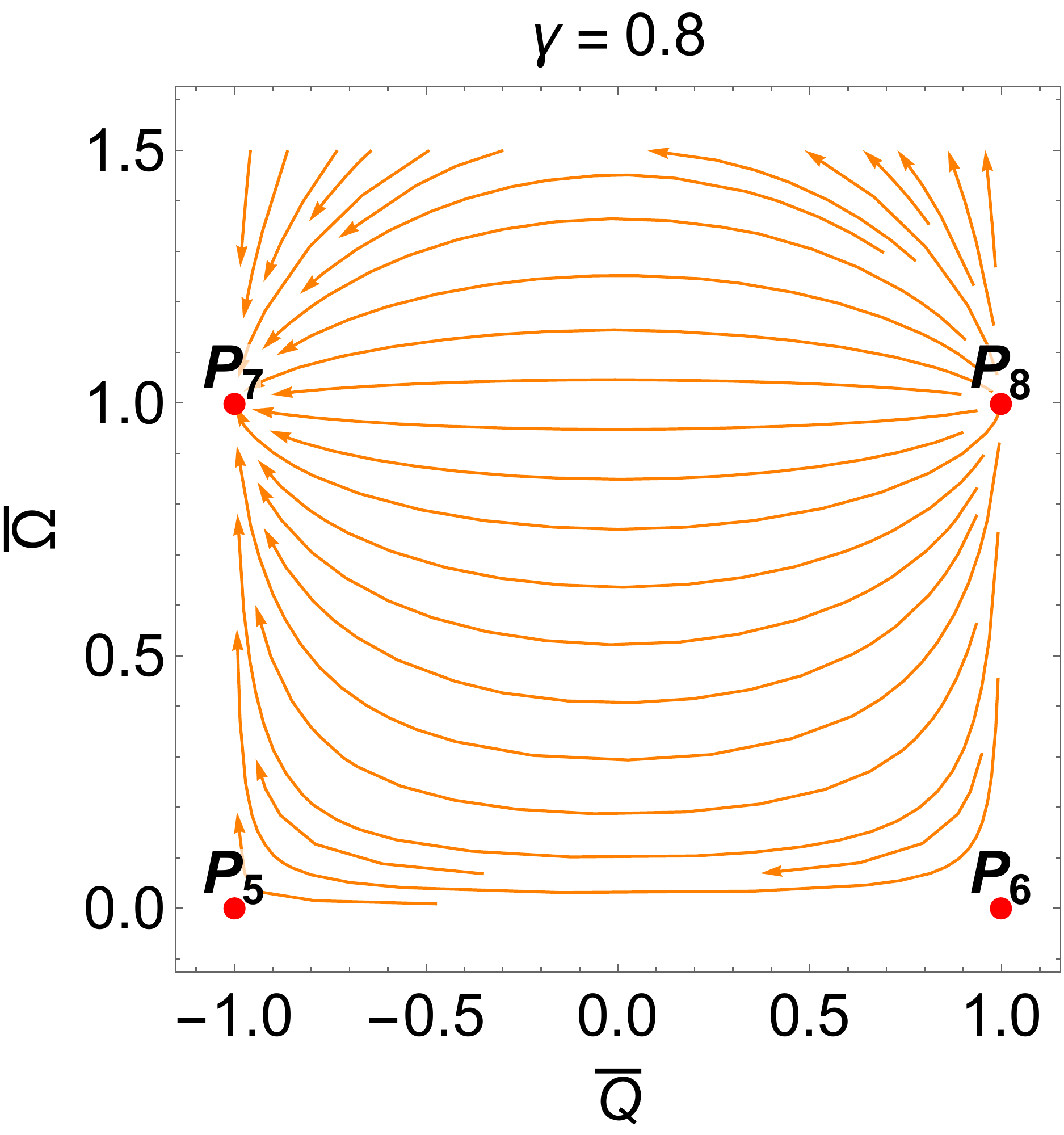}} 
    \subfigure[\label{fig:my_label2d}]{\includegraphics[width=0.4\textwidth]{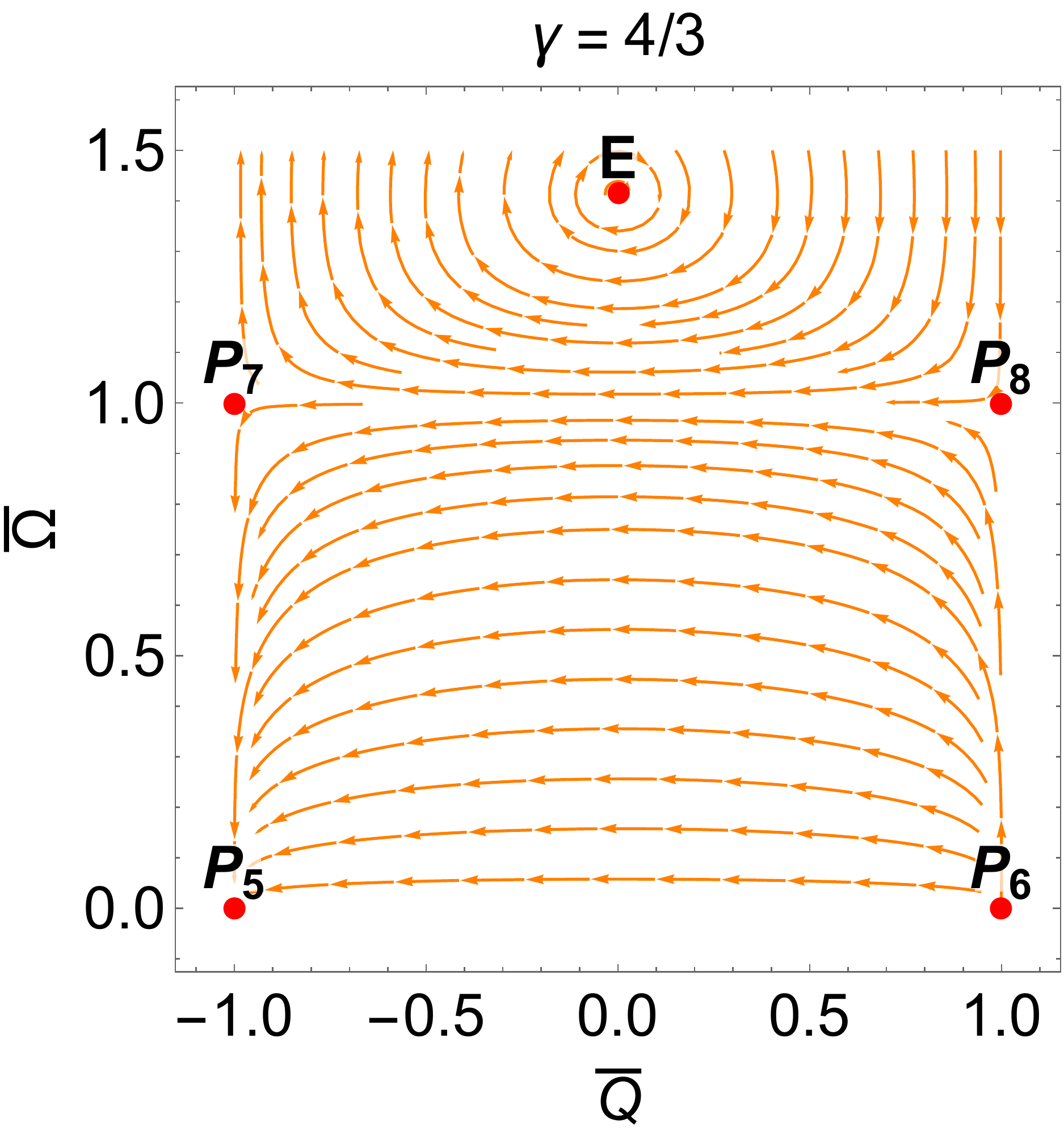}}
    \caption{Phase plane for system \eqref{eq76}, \eqref{eq77} corresponding to closed FLRW for different choices of $\gamma$.}
    \label{fig:my_label2}
\end{figure*}
Equilibrium points of \eqref{eq76} and \eqref{eq77}
are 
\begin{enumerate}
\item $P_5: (\bar{\Omega},  \bar{Q})=  (0,-1)$ with eigenvalues \newline $\left\{2-3 \gamma ,-\frac{3}{2} (\gamma -1)\right\}$.
\begin{enumerate}
    \item It is a source for $0\leq\gamma<\frac{2}{3}$.
    \item It is a saddle for $\frac{2}{3}<\gamma<1$.
    \item It is nonhyperbolic for $\gamma=\frac{2}{3}$ and $\gamma=1$.
    \item It is a sink for $1<\gamma\leq 2$.
\end{enumerate}
This equilibrium point is related to the isotropic point $P_5$ of KS. 
   The asymptotic metric at  $P_5$ is given by 
   \begingroup\makeatletter\def\f@size{8.5}\check@mathfonts
     \begin{align}
   &  ds^2= - dt^2 +  \frac{\left(1-\frac{3 \gamma  {D_0} t}{2}\right)^{\frac{4}{3 \gamma
   }}}{{c_1}^2} dr^2 \nonumber \\
   & + c_1 \left(1-\frac{3 \gamma  D_0 t}{2}\right)^{\frac{4}{3 \gamma }} \sin^2 r \; (d\y^2 + \sin^2 \y\, d\z^2). \label{metricP5}
\end{align} 
\endgroup
   The corresponding solution is a  flat matter--dominated FLRW contracting solution with $\bar{\Omega}_m=1$.
\item  $P_6: (\bar{\Omega},  \bar{Q})= (0,1)$ with eigenvalues \newline $\left\{\frac{3}{2}(\gamma -1),3 \gamma -2\right\}$.
\begin{enumerate}
    \item It is a sink for $0\leq \gamma<\frac{2}{3}$.
    \item It is a saddle for $\frac{2}{3}<\gamma<1$.
    \item It is nonhyperbolic for $\gamma=\frac{2}{3}$ and $\gamma=1$.
    \item It is a source for $1<\gamma\leq 2$.
\end{enumerate}
This equilibrium point is related to the isotropic point $P_6$ of KS.  The asymptotic metric at the equilibrium point $P_6$ is 
  \begingroup\makeatletter\def\f@size{8.5}\check@mathfonts
    \begin{align}
      &  ds^2= - dt^2 + \frac{\left(\frac{3 \gamma  {D_0} t}{2}+1\right)^{\frac{4}{3 \gamma
   }}}{{c_1}^2} dr^2 \nonumber \\
   & + c_1 \left(\frac{3 \gamma  D_0 t}{2}+1\right)^{\frac{4}{3 \gamma }} \sin^2 r \; (d\y^2 + \sin^2 \y\, d\z^2). \label{metricP6}
\end{align} 
\endgroup
   $P_6$ represents a  quintessence fluid or a zero-acceleration (Dirac-Milne) model for $\gamma=\frac{2}{3}$. 
In the limit $\gamma=0$ we have $\ell(t) \propto \left(\frac{3 \gamma  D_{0} t}{2}+1\right)^{\frac{2}{3
   \gamma }}\rightarrow  e^{D_0 t}$, i.e., the de Sitter solution. 
   
\item  $P_7: (\bar{\Omega},  \bar{Q})= (1,-1)$ with eigenvalues $\{-1,3 (\gamma -1)\}$.
\begin{enumerate}
    \item It is a sink for $0\leq \gamma<1$.
    \item It is  nonhyperbolic for $\gamma=1$.
    \item It is a saddle for $1<\gamma\leq 2$.
\end{enumerate}
This equilibrium point is related to the isotropic point $P_7$ of KS.
The line element \eqref{metric} becomes
\begin{align}
   &  ds^2= - dt^2 + c_1^{-2} {t^{\frac{4}{3}}} dr^2 \nonumber \\
   & +  {c_2^{-1}}{t^{\frac{4}{3}}} \sin^2 r \; (d\y^2 + \sin^2 \y\, d\z^2). \label{metricP7}
\end{align}
 Hence, the equilibrium point  can be associated with Einstein-de Sitter solution (\cite{WE}, Sec 9.1.1 (1)) with $\gamma= 1$. It is a contracting solution.  

\item  $P_8: (\bar{\Omega},  \bar{Q})= (1,1)$ with eigenvalues $\{3-3 \gamma ,1\}$.
\begin{enumerate}
    \item It is a source for $0\leq \gamma<1$.
    \item It is nonhyperbolic for $\gamma=1$.
    \item It is a saddle for $1<\gamma\leq 2$.
\end{enumerate}
This equilibrium point is related to the isotropic point $P_8$ of KS. 
The line element \eqref{metric} becomes
\begin{align}
   &  ds^2= - dt^2 + c_1^{-2} {t^{\frac{4}{3}}} dr^2 \nonumber \\
   &  +  {c_2^{-1}}{t^{\frac{4}{3}}} \sin^2 r \; (d\y^2 + \sin^2 \y\, d\z^2).
\end{align}  
 Hence, the equilibrium point  can be associated with  Einstein-de Sitter solution (\cite{WE}, Sec 9.1.1 (1)) with $\gamma= 1$. It is an expanding solution.  
\item  $E: (\bar{\Omega},  \bar{Q})= \left(\sqrt{\frac{3 \gamma -2}{3 \gamma -3}},0\right)$ with eigenvalues \newline $ \left\{-\frac{\sqrt{2-3 \gamma}}{\sqrt{2} },\frac{\sqrt{2-3 \gamma}}{\sqrt{2} }\right\}$. This point exists for $0\leq \gamma \leq \frac{2}{3}$ or $1<\gamma \leq 2$ and can be characterized as
\begin{enumerate}
    \item It is a saddle for $0\leq \gamma<\frac{2}{3}$.
    \item It is nonhyperbolic for $\gamma=\frac{2}{3}$ or $1<\gamma\leq 2$.
\end{enumerate}
This solution represents Einstein's static universe. It is characterized by $k= 1$, $\dot \ell = \ddot \ell=0$. It is usually viewed as a fluid model with a CC that is given {\it{a priori}} as a fixed universal constant \cite{Eddington:1930zz,Harrison:1967zza,Gibbons:1987jt,Gibbons:1988bm,Burd:1988ss,Noh:2020vnk,Barrow:2009sj,Barrow:2003ni}.
\end{enumerate}
 In the special case $\gamma=\frac{2}{3}$ there is one line of equilibrium points which are normally  hyperbolic, $M^{0}: (\bar{\Omega},  \bar{Q})= \left(0, \bar{Q}_c\right)$ with eigenvalues $\left\{-\frac{\bar{Q}_c}{2},0\right\}$, which is a sink for $\bar{Q}_c>0$ and a source for $\bar{Q}_c<0$.

The first equation of guiding system \eqref{eq72}, \eqref{eq73}, \eqref{eq74} becomes trivial up to second order in the $D$-expansion when $\gamma=1$. Using Taylor expansion up to the fourth order in $D$ the following time--averaged system is obtained
\begin{small}
\begin{subequations}
\label{eqaverg4D}
\begin{align}
& \frac{d\bar{\Omega}}{d \eta}=\frac{9 D^2 {\bar{Q}} \; \bar{\Omega}^5 \left(\omega ^2 - 2 \mu ^2\right)^3}{32 b^2 \mu ^6 \omega ^4}, \label{93a}
\\
&  \frac{d\bar{Q}}{d \eta} =-\frac{1}{32} \left(1-\bar{Q}^2\right) \left(\frac{9 D^2 \bar{\Omega}^4 \left(\omega ^2-2 \mu ^2\right)^3}{b^2 \mu ^6 \omega
   ^4}+16\right), \label{93b}
\\
& \frac{d \bar{\Phi}}{d \eta}= \frac{3 D \bar{\Omega}^2 \left(\omega ^2-2 \mu ^2\right)^3}{8 b^2 \mu ^6 \omega ^3} -\frac{3 D^3 \bar{\Omega}^4 \left(\omega ^2-2 \mu ^2\right)^5}{32 b^4 \mu ^{12} \omega ^5}, \label{93c}\\
& \frac{d D}{d \eta}=-\frac{3 D {\bar{Q}}}{2} - \frac{9 D^3 {\bar{Q}} \; \bar{\Omega}^4 \left(\omega ^2-2 \mu ^2\right)^3}{32 b^2 \mu ^6 \omega ^4}.
\end{align}
\end{subequations}
\end{small}
Given 
 \begin{align}
  & b\neq 0, \omega \neq 0,  \mu \neq 0, {\omega ^2-2
   \mu ^2} \neq 0,  {\bar{Q}}^2(\eta )\neq 1\nonumber\\
  &{-2 {\bar{Q}}'(\eta )+{\bar{Q}}^2(\eta )-1}\neq 0. \label{hypothesis} 
 \end{align}
From eq. \eqref{93b} we have  
\begin{small}
\begin{equation}
\label{94}
  \bar{\Omega}(\eta)= \frac{2 \sqrt{b} \mu ^{3/2} \omega  \sqrt[4]{{\bar{Q}}(\eta
   )^2-2 {\bar{Q}}'(\eta )-1}}{\sqrt{3} \sqrt{D} \left(\omega ^2-2 \mu ^2\right)^{3/4}
   \sqrt[4]{1-{\bar{Q}}^2(\eta )}}. 
\end{equation}
\end{small}
Substituting back eq. \eqref{94} in eq. \eqref{eqaverg4D} we have
\begin{small}
\begin{align}
D'(\eta)= D(\eta) {\bar{Q}}(\eta) \left(\frac{{\bar{Q}}'(\eta )}{1-{\bar{Q}}^2(\eta )}-1\right).
\end{align}
\end{small}
Using the method of the integrating factor we define 
\begin{equation}
  v=  e^{\int {\bar{Q}}(\eta ) d \eta} D(\eta),
\end{equation}
to obtain 
\begin{small}
\begin{align}
 v'(\eta)= &  D'(\eta ) e^{\int {\bar{Q}}(\eta ) \, d\eta }+D(\eta ) {\bar{Q}}(\eta ) e^{\int {\bar{Q}}(\eta )
   \, d\eta } \nonumber \\
    = &\frac{v(\eta) {\bar{Q}}(\eta ) {\bar{Q}}'(\eta )}{1-{\bar{Q}}^2(\eta )} \implies v(\eta)= \frac{v_0}{\sqrt{1-{\bar{Q}}^2(\eta )}}.
\end{align}
\end{small}
\noindent The integration constant can be absorbed in the indefinite integral. Then, we have  
\begin{small}
\begin{equation}
    D(\eta )=\frac{e^{-\int {\bar{Q}}(\eta ) \, d\eta }}{\sqrt{1-{\bar{Q}}^2(\eta )}}. \label{99}
\end{equation}
\end{small}
Substituting eq. \eqref{99} in eq. \eqref{94} we obtain
\begin{small}
\begin{align}
  & \bar{\Omega} (\eta )=\frac{2 \sqrt{b} \mu ^{3/2} \omega  e^{\frac{1}{2} \int
   {\bar{Q}}(\eta ) \, d\eta } }{\sqrt{3} \left(\omega ^2 - 2 \mu ^2\right)^{3/4}}   \sqrt[4]{{\bar{Q}}^2(\eta )-2 {\bar{Q}}'(\eta )-1}. \label{100}
\end{align}
\end{small}
Then, substituting eqs. \eqref{99} and \eqref{100} in eq. \eqref{93c} we obtain the quadrature
 \begingroup\makeatletter\def\f@size{8.5}\check@mathfonts
\begin{align}
      & \bar{\Phi} (\eta )= \frac{\left(\omega ^2-2 \mu
   ^2\right)^{3/2}}{2 b \mu ^3 \omega}   \bigints \frac{\sqrt{{\bar{Q}}^2(\eta )-2 {\bar{Q}}'(\eta )-1}}{\sqrt{1-{\bar{Q}}^2(\eta )}} d\eta  \nonumber \\
   & -\frac{\left(\omega ^2-2 \mu ^2\right)^2}{6 b^2 \mu ^6 \omega } \bigints \frac{D(\eta ) \left({\bar{Q}}^2(\eta )-2 {\bar{Q}}'(\eta )-1\right)}{ \left(1-{\bar{Q}}^2(\eta )\right)} d\eta. \label{101}
   \end{align}
\endgroup
   Substituting eqs. \eqref{99}, \eqref{100} in eq. \eqref{93a} we obtain under assumptions \eqref{hypothesis}  the differential equation
\begin{small}
 \begin{align} 
   & {\bar{Q}}''(\eta)={\bar{Q}(\eta)}\left(2 ({\bar{Q}(\eta)}^2-1) +{\bar{Q}}'(\eta)\left(\frac{4 {\bar{Q}}'(\eta)}{{\bar{Q}(\eta)}^2-1}-5\right)\right). \label{102}
\end{align}
\end{small}
\noindent
Summarizing,  system \eqref{eqaverg4D} admits the first integral 
\eqref{99}, \eqref{100}, \eqref{101}, where  ${\bar{Q}}$ satisfies the differential equation \eqref{102}. 

\noindent To analyze the asymptotic behavior of the solutions of eq. \eqref{102}  we introduce the variables $x={\bar{Q}}(\eta), \;  y= {\bar{Q}}'(\eta )$ to obtain  \begin{small}
\begin{subequations}
\label{system55}
\begin{align}
    & x'(\eta )=y(\eta ),  \\
    & y'(\eta )=x(\eta ) \left(2 (x^2(\eta )-1)  +y(\eta ) \left(\frac{4 y(\eta )}{x^2(\eta )-1}-5\right)\right).
\end{align}
\end{subequations}
\end{small}
     \begin{figure}[t!]
    \centering
    \includegraphics[width=0.45\textwidth]{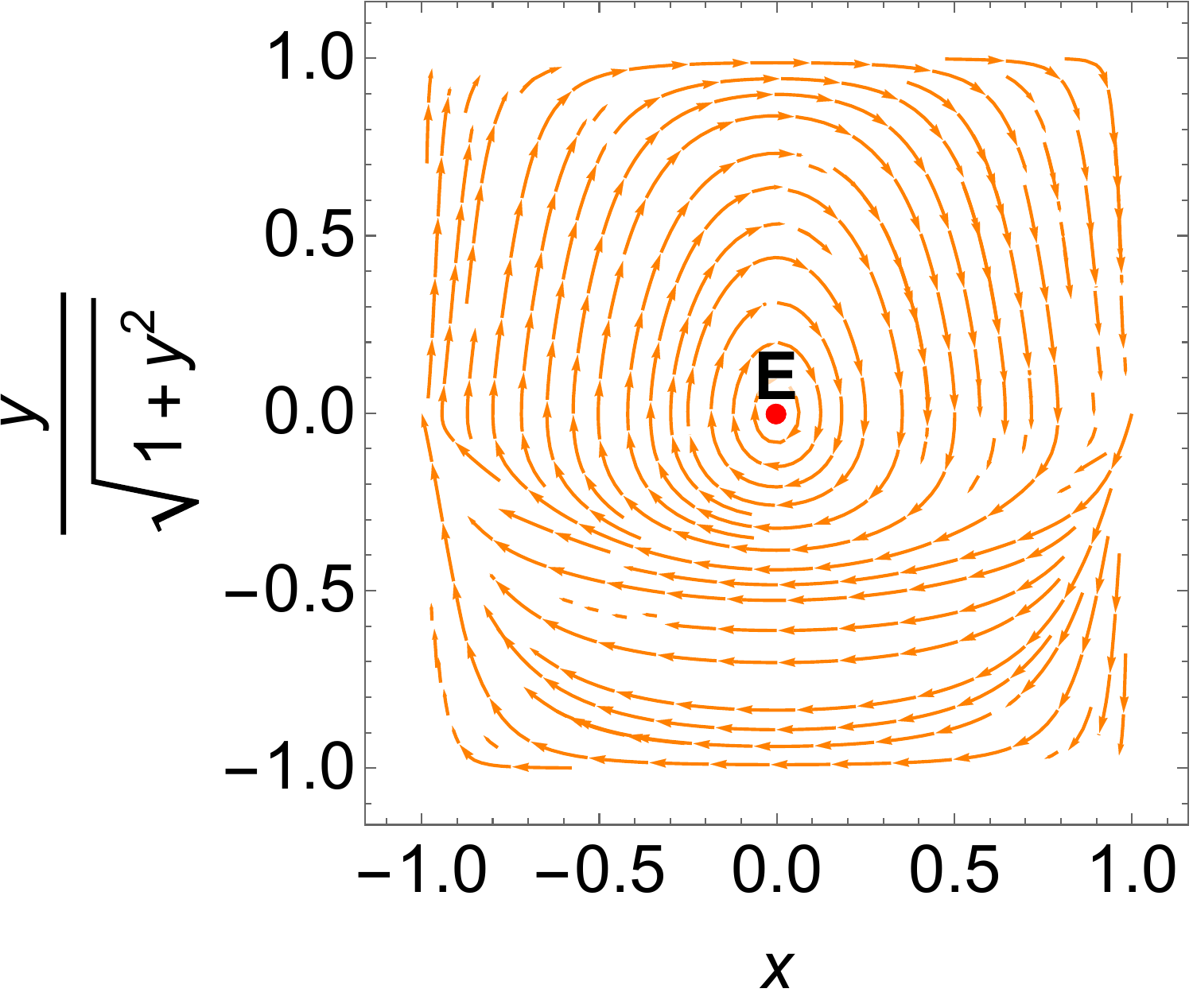}
    \caption{Compacted phase plane of system \eqref{system55}.}
    \label{fig:my_labelsyst55}
\end{figure}
The origin is an equilibrium point with eigenvalues \newline $\left\{i \sqrt{2},-i \sqrt{2}\right\}$. 
The dynamics of system \eqref{system55} in the coordinates  $(x, y/\sqrt{1+y^2})$ is presented in Figure \ref{fig:my_labelsyst55}  where the origin is a stable center. According to the analysis at first order of  the guiding system \eqref{eq72}, \eqref{eq73}, \eqref{eq74},   Einstein's static universe has coordinates $E: (\bar{\Omega},  \bar{Q})= \left(\sqrt{\frac{3 \gamma -2}{3 \gamma -3}},0\right)$. When $\gamma=1$, this point satisfies $\bar{\Omega} \rightarrow \infty$. However, by taking $\gamma=1$   the equation for $\bar{\Omega}$ is decoupled  and $\bar{\Omega}$ only takes  arbitrary constant values. On the other hand, since $\gamma=1$, we required higher order terms in Taylor expansion for obtaining system \eqref{eqaverg4D}. Since  extra $\bar{\Omega}$- coordinate is excluded from Figure \ref{fig:my_labelsyst55}, Einstein's static solution emerged. It is indeed a stable center in the $(Q, Q')$ phase space.  In Figure \ref{fig:my_labelsyst55}  the points  $(x,y)=(-1,0)$ and $(x,y)=(1,0)$ are saddles.

\noindent Now, we study the dynamics in the reduced phase space $\mathbf{x}=(\Omega,  Q)$, where  the effect of $D$ in the dynamics was neglected. The results from the linear stability analysis combined with Theorem \ref{KSLFZ11} for $\Sigma=0$ lead to the following local results. 
\begin{thm}
\label{thm12}
Late--time  attractors of full system \eqref{unperturbed1FLRWClosed} and time--averaged system  \eqref{avrgsystFLRWClosed} as $D\rightarrow 0$ for closed FLRW metric with positive curvature line element are 
\begin{enumerate}
  \item [(i)] The isotropic solution $P_5$ with metric \eqref{metricP5} 
       if  $1< \gamma \leq 2$.  The corresponding solution is a  flat matter--dominated FLRW contracting solution with $\bar{\Omega}_m=1$.
   
 \item[(ii)] The flat matter--dominated FLRW universe $P_6$    with metric \eqref{metricP6}
     if  $0< \gamma < \frac{2}{3}$. $P_6$ represents a  quintessence fluid or a zero-acceleration (Dirac-Milne) model for $\gamma=\frac{2}{3}$. 
In the limit $\gamma=0$ we have \newline $\ell(t) \propto \left(\frac{3 \gamma  D_{0} t}{2}+1\right)^{\frac{2}{3
   \gamma }}\rightarrow  e^{D_0 t}$, i.e.,  it corresponds to de Sitter solution. 
\item[(iii)]  The equilibrium point $P_7$ with metric \eqref{metricP7}
for $0 \leq\gamma<1$. The equilibrium point  can be associated with  Einstein-de Sitter solution. 
\end{enumerate}
\end{thm}
For global results when $D\in[0, \infty)$ see section \ref{SECT:5}. 

\section{Regular dynamical system on a compact phase space for Kantowski-Sachs and closed FLRW models}
\label{SECT:5}
According to Remark \ref{rem1}, Theorem \ref{KSLFZ11}  is valid on a finite time scale where $D$, given by eq. \eqref{D-KS}, remains close to zero, but at a critical time $t^*$ we have $D'(t^*)=0$ and $D$ becomes strictly increasing when $t>t^*$ such that  $\lim_{t\rightarrow \infty}D(t)= \infty.$  
\noindent
A lower bound for $t^*$ is estimated as 
$\bar{t}= \sup \{t>0: \Sigma(t)(1-Q^2(t) +3 Q(t) \Sigma(t))>0\}$. 
If $ \bar{t}= \infty$ this would mean that $\Sigma(t)(1-Q^2(t) +3 Q(t) \Sigma(t))>0$ is invariant for the flow and $D\rightarrow 0$ as $t\rightarrow \infty, \eta \rightarrow \infty$, in contradiction to  $\lim_{t\rightarrow \infty}D(t)= \infty$. Then, generically $\bar{t}<\infty$.  
 A similar result follows for closed FLRW for $D$ defined by eq.  \eqref{varsclosedFLRW}  and taking $\Sigma=\Sigma_0 = \bar{\Sigma}=0$. 

These results are  supported by the numerical evidence in  \ref{plotsKS} and  \ref{plotsclosedFLRW}. 

\subsection{Kantowski-Sachs metric}
In this section we analyze qualitatively the time--averaged system \eqref{avrgsystKS} as $D\rightarrow \infty$ by introducing the variable 
\begin{equation}
\label{compactD}
T=\frac{D}{1+ D}, 
\end{equation}
that maps $[0, \infty)$ to a finite interval $[0,1)$. Therefore, the  limit $D\rightarrow +\infty$ corresponds to $T=1$ and the limit   $D\rightarrow 0$ corresponds to $T=0$. Then, we have  the guiding system   \eqref{KSguidingC1}, \eqref{KSguidingC2}, \eqref{KSguidingC3} extended with equation  
\begingroup\makeatletter\def\f@size{8}\check@mathfonts
\begin{align}\label{avrgsystKSInfinity}
 & \frac{d T}{d \eta}=  \frac{1}{2} (1-T) T \left[3 \bar{Q} \left((\gamma -2) \bar{\Sigma }^2+(\gamma -1) \bar{\Omega }^2-\gamma \right)+2 \bar{Q}^2 \bar{\Sigma }-2 \bar{\Sigma }\right].
\end{align} 
\endgroup
We are interested in late--time or early-time  attractors, and in discussing relevant saddle  equilibrium points of the extended dynamical system   \eqref{KSguidingC1}, \eqref{KSguidingC2}, \eqref{KSguidingC3} \eqref{avrgsystKSInfinity}.  In this regard, we have the following results.  
\begin{enumerate}
 \item $P_1: (\bar{\Omega}, \bar{\Sigma}, \bar{Q}, T)=(0,-1,-1,0)$ with eigenvalues \newline $\left\{-2,-\frac{3}{2},-3 (2-\gamma), 3\right\}$. It is saddle. 
 
 \item $P_1^{\infty}: (\bar{\Omega}, \bar{\Sigma}, \bar{Q}, T)=(0,-1,-1,1)$ with eigenvalues \newline $\left\{-2,-\frac{3}{2},-3 (2-\gamma), -3\right\}$. It is a sink for $0\leq \gamma<2$. 
 
 \item $P_2:  (\bar{\Omega}, \bar{\Sigma}, \bar{Q}, T)=(0,1,1,0)$ with eigenvalues \newline $ \left\{2, \frac{3}{2}, 3(2 - \gamma), -3 \right\}$. It is saddle. 
 
 \item $P_2^{\infty}:  (\bar{\Omega}, \bar{\Sigma}, \bar{Q}, T)=(0,1,1,1)$ with eigenvalues \newline $ \left\{2, \frac{3}{2}, 3(2 - \gamma),3 \right\}$. It is a source for $0\leq \gamma<2$.  
 
 \item $P_3:  (\bar{\Omega}, \bar{\Sigma}, \bar{Q}, T)=(0,1,-1,0)$ with eigenvalues  \newline $ \left\{-6,-\frac{3}{2},-3 (2-\gamma),3\right\}$. It is saddle. 
 
 \item $P_3^{\infty}:  (\bar{\Omega}, \bar{\Sigma}, \bar{Q}, T)=(0,1,-1,1)$ with eigenvalues  \newline $ \left\{-6,-\frac{3}{2},-3 (2-\gamma),-3\right\}$. It is a sink for $0\leq \gamma<2$.  
 
 \item $P_4:  (\bar{\Omega}, \bar{\Sigma}, \bar{Q}, T)=(0,-1,1,0)$ with eigenvalues \newline $ \left\{6,\frac{3}{2},6-3 \gamma,-3\right\}$. It is saddle. 
 
 \item $P_4^{\infty}:  (\bar{\Omega}, \bar{\Sigma}, \bar{Q}, T)=(0,-1,1,1)$ with eigenvalues \newline $ \left\{6,\frac{3}{2},6-3 \gamma,3\right\}$. It is a source for $0\leq \gamma<2$. 
 
  \item $P_5:  (\bar{\Omega}, \bar{\Sigma}, \bar{Q}, T)=(0,0,-1,0)$ with eigenvalues \newline $\left\{2-3 \gamma ,-\frac{3}{2} (\gamma -1),3-\frac{3 \gamma }{2},\frac{3 \gamma }{2}\right\}$. It is a source for $0<\gamma <\frac{2}{3}$. 
  
 \item $P_5^{\infty}:  (\bar{\Omega}, \bar{\Sigma}, \bar{Q}, T)=(0,0,-1,1)$ with eigenvalues \newline $\left\{2-3 \gamma ,-\frac{3}{2} (\gamma -1),3-\frac{3 \gamma }{2},-\frac{3 \gamma }{2}\right\}$.  It is a saddle. 
 
  \item $P_6:  (\bar{\Omega}, \bar{\Sigma}, \bar{Q}, T)=(0,0,1,0)$ with eigenvalues \newline $ \left\{\frac{3 (\gamma -2)}{2},\frac{3 (\gamma -1)}{2},3 \gamma -2, -\frac{3 \gamma }{2}\right\}$. It is a sink for $0<\gamma <\frac{2}{3}$.

 \item $P_6^{\infty}:  (\bar{\Omega}, \bar{\Sigma}, \bar{Q}, T)=(0,0,1,1)$ with eigenvalues \newline $ \left\{\frac{3 (\gamma -2)}{2},\frac{3 (\gamma -1)}{2},3 \gamma -2,\frac{3 \gamma }{2}\right\}$. It is a saddle. 
 
  \item $P_7: (\bar{\Omega}, \bar{\Sigma}, \bar{Q}, T)=(1,0,-1,0)$ with eigenvalues \newline $ \left\{-1,3 (\gamma -1),\frac{3}{2}, \frac{3}{2}\right\}$. It is saddle. 
 
 \item $P_7^{\infty}: (\bar{\Omega}, \bar{\Sigma}, \bar{Q}, T)=(1,0,-1,1)$ with eigenvalues \newline $ \left\{-1,3 (\gamma -1),\frac{3}{2}, -\frac{3}{2}\right\}$. It is a saddle. 
 
  \item $P_8:  (\bar{\Omega}, \bar{\Sigma}, \bar{Q}, T)=(1,0,1,0)$ with eigenvalues \newline $ \left\{1,3(1- \gamma), -\frac{3}{2}, -\frac{3}{2}\right\} $. It is saddle.

 \item $P_8^{\infty}:  (\bar{\Omega}, \bar{\Sigma}, \bar{Q}, T)=(1,0,1,1)$ with eigenvalues \newline $ \left\{1,3(1- \gamma), -\frac{3}{2}, \frac{3}{2}\right\} $. It is a saddle.

  \item $P_9:  (\bar{\Omega}, \bar{\Sigma}, \bar{Q}, T)=\left(0,\frac{2-3 \gamma }{\left| 4-3 \gamma \right| },-\frac{2}{\left| 4-3 \gamma \right| },0\right)$ with eigenvalues \newline \begingroup\makeatletter\def\f@size{7.5}\check@mathfonts $ \Bigg\{-\frac{3 \left(\gamma
   +\sqrt{(2-\gamma ) (\gamma  (24 \gamma -41)+18)}-2\right)}{2 \left| 4-3 \gamma \right| },\frac{3 \left(-\gamma +\sqrt{(2-\gamma ) (\gamma 
   (24 \gamma -41)+18)}+2\right)}{2 \left| 4-3 \gamma \right| }$, \newline$\frac{3-3 \gamma }{\left| 4-3 \gamma \right| }, \frac{3 \gamma }{\left| 4-3
   \gamma \right| }\Bigg\}$
   \endgroup. Exists for $0\leq \gamma\leq \frac{2}{3}$ or $\gamma=2$ and is a saddle. 
 
 \item $P_9^{\infty}:  (\bar{\Omega}, \bar{\Sigma}, \bar{Q}, T)=\left(0,\frac{2-3 \gamma }{\left| 4-3 \gamma \right| },-\frac{2}{\left| 4-3 \gamma \right| },1\right)$ with eigenvalues \newline \begingroup\makeatletter\def\f@size{7.5}\check@mathfonts $ \Bigg\{-\frac{3 \left(\gamma
   +\sqrt{(2-\gamma ) (\gamma  (24 \gamma -41)+18)}-2\right)}{2 \left| 4-3 \gamma \right| },\frac{3 \left(-\gamma +\sqrt{(2-\gamma ) (\gamma 
   (24 \gamma -41)+18)}+2\right)}{2 \left| 4-3 \gamma \right| }$, \newline$\frac{3(1- \gamma) }{\left| 4-3 \gamma \right| }, -\frac{3 \gamma }{\left| 4-3
   \gamma \right| }\Bigg\}$
   \endgroup. Exists for $0\leq \gamma\leq \frac{2}{3}$ or $\gamma=2$ and is a saddle.
   
   \item $P_{10}:  (\bar{\Omega}, \bar{\Sigma}, \bar{Q}, T)= \left(0,\frac{3 \gamma -2}{\left| 4-3 \gamma \right| },\frac{2}{\left| 4-3 \gamma \right| }, 0\right)$ with eigenvalues \newline \begingroup\makeatletter\def\f@size{7.5}\check@mathfonts$ \Bigg\{\frac{3 \left(\gamma +\sqrt{(2-\gamma ) (\gamma  (24
   \gamma -41)+18)}-2\right)}{2 \left| 4-3 \gamma \right| }, -\frac{3 \left(-\gamma
   +\sqrt{(2-\gamma ) (\gamma  (24 \gamma -41)+18)}+2\right)}{2 \left| 4-3 \gamma \right| }$,\newline $-\frac{3 (1-\gamma)}{\left| 4-3 \gamma \right| },-\frac{3 \gamma }{\left| 4-3 \gamma
   \right| }\Bigg\}$\endgroup. Exists for $0\leq \gamma\leq \frac{2}{3}$ or $\gamma=2$ and is saddle. 
   
   \item $P_{10}^{\infty}:  (\bar{\Omega}, \bar{\Sigma}, \bar{Q}, T)= \left(0,\frac{3 \gamma -2}{\left| 4-3 \gamma \right| },\frac{2}{\left| 4-3 \gamma \right| },1\right)$ with eigenvalues \newline \begingroup\makeatletter\def\f@size{7.5}\check@mathfonts$ \Bigg\{\frac{3 \left(\gamma +\sqrt{(2-\gamma ) (\gamma  (24
   \gamma -41)+18)}-2\right)}{2 \left| 4-3 \gamma \right| }, -\frac{3 \left(-\gamma
   +\sqrt{(2-\gamma ) (\gamma  (24 \gamma -41)+18)}+2\right)}{2 \left| 4-3 \gamma \right| }$,\newline $-\frac{3 (1-\gamma)}{\left| 4-3 \gamma \right| },\frac{3 \gamma }{\left| 4-3 \gamma
   \right| }\Bigg\}$\endgroup. Exists for $0\leq \gamma\leq \frac{2}{3}$ or $\gamma=2$ and is saddle. 
   \end{enumerate}
In the special case $\gamma=0$, there exist four lines of equilibrium points which are normally  hyperbolic:
\begin{enumerate}
    \item $I_{-}: (\bar{\Omega}, \bar{\Sigma}, \bar{Q}, T)= \left(0 ,0 , -1, T_c\right)$ with eigenvalues \newline $\left\{3,2,\frac{3}{2},0\right\}$ is a source. 
    \item $I_{+}: (\bar{\Omega}, \bar{\Sigma}, \bar{Q}, T)= \left(0 ,0 , 1, T_c\right)$  with eigenvalues \newline $\left\{-3,-2,-\frac{3}{2},0\right\}$ is a sink. 
    \item $J_{-}:  (\bar{\Omega}, \bar{\Sigma}, \bar{Q}, T)= \left(0, \frac{1}{2}, -\frac{1}{2}, T_c \right)$ with eigenvalues $\left\{3,-\frac{3}{2},\frac{3}{4},0\right\}$ is a saddle. 
    \item $J_{+}:(\bar{\Omega}, \bar{\Sigma}, \bar{Q}, T)= \left(0, -\frac{1}{2}, \frac{1}{2}, T_c \right)$ with eigenvalues $\left\{-3,\frac{3}{2},-\frac{3}{4},0\right\}$ is a saddle. 
   \end{enumerate}

In the special case $\gamma=1$  there exist four lines of equilibrium points which are normally  hyperbolic, say, 
\begin{enumerate}
    \item $K_{-}: (\bar{\Omega}, \bar{\Sigma}, \bar{Q}, T)=(\bar{\Omega}_c, 0, -1, 0)$ with eigenvalues $\left\{\frac{3}{2},\frac{3}{2},-1,0\right\}$ is a saddle. 
    
    \item $K_{+}: (\bar{\Omega}, \bar{\Sigma}, \bar{Q}, T)=(\bar{\Omega}_c, 0, 1, 0)$ with eigenvalues \newline $\left\{-\frac{3}{2},-\frac{3}{2},1,0\right\}$ is a saddle. 
    
   \item $K_{-}^{\infty}: (\bar{\Omega}, \bar{\Sigma}, \bar{Q}, T)=(\bar{\Omega}_c, 0, -1, 1)$ with eigenvalues \newline  $\left\{-\frac{3}{2},\frac{3}{2},-1,0\right\}$ is a saddle. 
   
    \item $K_{+}^{\infty}:(\bar{\Omega}, \bar{\Sigma}, \bar{Q}, T)=(\bar{\Omega}_c, 0, 1, 1)$ with eigenvalues \newline $\left\{-\frac{3}{2},\frac{3}{2},1,0\right\}$ is a saddle.
\end{enumerate}

In the special case $\gamma=2$  there exist four lines of equilibrium points which are normally  hyperbolic, say, 
\begin{enumerate}
    \item $L_{-}: (\bar{\Omega}, \bar{\Sigma}, \bar{Q}, T)=(0, \bar{\Sigma}_c, -1, 0)$ with eigenvalues \newline $\left\{3,-\frac{3}{2},0,-2 \bar{\Sigma }_c-4\right\}$ is a saddle. 
    \item $L_{+}: (\bar{\Omega}, \bar{\Sigma}, \bar{Q}, T)=(0, \bar{\Sigma}_c, 1, 0)$ with eigenvalues \newline $\left\{-3,\frac{3}{2},0,4-2 \bar{\Sigma }_c\right\}$ is a saddle.
    
   \item $L_{-}^{\infty}: (\bar{\Omega}, \bar{\Sigma}, \bar{Q}, T)=(0, \bar{\Sigma}_c, -1, 1)$ with eigenvalues \newline $\left\{-3,-\frac{3}{2},0,-2 \bar{\Sigma }_c-4\right\}$ is a sink for $\bar{\Sigma }_c>-2$.
   
    \item $L_{+}^{\infty}:(\bar{\Omega}, \bar{\Sigma}, \bar{Q}, T)=(0, \bar{\Sigma}_c, 1, 1)$ with eigenvalues \newline $\left\{3,\frac{3}{2},0,4-2 \bar{\Sigma }_c\right\}$ is a source for $\bar{\Sigma }_c<2$.   
\end{enumerate}
As before, the subindex $\pm$ indicates if they correspond to contracting (``$-$'') or to expanding  (``$+$'') solutions. Superindex $\infty$ refers to solutions with $D\rightarrow \infty$. 

\begin{figure}
    \centering
    \subfigure[\label{fig:AKSAlternativePhaseCC} Cosmological constant $\gamma=0$.]{\includegraphics[scale = 0.27]{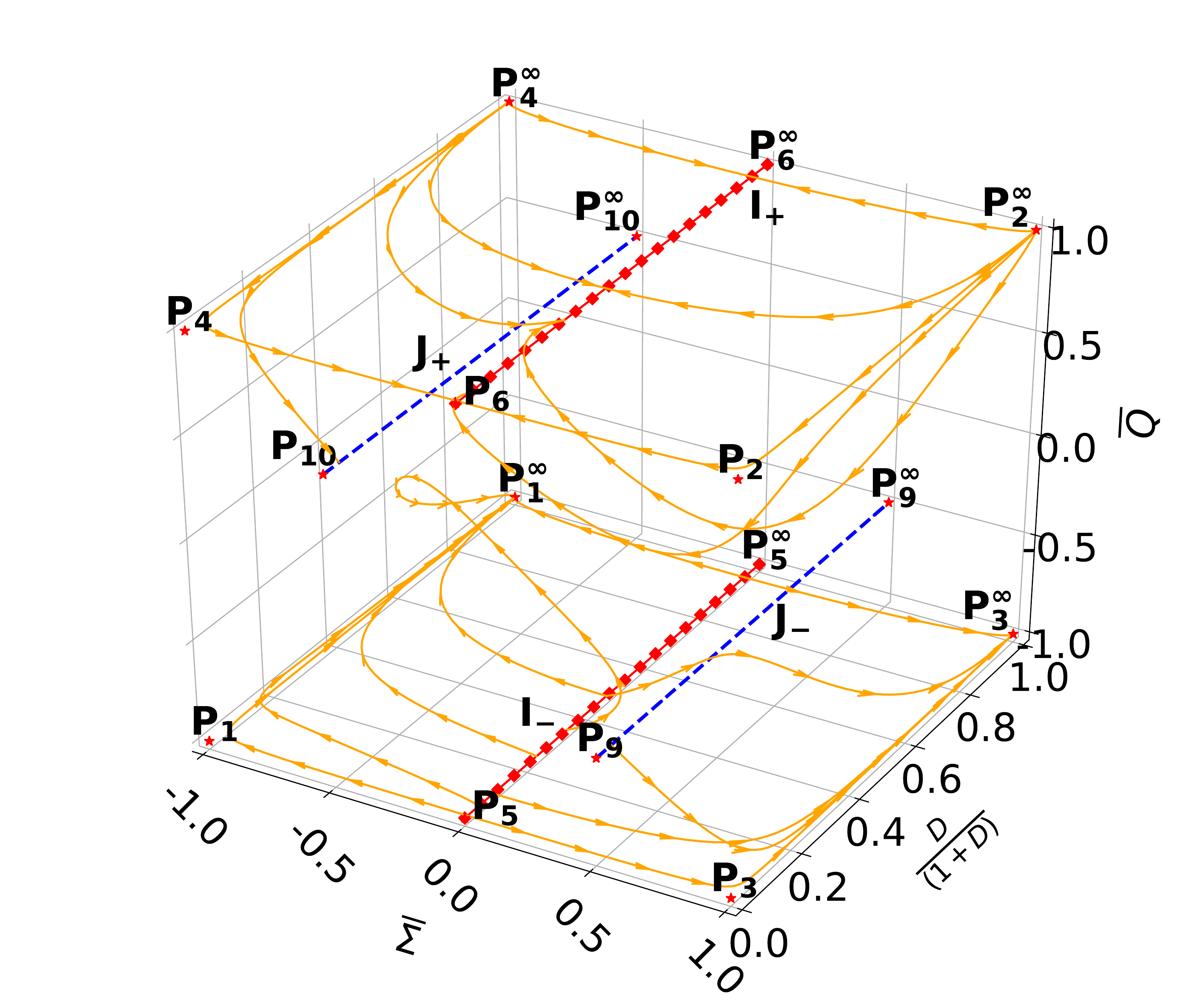}}
    \subfigure[\label{fig:AKSAlternativePhaseStiff} Stiff matter $\gamma=2$.]{\includegraphics[scale = 0.27]{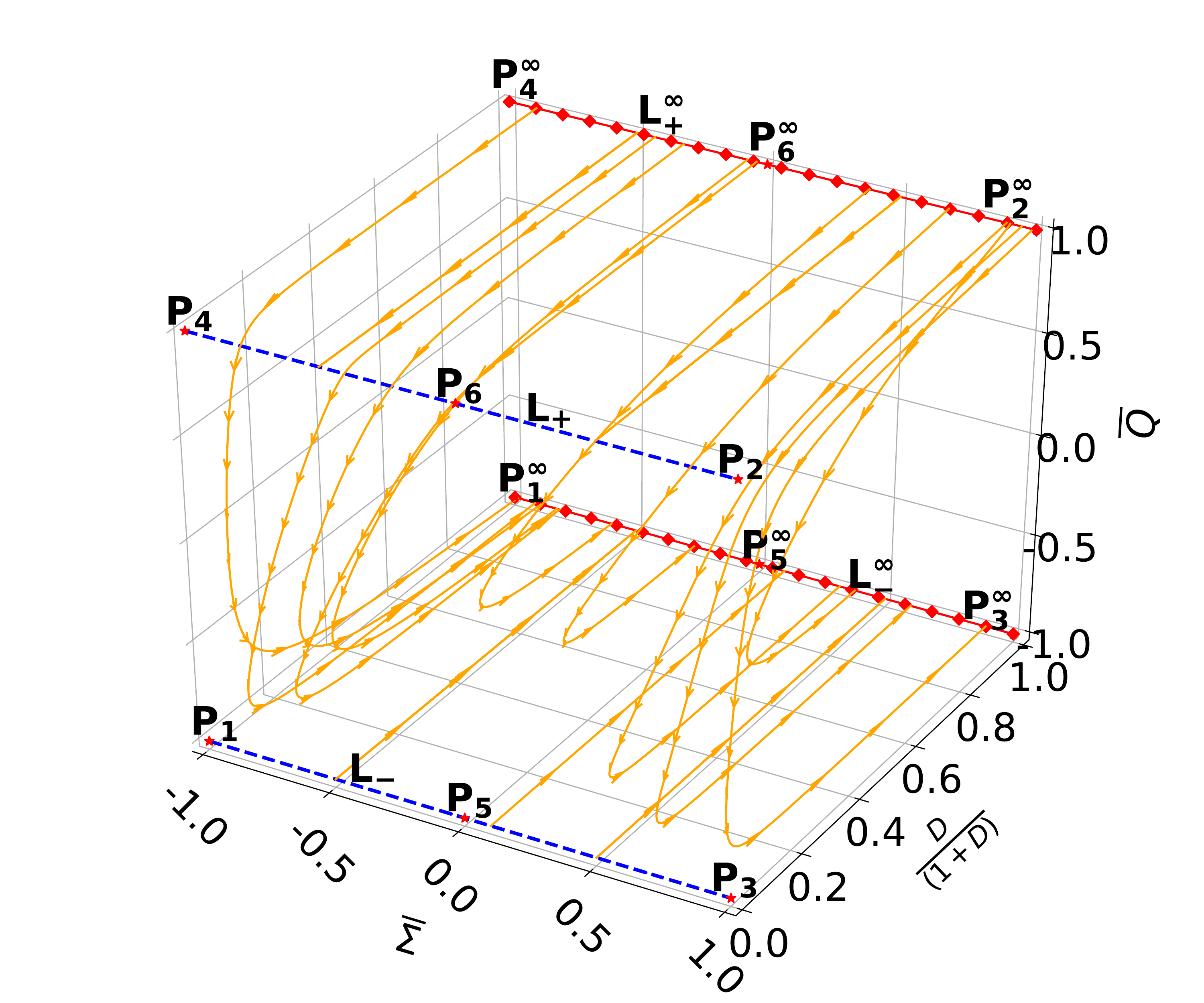}}
    \caption{\label{KSAlternativePhase} Dynamics of the averaged system \eqref{avrgsystKS} at the invariant set $\bar{\Omega}=0$ for $\gamma=0$ and $2$, represented in the compact space $(\bar{\Sigma}, D/(1+D), \bar{Q})$.}
\end{figure}

\noindent For KS model the extended phase is four dimensional. To illustrate the stability of the aforementioned normally hyperbolic lines we examine numerically some invariant sets. In Figure \ref{KSAlternativePhase} the dynamics at the invariant set $\bar{\Omega}=0$  corresponding to vacuum solutions is represented in the compact space $(\bar{\Sigma}, D/(1+D), \bar{Q})$ for $\gamma=0$ and $\gamma=2$. 

\noindent For $\gamma=1$ and $\bar{\Sigma}=0$, the model is reduced to the closed FLRW metric with dust. The stability of the lines $K_{\pm}$ and $K_{\pm}^{\infty}$ on this invariant set  corresponding  to isotropic solutions  is illustrated in Figure \ref{fig:ClosedFLRWPhaseDust}.

\noindent  Also, we refer to  Figures \ref{Figure8}, \ref{Figure9}, \ref{Figure10}, \ref{Figure11}, \ref{Figure12}, \ref{Figure13}, \ref{Figure14}, and  \ref{Figure15},  where projections of some orbits of the full system and the averaged system are presented. In these figures it is numerically confirmed the result of Theorem \ref{KSLFZ11} for KS metric. That is to say,   solutions of the full system (blue lines) follow the track of the solutions of the averaged system (orange lines) for the whole $D$-range. 

\subsection{FLRW metric with positive curvature}
\begin{figure}
    \centering
    \subfigure[\label{fig:ClosedFLRWPhaseCC} Cosmological constant $\gamma=0$.]{\includegraphics[scale = 0.26]{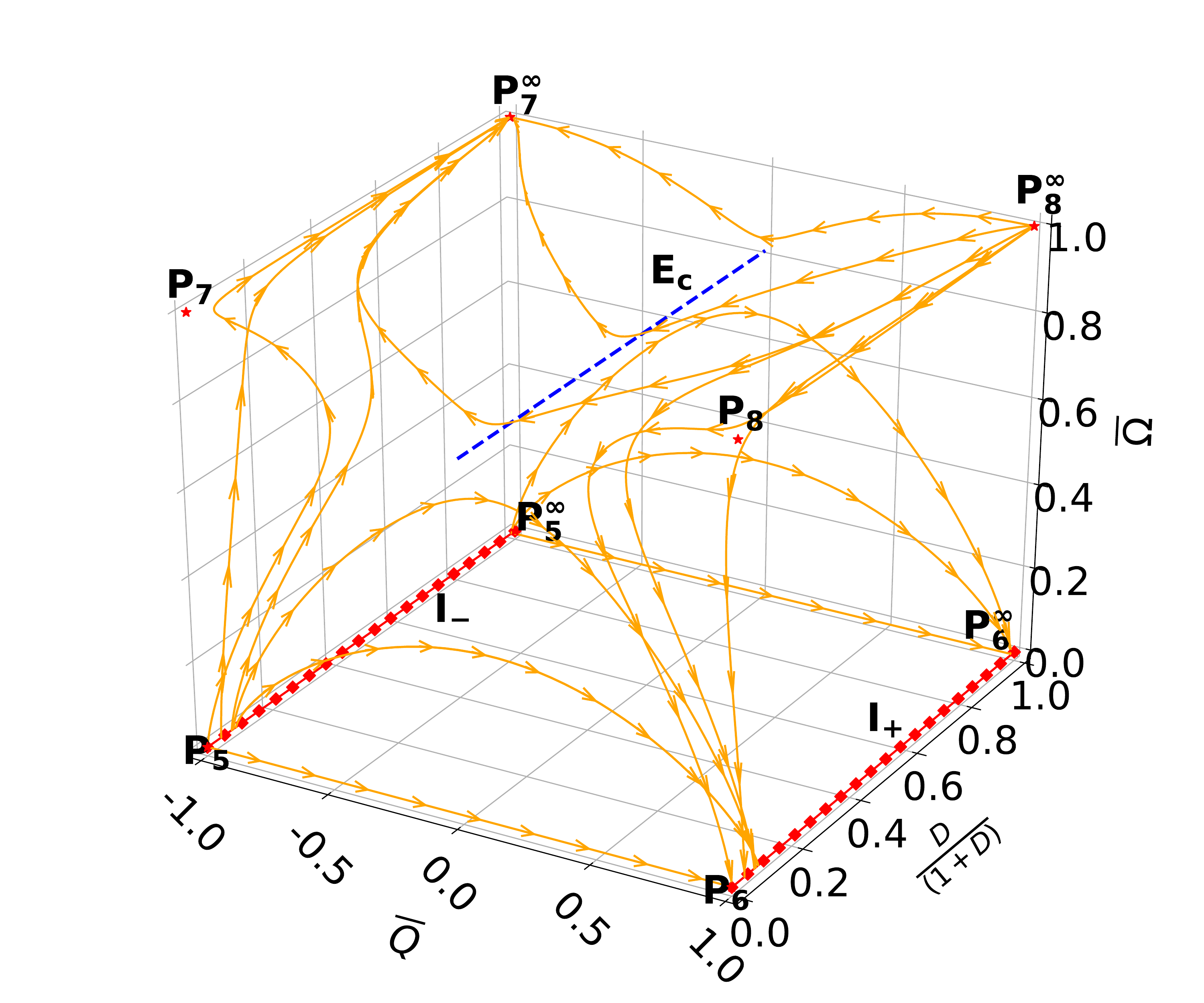}}
    \subfigure[\label{fig:ClosedFLRWPhaseBif} Zero acceleration $\gamma=\frac{2}{3}$.]{\includegraphics[scale = 0.26]{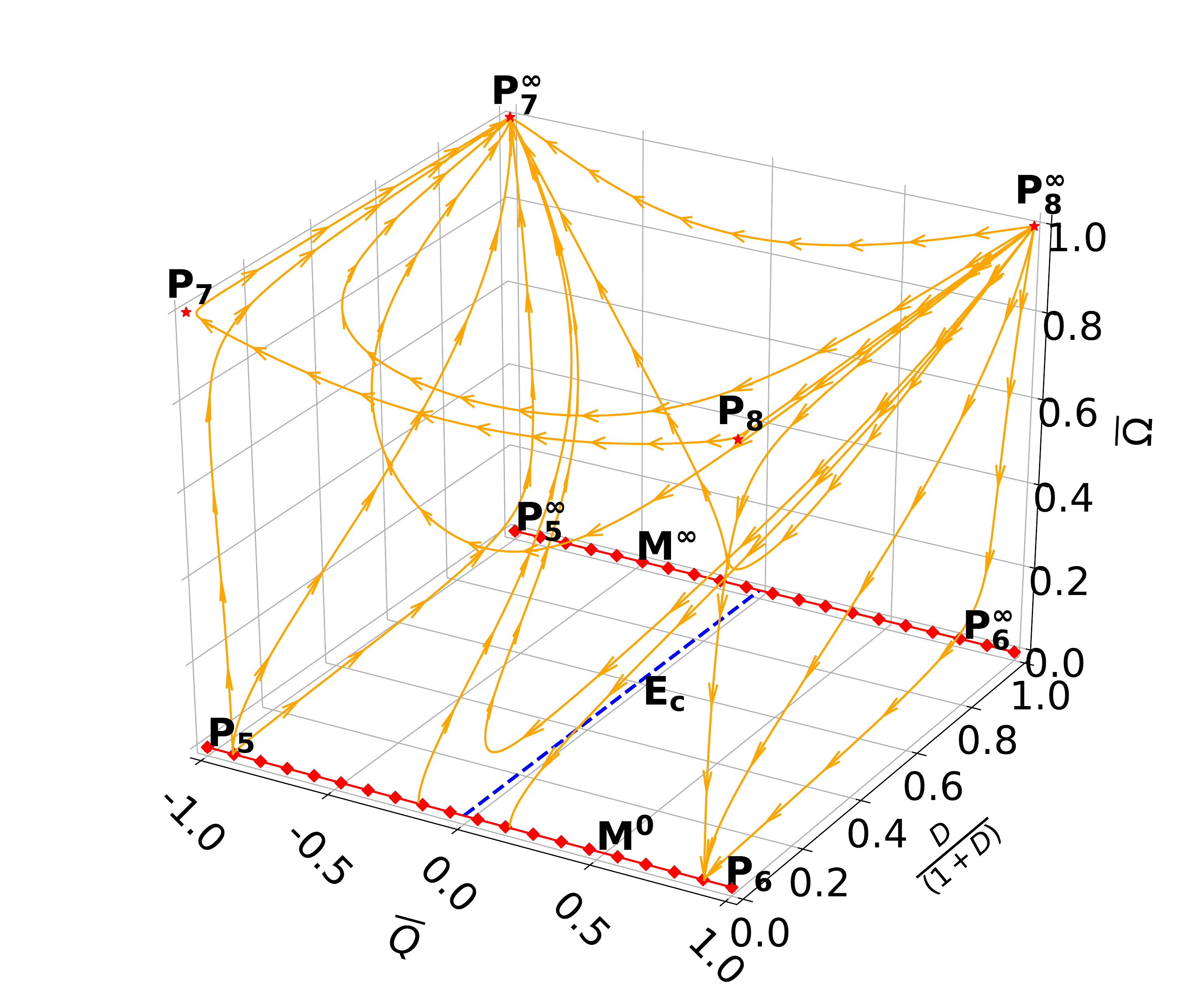}}
    \subfigure[\label{fig:ClosedFLRWPhaseDust} Dust $\gamma=1$.]{\includegraphics[scale = 0.26]{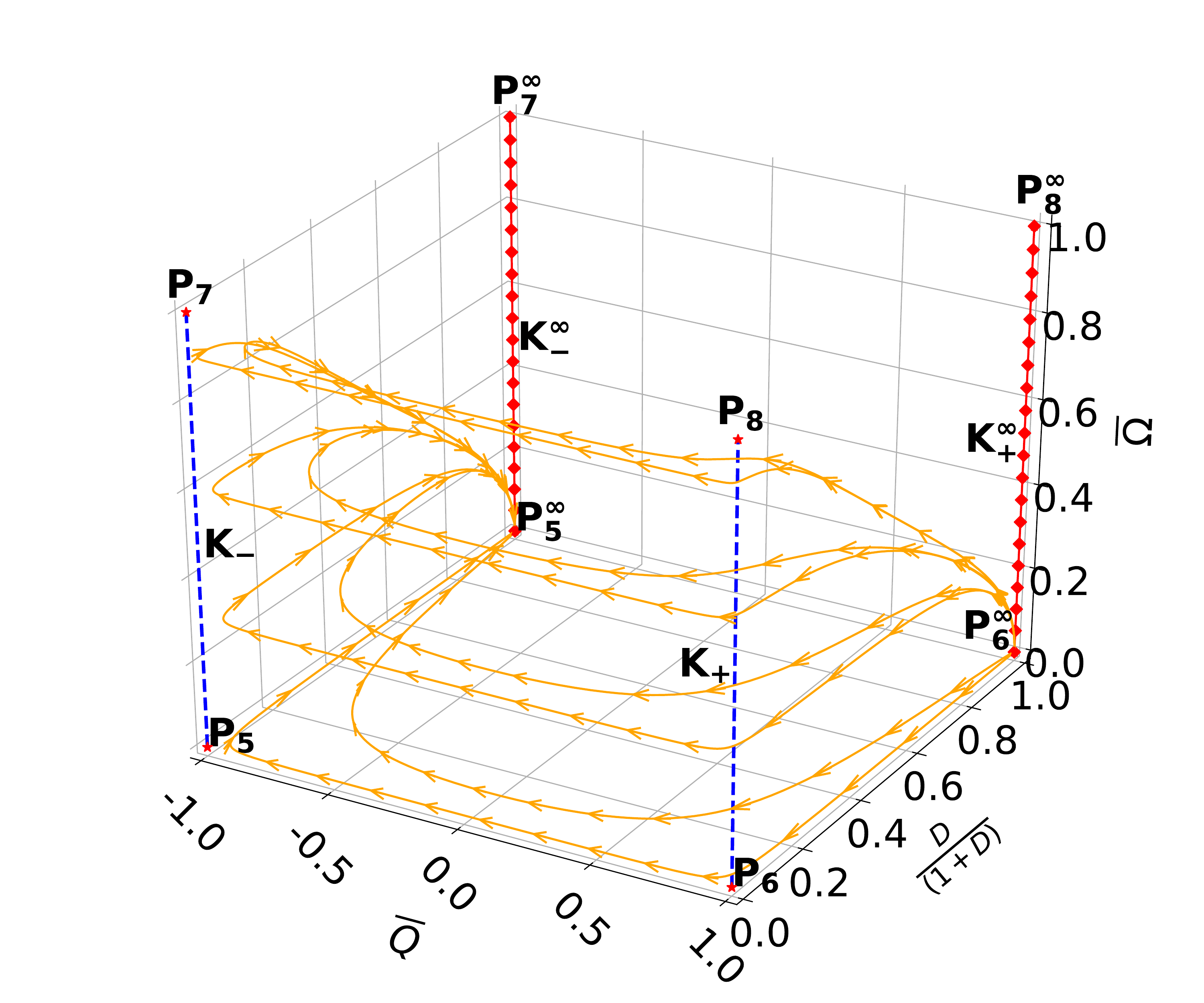}}
    \caption{\label{ClosedFLRWPhase} Dynamics of the averaged system \eqref{avrgsystFLRWClosed} for $\gamma=0$ and $\frac{2}{3}$, and of the averaged system \eqref{eqaverg4D} for $\gamma=1$, represented in the compact space \ $(\bar{Q}, D/(1+D), \bar{\Omega})$.}
\end{figure}
Using the variable \eqref{compactD}  the time--averaged system for $\gamma\neq 1$ \eqref{avrgsystFLRWClosed} becomes the guiding system  \eqref{eq76} and \eqref{eq77} extended with equation
\begin{small}
\begin{align}
\label{avrgsystFLRWClosedInfinity}
& \frac{d T}{d \eta}= \frac{3}{2} (1-T) T \bar{Q} \left[(\gamma -1) \bar{\Omega }^2-\gamma \right].
\end{align}
\end{small}
We are interested in late--time or early-time  attractors and  in discussing the relevant saddle  equilibrium points of the extended system  \eqref{eq76}, \eqref{eq77} and \eqref{avrgsystFLRWClosedInfinity}.  In this regard, we have the following results.  
\begin{enumerate}
   \item $P_5: (\bar{\Omega}, \bar{Q}, T)= (0,  -1, 0)$ with eigenvalues  \newline $ \left\{2-3 \gamma ,-\frac{3}{2} (\gamma -1),\frac{3 \gamma }{2}\right\}$ is a source for $0<\gamma <\frac{2}{3}$. 
   
   \item $P_5^{\infty}: (\bar{\Omega}, \bar{Q}, T)= (0,  -1, 1)$ with eigenvalues  \newline $ \left\{2-3 \gamma ,-\frac{3}{2} (\gamma -1),-\frac{3 \gamma }{2}\right\}$ is a sink for $1<\gamma\leq 2$. 
   
   \item $P_6: (\bar{\Omega}, \bar{Q}, T)=(0, 1,  0)$ with eigenvalues  \newline $ \left\{\frac{3 (\gamma -1)}{2},3 \gamma -2,-\frac{3 \gamma }{2}\right\}$ is a sink  $0<\gamma <\frac{2}{3}$.  
   
   \item $P_6^{\infty}: (\bar{\Omega}, \bar{Q}, T)=(0, 1,  1)$ with eigenvalues  \newline $ \left\{\frac{3 (\gamma -1)}{2},3 \gamma -2,\frac{3 \gamma }{2}\right\}$ is a source for $1<\gamma\leq 2$.  
   
  \item $P_7: (\bar{\Omega}, \bar{Q}, T)=(1,  -1, 0)$ with eigenvalues  \newline $ \left\{-1,3 (\gamma -1),\frac{3}{2}\right\}$. It is saddle. 
 
  \item $P_7^{\infty}: (\bar{\Omega}, \bar{Q}, T)=(1,  -1,  1)$ with eigenvalues  \newline $ \left\{-1,3 (\gamma -1), -\frac{3}{2}\right\}$ is a sink for $0\leq \gamma<1$.
   
  \item $P_8: (\bar{\Omega}, \bar{Q}, T)=(1,  1,  0)$ with eigenvalues  \newline $\left\{3-3 \gamma,1,-\frac{3}{2}\right\}$ is a source for $0\leq \gamma<1$.  It is saddle. 
   
   \item $P_8^{\infty}: (\bar{\Omega}, \bar{Q}, T)=(1,  1,  1)$ with eigenvalues  \newline $\left\{3-3 \gamma,1,\frac{3}{2}\right\}$ is a source for $0\leq \gamma<1$.  
   \item $E_c: (\bar{\Omega}, \bar{Q}, T)=\left(\sqrt{\frac{{3 \gamma -2}}{{3 \gamma -3}}},  0, T_c\right)$ with eigenvalues \newline $\left\{0,-\sqrt{\frac{(5-3 \gamma ) \gamma -2}{2
   (\gamma -1)}},\sqrt{\frac{(5-3 \gamma ) \gamma -2}{2
   (\gamma -1)}}\right\}$. This line of equilibrium points exists for $0\leq \gamma \leq \frac{2}{3}$ or $1<\gamma \leq 2$ and can be characterized as a nonhyperbolic saddle for $0\leq \gamma <\frac{2}{3}$. For $1<\gamma \leq 2$  two eigenvalues are purely imaginary and one eigenvalue is zero. Then, it can be a center or  a focus. 
\end{enumerate}

In the special case $\gamma=0$  there exist two  lines of equilibrium points which are normally  hyperbolic, say, 
\begin{enumerate}
    \item $I_{-}: (\bar{\Omega},  \bar{Q}, T)= \left(0, -1, T_c\right)$ with eigenvalues \newline $\left\{2,\frac{3}{2},0\right\}$  is a source. 
    
    \item $I_{+}: (\bar{\Omega}, \bar{Q}, T)= \left(0, 1, T_c\right)$ with eigenvalues \newline $\left\{-2,-\frac{3}{2},0\right\}$ is a sink.   
  \end{enumerate}
  
  In the special case $\gamma=\frac{2}{3}$ there exists two lines of equilibrium points which are normally  hyperbolic, say, 
\begin{enumerate}
    \item $M^{0}: (\bar{\Omega},  \bar{Q}, T)= \left(0, \bar{Q}_c, 0\right)$ with eigenvalues \newline $\left\{-\bar{Q}_c,-\frac{\bar{Q}_c}{2},0\right\}$ is a sink for $\bar{Q}_c>0$ and a source for $\bar{Q}_c<0$. 
    
    \item $M^{\infty}: (\bar{\Omega}, \bar{Q}, T)= \left(0, \bar{Q}_c, 1\right)$ with eigenvalues \newline $\left\{\bar{Q}_c,-\frac{\bar{Q}_c}{2},0\right\}$ is a saddle.  
  \end{enumerate}
In the special case $\gamma=1$  there exist four lines of equilibrium points which are normally  hyperbolic, say, 
\begin{enumerate}
    \item $K_{-}: (\bar{\Omega},  \bar{Q}, T)=(\bar{\Omega}_c,   -1, 0)$ with eigenvalues  \newline $\left\{\frac{3}{2},-1,0\right\}$ is a saddle. 
    \item $K_{+}: (\bar{\Omega},   \bar{Q}, T)=(\bar{\Omega}_c,   1, 0)$ with eigenvalues \newline $\left\{-\frac{3}{2},1,0\right\}$ is a saddle. 
   \item $K_{-}^{\infty}: (\bar{\Omega},  \bar{Q}, T)=(\bar{\Omega}_c,   -1, 1)$ with eigenvalues  \newline $\left\{-\frac{3}{2},-1,0\right\}$ is a sink. 
    \item $K_{+}^{\infty}:(\bar{\Omega},  \bar{Q}, T)=(\bar{\Omega}_c,   1, 1)$ with eigenvalues \newline  $\left\{\frac{3}{2},1,0\right\}$ is a source. 
\end{enumerate}
In Figure \ref{ClosedFLRWPhase} the dynamics of the averaged system \eqref{avrgsystFLRWClosed} for $\gamma=0$, $\frac{2}{3}$ and of the averaged system \eqref{eqaverg4D} for $\gamma=1$ is represented in the compact space \ $(\bar{Q}, D/(1+D), \bar{\Omega})$.
In Figures \ref{Figure8}, \ref{Figure9}, \ref{Figure10}, \ref{Figure11}, \ref{Figure12}, \ref{Figure13}, \ref{Figure14}, and  \ref{Figure15} projections of some solutions of the full system \eqref{unperturbed1KS} and time--averaged system \eqref{avrgsystKS} in the $(\Sigma, D/(1+D), \Omega^{2})$ and $(Q, D, \Omega^{2})$ space are presented with their respective projections when $D=0$.
In figures \ref{Figure16}, \ref{Figure17}, \ref{Figure18}, and \ref{Figure19} projections of some solutions of the full system \eqref{unperturbed1FLRWClosed} and time--averaged system \eqref{avrgsystFLRWClosed} for $\gamma\neq 1$ and of system \eqref{eqaverg4D} for $\gamma=1$ in the $(Q, D/(1+D), \Omega^{2})$ space  are presented with their respective projection when $D=0$.  Figures \ref{Figure16} and \ref{FIGURE25} show how the solutions of the full system (blue lines) follow the track of the solutions of the averaged system (orange lines) for the whole $D$-range. Figures \ref{Figure17}, \ref{Figure18}, \ref{Figure19}, and \ref{FIGURE27} are evidence that the main theorem presented in Section \ref{SECT:II} is fulfilled for the FLRW metric with positive curvature ($k=+1$) only when $D$ is bounded. Precisely, the solutions of the full system (blue lines) follow the track of the solutions of the averaged system (orange lines) for the time interval $t D =\mathcal{O}(1)$. However, when $D$ becomes infinite ($T\rightarrow 1$) and for $\gamma\geq 1$  we have the following conjecture. {\emph{For closed Friedmann--Lemaître-Robertson--Walker (FLRW) and $1\leq\gamma\leq2$ the  solutions of the full system depart from the solutions of the averaged system}}.

\subsubsection{Closed FLRW: regime $\Omega>1$} 
\noindent

Now, we describe the regime $\Omega>1$. We define 
\begin{small}
\begin{equation}\label{eqU}
U(\phi)=  2 f^2\left(1-\cos \left(\frac{\phi
   }{f}\right)\right) - {\phi ^2}.     
\end{equation}
\end{small}
Substituting the form of  potential  \eqref{pot_v2} in   constraint  \eqref{GaussFLRW} we obtain for closed FLRW models
\begin{small}
\begin{align}
3 D^2 := 3H^2+ \frac{3}{\ell^2}= & \rho_m+ \frac{1}{2} \dot{\phi}^2 + \frac{\omega^2}{2}\phi^2  +  \frac{\left(\omega ^2-2 \mu ^2\right)}{2}U(\phi).
\end{align}
\end{small}
\noindent Then, 
\begin{small}
\begin{align}
 & 1= \frac{\rho_m}{3D^2} + \frac{\dot{\phi}^2}{6 D^2}  + \frac{\omega^2\phi^2}{6 D^2}  + \frac{\left(\omega ^2-2 \mu ^2\right)}{6 D^2} U(\phi).  
\end{align}
\end{small}
Using definition \eqref{varsclosedFLRW} we obtain 
\begin{small}
\begin{align}
 & 1= \frac{\rho_m}{3D^2} + \Omega^2 + \frac{\left(\omega ^2-2 \mu ^2\right)}{6 D^2} U(\phi).  \label{103}
\end{align}
\end{small}
\begin{figure}
    \centering
    \subfigure[\label{fig:AClosedFLRWRad3D} Projections in the space $(Q, D/(1+D), \Omega^2/(1+\Omega^2))$.]{\includegraphics[scale = 0.27]{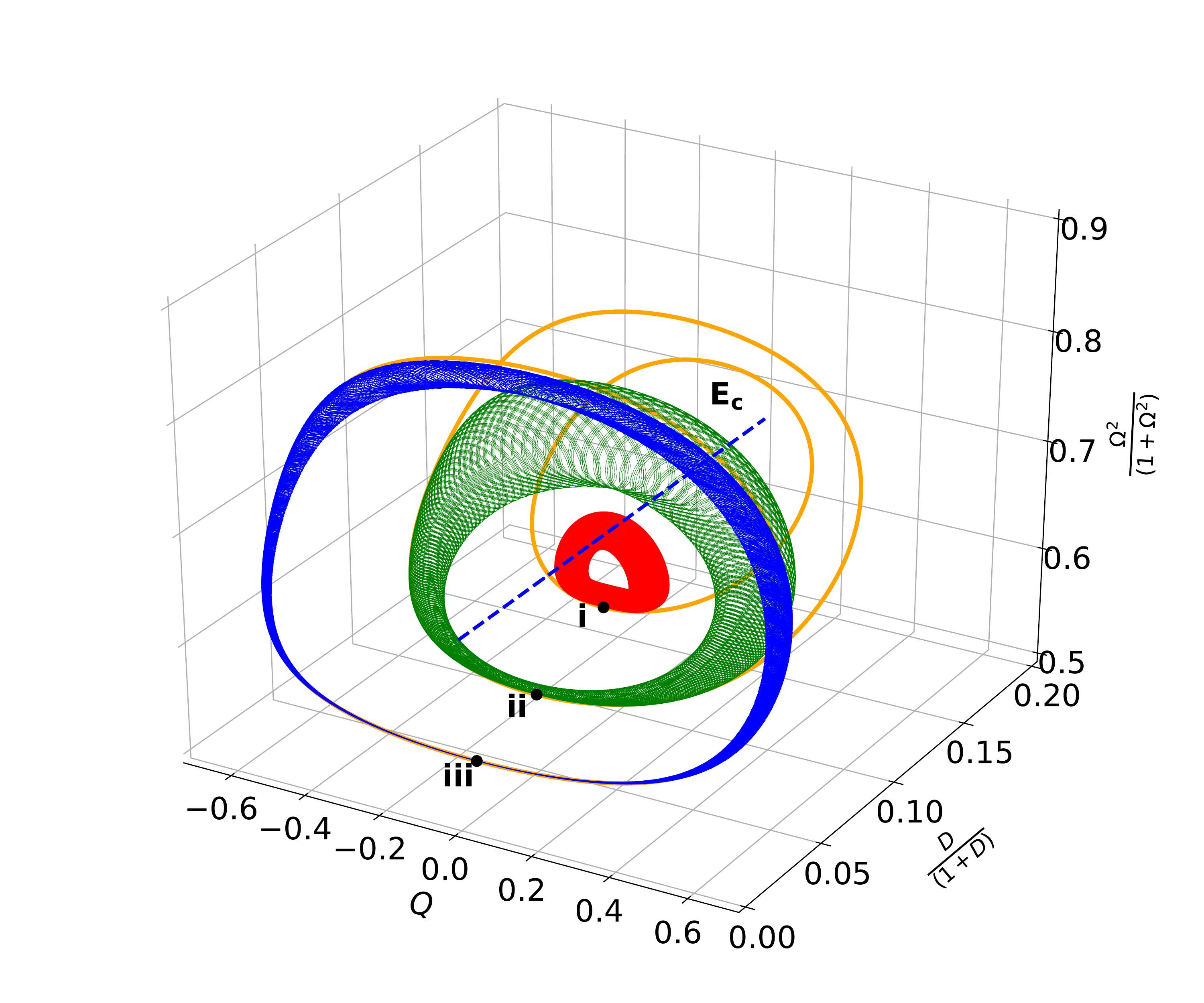}}
    \subfigure[\label{fig:AClosedFLRWRad2D} Projection in the space $(Q, \Omega^2/(1+\Omega^2))$.]{\includegraphics[scale = 0.55]{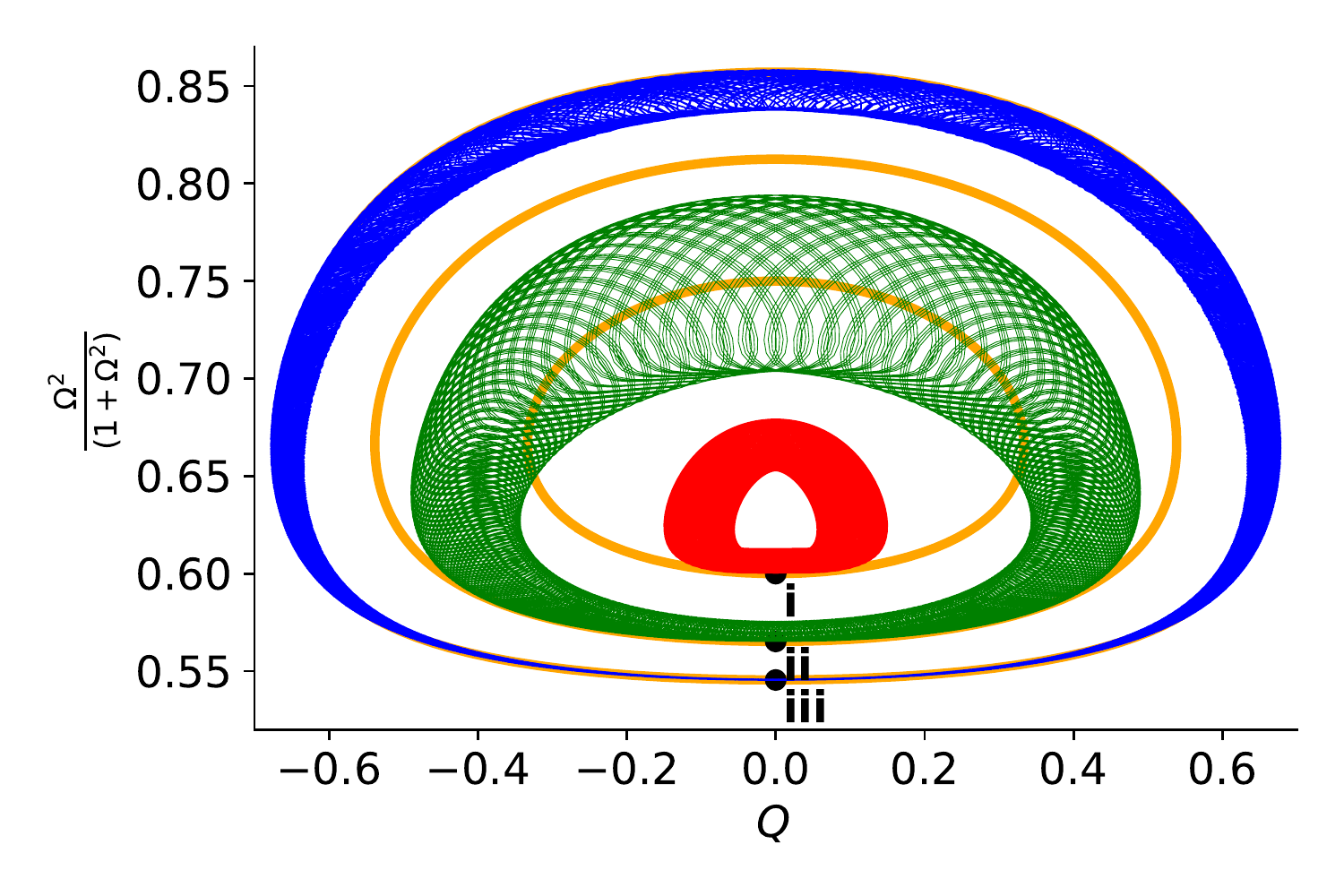}}
    \caption{Some solutions of the full system \eqref{unperturbed1FLRWClosed} (blue) and time--averaged system \eqref{avrgsystFLRWClosed} (orange) for the FLRW metric with positive curvature ($k=+1$) when $\gamma=\frac{4}{3}$. We have used for both systems the initial data sets presented in Table \ref{Tab5b}. \label{Figure20}}
\end{figure}
\begin{figure}
    \centering
    \subfigure[\label{fig:AClosedFLRWStiff3D} Projections in the space $(Q, D/(1+D), \Omega^2/(1+\Omega^2))$.]{\includegraphics[scale = 0.27]{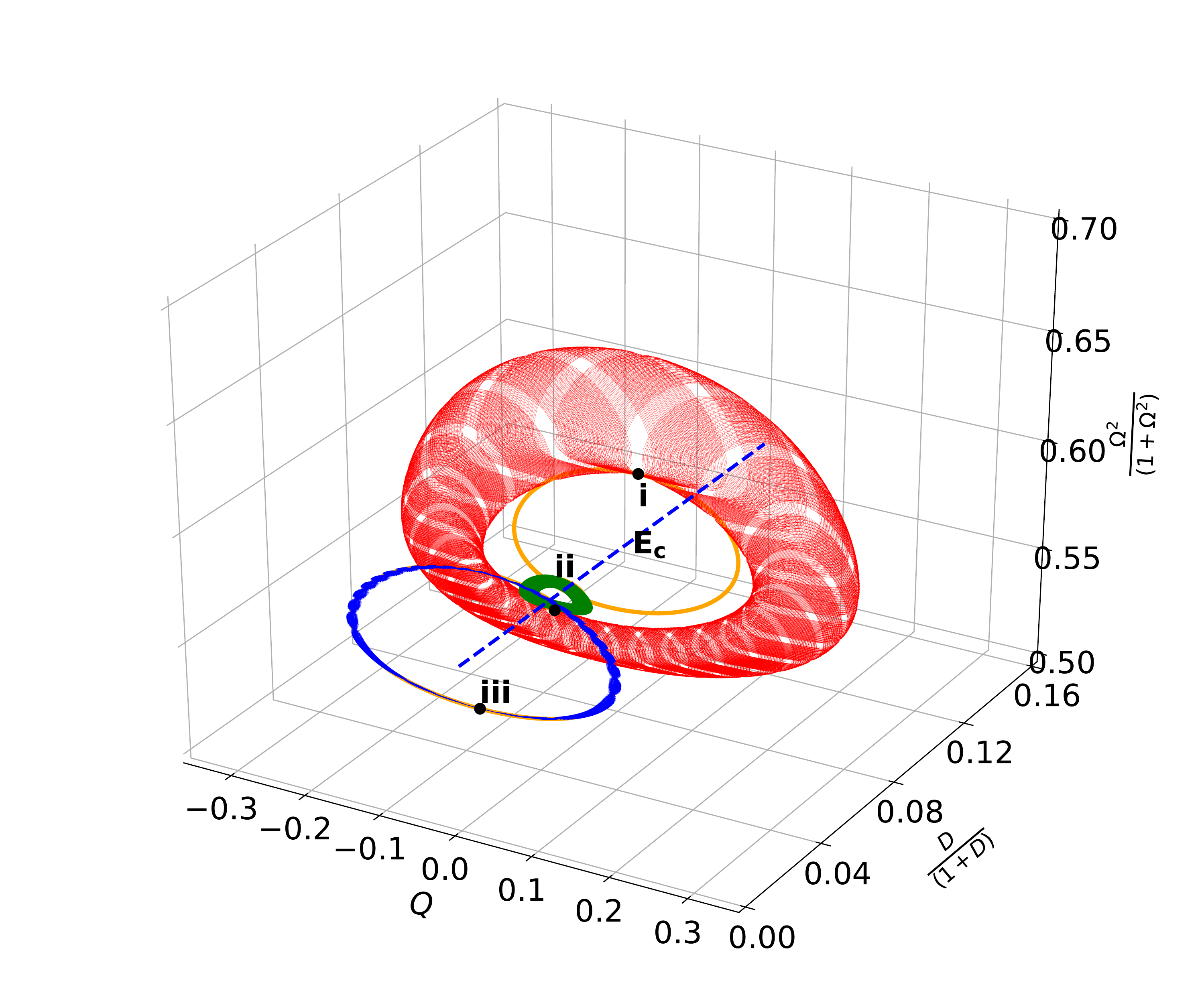}}
    \subfigure[\label{fig:AClosedFLRWStiff2D} Projection in the space $(Q, \Omega^2/(1+\Omega^2))$.]{\includegraphics[scale = 0.55]{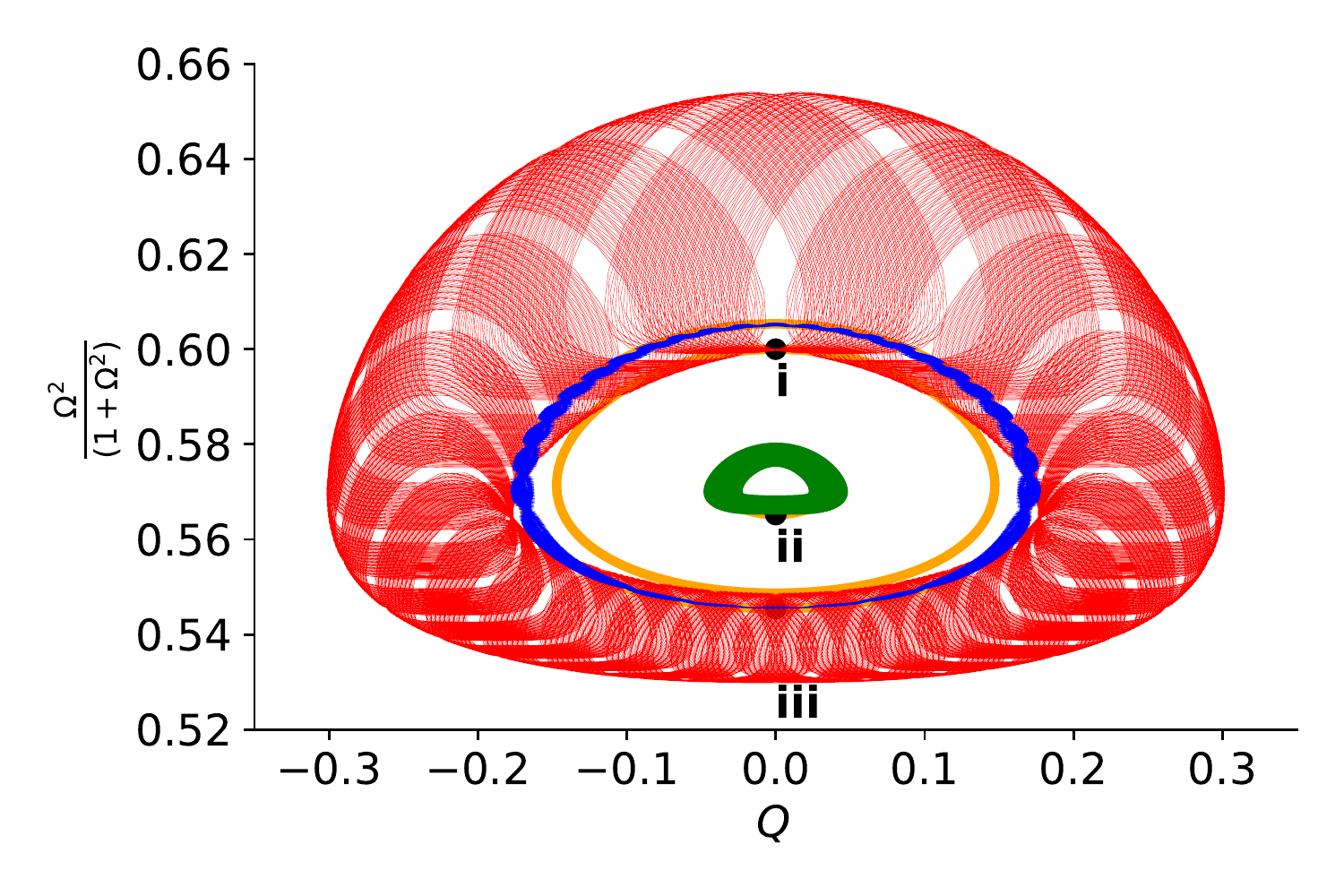}}
    \caption{Some solutions of the full system \eqref{unperturbed1FLRWClosed} (blue) and time--averaged system \eqref{avrgsystFLRWClosed} (orange) for the FLRW metric with positive curvature ($k=+1$) when $\gamma=2$. We have used for both systems the initial data sets presented in Table \ref{Tab5b}. \label{Figure21}}
\end{figure}
\noindent 
\begin{table}[t!]
\caption{\label{Tab5b} Here we present three initial data sets for the simulation of full system \eqref{unperturbed1FLRWClosed} and time--averaged system  \eqref{avrgsystFLRWClosed} for FLRW metric with positive curvature ($k=+1$) and for $\gamma=\frac{4}{3}$ (radiation) and $\gamma=2$ (stiff fluid). All the conditions are chosen in order to fulfill the inequalities $\Omega>1$ and $0\leq Q\leq 1$.}
\footnotesize\setlength{\tabcolsep}{9pt}
    \begin{tabular}{lcccccc}\hline
Sol.  & \multicolumn{1}{c}{$D(0)$} & \multicolumn{1}{c}{$\Omega^2(0)$} & \multicolumn{1}{c}{$Q(0)$} &  \multicolumn{1}{c}{$\varphi(0)$}  & \multicolumn{1}{c}{$t(0)$}  \\\hline
        i & $0.1$ & $1.5$ & $0$ & $0$ & $0$ \\
        ii & $0.05$ & $1.3$ & $0$ & $0$ & $0$ \\
        iii & $0.01$ & $1.2$ & $0$ & $0$ & $0$ \\ \hline
    \end{tabular}
\end{table}
The function $U(\phi)$ defined by eq. \eqref{eqU} 
satisfies
$U'(0)=U''(0)=U'''(0)=0$, and 
$U^{(iv)}(0)=-\frac{2}{f^2}$, which implies that $\phi=0$ is a global degenerated maximum of order 2 for $U(\phi)$. Therefore, $U(\phi)\leq 0$. 
Then, from  $\omega^2> 2 \mu^2$ and \eqref{103} it follows that $\Omega$, $\Sigma^2$ and $\frac{\rho_m}{3 D^2}$ can be greater than 1.  This implies that we can have solutions with $\Omega>1$ preserving the  non-negativity of the energy densities. 
That is, even if $U(\phi)\leq 0$,  we have $ \frac{\omega^2}{2}\phi^2  +  \frac{\left(\omega ^2-2 \mu ^2\right)}{2}U(\phi)\geq 0$ because the sign of term $\frac{\omega^2}{2}\phi^2$ is dominant over the last  term in eq. \eqref{103}.

\noindent
This  interesting behavior when $\Omega>1$ can be seen in Figures \ref{fig:my_label2d}, \ref{fig:my_labelsyst55}, \ref{Figure20} and \ref{Figure21}, where we show the solutions of the full system \eqref{unperturbed1FLRWClosed} and the time--averaged system \eqref{avrgsystFLRWClosed} considering the values $\gamma=\frac{4}{3}$ and $\gamma=2$ and using as initial conditions the three data sets presented in Table \ref{Tab5b}. 

\noindent This dynamical behavior related to spiral tubes  has been presented before in the literature in \cite{Heinzle:2004sr} and it is related to the  fact that the line of equilibrium points  $E_c$ (representing Einstein's static universes) has purely imaginary eigenvalues. 
In  \cite{Heinzle:2004sr} a comprehensive dynamical description  for closed cosmologies when the
matter source admits Einstein’s static model was presented. Moreover, theorems about the global asymptotic
behavior of solutions were established. Results in  \cite{Heinzle:2004sr} and  \cite{Heinzle:2006zy}  disprove   claims of non-predictability and chaos for models
close to   Einstein's model given in \cite{deOliveira:1997ej,Barguine:2001rb,DeOliveira:2002ih,Soares:2005fn}.

\noindent
To illustrate the existence of spiral tubes  we integrate the full system \eqref{unperturbed1FLRWClosed} and the time--averaged system \eqref{avrgsystFLRWClosed} using the fixed constants $\mu=\sqrt{2}/2$, $b=\sqrt{2}/5$, $\omega=\sqrt{2}$,  $f=\frac{b \mu ^3}{\omega ^2-2 \mu ^2}=1/10 \geq 0$. We select  $\gamma=\frac{4}{3}$ and $\gamma=2$ for the barotropic index (cases where  $E_c$ exists), and we use as initial conditions the three data sets presented in Table \ref{Tab5b}. In Figures \ref{Figure20} and \ref{Figure21} projections of the orbits showing this behavior in the $(Q, D/(1+D), \Omega^{2})$ space are presented with their respective projection when $D=0$. Figures \ref{fig:AClosedFLRWRad3D} and \ref{fig:AClosedFLRWRad2D} show solutions for a fluid corresponding to radiation ($\gamma=\frac{4}{3}$). Figures \ref{fig:AClosedFLRWStiff3D} and \ref{fig:AClosedFLRWStiff2D} show solutions for a fluid corresponding to stiff fluid ($\gamma=2$).  
\section{Conclusions}
\label{Conclusions}
This is the last paper of the ``Averaging generalized scalar field cosmologies'' research program. We  have used asymptotic methods and averaging theory to explore the solutions space of  scalar field cosmologies with generalized harmonic potential \eqref{pot} in vacuum or minimally coupled to matter. Different from references \cite{Leon:2021lct,Leon:2021rcx}, here we have studied systems where  Hubble scalar is not monotonic, but the systems admit a function $D$ given by eq. \eqref{GEN-D}  playing the role of a time-depending perturbation parameter, which is   decreasing for a finite time scale $t< t^*$ where $D'(t^*)=0$. 
For $t> t^*$  monotony of the quantity $D$ changes and  this parameter increases without bound. 

\noindent We have proved Theorem   \ref{KSLFZ11} which states that late--time  attractors of full and time--averaged systems are the same  when the quantity $D$ tends to zero. More specifically, according to Theorem  \ref{KSLFZ11} for KS metrics and positively curved FLRW models, the quantity $D$ controls the magnitude of error between full time--dependent and time--averaged solutions as $D\rightarrow 0$. Therefore, the analysis of the system  is  reduced to study the corresponding time--averaged equation as $D\rightarrow 0$. 
However, for  KS metric the  initial region $0<Q<1, \Sigma^2+\Omega^2<1, \Sigma(1-Q^2 +3 Q \Sigma)>0$ (and for closed FLRW the  initial region $Q>0$, respectively) is not invariant for the full system \eqref{unperturbed1KS} and for the time--averaged equations \eqref{eq65a}, \eqref{eq65b}, \eqref{eq65c}, \eqref{eq65d}, and \eqref{eq65e}.
According to Remark \ref{rem1}, Theorem \ref{KSLFZ11}  is valid on a time scale $t <t^*$  where $D$ remains close to zero, but  $D(t)$ changes its monotony at a critical time $t^*$ becoming monotonic increasing  without bound. 

\noindent We have formulated Theorems \ref{thm11} and \ref{thm12} concerning the late--time behavior of our model valid when the evolution equation for $D$ is decoupled, whose proofs are based on Theorem \ref{KSLFZ11} and linear stability analysis. Hence,  we can  establish the stability of a periodic solution as it exactly matches the stability of the stationary solution of the time--averaged equation. \noindent In particular,  for KS metric the local late--time  attractor of full system \eqref{unperturbed1KS} and time--averaged system  \eqref{avrgsystKS} (where the evolution equation for $D$ is decoupled) are the following. 
\begin{enumerate}
 \item[(i)]   The anisotropic solution $P_1$  with $\bar{\Omega}_m=0$ if    $0\leq \gamma < 2$, which represents a non--flat LRS Kasner ($p_1=-\frac{1}{3}, p_2= p_3= \frac{2}{3}$)  contracting solution with $H<0$. This solution is singular at finite time $t_1=\frac{1}{3D_0}$ and is valid  for $t>t_1$.  

 \item[(ii)]   The anisotropic solution $P_3$  with $\bar{\Omega}_m=0$ if    $0\leq \gamma < 2$, which represents a Taub (flat LRS Kasner) contracting solution ($p_1=1, p_2= 0, p_3= 0$)
\cite{WE}.

    \item[(iii)]  The flat matter--dominated FLRW universe $P_6$ if  $0\leq \gamma < \frac{2}{3}$. $P_6$ represents a  quintessence fluid or a zero-acceleration (Dirac-Milne) model for $\gamma=\frac{2}{3}$. 
In the limit $\gamma=0$ we have the de Sitter solution.  
 \end{enumerate}

We have commented that, although for $t < t^*$, $D(t)$ remains close to zero, once the orbit crosses the  initial region, $D$ changes its monotony, and it becomes strictly increasing without bound. Hence, Theorem \ref{KSLFZ11}  is valid on a time scale $t D =\mathcal{O}(1)$. To investigate the region $D\rightarrow \infty$ we have used the transformation of coordinates \eqref{compactD}
that maps $[0, \infty)$ to a finite interval $[0,1)$. Therefore, the  limit $D\rightarrow +\infty$ corresponds to $T=1$ and the limit   $D\rightarrow 0$ corresponds to $T=0$. This defines a regular dynamical system over a compact phase space that allows to obtain global results. We have studied the stability of the fixed points in a compactified phase space.  \noindent These numerical results support the claim that late--time attractors in the extended phase space  $(\mathbf{x}, T)$, where $T=\frac{D}{1+D}$ and $\mathbf{x}= (\Omega, \Sigma, Q)$  for both the original system and the time--averaged  are the same for KS. When the stability of the equilibrium point of the time--averaged is analyzed in extended phase space, we find for KS metric that the extra variable $T$ introduces equilibrium points ``at infinity'', $P_1^{\infty}$ which is a non--flat LRS Kasner solution and $P_3^{\infty}$ which is Taub (flat LRS Kasner). They are contracting solutions and sink for $0\leq \gamma<2$ in the extended (global) phase space.   Their analogous points $P_1$ and $P_3$ (with $D=H=T=0$) become saddle along the $T$-axis in the extended phase space. The only equilibrium point that remains a sink for KS for $0\leq \gamma < \frac{2}{3}$ in the extended phase space is $P_6$.

\noindent Figures \ref{Figure8}, \ref{Figure9}, \ref{Figure10}, \ref{Figure11}, \ref{Figure12}, \ref{Figure13}, \ref{Figure14}, and  \ref{Figure15} are a numerical confirmation that main Theorem \ref{KSLFZ11} presented in Section \ref{SECT:II} is fulfilled for the KS metric.  That is to say, the solutions of the full system (blue lines) follow the track of the solutions of the averaged system (orange lines) for the whole $D$-range.

\noindent On the other hand, local late--time  attractors of full system \eqref{unperturbed1FLRWClosed} and time--averaged system  \eqref{avrgsystFLRWClosed} (where the evolution equation for $D$ is decoupled)  for closed FLRW metric with positive curvature are the following. 
\begin{enumerate}
  \item [(i)] The isotropic solution $P_5$  if  $1< \gamma \leq 2$. The corresponding solution is a  flat matter--dominated FLRW contracting solution with $\bar{\Omega}_m=1$.
   
 \item[(ii)] The flat matter--dominated FLRW universe $P_6$   if  $0 \leq \gamma < \frac{2}{3}$. $P_6$ represents a  quintessence fluid or a zero-acceleration (Dirac-Milne) model for $\gamma=\frac{2}{3}$. 
In the limit $\gamma=0$ we have the de Sitter solution. 
\item[(iii)]  The equilibrium point $P_7$ if $0\leq \gamma<1$. The equilibrium point  can be associated with Einstein-de Sitter solution. 
\end{enumerate}
When  the stability of the equilibrium point of the time--averaged extended phase space for closed FLRW metric is analyzed in the extended phase space  $(\mathbf{x}, T)$, where $\mathbf{x}= (\Omega, Q)$, for closed FLRW we find  that the extra variable $T$ introduces equilibrium points ``at infinity'', $P_5^{\infty}$ which is a sink for $1<\gamma\leq 2$  and $P_7^{\infty}$ which is a sink for $0 \leq\gamma<1$. As for KS, the only equilibrium point that remains a sink for KS for $0\leq \gamma < \frac{2}{3}$ in the extended phase space is $P_6$. 

\noindent In Figures \ref{Figure16}, \ref{Figure17}, \ref{Figure18}, and \ref{Figure19} we have presented projections of some solutions of the full system \eqref{unperturbed1FLRWClosed} and time--averaged system \eqref{avrgsystFLRWClosed} for $\gamma\neq 1$. Also, system \eqref{eqaverg4D} for $\gamma=1$ in the $(Q, D/(1+D), \Omega^{2})$ space  are presented with its respective projection when $D=0$. 
Figures \ref{Figure16}  and \ref{FIGURE25} show how solutions of the full system (blue lines) follow the track of solutions of the averaged system (orange lines) for the whole $D$-range. Figures \ref{Figure17}, \ref{Figure18}, \ref{Figure19}, and  \ref{FIGURE27} are evidence that  main theorem presented in Section \ref{SECT:II} is fulfilled for FLRW metric with positive curvature ($k=+1$) only when $D$ is bounded. Namely, the solutions of the full system (blue lines) follow the track of solutions of the averaged system (orange lines) for the time interval $t D =\mathcal{O}(1)$. However, when $D$ becomes infinite ($T\rightarrow 1$) and for $\gamma\geq 1$, the solutions of full system (blue lines) depart from solutions of  averaged system (orange lines) as $D$ becomes large. Then, different from KS, for the full system and given $\gamma\geq 1$ the orbits (blue lines) do not follow the track of solutions of the averaged system, while $P_8^{\infty}$ and $P_7^{\infty}$ are  early and late--time attractor, respectively, as $D\rightarrow \infty$. This is a rather different  behavior from time--averaged system, where they are saddle. 
This can be anticipated because when $D$ becomes large, the approximation obtained under the assumption of small $D$  fails.

\noindent
Additionally, for closed FLRW we have found by numerical tools the existence of spiral tubes confined in  a finite region of the phase space when  the line of equilibrium points  $E_c$ (representing Einstein's static universes)  exist. This kind of dynamical structures have been presented before in reference \cite{Heinzle:2004sr} and they exist for any matter source that admits Einstein’s static model.
Results in the line of  \cite{Heinzle:2004sr} are of interest since they disprove the claims of non-predictability and chaos for models
close to   Einstein's model related to the existence of infinitely many homoclinic orbits whose
$\alpha$ and $\omega$- limits are the same periodic orbit producing chaotic sets in the whole state space. Thus, a set of models in a neighborhood of Einstein’s model were claimed to be unpredictable and characterized by ``homoclinic chaos'' \cite{deOliveira:1997ej,Barguine:2001rb,DeOliveira:2002ih,Soares:2005fn}. However, the asserted ``homoclinic phenomena'', if they occur at all, must be confined to narrow regions of the phase space  \cite{Heinzle:2004sr} (see also \cite{Heinzle:2006zy}).

\noindent Now, we summarize the results of the ``Averaging generalized scalar field cosmologies'' research  program. In reference \cite{Leon:2021lct} it was proved that for LRS Bianchi III the late--time attractors are  a  matter--dominated flat FLRW universe  if  $0\leq \gamma \leq \frac{2}{3}$ (mimicking de Sitter, quintessence or zero acceleration solutions), a matter--curvature scaling solution  if  $\frac{2}{3}<\gamma <1$, and Bianchi III flat spacetime  for $1\leq \gamma\leq 2$.  For FLRW metric with $k=-1$ the late--time attractors are  flat matter--dominated FLRW universe  if  $0\leq \gamma \leq \frac{2}{3}$ and Milne solution if $\frac{2}{3}<\gamma <2$.  In all metrics, matter--dominated flat  FLRW universe represents quintessence fluid if $0< \gamma < \frac{2}{3}$.
In reference \cite{Leon:2021rcx} for flat FLRW and LRS Bianchi I metrics it was obtained that late--time  attractors of full and time--averaged systems are given by flat  matter--dominated  FLRW solution and Einstein-de Sitter solution.  It is interesting to note that for  FLRW with negative or zero curvature and  for Bianchi I metric when the matter fluid corresponds to a CC,  $H$ asymptotically tends  to constant values depending on  initial conditions. That is consistent with de Sitter expansion. 
In addition, for FLRW models with negative curvature for any $\gamma<\frac{2}{3}$ and $\Omega_k>0$, $\Omega_k \rightarrow 0$, or when $\gamma>\frac{2}{3}$ and $\Omega_k>0$ the universe becomes  curvature dominated  ($\Omega_k \rightarrow 1$). 
For flat FLRW and dust background from the qualitative analysis performed in paper \cite{Leon:2021rcx}  we have that
$\lim_{\tau\rightarrow +\infty}\bar{\Omega}(\tau)= \text{const.}$,  and $\lim_{\tau\rightarrow +\infty}H(\tau)=0$. Also, it was numerically proved that as $H\rightarrow 0$, the values of $\bar{\Omega}$ give an upper bound for the values $\Omega$ of the original system. Therefore, by controlling the error of the time--averaged higher order system the error of the original system can also be controlled.

\noindent Finally, in the present research we have proved that in KS metric the global late--time  attractors  of  full and time--averaged systems are  two anisotropic contracting solutions, $P_1^{\infty}$ which is a non--flat LRS Kasner and  $P_3^{\infty}$ which is a  Taub (flat LRS Kasner) for $0\leq \gamma<2$, and $P_6$ which is a  matter--dominated flat FLRW universe  if  $0\leq \gamma \leq \frac{2}{3}$ (mimicking de Sitter, quintessence or zero
acceleration solutions). For FLRW metric with $k=+1$  global late--time  attractors  of time--averaged system are  $P_5^{\infty}$ which is a flat matter--dominated contracting solution that is a sink for $1<\gamma\leq 2$,    $P_6$ which is a  matter--dominated flat FLRW universe mimicking de Sitter, quintessence or zero
acceleration solutions if  $0\leq \gamma \leq \frac{2}{3}$, and the point $P_7^{\infty}$ which is an Einstein-de Sitter solution for $0\leq\gamma<1$ and large $t$. 
However, when $D$ becomes infinite ($T\rightarrow 1$) and for $\gamma\geq 1$  solutions of the  full system (blue lines) depart from solutions of averaged system (orange lines) as $D$ is large. Then,  different from KS, for the full system and given $\gamma>1$ the orbits (blue lines) do not follow the track of the averaged system and $P_8^{\infty}$ and $P_7^{\infty}$ are the early and late--time attractor, respectively, as $D\rightarrow \infty$. This is a rather different  behavior from the time--averaged system, where they are saddle. Therefore, this analysis completes the characterization of the full class of homogeneous but anisotropic solutions and their isotropic limits with exception of LRS Bianchi V. 
Our analytical results were strongly supported by numerical results in \ref{numerics}. We have shown that asymptotic methods and averaging theory are powerful tools to investigate scalar field cosmologies with generalized harmonic potential. 

\section*{Acknowledgements}

This research was funded by  Agencia Nacional de Investigaci\'on y Desarrollo- ANID  through the program FONDECYT Iniciaci\'on grant no.
11180126 and by Vicerrector\'{\i}a de Investigaci\'on y Desarrollo Tecnol\'ogico at
Universidad Cat\'olica del Norte.  Genly Leon thanks Bertha Cuadros-Melgar for her useful comments. Ellen de los Milagros Fern\'andez Flores  is acknowledged for proofreading this manuscript and for improving the English. We thank anonymous referee for his/her valuable comments which  helped   improve our work.

\appendix

\section{Proof of Theorem \ref{KSLFZ11}}
\label{gLKSFZ11}

\begin{lem}[\textbf{Gronwall's Lemma (Integral form)}]
\label{Gronwall}
 Let $\xi(t)$ be  a nonnegative function  summable over  $[0,T]$, which satisfies almost everywhere the integral inequality $$\xi(t)\leq C_1 \int_0^t \xi(s)ds +C_2, \;  C_1, C_2\geq 0.$$
       Then, 
      $\xi(t)\leq C_2  e^{C_1 t},$
        almost everywhere for $t$ in $0\leq t\leq T$. In particular, if    
     $$\xi(t)\leq C_1 \int_0^t \xi(s)ds, \;  C_1\geq 0,$$
        almost everywhere for $t$ in $0\leq t\leq T$. Then,  $
           \xi \equiv 0$  
        almost everywhere for $t$ in $0\leq t\leq T$.
 \end{lem}
\begin{lem}[Mean value Theorem]
\label{lemma6}
 Let $U \subset \mathbb{R}^n$ be open, $\mathbf{f}: U \rightarrow \mathbb{R}^m$ continuously differentiable, and $\mathbf{x}\in U$, $\mathbf{h}\in \mathbb{R}^m$ vectors such that the line segment $\mathbf{x}+z \; \mathbf{h}$,  $0 \leq z \leq 1$ remains in $U$. Then, we have
\begin{equation}
    \mathbf{f}(\mathbf{x}+\mathbf{h})-\mathbf{f}(\mathbf{x}) = \left (\int_0^1 \mathbb{D}\mathbf{f}(\mathbf{x}+z \; \mathbf{h})\,dz\right)\cdot \mathbf{h},
\end{equation} where  $\mathbb{D} \mathbf{f}$ denotes the Jacobian matrix  of $\mathbf{f}$ and the integral of a matrix is understood as a componentwise.
\end{lem}

\subsection{Proof of Theorem \ref{KSLFZ11}.}
From eq.  \eqref{EQ:96b}
$D$ is a  monotonic decreasing function of  $t$ whenever $0<Q<1, \Sigma^2+\Omega^2<1, \Sigma(1-Q^2 +3 Q \Sigma)>0$. 
The last restriction holds, in particular, by choosing $Q>0, \Sigma>0$. In the last case  we  define the bootstrapping sequences 
\begin{align}
    & \left\{\begin{array}{c}
       t_0=t_0,   \\\\
        D_0=D(t_0) 
    \end{array}\right.,  \nonumber\\ & \left\{\begin{array}{c}
       {t_{n+1}}= {t_{n}} +\frac{1}{D_n}   \\ \\
       D_{n+1}= D(t_{n+1})  
    \end{array}\right..
\end{align}
This sequence, however, is finite; that is,  $t_n < t^*$  with $t^*$ such that  $\lim_{t\rightarrow t^*}D'(t)=0.$ 
If $Q(t_n)< 0$ or $\Sigma(t_n)< 0$ for some $n$ we stop the integration  because $D$ changes its monotony to become a monotonic increasing function. However,  $(t, \eta, \Sigma, Q, \Phi) \mapsto (-t, -\eta, -\Sigma, -Q, -\Phi)$ leaves invariant the system. Therefore, the solution is completed by using the above symmetry. 

Given  expansions \eqref{AppKSquasilinear211}, 
system \eqref{eqT59} becomes 
\begingroup\makeatletter\def\f@size{7.5}\check@mathfonts
\begin{align}
\dot{\Omega_0}= &  \frac{1}{2} D {Q_0} {\Omega_0} \Bigg(3 \gamma  \left(1-{\Sigma_0}^2-{\Omega_0}^2\right)    -2 {Q_0}
   {\Sigma_0}+6 {\Sigma_0}^2+3 {\Omega_0}^2-3\Bigg)  \nonumber \\
   & +\frac{3}{2} D {Q_0} \left({\Omega_0}^2-1\right) {\Omega_0} \cos (2 ({\Phi_0}-t \omega ))   \nonumber \\
   &   +D {\Sigma_0} {\Omega_0}  -D \frac{\partial g_1}{\partial t}+  \mathcal{O}(D^2),
\end{align}
\begin{align}
\dot{\Sigma_0}= &  \frac{1}{2} D \Bigg(\left({\Sigma_0}^2-1\right) \left(-2 {Q_0}^2-3 (\gamma -2) {Q_0} {\Sigma_0}+2\right)   -3  (\gamma -1) {Q_0} {\Sigma_0} {\Omega_0}^2\Bigg)   \nonumber \end{align}
\begin{align}
   &  +\frac{3}{2} D {Q_0} {\Sigma_0} {\Omega_0}^2 \cos (2 ({\Phi_0}-t \omega ))    -D \frac{\partial g_2}{\partial t}+  \mathcal{O}(D^2), 
\end{align}
\begin{align}
\dot{Q_0}= & \frac{1}{2} D \left(1-{Q_0}^2\right) \Bigg(3 \gamma  \left({\Sigma_0}^2+{\Omega_0}^2-1\right)    +2 {\Sigma_0} ({Q_0}-3 {\Sigma_0})-3
   {\Omega_0}^2+2\Bigg)    \nonumber \\
   &  +\frac{3}{2} D \left({Q_0}^2-1\right) {\Omega_0}^2 \cos (2 ({\Phi_0}-t \omega ))      -D \frac{\partial g_3}{\partial t}+  \mathcal{O}(D^2), 
\end{align}
\begin{align}
 \dot{\Phi_0}=  &\left(\frac{3}{2} {Q_0} \sin (2 ({\Phi_0}-t \omega
   ))-\frac{\partial g_4}{\partial t}\right) D+\mathcal{O}\left(D^2\right).
\end{align}
\endgroup
Let $\Delta\mathbf{x}_0(t)= (\Omega_0-\bar{\Omega}, {\Sigma_0}-  \bar{\Sigma}, {Q_0}-  \bar{Q})^T$  and  $\Delta \Phi_0= \Phi_0 - \bar{\Phi}$. Taking the initial conditions  $\Omega_0(t_n)=\bar{\Omega}(t_n)= {\Omega_{n}}, \;   \Sigma_0(t_n)=\bar{\Sigma}(t_n)= {\Sigma}_{n},$ $ Q_0(t_n)=\bar{Q}(t_n)= {Q}_{n}, \; \Phi_0(t_n)=\bar{\Phi}(t_n)= {\Phi}_{n},$ such that
$$0<Q_n<1, \Sigma_n^2+\Omega_n^2<1, \Sigma_n(1-Q_n^2 +3 Q_n \Sigma_n)>0,$$
Eqs. \eqref{eqT602} become
\begin{subequations}
\label{EDOgLKSFZ11}
  \begin{align}
   & \frac{\partial g_1}{\partial t}=  \frac{3}{2} {Q_0} {\Omega_0} \left({\Omega_0}^2-1\right) \cos (2 ({\Phi_0}-t \omega )),
\\
   & \frac{\partial g_2}{\partial t}=  \frac{3}{2} {Q_0} {\Sigma_0} {\Omega_0}^2 \cos (2 ({\Phi_0}-t \omega )), \\
   &\frac{\partial g_3}{\partial t}=  \frac{3}{2} \left({Q_0}^2-1\right) {\Omega_0}^2 \cos (2
   ({\Phi_0}-t \omega )), 
\\
   &\frac{\partial g_4}{\partial t}= \frac{3}{2} {Q_0}
   \sin (2 ({\Phi_0}-t \omega )),
  \end{align} 
    \end{subequations}
Explicit expressions for $g_i$  obtained   by integration of \eqref{EDOgLKSFZ11} are given by  
\begingroup\makeatletter\def\f@size{8.0}\check@mathfonts
\begin{align}
    & g_1(D, \Omega_{0}, \Sigma_{0}, Q_{0}, \Phi_{0}, t)  =   \frac{3}{4 \omega} {Q_0} {\Omega_0} \left(1-{\Omega_0}^2\right) \sin (2 ({\Phi_0}-t \omega )), 
\\
    & g_2(D, \Omega_{0}, \Sigma_{0}, Q_{0}, \Phi_{0}, t) =   -\frac{3}{4 \omega}  {Q_0} {\Sigma_0} {\Omega_0}^2 \sin (2 ({\Phi_0}-t \omega )), 
\\
    & g_3(D, \Omega_{0}, \Sigma_{0}, Q_{0}, \Phi_{0}, t)  =  \frac{3}{4 \omega } \left(1-{Q_0}^2\right) {\Omega_0}^2 \sin (2 ({\Phi_0}-t \omega )), 
\\
    &g_4(D, \Omega_{0}, \Sigma_{0}, Q_{0}, \Phi_{0}, t)   =   \frac{3 Q_0 \cos (2 (\Phi_0-t \omega ))}{4 \omega }, 
\end{align}
\endgroup
where we have set four integration functions \newline $C_i(D, \Omega_{0}, \Sigma_{0}, Q_{0}, \Phi_{0}), i=1, 2, 3, 4$ to zero. 
\noindent
The functions  $g_i, i=1, 2, 3, 4$ are continuously differentiable such that their partial derivatives are bounded on $t\in [t_n, t_{n+1}]$. 

\noindent
The second order expansion around $D=0$ of system \eqref{EqY602} is written as
\begin{widetext}
\begingroup\makeatletter\def\f@size{7.5}\check@mathfonts
 \begin{align}
 & \dot{\Delta\Omega_0}= \frac{1}{2} D \Bigg(\bar{\Omega } \left(3 \bar{Q} \left((\gamma -2) \bar{\Sigma }^2+(\gamma -1) \left(\bar{\Omega }^2-1\right)\right)+2 \bar{Q}^2 \bar{\Sigma }-2 \bar{\Sigma }\right)   +\Omega_0 \left(Q_0 \left(-3
   \gamma  \left(\Sigma_0^2+\Omega_0^2-1\right)-2 Q_0 \Sigma_0+6 \Sigma_0^2+3 \Omega_0^2-3\right)+2 \Sigma_0\right)\Bigg) \nonumber \\
   & +D^2 \sin (2 (\Phi_0-t \omega ))
   \Bigg( \frac{\Omega_0 \cos (2 (\Phi_0-t \omega )) \left(-9 b^2 \mu ^6 \omega ^2 \left(\Omega_0^2-1\right)
   \left(\left(5 Q_0^2-1\right) \Omega_0^2+Q_0^2\right)-2 \Omega_0^2 \left(2 \mu ^2-\omega ^2\right)^3\right)}{8 b^2 \mu ^6 \omega ^3} +\frac{\Omega_0^3 \left(2 \mu ^2-\omega ^2\right)^3}{4 b^2 \mu ^6 \omega ^3} \nonumber \\
   & +\frac{3 \Omega_0 \left(-3 \gamma  \Sigma_0^2+3
   \gamma +\Omega_0^2 \left(-3 \gamma +Q_0 \left(2 \left(Q_0^2-1\right) \Sigma_0+3 (\gamma -2) Q_0 \Sigma_0^2-3 \gamma  Q_0+Q_0\right)+2\right)+3 (\gamma -1) Q_0^2
   \Omega_0^4+2 Q_0^2+6 \Sigma_0^2-2\right)}{8 \omega }\Bigg), \label{firsteq}
\\
   & \dot{\Delta\Sigma_0}=\frac{1}{2} D \left(\left(\bar{Q}^2-1\right) \left(3 (\gamma -2) \bar{\Sigma }^2+3 (\gamma -1) \bar{\Omega }^2+2 \bar{Q} \bar{\Sigma }-3 \gamma +2\right)+\left(\Sigma_0^2-1\right) \left(-2 Q_0^2-3 (\gamma -2)
   Q_0 \Sigma_0+2\right)-3 (\gamma -1) Q_0 \Sigma_0 \Omega_0^2\right) \nonumber\\
   & +D^2 \sin (2 (\Phi_0-t \omega )) \Bigg(\frac{3 \Omega_0^2 \left(Q_0 \left(2 Q_0^2
   \left(\Sigma_0^2-1\right)+Q_0 \Sigma_0 \left(3 \gamma  \left(\Sigma_0^2+\Omega_0^2-1\right)-6 \Sigma_0^2-3 \Omega_0^2+4\right)-2 \Sigma_0^2+2\right)-4
   \Sigma_0\right)}{8 \omega } \nonumber \\
   & \quad \quad \quad \quad \quad \quad \quad \quad \quad \quad   +\frac{9 \left(1-5 Q_0^2\right) \Sigma_0 \Omega_0^4 \cos (2 (\Phi_0-t \omega ))}{8 \omega }\Bigg), \label{secondeq}
\\
 & \dot{\Delta Q_0}=\frac{1}{2} D \left(3 (\gamma -1) \bar{Q} \bar{\Sigma } \bar{\Omega }^2-\left(\bar{\Sigma }^2-1\right) \left(-3 (\gamma -2) \bar{Q} \bar{\Sigma }-2 \bar{Q}^2+2\right)+\left(1-Q_0^2\right) \left(3 \gamma  \left(\Sigma_0^2+\Omega_0^2-1\right)+2 \Sigma_0 (Q_0-3 \Sigma_0)-3 \Omega_0^2+2\right)\right) \nonumber \\
 & +D^2 \sin (2 (\Phi_0-t \omega )) \left(\frac{3 Q_0 \left(Q_0^2-1\right)
   \Omega_0^2 \left(3 \gamma  \left(\Sigma_0^2+\Omega_0^2-1\right)+2 \Sigma_0 (Q_0-3 \Sigma_0)-3 \Omega_0^2\right)}{8 \omega }-\frac{45 Q_0
   \left(Q_0^2-1\right) \Omega_0^4 \cos (2 (\Phi_0-t \omega ))}{8 \omega }\right), \label{thirdeq}
\\
   & \dot{\Delta\Phi_0}=D^2 \Bigg(\frac{3 \bar{\Omega }^2 \left(2 \mu ^2-\omega ^2\right)^3}{8 b^2 \mu ^6 \omega ^3}+\frac{\cos (2 (\Phi_0-t \omega )) \left(3 b^2 \mu ^6 \omega ^2 \left(-3 \gamma  \left(\Sigma_0^2+ \Omega_{0}^2-1\right)+2 Q_0^2+6 \Sigma_0^2+3 \Omega_0^2-2\right)+4 \Omega_0^2 \left(2 \mu ^2-\omega ^2\right)^3\right)}{8 b^2 \mu ^6 \omega ^3} \nonumber \\
   & +\frac{9 b^2 \mu ^6 \omega ^2 \left(2
   Q_0^2+\Omega_0^2\right)+\cos (4 (\Phi_0-t \omega )) \left(9 b^2 \mu ^6 \omega ^2 \left(2 Q_0^2 \left(\Omega_0^2+1\right)-\Omega_0^2\right)-2 \Omega_0^2 \left(2 \mu
   ^2-\omega ^2\right)^3\right)-6 \Omega_0^2 \left(2 \mu ^2-\omega ^2\right)^3}{16 b^2 \mu ^6 \omega ^3}\Bigg). \label{lasteq}
  \end{align}
  \endgroup
Denoting $\mathbf{x}_0=(\Omega_0, \Sigma_0, Q_0)^T$, $\bar{\mathbf{x}}=(\bar{\Omega}, \bar{\Sigma}, \bar{Q})^T$, and $\Delta\mathbf{x}_0(t)= (\Omega_0-\bar{\Omega}, {\Sigma_0}-  \bar{\Sigma}, {Q_0}-  \bar{Q})^T$ with $0\leq |\Delta\mathbf{x}_0|:=\max \left\{|\Omega_0-\bar{\Omega}|, |{\Sigma_0}-  \bar{\Sigma}|, |{Q_0}-  \bar{Q}| \right\}< \infty$ in the closed interval $[t_n,t]$, Eqs. \eqref{EqY602} are reduced to a 3-dimensional system  
\begin{align}
 & \dot{\Delta\mathbf{x}_0}= D \left(\bar{\mathbf{f}}( {\mathbf{x}}_0)-\bar{\mathbf{f}}(\bar{\mathbf{x}})\right) +   \mathcal{O}(D^2), 
  \end{align} 
plus eq. \eqref{lasteq},  
where the vector function $\bar{\mathbf{f}}$ in eq.   \eqref{EqY602}
is explicitly given by  
\begin{align*}
    &\bar{\mathbf{f}}(y_1, y_2, y_3) = \left(
\begin{array}{c}
-\frac{1}{2} y_1 \left(3 (\gamma -1)
   \left(y_1^2-1\right) y_3+3 (\gamma -2)
   y_2^2 y_3+2 y_2
   \left(y_3^2-1\right)\right) \\
\frac{1}{2}
   \left(\left(y_2^2-1\right) \left(-3 (\gamma -2)
   y_2 y_3-2 y_3^2+2\right)-3 (\gamma -1)
   y_1^2 y_2 y_3\right)\\
-\frac{1}{2}
   \left(y_3^2-1\right) \left(-3 \gamma +3 (\gamma -1)
   y_1^2+3 (\gamma -2) y_2^2+2 y_2
   y_3+2\right)\\
\end{array}
\right).
\end{align*}
The last row corresponding to $\dot{\Delta\Phi_0}$ was omitted. 
This vector function with polynomial components in variables $(y_1, y_2, y_3)$ is continuously differentiable in all its components.

Using the same initial conditions for $\mathbf{x}_0$ and $\bar{\mathbf{x}}$, we obtain by integration 
\begingroup\makeatletter\def\f@size{8}\check@mathfonts
\begin{align}
\label{int}
 & \Delta\mathbf{x}_0(t) = \int_{t_n}^t \dot{\Delta\mathbf{x}_0} d s  =  \int_{t_n}^t \left(D \left(\bar{\mathbf{f}}( {\mathbf{x}}_0)-\bar{\mathbf{f}}(\bar{\mathbf{x}})\right) +   \mathcal{O}(D^2)\right) ds. 
\end{align}
\endgroup
The terms of order $\mathcal{O}(D^2)$ under the integral sign in eq.   \eqref{int} come from the second order terms in the series expansion centered in $D=0$ of $\dot \Delta \Omega_0$, $\dot \Delta \Sigma_0$, and  $\dot \Delta Q_0$ in eqs. \eqref{firsteq}, \eqref{secondeq}, and \eqref{thirdeq}. These terms are bounded in the interval $[t_n, t_{n+1}]$ by $M_1 D_n^2$, where 
\begingroup\makeatletter\def\f@size{8}\check@mathfonts
\begin{align*}
& M_1= \max_{t\in[t_{n},t_{n+1}]}  \Bigg\{ \Bigg{|} \frac{\Omega_0 \cos (2 (\Phi_0-t \omega )) \left(-9 b^2 \mu ^6 \omega ^2 \left(\Omega_0^2-1\right)
   \left(\left(5 Q_0^2-1\right) \Omega_0^2+Q_0^2\right)-2 \Omega_0^2 \left(2 \mu ^2-\omega ^2\right)^3\right)}{8 b^2 \mu ^6 \omega ^3} +\frac{\Omega_0^3 \left(2 \mu ^2-\omega ^2\right)^3}{4 b^2 \mu ^6 \omega ^3} \nonumber \\
   & +\frac{3 \Omega_0 \left(-3 \gamma  \Sigma_0^2+3
   \gamma +\Omega_0^2 \left(-3 \gamma +Q_0 \left(2 \left(Q_0^2-1\right) \Sigma_0+3 (\gamma -2) Q_0 \Sigma_0^2-3 \gamma  Q_0+Q_0\right)+2\right)+3 (\gamma -1) Q_0^2
   \Omega_0^4+2 Q_0^2+6 \Sigma_0^2-2\right)}{8 \omega }\Bigg{|}, \nonumber \\
   & \Bigg{|}\frac{3 \Omega_0^2 \left(Q_0 \left(2 Q_0^2
   \left(\Sigma_0^2-1\right)+Q_0 \Sigma_0 \left(3 \gamma  \left(\Sigma_0^2+\Omega_0^2-1\right)-6 \Sigma_0^2-3 \Omega_0^2+4\right)-2 \Sigma_0^2+2\right)-4
   \Sigma_0\right)}{8 \omega } +\frac{9 \left(1-5 Q_0^2\right) \Sigma_0 \Omega_0^4 \cos (2 (\Phi_0-t \omega ))}{8 \omega }\Bigg{|}, \nonumber \\
   & \Bigg{|}\frac{3 Q_0 \left(Q_0^2-1\right)
   \Omega_0^2 \left(3 \gamma  \left(\Sigma_0^2+\Omega_0^2-1\right)+2 \Sigma_0 (Q_0-3 \Sigma_0)-3 \Omega_0^2\right)}{8 \omega }-\frac{45 Q_0
   \left(Q_0^2-1\right) \Omega_0^4 \cos (2 (\Phi_0-t \omega ))}{8 \omega }\Bigg{|} \Bigg\},
\end{align*}
\endgroup
\noindent
is finite by continuity of $\bar{\Omega}, \Omega_0, \bar{\Sigma}, \Sigma_0, \bar{Q}, Q_0, \Phi_0$ in the interval $[t_{n},t_{n+1}]$.  Taking the sup norm 
$ |\Delta\mathbf{x}_0|=\max \left\{|\Omega_0-\bar{\Omega}|, |{\Sigma_0}-  \bar{\Sigma}|, |{Q_0}-  \bar{Q}| \right\}$ we have  
\begingroup\makeatletter\def\f@size{10}\check@mathfonts
\begin{align*}
 & \Big{|}\Delta\mathbf{x}_0(t) \Big{|} = \Big{|} \int_{t_n}^t \dot{\Delta\mathbf{x}_0} d s \Big{|}   = \Bigg{|} \int_{t_n}^t \Bigg(D \left(\bar{\mathbf{f}}( {\mathbf{x}}_0)-\bar{\mathbf{f}}(\bar{\mathbf{x}})\right) +   \mathcal{O}(D^2)\Bigg) ds  \Bigg{|}  \nonumber \\
 & \leq D_n \int_{t_n}^t  \Big{|} \bar{\mathbf{f}}( {\mathbf{x}}_0)-\bar{\mathbf{f}}(\bar{\mathbf{x}})\Big{|} ds +  M_1 D_n^2 (t-t_n),
\end{align*}
\endgroup
for all $t\in[t_n, t_{n+1}]$. \newline 
Using Lemma \ref{lemma6} we have 
\begin{align}
   & \bar{\mathbf{f}}( {\mathbf{x}}_0(s))-\bar{\mathbf{f}}(\bar{\mathbf{x}}(s)) =  {\mathbb{A}(s)}\cdot \left({\mathbf{x}}_0(s) - \bar{\mathbf{x}}(s)\right),
\end{align} 
where 
\begin{equation}
\mathbb{A}(s)=    \left (\int_0^1 \mathbb{D}   \bar{\mathbf{f}}\left(\bar{\mathbf{x}}(s)+\tau \left({\mathbf{x}}_0(s) - \bar{\mathbf{x}}(s)\right)\right)\,d\tau\right),
\end{equation} $\mathbb{D}\bar{\mathbf{f}}$ denotes the Jacobian matrix  of $\bar{\mathbf{f}}$ and the integral of a matrix is  understood as a componentwise.
We denote the matrix elements of $\mathbb{A}$ as
\begin{equation}
    \mathbb{A}(s)= \left(
\begin{array}{ccc}
 a_{1 1}(s) & a_{1 2}(s) & a_{1 3}(s)\\
 a_{2 1}(s) & a_{2 2}(s) & a_{2 3}(s)\\
 a_{3 1}(s) & a_{3 2}(s) & a_{3 3}(s)\\
\end{array}
\right),
\end{equation}
where $a_{i j}$ are polynomial functions of  $\bar{\Omega}, \Omega_0, \bar{\Sigma}, \Sigma_0, \bar{Q}, Q_0$, and they are explicitly given by 
\begin{small}
\begin{subequations}
\label{aij}
\begin{align}
    a_{1 1}(s)=&\frac{1}{24} \bar{Q} \left(-6 (\gamma -2) \Sigma_0
   \bar{\Sigma }-9 (\gamma -2) \bar{\Sigma }^2-9 (\gamma -1)
   \left(2 \Omega_0 \bar{\Omega }+3 \bar{\Omega
   }^2+\Omega_0^2-2\right)-3 (\gamma -2)
   \Sigma_0^2\right)\nonumber\\& +Q_0 \left(\frac{1}{24}
   \left(-6 (\gamma -2) \Sigma_0 \bar{\Sigma }-3
   (\gamma -2) \bar{\Sigma }^2-9 (\gamma -1) \left(2
   \Omega_0 \bar{\Omega }+\bar{\Omega }^2+3
   \Omega_0^2-2\right)-9 (\gamma -2) \Sigma_0^2\right)+\frac{1}{6} \bar{Q} \left(-\bar{\Sigma
   }-\Sigma_0\right)\right)\nonumber\\& +\frac{1}{12} \bar{Q}^2
   \left(-3 \bar{\Sigma }-\Sigma_0\right)+\frac{1}{12} Q_0^2 \left(-\bar{\Sigma }-3
   \Sigma_0\right)+\frac{1}{2} \left(\bar{\Sigma
   }+\Sigma_0\right),
\\
 a_{1 2}(s)=&-\frac{1}{4} (\gamma -2) \bar{Q} \left(\bar{\Sigma } \left(3
   \bar{\Omega }+\Omega_0\right)+\Sigma_0
   \left(\bar{\Omega }+\Omega_0\right)\right)+Q_0 \left(\frac{1}{6} \bar{Q}
   \left(-\bar{\Omega }-\Omega_0\right)-\frac{1}{4}
   (\gamma -2) \left(\bar{\Sigma } \left(\bar{\Omega
   }+\Omega_0\right)+\Sigma_0
   \left(\bar{\Omega }+3 \Omega_0\right)\right)\right)\nonumber \\ & +\frac{1}{12} \bar{Q}^2 \left(-3
   \bar{\Omega }-\Omega_0\right)+\frac{1}{12}
   Q_0^2 \left(-\bar{\Omega }-3 \Omega_0\right)+\frac{1}{2} \left(\bar{\Omega }+\Omega_0\right),
\\
 a_{1 3}(s)=&\frac{1}{8} \Bigg(-2 (\gamma -2) \Sigma_0
   \bar{\Sigma } \left(\bar{\Omega }+\Omega_0\right)-(\gamma -2) \bar{\Sigma }^2 \left(3 \bar{\Omega
   }+\Omega_0\right)+\bar{\Omega } \left(-3 (\gamma
   -1) \bar{\Omega } \left(\bar{\Omega }+\Omega_0\right)-\gamma  \left(\Sigma_0^2+3
   \Omega_0^2-6\right)+2 \Sigma_0^2+3
   \Omega_0^2-6\right)\nonumber \\ & +3 \Omega_0
   \left(-\gamma  \left(\Sigma_0^2+\Omega_0^2-2\right)+2 \Sigma_0^2+\Omega_0^2-2\right)\Bigg) +\frac{1}{6} \bar{Q}
   \left(-\bar{\Sigma } \left(3 \bar{\Omega }+\Omega_0\right)-\Sigma_0 \left(\bar{\Omega
   }+\Omega_0\right)\right)       \nonumber \\ & +\frac{1}{6} Q_0
   \left(-\bar{\Sigma } \left(\bar{\Omega }+\Omega_0\right)-\Sigma_0 \left(\bar{\Omega }+3
   \Omega_0\right)\right),
\\
 a_{2 1}(s)=&-\frac{1}{4} (\gamma -1) \bar{Q} \left(\bar{\Sigma }
   \left(3 \bar{\Omega }+\Omega_0\right)+\Sigma_0 \left(\bar{\Omega
   }+\Omega_0\right)\right)-\frac{1}{4} (\gamma -1)
   Q_0 \left(\bar{\Sigma } \left(\bar{\Omega
   }+\Omega_0\right)+\Sigma_0
   \left(\bar{\Omega }+3 \Omega_0\right)\right),
\\
 a_{2 2}(s)=&\frac{1}{8} \bar{Q} \left(-6 (\gamma -2)
   \Sigma_0 \bar{\Sigma }-9 (\gamma -2) \bar{\Sigma
   }^2-(\gamma -1) \bar{\Omega } \left(3 \bar{\Omega }+2
   \Omega_0\right)-3 (\gamma -2) \left(\Sigma_0^2-2\right)-(\gamma -1) \Omega_0^2\right)\nonumber \\ & +Q_0 \left(\frac{1}{8} \left(-6 (\gamma
   -2) \Sigma_0 \bar{\Sigma }-3 (\gamma -2)
   \bar{\Sigma }^2-(\gamma -1) \bar{\Omega }
   \left(\bar{\Omega }+2 \Omega_0\right)-3 (\gamma
   -2) \left(3 \Sigma_0^2-2\right)-3 (\gamma -1)
   \Omega_0^2\right)+\frac{1}{3} \bar{Q}
   \left(-\bar{\Sigma }-\Sigma_0\right)\right)\nonumber\\& +\frac{1}{6} \bar{Q}^2 \left(-3
   \bar{\Sigma }-\Sigma_0\right)+\frac{1}{6}
   Q_0^2 \left(-\bar{\Sigma }-3 \Sigma_0\right)+\bar{\Sigma }+\Sigma_0,
\\
 a_{2 3}(s)=&\frac{1}{8}
   \Bigg(\bar{\Sigma } \left(-3 (\gamma -2) \bar{\Sigma }
   \left(\bar{\Sigma }+\Sigma_0\right)-3 (\gamma -2)
   \left(\Sigma_0^2-2\right)-(\gamma -1)
   \Omega_0^2\right)\nonumber \\ & -2 (\gamma -1) \Omega_0
   \bar{\Omega } \left(\bar{\Sigma }+\Sigma_0\right)-(\gamma -1) \bar{\Omega }^2 \left(3 \bar{\Sigma
   }+\Sigma_0\right)-3 (\gamma -2) \Sigma_0
   \left(\Sigma_0^2-2\right)-3 (\gamma -1)
   \Sigma_0 \Omega_0^2\Bigg)\nonumber \\ &+\frac{1}{6}
   \bar{Q} \left(-2 \Sigma_0 \bar{\Sigma }-3
   \bar{\Sigma }^2-\Sigma_0^2+6\right)+\frac{1}{6}
   Q_0 \left(-\bar{\Sigma } \left(\bar{\Sigma }+2
   \Sigma_0\right)-3 \Sigma_0^2+6\right),
\\
 a_{3 1}(s)=&\frac{3}{2} (\gamma -1) \left(\bar{\Omega
   }+\Omega_0\right)-\frac{1}{4} (\gamma -1)
   \bar{Q}^2 \left(3 \bar{\Omega }+\Omega_0\right)-\frac{1}{2} (\gamma -1) Q_0 \bar{Q}
   \left(\bar{\Omega }+\Omega_0\right)-\frac{1}{4}
   (\gamma -1) Q_0^2 \left(\bar{\Omega }+3
   \Omega_0\right),
\\
 a_{3 2}(s)=&\frac{3}{2} (\gamma -2)
   \left(\bar{\Sigma }+\Sigma_0\right)-\frac{1}{4}
   (\gamma -2) \bar{Q}^2 \left(3 \bar{\Sigma }+\Sigma_0\right) +Q_0^2 \left(-\frac{1}{4} (\gamma -2)
   \left(\bar{\Sigma }+3 \Sigma_0\right)-\frac{\bar{Q}}{4}\right)\nonumber \\& +Q_0
   \left(-\frac{1}{2} (\gamma -2) \bar{Q} \left(\bar{\Sigma
   }+\Sigma_0\right)-\frac{\bar{Q}^2}{4}+\frac{1}{2}\right) -\frac{\bar{Q}^3}{4}+\frac{\bar{Q}}{2}-\frac{Q_0^3}{4}, 
\\
 a_{3 3}(s)=&\frac{
   1}{4} \bar{Q} \left(-2 (\gamma -2) \Sigma_0
   \bar{\Sigma }-3 (\gamma -2) \bar{\Sigma }^2-(\gamma -1)
   \bar{\Omega } \left(3 \bar{\Omega }+2 \Omega_0\right)-\gamma  \left(\Sigma_0^2+\Omega_0^2-6\right)+2 \Sigma_0^2+\Omega_0^2-4\right)\nonumber \\ & +Q_0 \left(\frac{1}{4} \left(-2
   (\gamma -2) \Sigma_0 \bar{\Sigma }-(\gamma -2)
   \bar{\Sigma }^2-(\gamma -1) \bar{\Omega }
   \left(\bar{\Omega }+2 \Omega_0\right)-3 \gamma 
   \left(\Sigma_0^2+\Omega_0^2-2\right)+6
   \Sigma_0^2+3 \Omega_0^2-4\right)+\frac{1}{2} \bar{Q} \left(-\bar{\Sigma
   }-\Sigma_0\right)\right)\nonumber \\ & +\frac{1}{4} \bar{Q}^2
   \left(-3 \bar{\Sigma }-\Sigma_0\right)+\frac{1}{4} Q_0^2 \left(-\bar{\Sigma }-3
   \Sigma_0\right)+\frac{1}{2} \left(\bar{\Sigma
   }+\Sigma_0\right).
\end{align}
\end{subequations}
\end{small}
Furthermore, $|\mathbb{A} \cdot \Delta\mathbf{x}_0 | \leq 3  |\mathbb{A} | |\Delta\mathbf{x}_0 |$
where the sup norm of a matrix
$ {|} \left(
\begin{array}{cc}
a_{i j}
\end{array}
\right) {|}$ is defined by $\max\{|a_{i j}|, i=1,2,3, j=1,2,3\}$ with $a_{i j}$ given by eqs. \eqref{aij}. 
Define $L_1= 3 \max_{s\in[t_n,t_{n+1}]}|
\mathbb{A}(s)|$, 
which is constant by continuity of $\bar{\Omega}, \Omega_0, \bar{\Sigma}, \Sigma_0, \bar{Q}, Q_0$  in $[t_n, t_{n+1}]$. 
Therefore, 
\begin{small}
\begin{align*}
 & \Big{|}\Delta\mathbf{x}_0(t) \Big{|}  \leq D_n \int_{t_n}^t  \Big{|} \mathbb{A} (s) \cdot \Delta\mathbf{x}_0(s) \Big{|} ds  +  M_1 D_n^2 (t-t_n) \nonumber \\
 & \leq  L_1 D_n \int_{t_n}^t  \Big{|}  \Delta\mathbf{x}_0(s) \Big{|} ds +   M_1 D_n.
\end{align*}
\end{small}
Using   Gronwall's Lemma \ref{Gronwall}  we have for $t \in[t_n, t_{n+1}]$, 
\begin{align*}
 & \Big{|} \Delta \mathbf{x}_0(t)  \Big{|} \leq   M_1  D_n   e^{L_1  D_n(t-t_n)} \leq    M_1  {D_n}e^{L_1},
 \end{align*} 
due to $t-t_n\leq {t_{n+1}}- {t_{n}} =\frac{1}{D_n}$.
 Then, 
 \begin{small}
 \begin{align*}
& \Big{|} \Delta \Omega_0(t) \Big{|} \leq    M_1 e^{L_1} {D_n}, \nonumber \\ 
& \Big{|} \Delta \Sigma_0(t) \Big{|} \leq     M_1 e^{L_1} {D_n}, \\ &  \Big{|} \Delta \Omega_{k0}(t) \Big{|} \leq   M_1 e^{L_1} {D_n}.
\end{align*} 
 \end{small}
Furthermore,  defining 
\begingroup\makeatletter\def\f@size{8}\check@mathfonts
\begin{align*}
M_2= & \max_{t\in[t_{n},t_{n+1}]}\Bigg{|} \frac{3 \bar{\Omega }^2 \left(2 \mu ^2-\omega ^2\right)^3}{8 b^2 \mu ^6 \omega ^3}  +\frac{\cos (2 (\Phi_0-t \omega )) \left(3 b^2 \mu ^6 \omega ^2 \left(-3 \gamma  \left(\Sigma_0^2+ \Omega_{0}^2-1\right)+2 Q_0^2+6 \Sigma_0^2+3 \Omega_0^2-2\right)+4 \Omega_0^2 \left(2 \mu ^2-\omega ^2\right)^3\right)}{8 b^2 \mu ^6 \omega ^3} \nonumber \\
   & +\frac{9 b^2 \mu ^6 \omega ^2 \left(2
   Q_0^2+\Omega_0^2\right)+\cos (4 (\Phi_0-t \omega )) \left(9 b^2 \mu ^6 \omega ^2 \left(2 Q_0^2 \left(\Omega_0^2+1\right)-\Omega_0^2\right)-2 \Omega_0^2 \left(2 \mu
   ^2-\omega ^2\right)^3\right)-6 \Omega_0^2 \left(2 \mu ^2-\omega ^2\right)^3}{16 b^2 \mu ^6 \omega ^3} \Bigg{|},
\end{align*}
\endgroup 
\end{widetext}
\noindent which is finite by continuity of $\bar{\Omega}, \Omega_0, \bar{\Sigma}, \Sigma_0,  \Phi_0$ in the closed interval $[t_{n},t_{n+1}]$. We obtain from eq.   \eqref{lasteq} that
\begin{align*}
& |\Delta \Phi_0(t)| = \Big{|} \int_{t_n}^{t} \dot{\Delta\Phi_0}(s)  d s\Big{|}   \leq  M_2 D_n^2 (t-t_n) \leq  M_2 D_n,
\end{align*}
due to $t-t_n\leq {t_{n+1}}- {t_{n}} =\frac{1}{D_n}$.
Finally, it follows that the functions $\Omega_{0}, \Sigma_{0}, Q_0, \Phi_0$ and  $\bar{\Omega},  \bar{\Sigma}, \bar{Q}, \bar{\Phi}$  have the same asymptotics as $D_n \rightarrow 0$.
$\square$

\section{Numerical simulation}
\label{numerics}

In this section we present the numerical evidence that supports the main theorem presented in  section \ref{SECT:II}. Precisely, we numerically solve full and time--averaged systems obtained for KS and FLRW with positive curvature metrics. 
\newline 
To this purpose  an algorithm in the programming language \textit{Python} was elaborated. This program solves the systems of differential equations using the \textit{solve\_ivp} code provided by the \textit{SciPy} open-source \textit{Python}-based ecosystem. The used integration method was \textit{Radau}, which is an implicit Runge-Kutta method of the Radau IIa family of order $5$  with  relative and absolute tolerances of $10^{-4}$ and $10^{-7}$, respectively. All systems of differential equations were integrated with respect to $\eta$, instead of $t$,  with an integration range of $-40\leq\eta\leq 10$ for the original system and $-40\leq\eta\leq 100$ for the time--averaged system. All of them were partitioned in $10000$ data points.  
For numerical simulations in Figures  \ref{FIGURE25} and \ref{FIGURE27} we have reduced the integration range to $-30\leq\eta\leq 10$ for the original system and $-30\leq\eta\leq 30$ for the time--averaged system, partitioned in $5000$ data points.
\newline 
Furthermore, each full and time--averaged systems were solved considering only one matter component. The chosen components were CC ($\gamma=0$), non relativistic matter or dust ($\gamma=1$), radiation ($\gamma=\frac{4}{3}$), and stiff fluid ($\gamma=2$). Vacuum solutions correspond to those where $\Omega=\Omega_m\equiv 0$ and  solutions without matter correspond to $\Omega_m\equiv 0$. 
\newline 
Finally, we have considered the fixed constants $\mu=\sqrt{2}/2$, $b=\sqrt{2}/5$ and $\omega=\sqrt{2}$,  $f=\frac{b \mu ^3}{\omega ^2-2 \mu ^2}=1/10 \geq 0$,  such that the  generalized harmonic potential becomes $V(\phi)=\frac{\phi ^2}{2}+\frac{1}{100}(1-\cos(10\phi))$.

In Figures \ref{Figure20} and \ref{Figure21} projections of the orbits showing this behavior in the $(Q, D/(1+D), \Omega^{2})$ space are presented with their respective projection when $D=0$. Figures \ref{fig:AClosedFLRWRad3D} and \ref{fig:AClosedFLRWRad2D} show solutions for a fluid corresponding to radiation ($\gamma=\frac{4}{3}$). Figures \ref{fig:AClosedFLRWStiff3D} and \ref{fig:AClosedFLRWStiff2D} show solutions for a fluid corresponding to stiff fluid ($\gamma=2$).  

\noindent
This dynamical behavior related to spiral tubes  has been presented before in the literature in \cite{Heinzle:2004sr} and it is related to the  fact that the line of equilibrium points  $E_c$ (representing Einstein's static universes) has purely imaginary eigenvalues. 
 We present numerical results in the same line of \cite{Heinzle:2004sr} in Figures \ref{Figure20} and \ref{Figure21}. 

\subsection{Kantowski-Sachs}
\label{plotsKS}
\begin{figure*}
    \centering
    \subfigure[\label{fig:KSCC3DS} Projections in the space $(\Sigma, D/(1+D), \Omega^2)$. The surface is given by the constraint $\Omega^{2}=1-\Sigma^{2}$.]{\includegraphics[scale = 0.4]{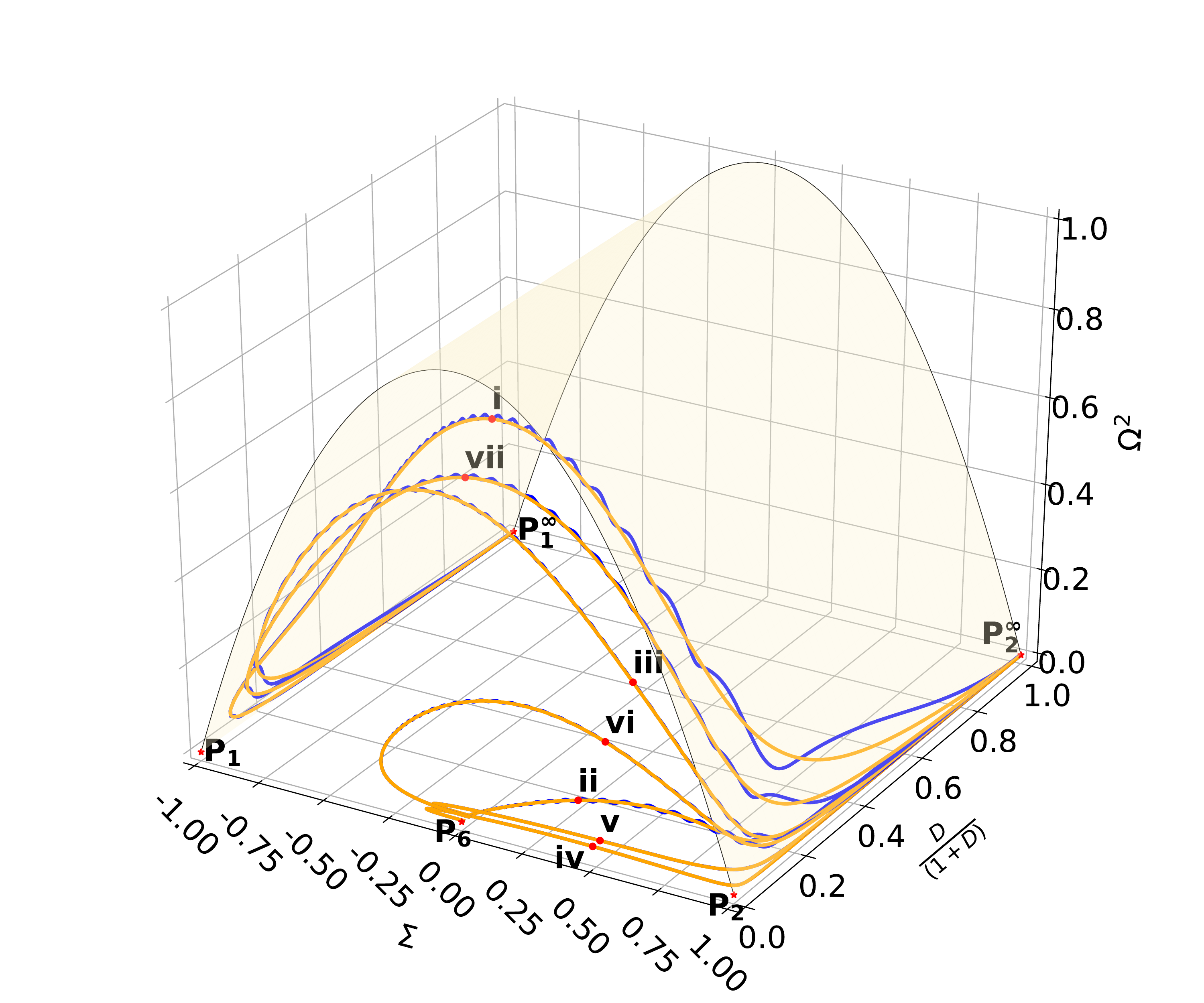}}
    \subfigure[\label{fig:KSCC2DS} Projection in the space $(\Sigma, \Omega^2)$. The black line represent the constraint $\Omega^{2}=1-\Sigma^{2}$.]{\includegraphics[scale = 0.5]{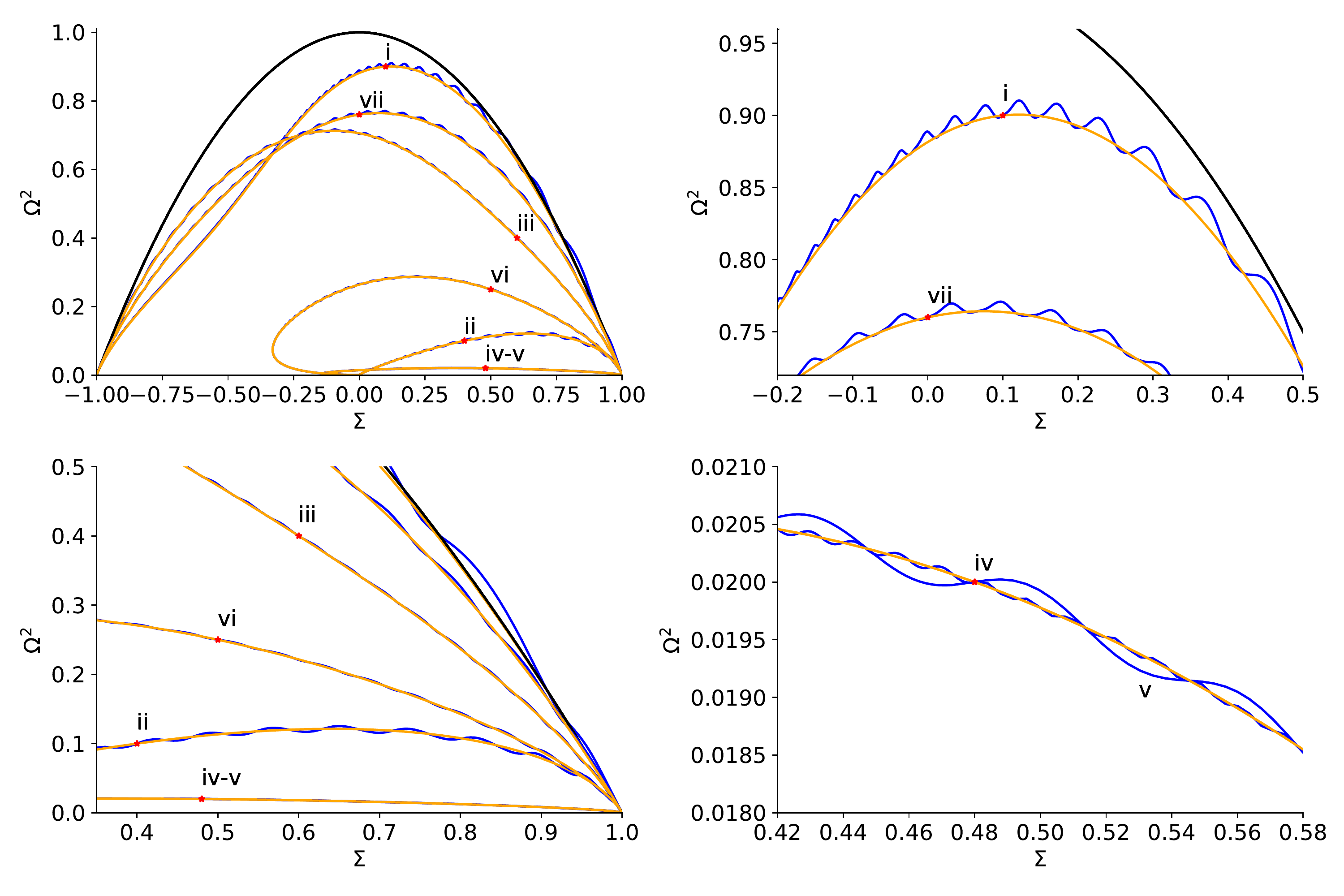}}
    \caption{Some solutions of the full system \eqref{unperturbed1KS} (blue) and time--averaged system \eqref{avrgsystKS} (orange) for the KS metric when $\gamma=0$  in the projection $Q=0$. We have used for both systems the initial data sets presented in Table \ref{Tab4}. \label{Figure8}}
\end{figure*}

\begin{figure*}
    \centering
    \subfigure[\label{fig:KSCC3DQ} Projections in the space $(Q, D/(1+D), \Omega^2)$. The surface is given by the constraint $Q=0$.]{\includegraphics[scale = 0.4]{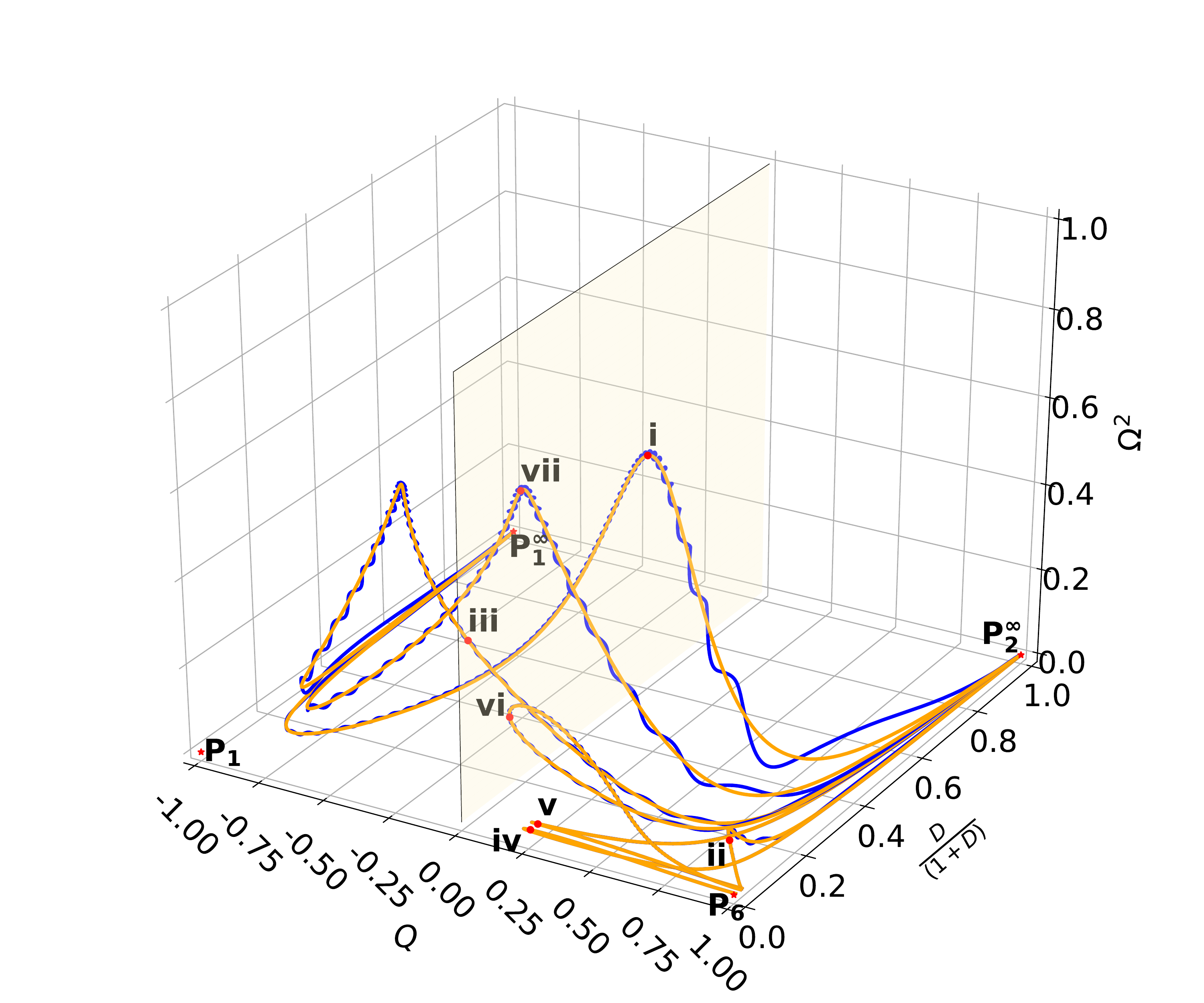}}
    \subfigure[\label{fig:KSCC2DQ} Projection in the space $(Q, \Omega^2)$. The black line represent the constraint $Q=0$.]{\includegraphics[scale = 0.5]{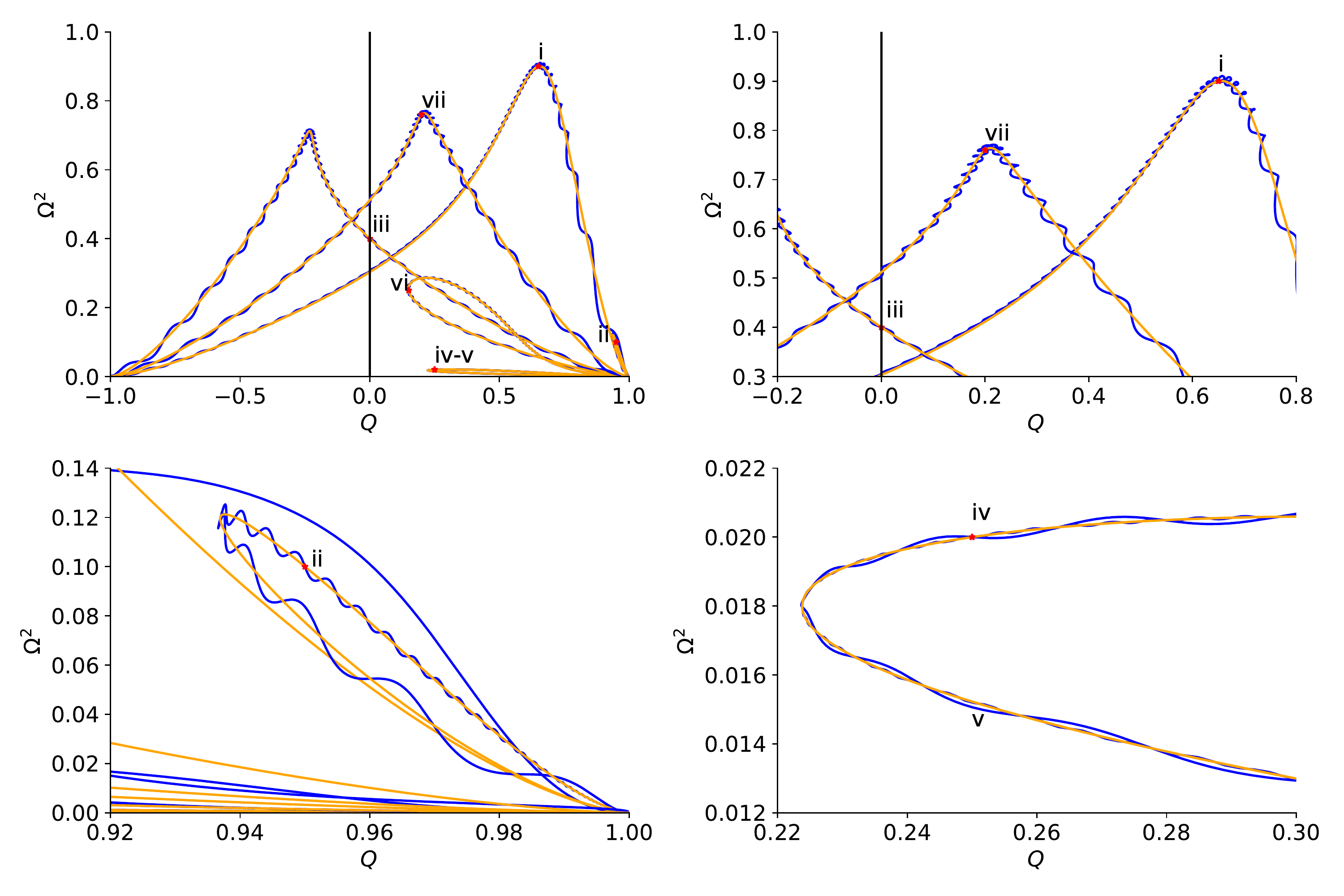}}
    \caption{Some solutions of the full system \eqref{unperturbed1KS} (blue) and time--averaged system \eqref{avrgsystKS} (orange) for the KS metric when $\gamma=0$  in the projection $\Sigma=0$. We have used for both systems the initial data sets presented in Table \ref{Tab4}. \label{Figure9}}
\end{figure*}

\begin{figure*}
    \centering
    \subfigure[\label{fig:KSDust3DS} Projections in the space $(\Sigma, D/(1+D), \Omega^2)$. The surface is given by the constraint $\Omega^{2}=1-\Sigma^{2}$.]{\includegraphics[scale = 0.4]{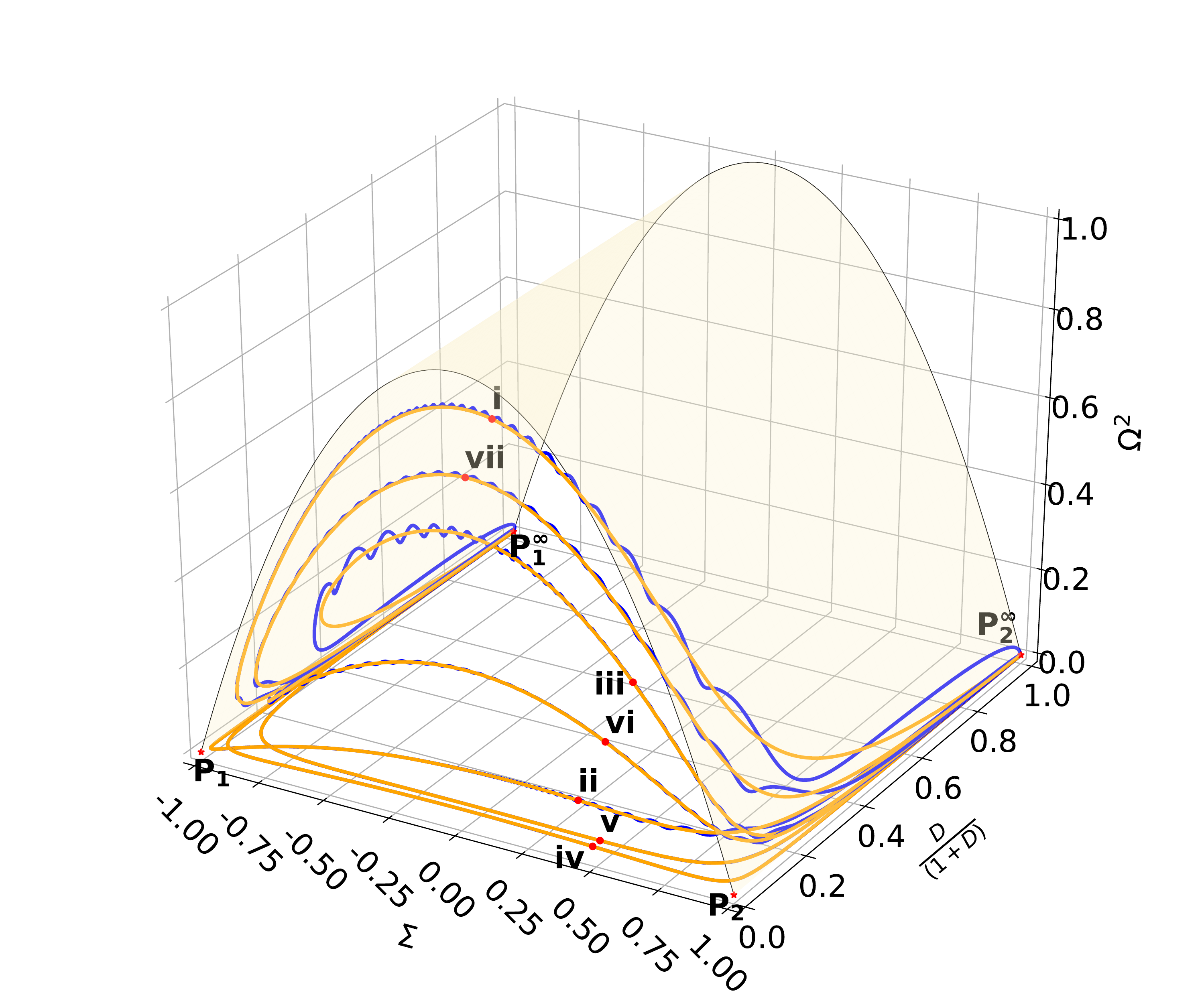}}
    \subfigure[\label{fig:KSDust2DS} Projection in the space $(\Sigma, \Omega^2)$. The black line represent the constraint $\Omega^{2}=1-\Sigma^{2}$.]{\includegraphics[scale = 0.5]{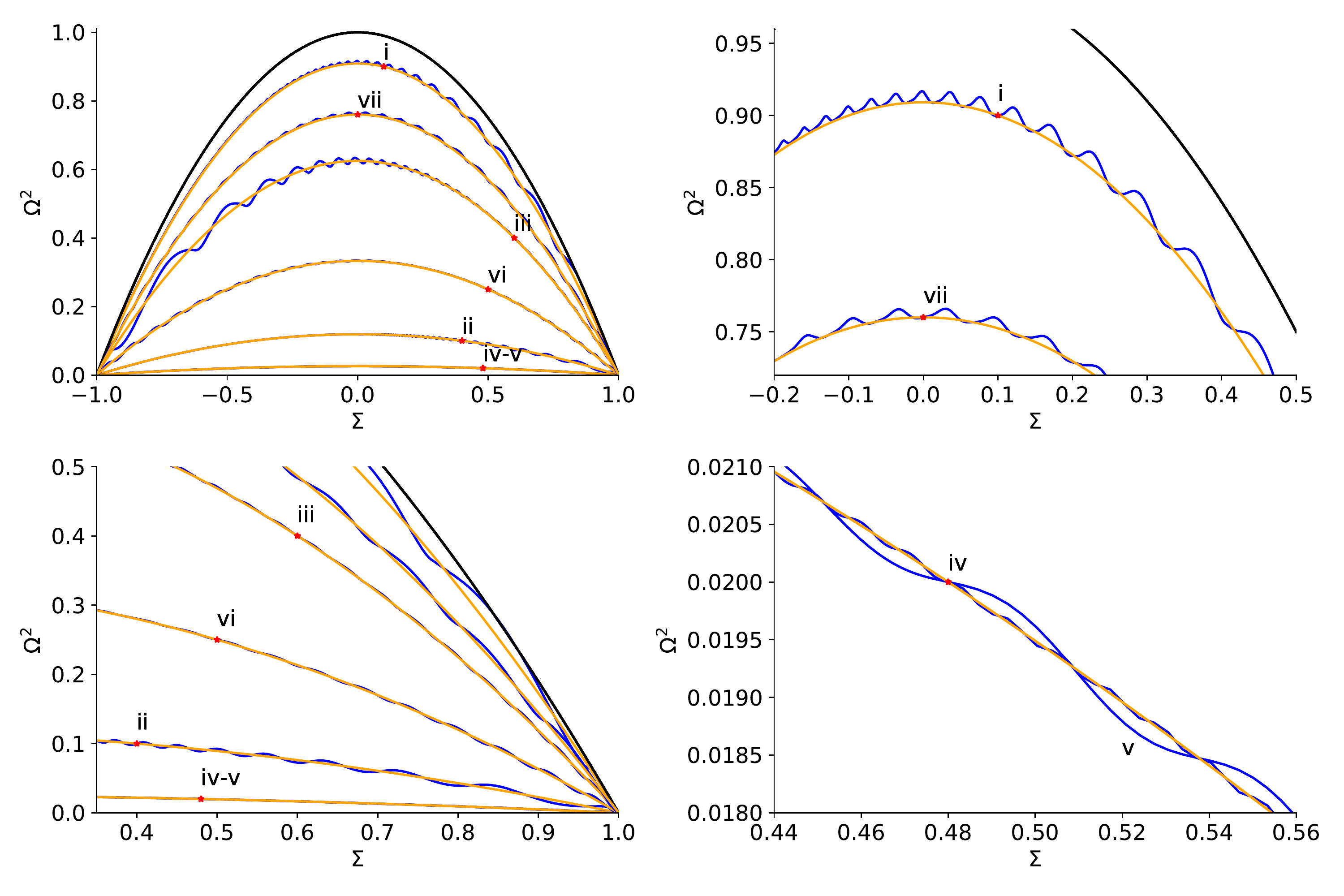}}
    \caption{Some solutions of the full system \eqref{unperturbed1KS} (blue) and time--averaged system \eqref{avrgsystKS} (orange) for the KS metric when $\gamma=1$  in the projection $Q=0$. We have used for both systems the initial data sets presented in Table \ref{Tab4}. \label{Figure10}}
\end{figure*}

\begin{figure*}
    \centering
    \subfigure[\label{fig:KSDust3DQ} Projections in the space $(Q, D/(1+D), \Omega^2)$. The surface is given by the constraint $Q=0$.]{\includegraphics[scale = 0.4]{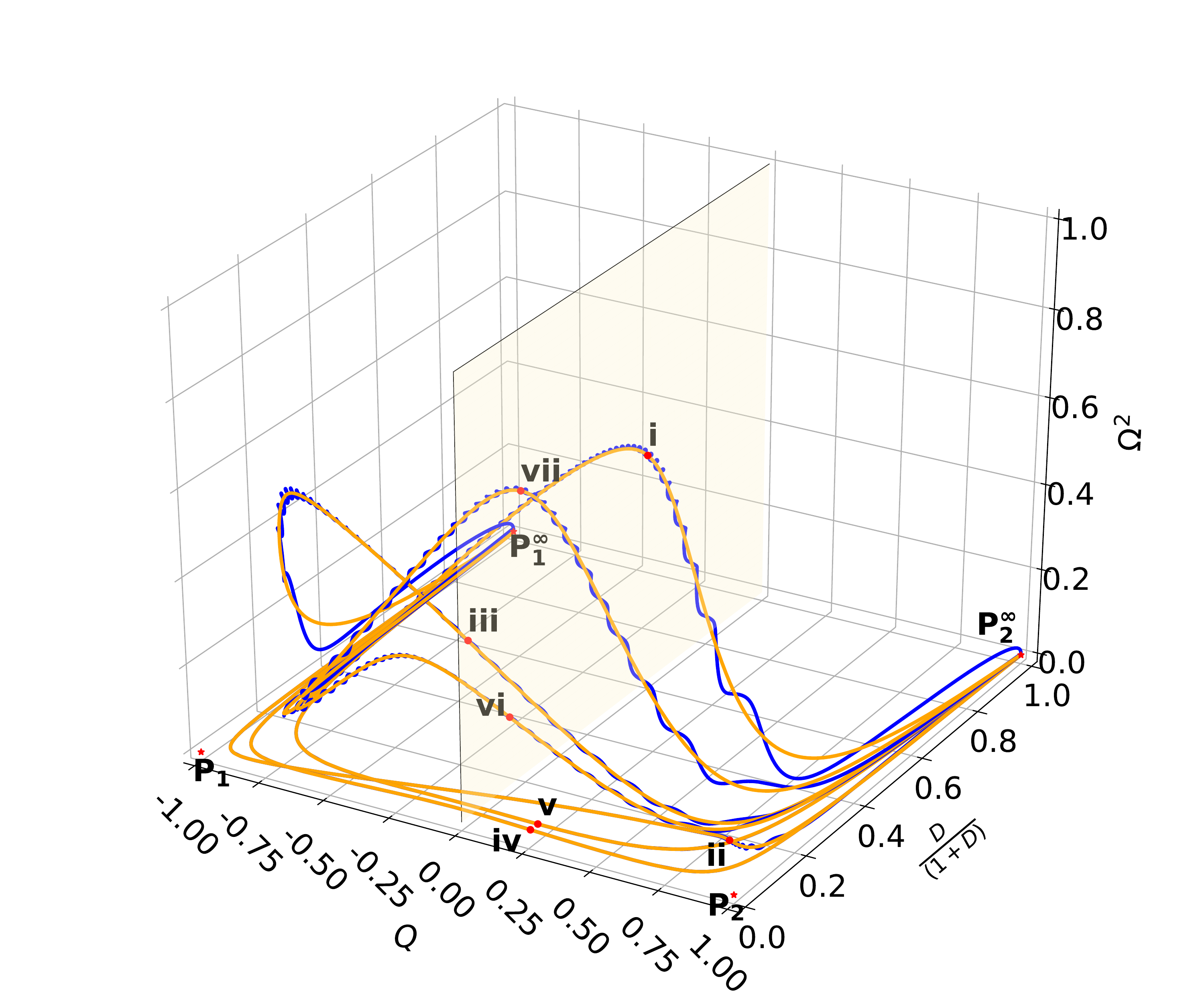}}
    \subfigure[\label{fig:KSDust2DQ} Projection in the space $(Q, \Omega^2)$. The black line represent the constraint $Q=0$.]{\includegraphics[scale = 0.5]{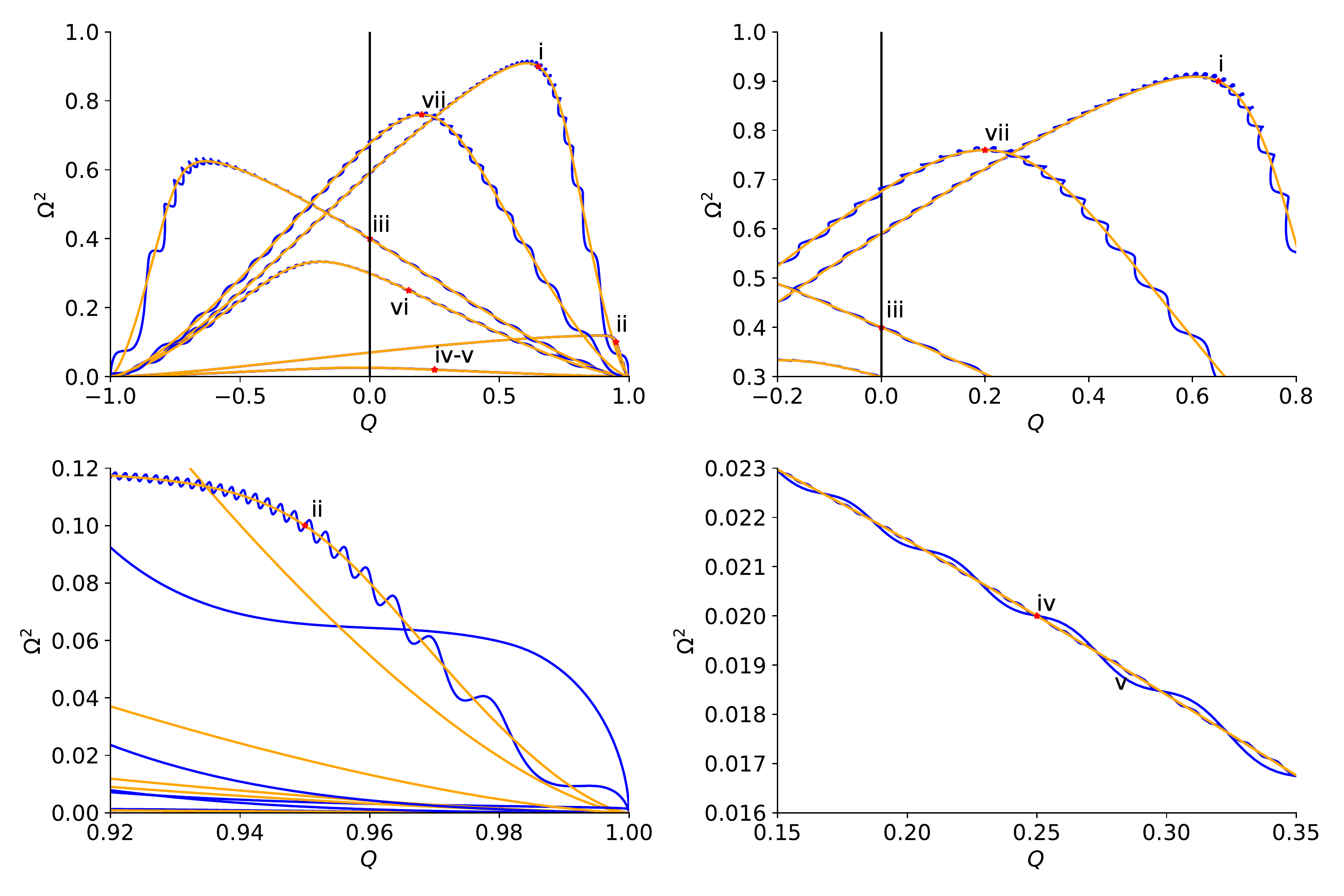}}
    \caption{Some solutions of the full system \eqref{unperturbed1KS} (blue) and time--averaged system \eqref{avrgsystKS} (orange) for the KS metric when $\gamma=1$  in the projection $\Sigma=0$. We have used for both systems the initial data sets presented in Table \ref{Tab4}. \label{Figure11}}
\end{figure*}

\begin{figure*}
    \centering
    \subfigure[\label{fig:KSRad3DS} Projections in the space $(\Sigma, D/(1+D), \Omega^2)$. The surface is given by the constraint $\Omega^{2}=1-\Sigma^{2}$.]{\includegraphics[scale = 0.4]{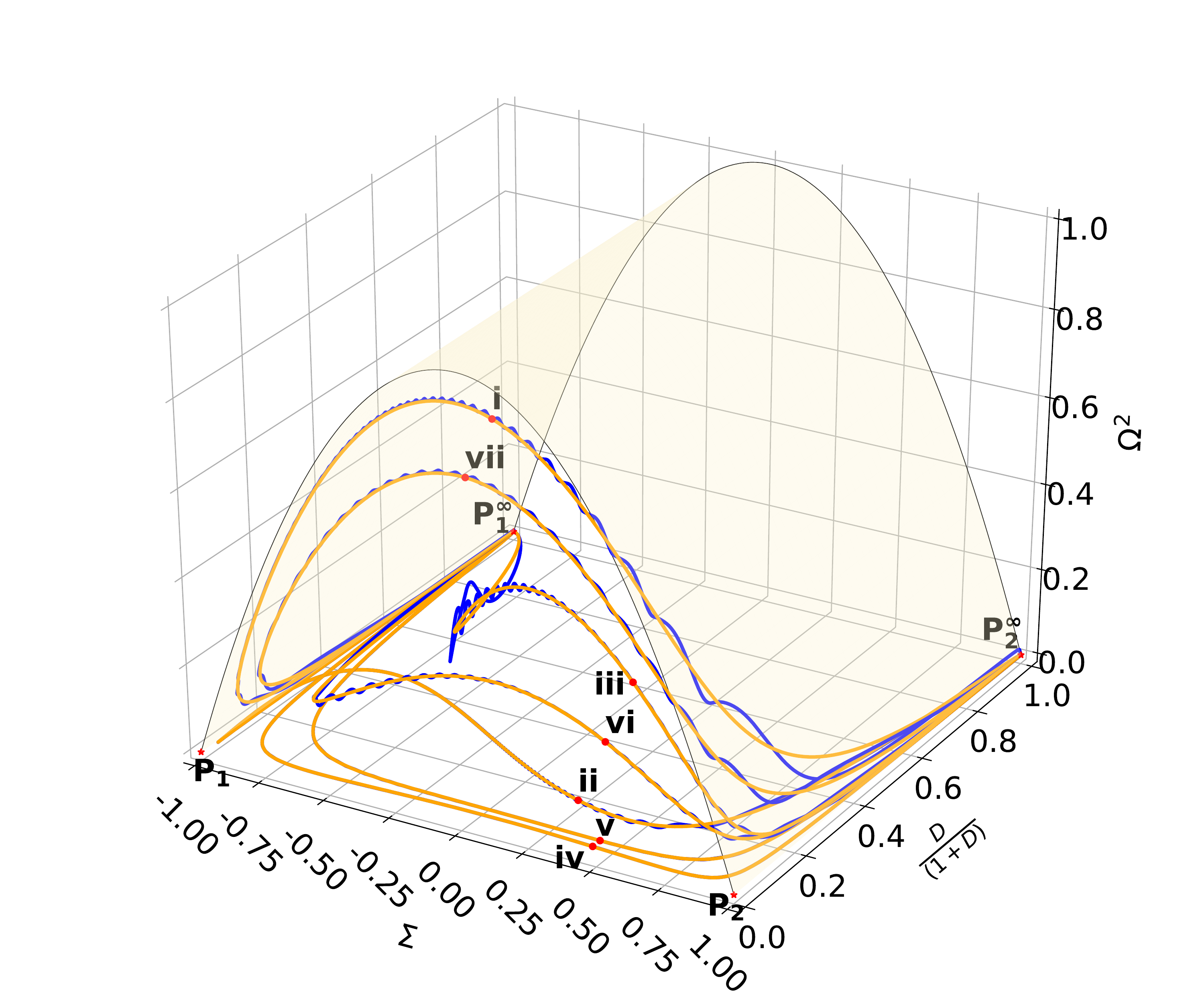}}
    \subfigure[\label{fig:KSRad2DS} Projection in the space $(\Sigma, \Omega^2)$. The black line represent the constraint $\Omega^{2}=1-\Sigma^{2}$.]{\includegraphics[scale = 0.5]{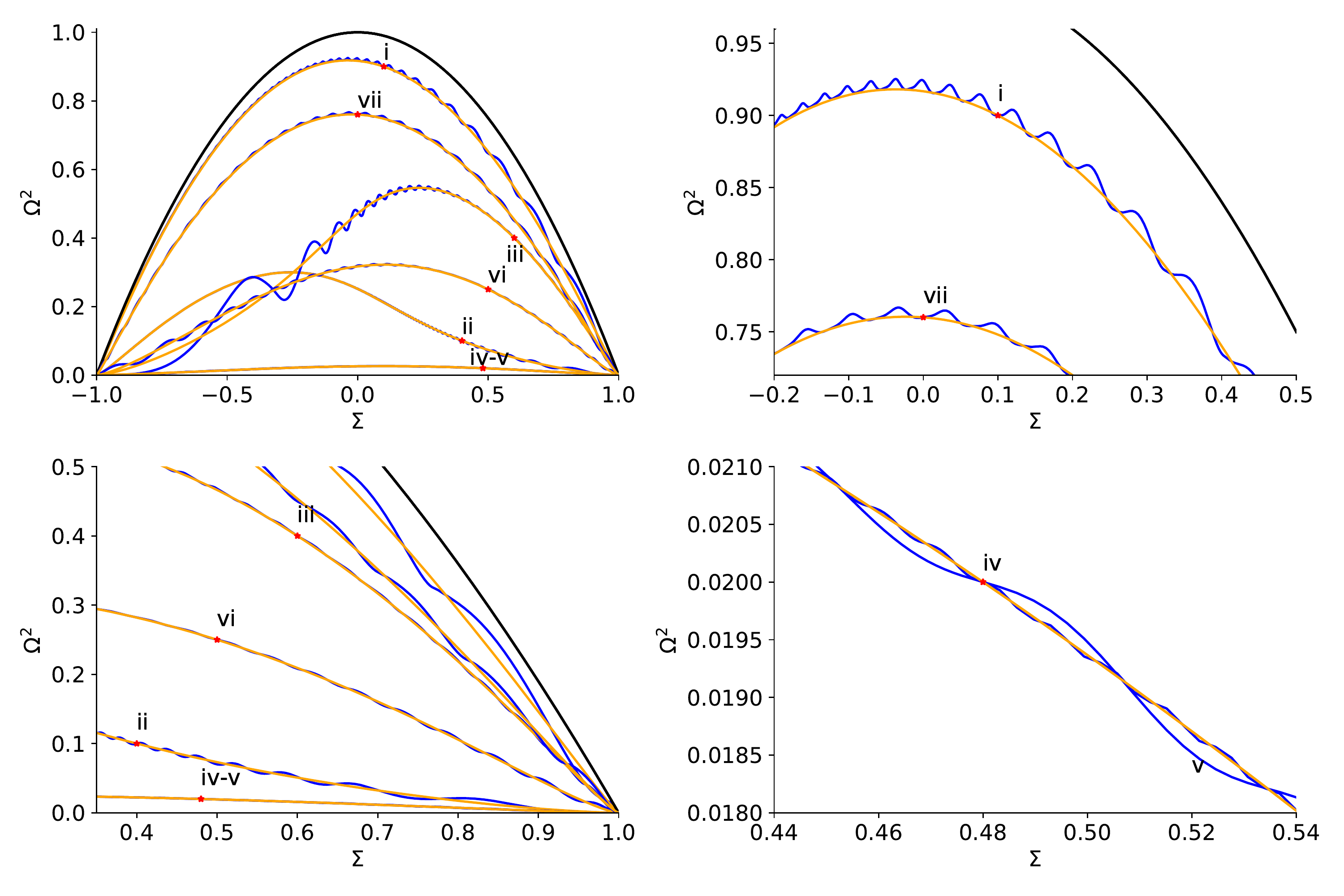}}
    \caption{Some solutions of the full system \eqref{unperturbed1KS} (blue) and time--averaged system \eqref{avrgsystKS} (orange) for the KS metric when $\gamma=\frac{4}{3}$  in the projection $Q=0$. We have used for both systems the initial data sets presented in Table \ref{Tab4}. \label{Figure12}}
\end{figure*}

\begin{figure*}
    \centering
    \subfigure[\label{fig:KSRad3DQ} Projections in the space $(Q, D/(1+D), \Omega^2)$. The surface is given by the constraint $Q=0$.]{\includegraphics[scale = 0.4]{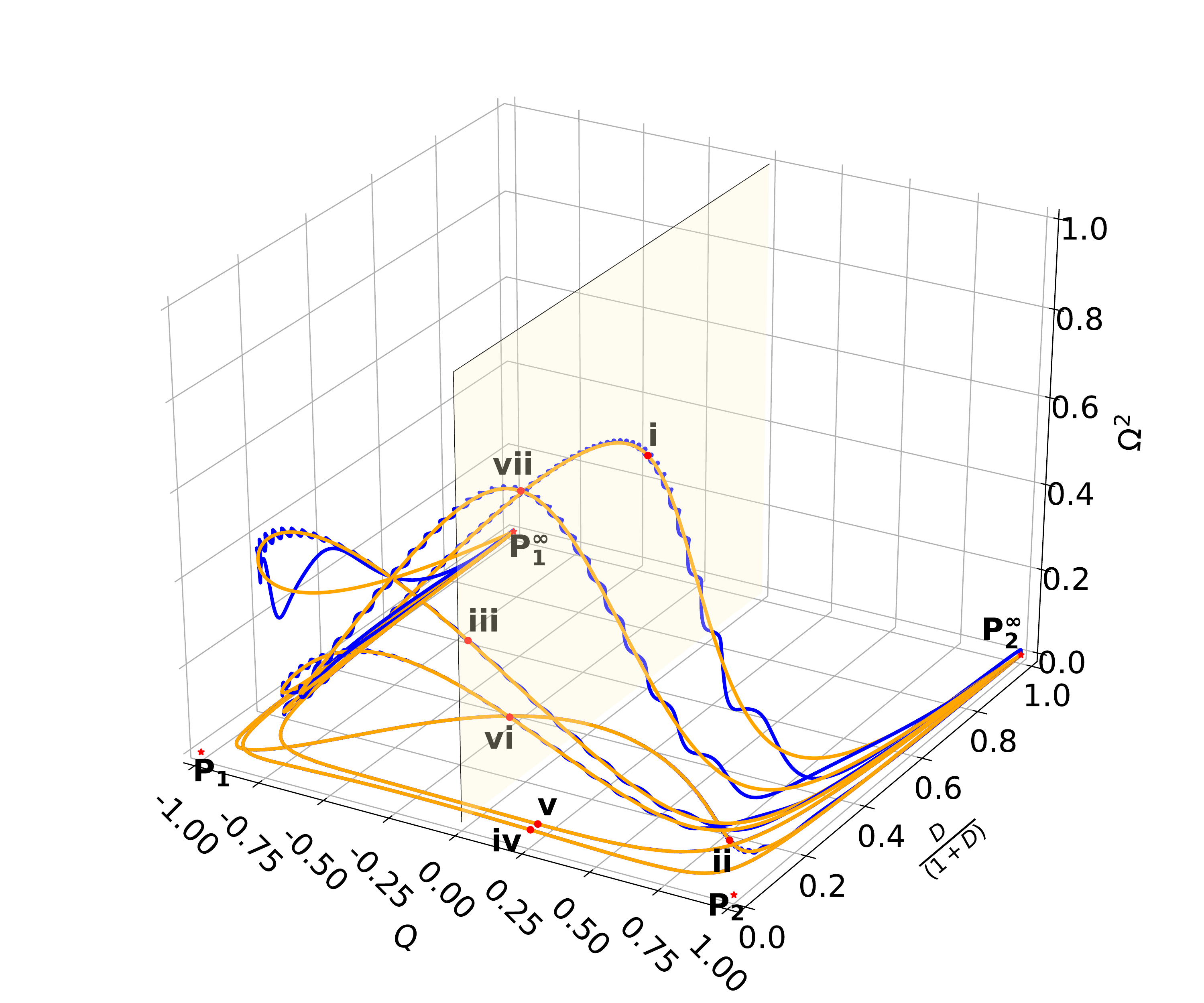}}
    \subfigure[\label{fig:KSRad2DQ} Projection in the space $(Q, \Omega^2)$. The black line represent the constraint $Q=0$.]{\includegraphics[scale = 0.5]{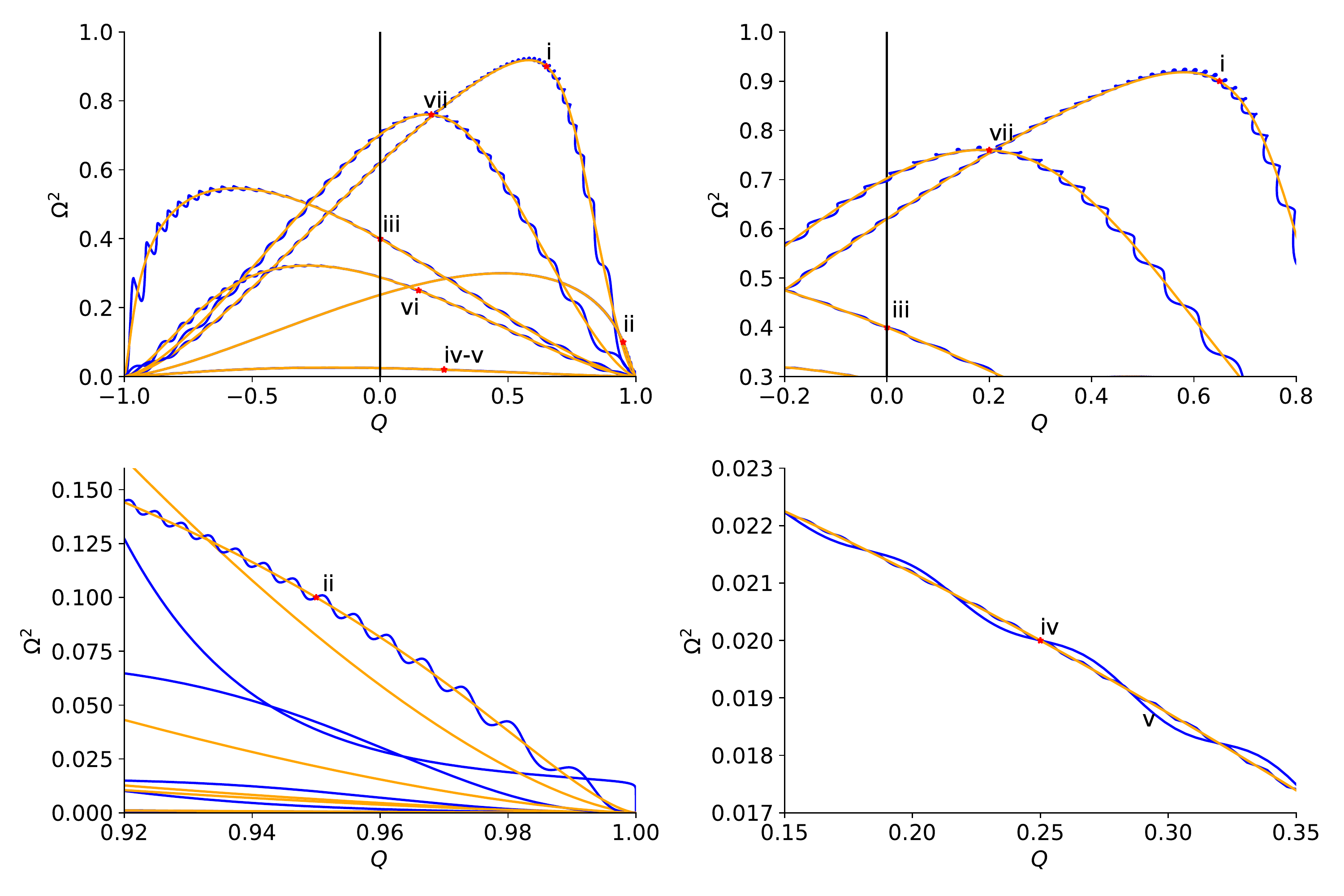}}
    \caption{Some solutions of the full system \eqref{unperturbed1KS} (blue) and time--averaged system \eqref{avrgsystKS} (orange) for the KS metric when $\gamma=\frac{4}{3}$  in the projection $\Sigma=0$. We have used for both systems the initial data sets presented in Table \ref{Tab4}. \label{Figure13}}
\end{figure*}

\begin{figure*}
    \centering
    \subfigure[\label{fig:KSStiff3DS} Projections in the space $(\Sigma, D/(1+D), \Omega^2)$. The surface is given by the constraint $\Omega^{2}=1-\Sigma^{2}$.]{\includegraphics[scale = 0.4]{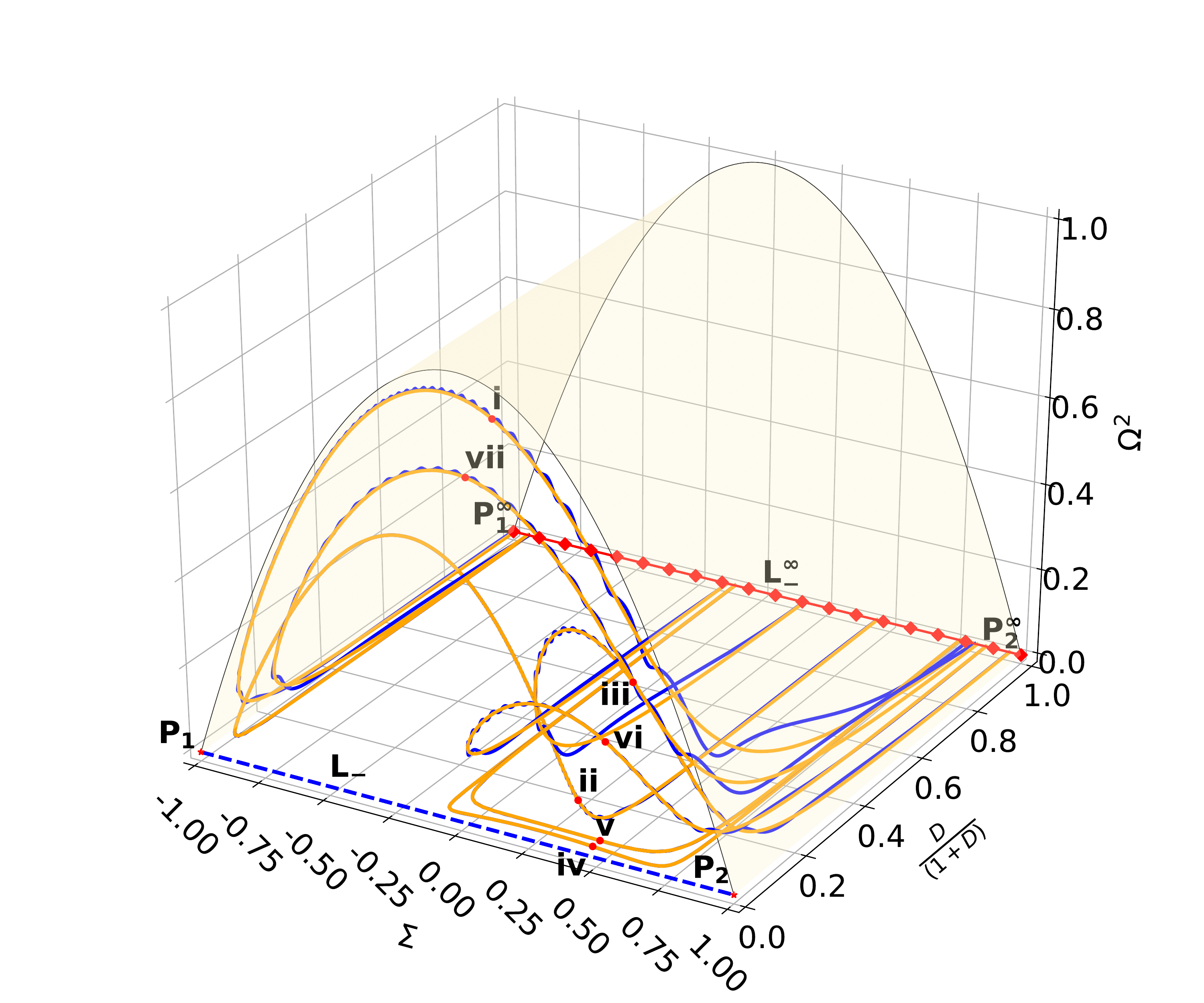}}
    \subfigure[\label{fig:KSStiff2DS} Projection in the space $(\Sigma, \Omega^2)$. The black line represent the constraint $\Omega^{2}=1-\Sigma^{2}$.]{\includegraphics[scale = 0.5]{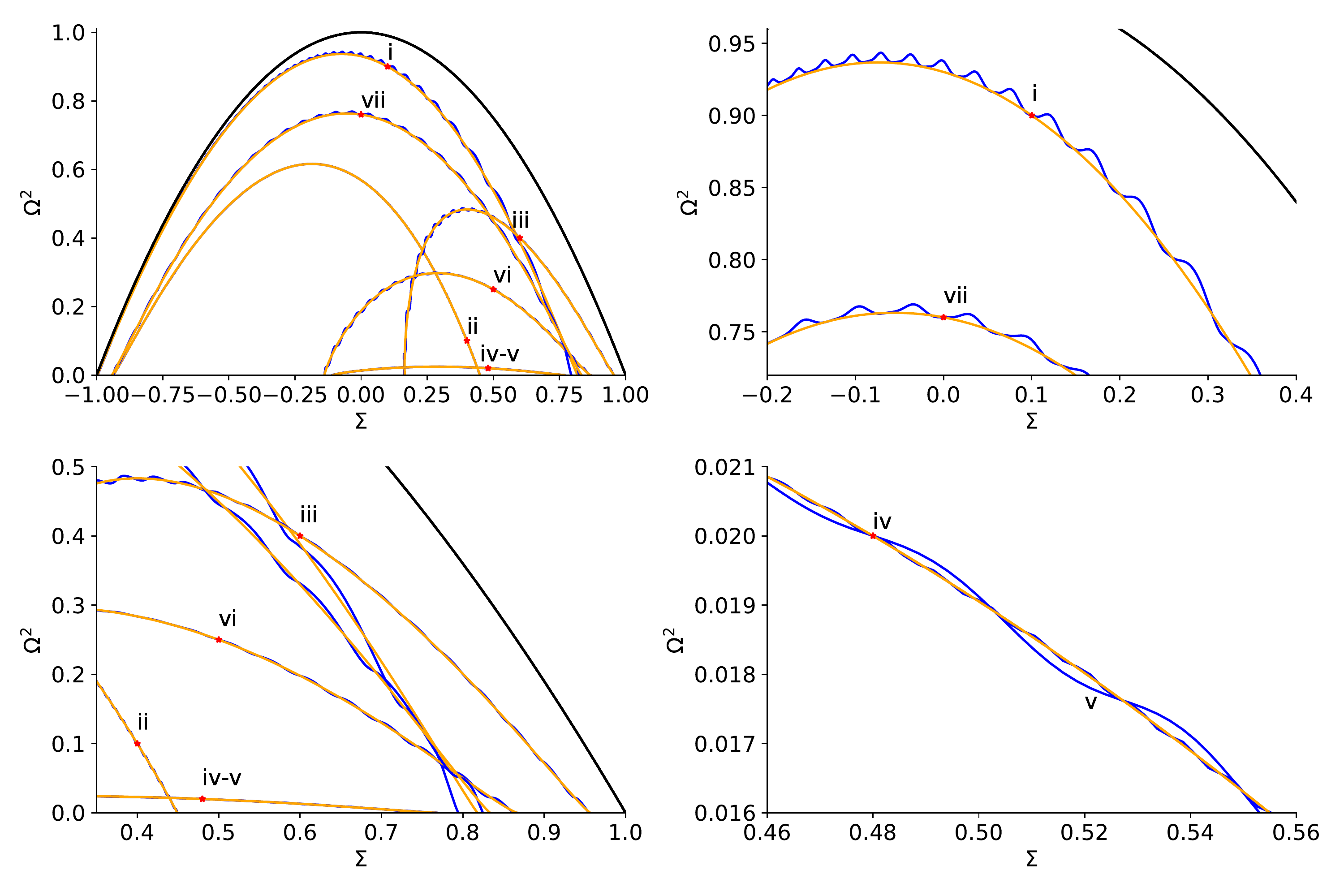}}
    \caption{Some solutions of the full system \eqref{unperturbed1KS} (blue) and time--averaged system \eqref{avrgsystKS} (orange) for the KS metric when $\gamma=2$  in the projection $Q=0$. We have used for both systems the initial data sets presented in Table \ref{Tab4}. \label{Figure14}}
\end{figure*}

\begin{figure*}
    \centering
    \subfigure[\label{fig:KSStiff3DQ} Projections in the space $(Q, D/(1+D), \Omega^2)$. The surface is given by the constraint $Q=0$.]{\includegraphics[scale = 0.4]{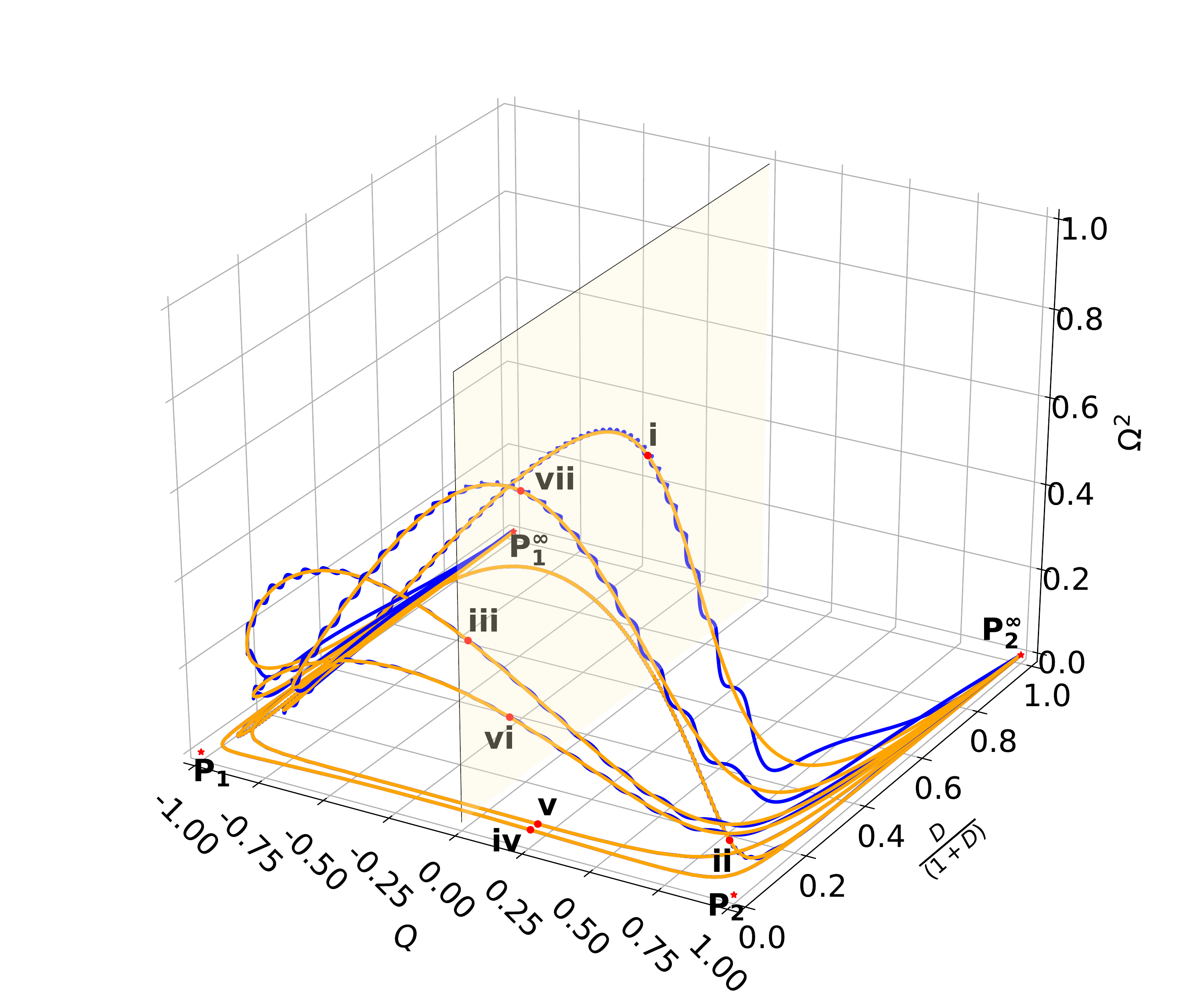}}
    \subfigure[\label{fig:KSStiff2DQ} Projection in the space $(Q, \Omega^2)$. The black line represent the constraint $Q=0$.]{\includegraphics[scale = 0.5]{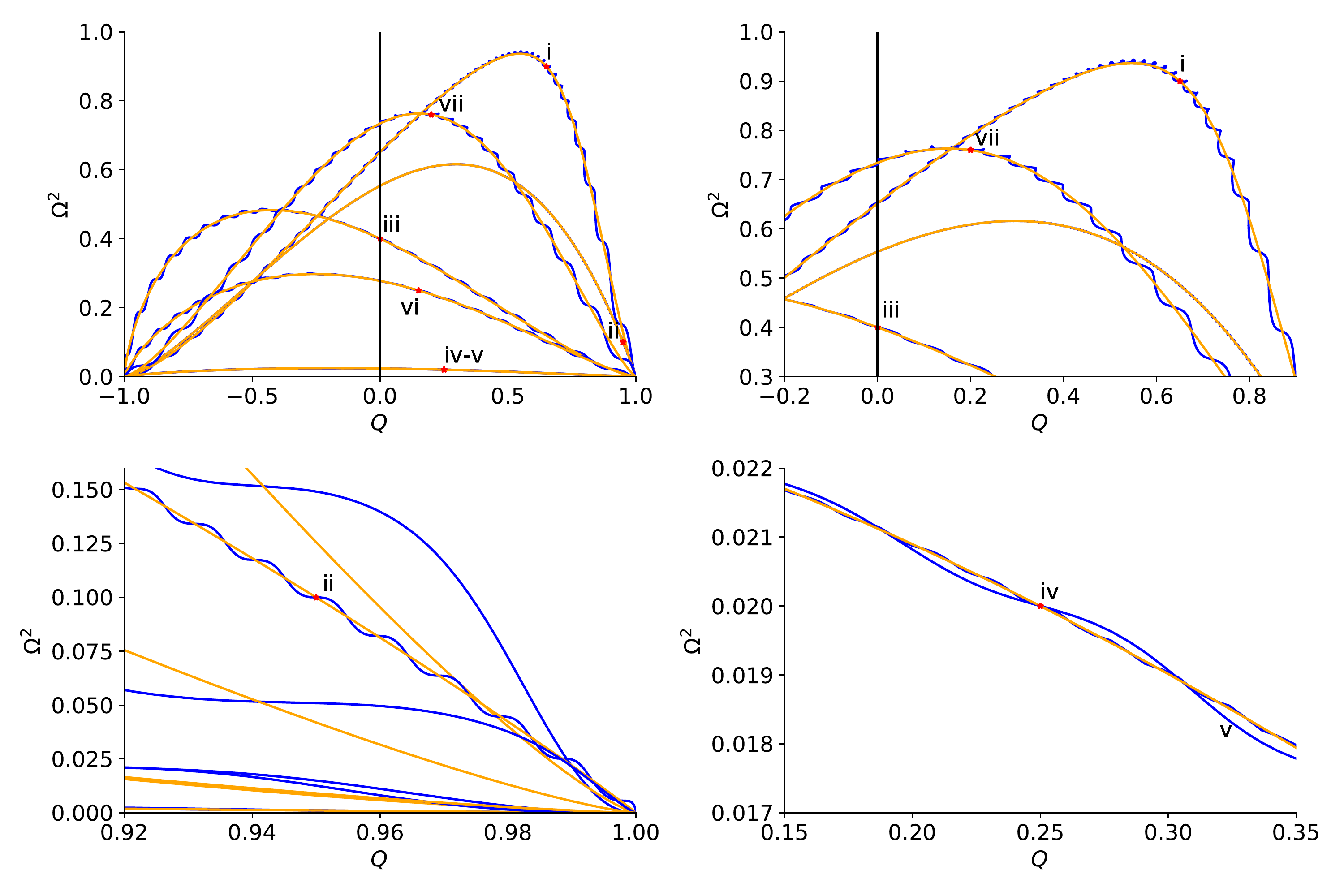}}
    \caption{Some solutions of the full system \eqref{unperturbed1KS} (blue) and time--averaged system \eqref{avrgsystKS} (orange) for the KS metric when $\gamma=2$  in the projection $\Sigma=0$. We have used for both systems the initial data sets presented in Table \ref{Tab4}. \label{Figure15}}
\end{figure*}

For  KS metric we integrate
\begin{enumerate}
\item the full system given by  \eqref{unperturbed1KS}. 
\item The time--averaged system \eqref{avrgsystKS}.
\end{enumerate}
As  initial conditions we use the seven data set presented in Table \ref{Tab4} for a better comparison of both systems.
\begin{table}
\caption{\label{Tab4} Here we present seven initial data sets for the simulation of the full system \eqref{unperturbed1KS} and time--averaged system  \eqref{avrgsystKS} for the KS metric. All the conditions are chosen in order to fulfill the inequalities $\Sigma^{2}(0)+\Omega^{2}(0)\leq 1$ and $0\leq Q\leq 1$.}
\footnotesize\setlength{\tabcolsep}{7pt}
    \begin{tabular}{lccccccc}\hline
Sol.  & \multicolumn{1}{c}{$D(0)$} & \multicolumn{1}{c}{$\Sigma(0)$} & \multicolumn{1}{c}{$\Omega^2(0)$} & \multicolumn{1}{c}{$Q(0)$} &  \multicolumn{1}{c}{$\varphi(0)$}  & \multicolumn{1}{c}{$t(0)$}  \\\hline
        i & $0.03$ & $0.1$ & $0.9$ & $0.65$ & $0$ &  $0$  \\
        ii & $0.03$ & $0.4$ & $0.1$ & $0.95$ & $0$ &  $0$ \\
        iii & $0.03$ & $0.6$ & $0.4$ & $0$ & $0$ & $0$ \\
        iv & $0.006$ & $0.48$ & $0.02$ & $0.25$ & $0$ & $0$ \\
        v & $0.03$ & $0.48$ & $0.02$ & $0.25$ & $0$ & $0$ \\
        vi & $0.03$ & $0.5$ & $0.25$ & $0.15$ & $0$ & $0$ \\
        vii & $0.03$ & $0$ & $0.76$ & $0.2$ & $0$ & $0$ \\ \hline
    \end{tabular}
\end{table}
In Figures \ref{Figure8}, \ref{Figure9}, \ref{Figure10}, \ref{Figure11}, \ref{Figure12}, \ref{Figure13}, \ref{Figure14}, and  \ref{Figure15} projections of some solutions of the full system \eqref{unperturbed1KS} and time--averaged system \eqref{avrgsystKS} in the $(\Sigma, D/(1+D), \Omega^{2})$ and $(Q, D/(1+D), \Omega^{2})$ space are presented with their respective projections when $D=0$. Figures \ref{Figure8} and \ref{Figure9} show solutions for a fluid corresponding to CC ($\gamma=0$). Figures \ref{Figure10} and \ref{Figure11} show solutions for a fluid corresponding to dust ($\gamma=1$). Figures \ref{Figure12} and \ref{Figure13} show solutions for a fluid corresponding to radiation ($\gamma=\frac{4}{3}$). Figures \ref{Figure14} and \ref{Figure15} show solutions for a fluid corresponding to stiff fluid ($\gamma=2$). Figures \ref{Figure8}, \ref{Figure9}, \ref{Figure10}, \ref{Figure11}, \ref{Figure12}, \ref{Figure13}, \ref{Figure14}, and  \ref{Figure15} are a numerical confirmation that  main theorem \ref{KSLFZ11} presented in Section \ref{SECT:II} is fulfilled for the KS metric.  That is,   solutions of   full system (blue lines) follow the track of   solutions of  averaged system (orange lines) for the whole $D$-range.

\subsection{FLRW metric with positive curvature}
\label{plotsclosedFLRW}
\begin{figure*}
    \centering
    \subfigure[\label{fig:ClosedFLRWCC3D} Projections in the space $(Q, D/(1+D), \Omega^2/(1+\Omega^2))$. The surface is given by the constraint $Q=0$.]{\includegraphics[scale = 0.4]{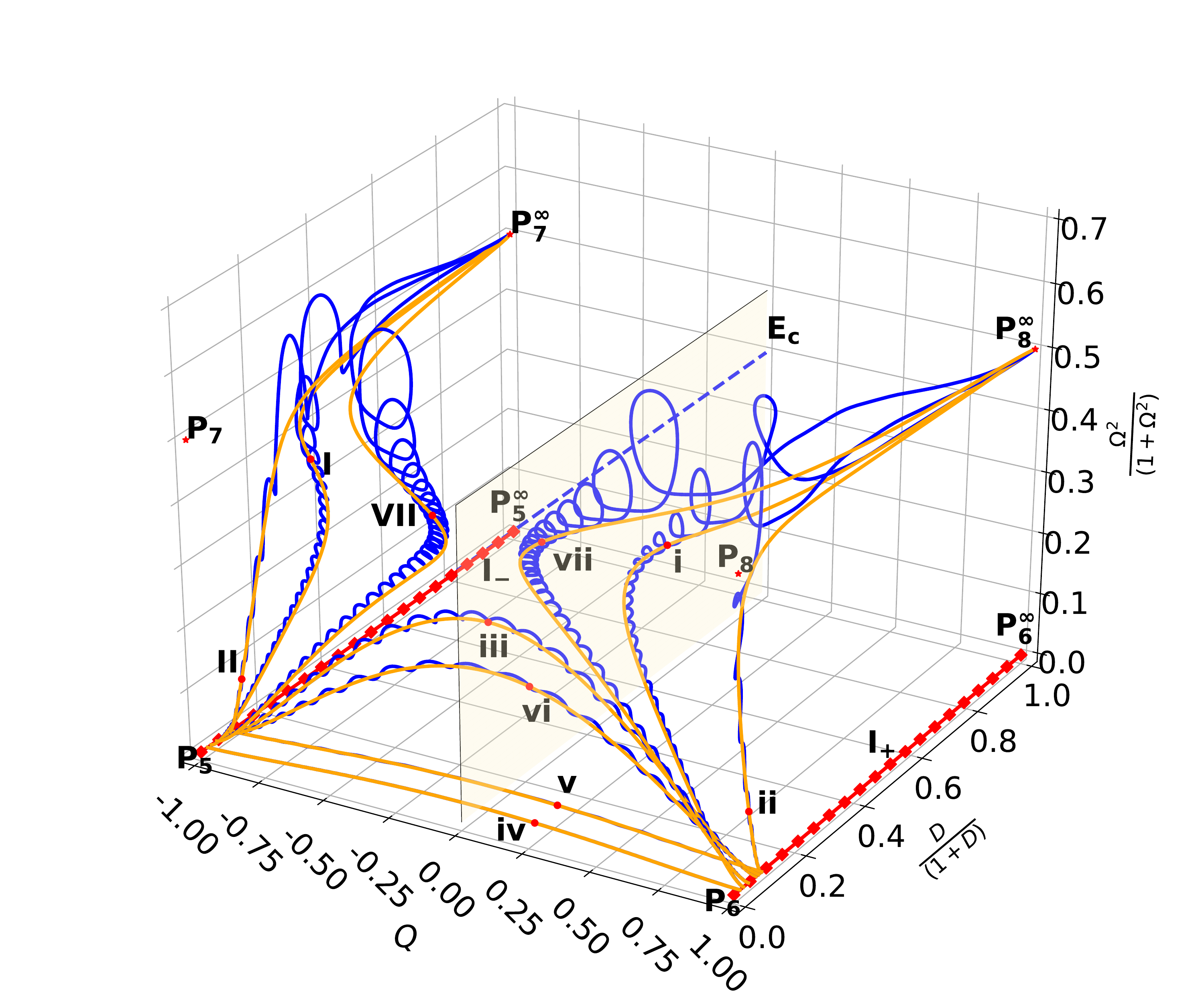}}
    \subfigure[\label{fig:ClosedFLRWCC2D} Projection in the space $(Q, \Omega^2)$. The black line represent the constraint $Q=0$.]{\includegraphics[scale = 0.5]{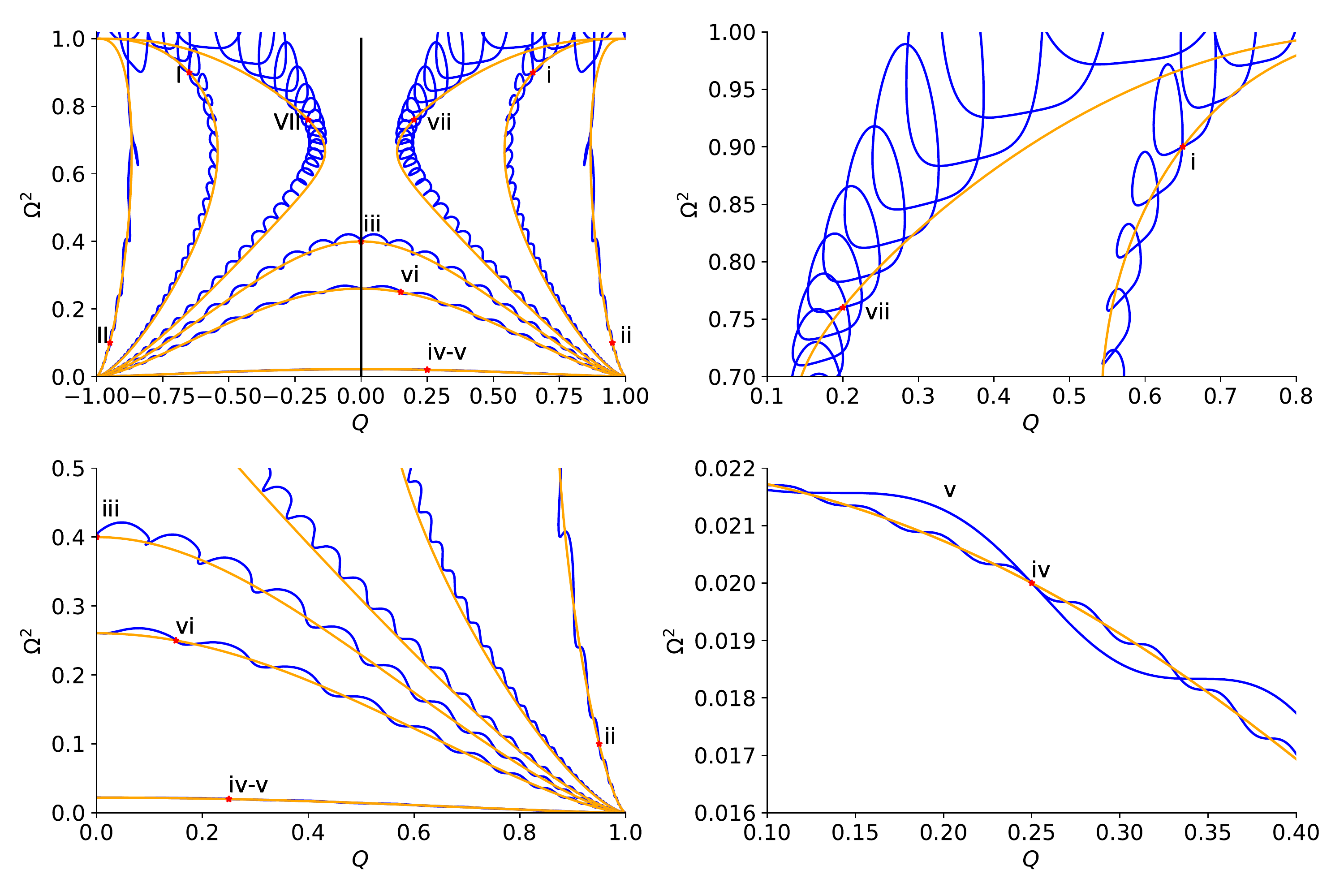}}
    \caption{Some solutions of the full system \eqref{unperturbed1FLRWClosed} (blue) and time--averaged system \eqref{avrgsystFLRWClosed} (orange) for the FLRW metric with positive curvature ($k=+1$) when $\gamma=0$. We have used for both systems the initial data sets presented in Table \ref{Tab5a}. \label{Figure16}}
\end{figure*}

\begin{figure*}
    \centering
    \subfigure[\label{fig:ClosedFLRWDust3D} Projections in the space $(Q, D/(1+D), \Omega^2/(1+\Omega^2))$. The surface is given by the constraint $Q=0$.]{\includegraphics[scale = 0.4]{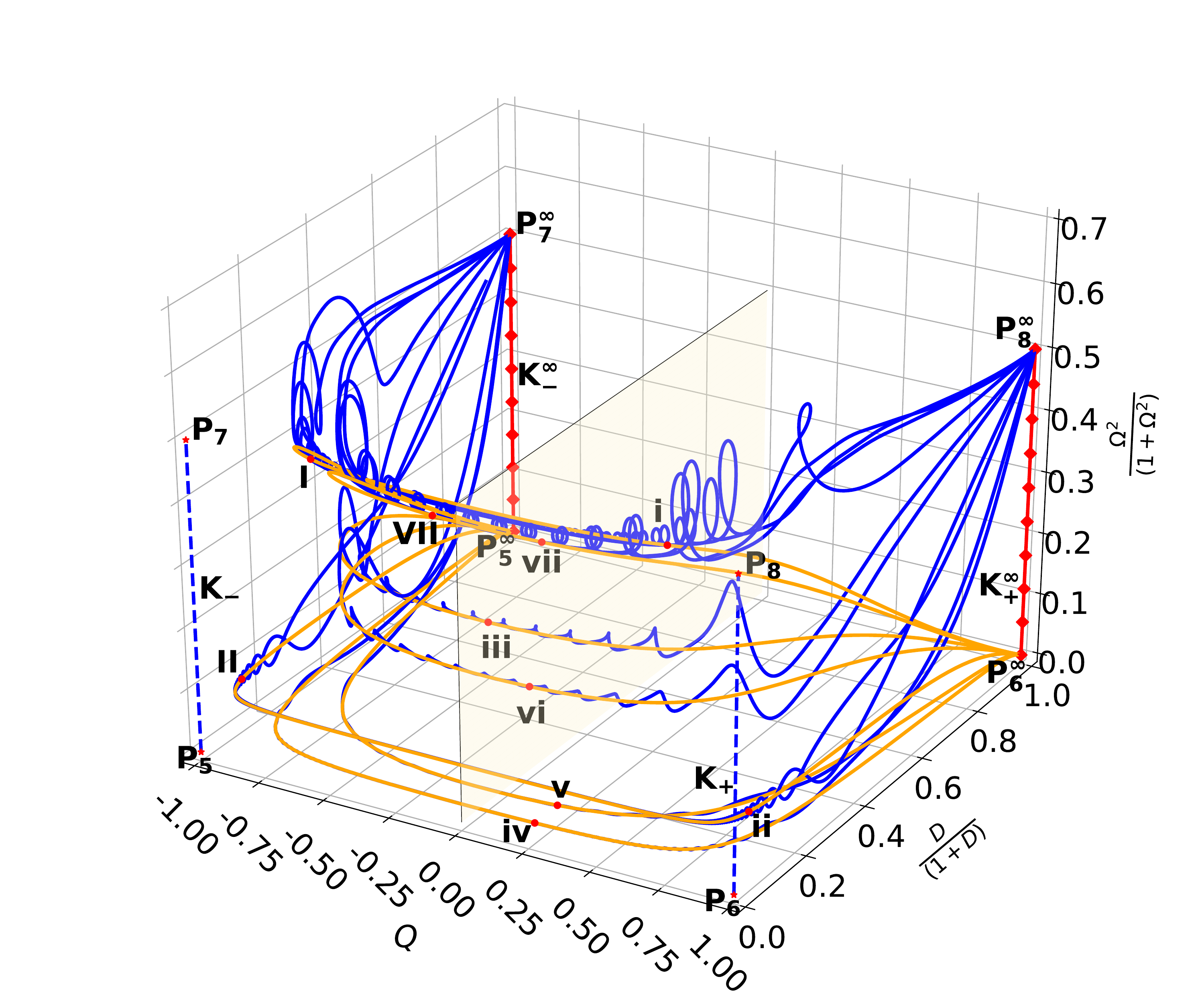}}
    \subfigure[\label{fig:ClosedFLRWDust2D} Projection in the space $(Q, \Omega^2)$. The black line represent the constraint $Q=0$.]{\includegraphics[scale = 0.5]{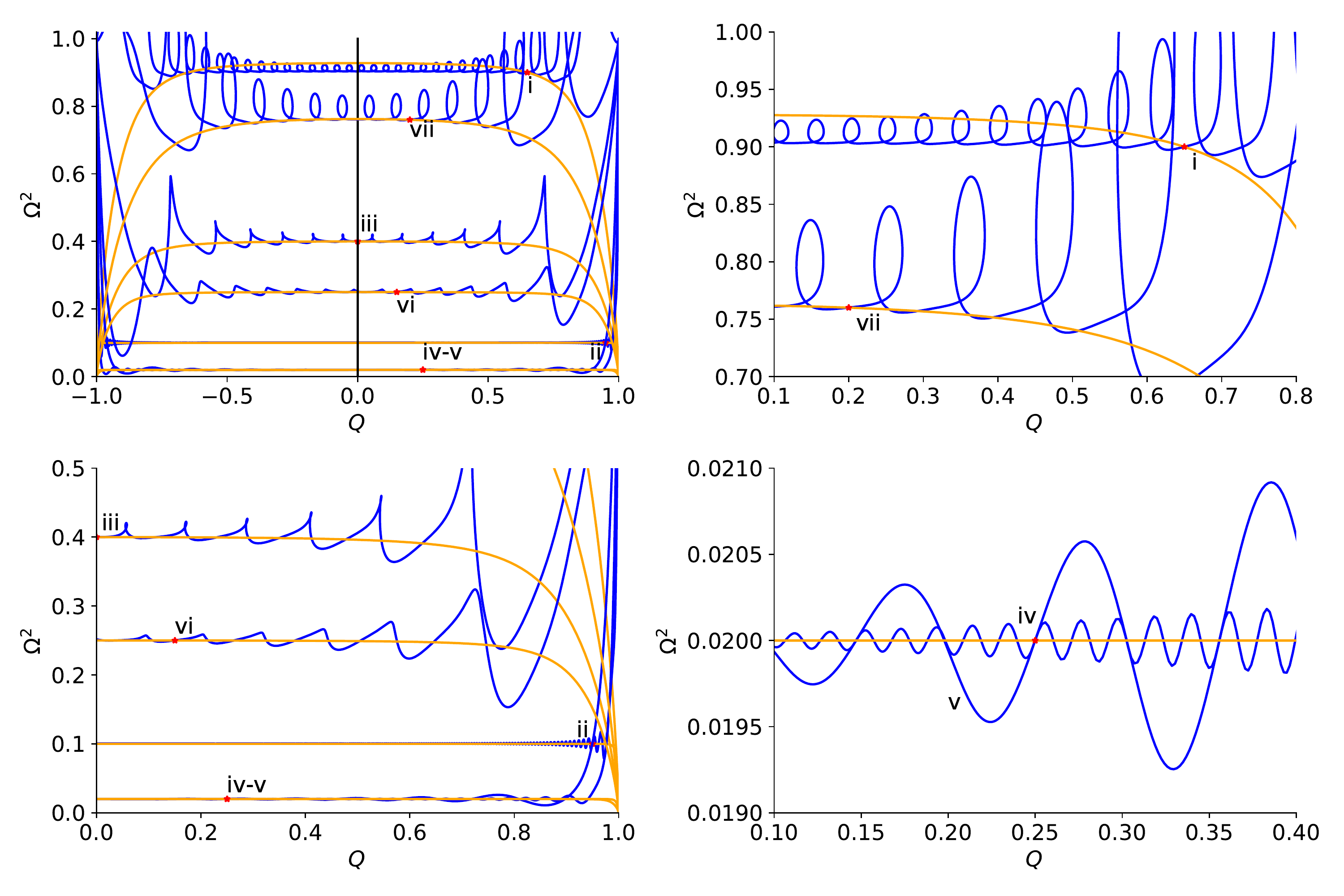}}
    \caption{Some solutions of the full system \eqref{unperturbed1FLRWClosed} (blue) and time--averaged system \eqref{eqaverg4D} (orange) for the FLRW metric with positive curvature ($k=+1$) when $\gamma=1$. We have used for both systems the initial data sets presented in Table \ref{Tab5a}. \label{Figure17}}
\end{figure*}

\begin{figure*}
    \centering
    \subfigure[\label{fig:ClosedFLRWRad3D} Projections in the space $(Q, D/(1+D), \Omega^2/(1+\Omega^2))$. The surface is given by the constraint $Q=0$.]{\includegraphics[scale = 0.4]{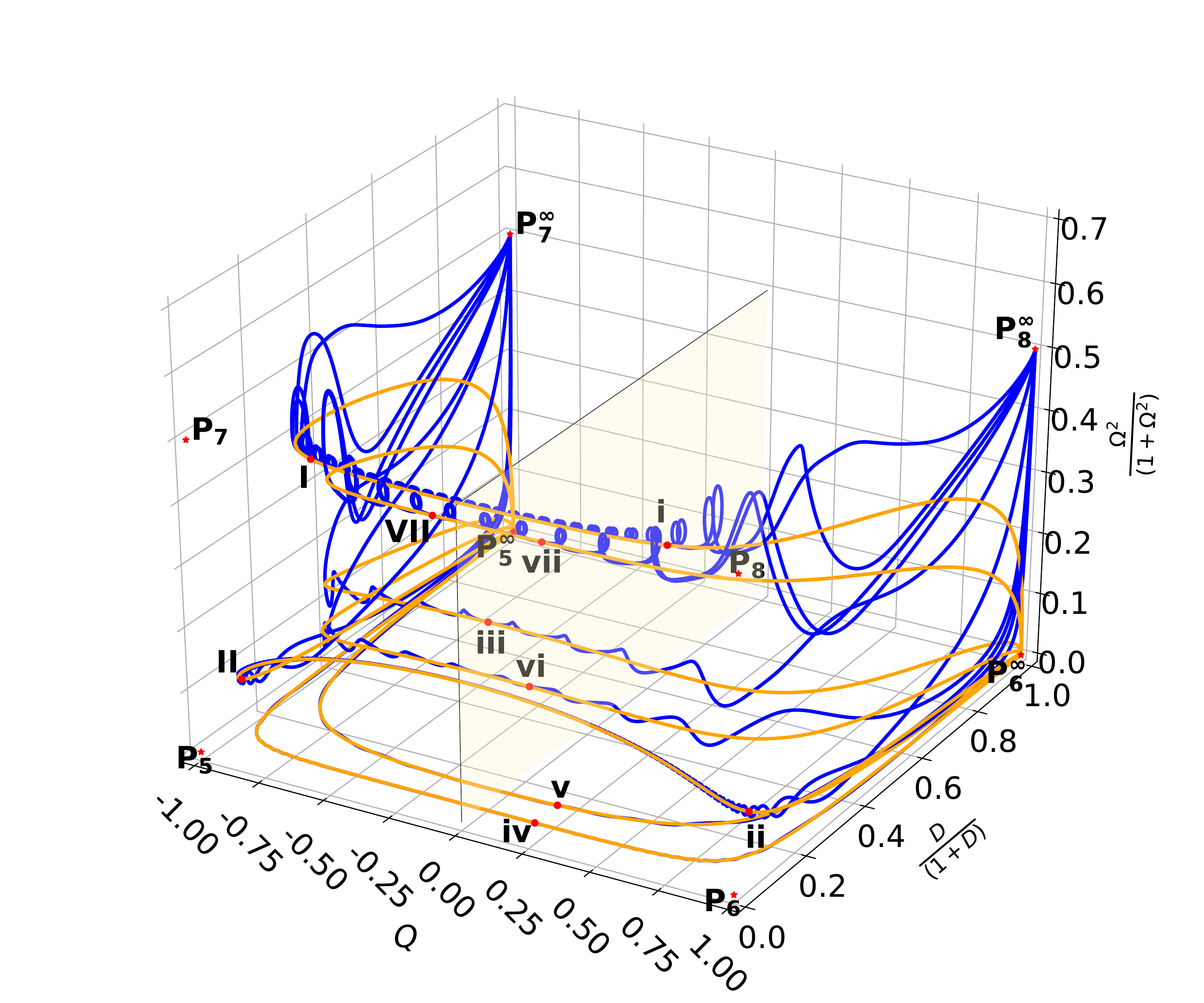}}
    \subfigure[\label{fig:ClosedFLRWRad2D} Projection in the space $(Q, \Omega^2)$. The black line represent the constraint $Q=0$.]{\includegraphics[scale = 0.5]{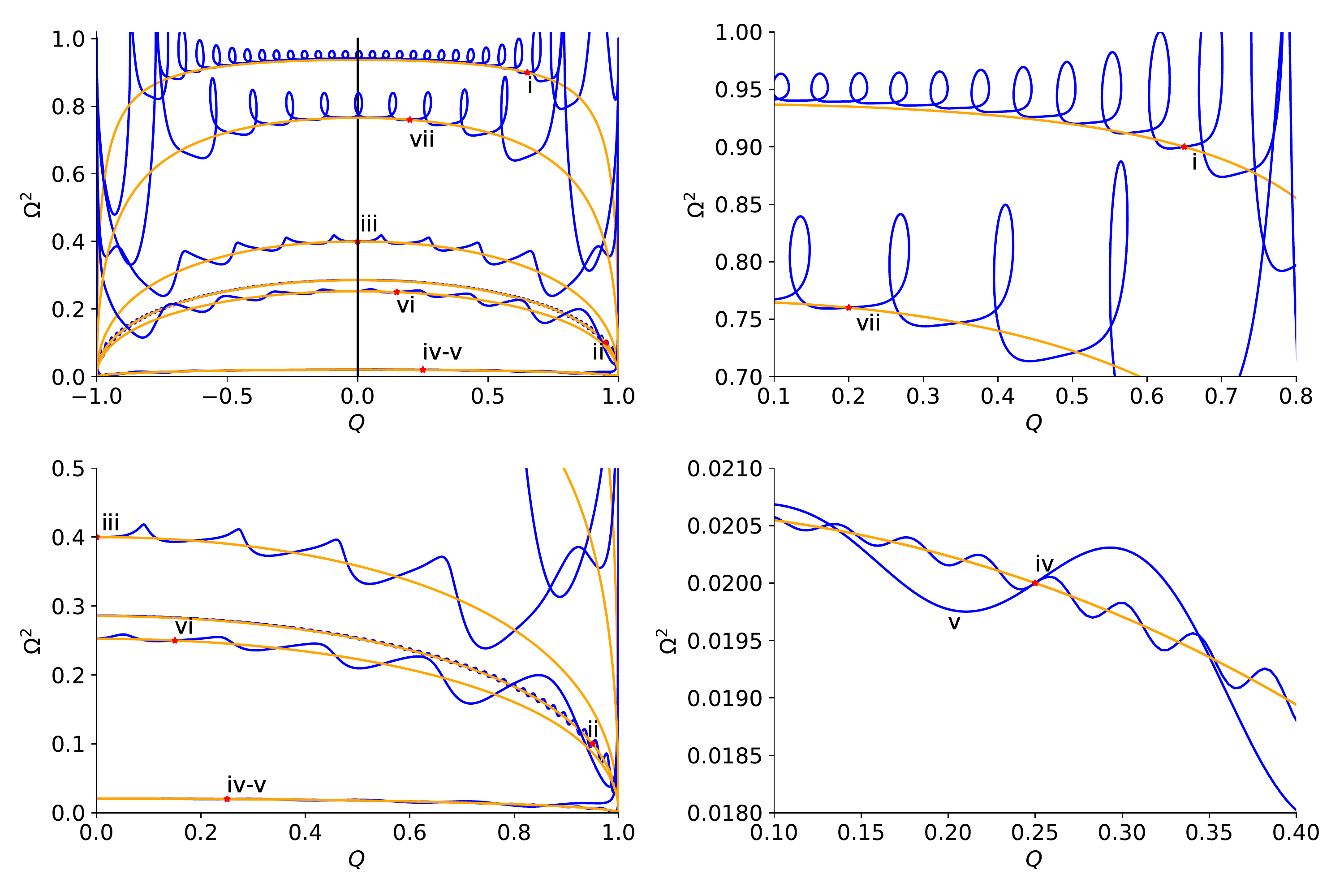}}
    \caption{Some solutions of the full system \eqref{unperturbed1FLRWClosed} (blue) and time--averaged system \eqref{avrgsystFLRWClosed} (orange) for the FLRW metric with positive curvature ($k=+1$) when $\gamma=\frac{4}{3}$. We have used for both systems the initial data sets presented in Table \ref{Tab5a}. \label{Figure18}}
\end{figure*}

\begin{figure*}
    \centering
    \subfigure[\label{fig:ClosedFLRWStiff3D} Projections in the space $(Q, D/(1+D), \Omega^2/(1+\Omega^2))$. The surface is given by the constraint $Q=0$.]{\includegraphics[scale = 0.4]{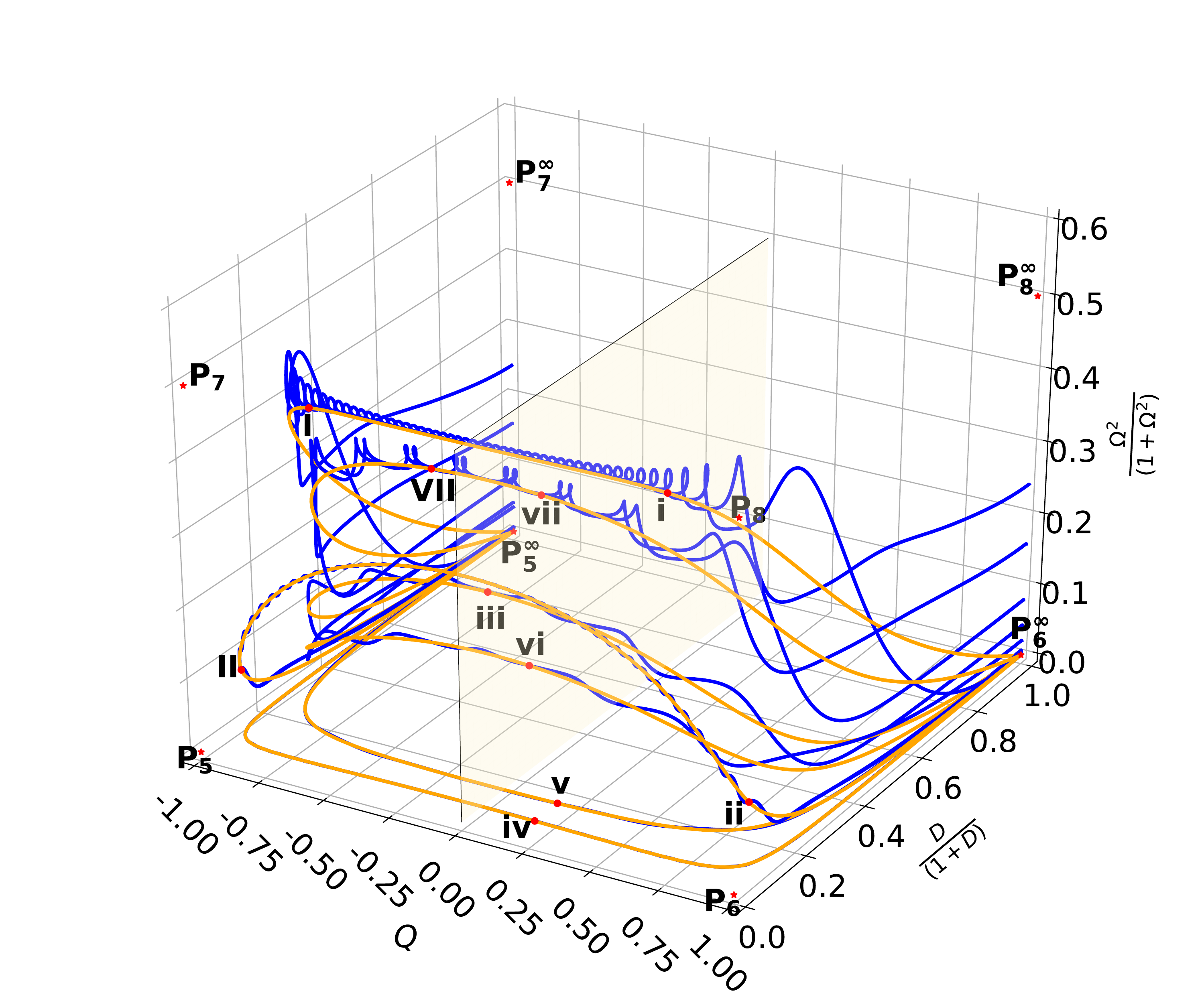}}
    \subfigure[\label{fig:ClosedFLRWStiff2D} Projection in the space $(Q, \Omega^2)$. The black line represent the constraint $Q=0$.]{\includegraphics[scale = 0.5]{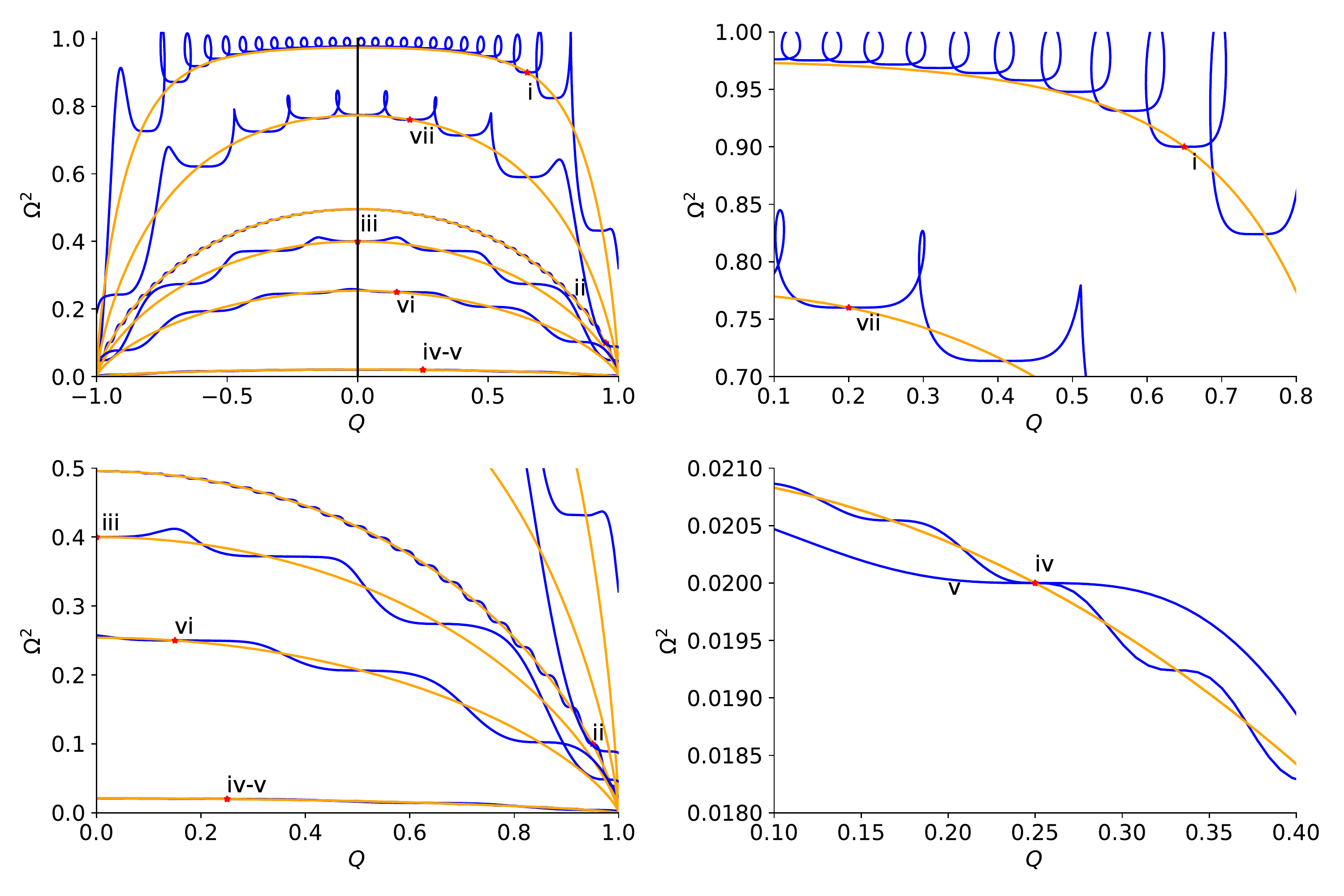}}
    \caption{Some solutions of the full system \eqref{unperturbed1FLRWClosed} (blue) and time--averaged system \eqref{avrgsystFLRWClosed} (orange) for the FLRW metric with positive curvature ($k=+1$) when $\gamma=2$. We have used for both systems the initial data sets presented in Table \ref{Tab5a}. \label{Figure19}}
\end{figure*}

\begin{figure*}[h!]
    \centering
      \subfigure[ $\gamma=0.1$.]{\includegraphics[scale=0.25]{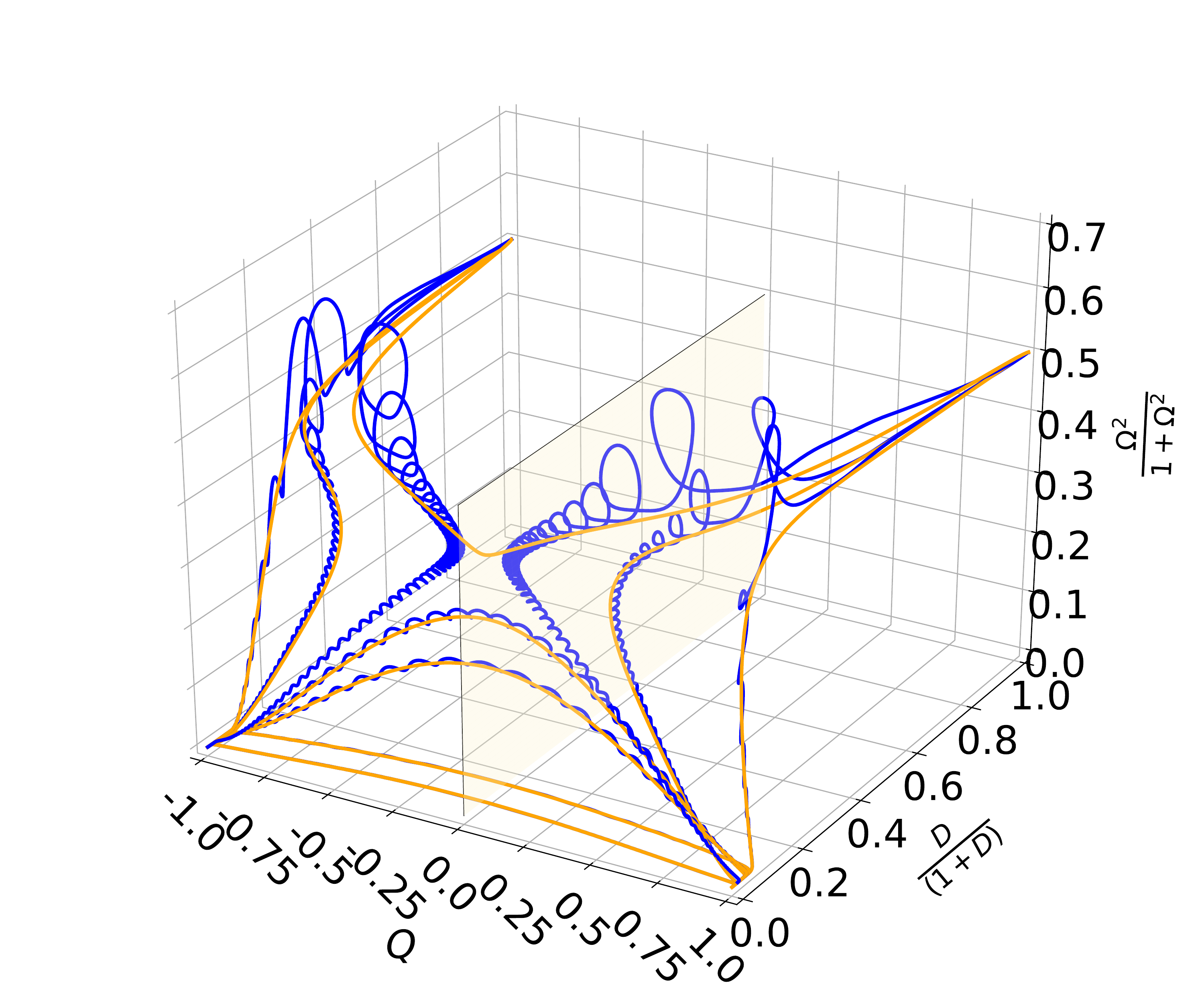}}
      \subfigure[ $\gamma=0.3$.]{\includegraphics[scale=0.25]{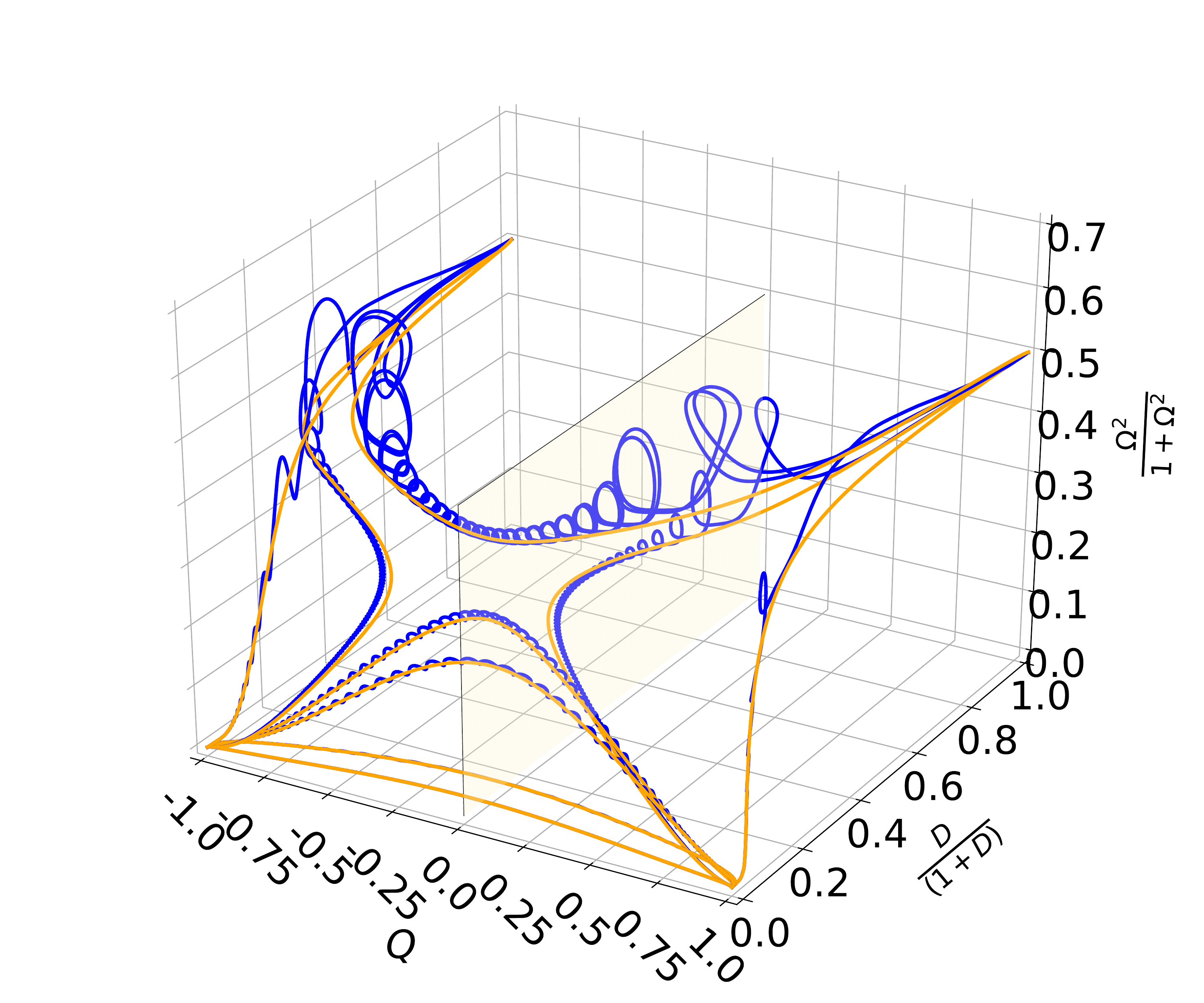}}
    \subfigure[ $\gamma=0.4$.]{\includegraphics[scale=0.25]{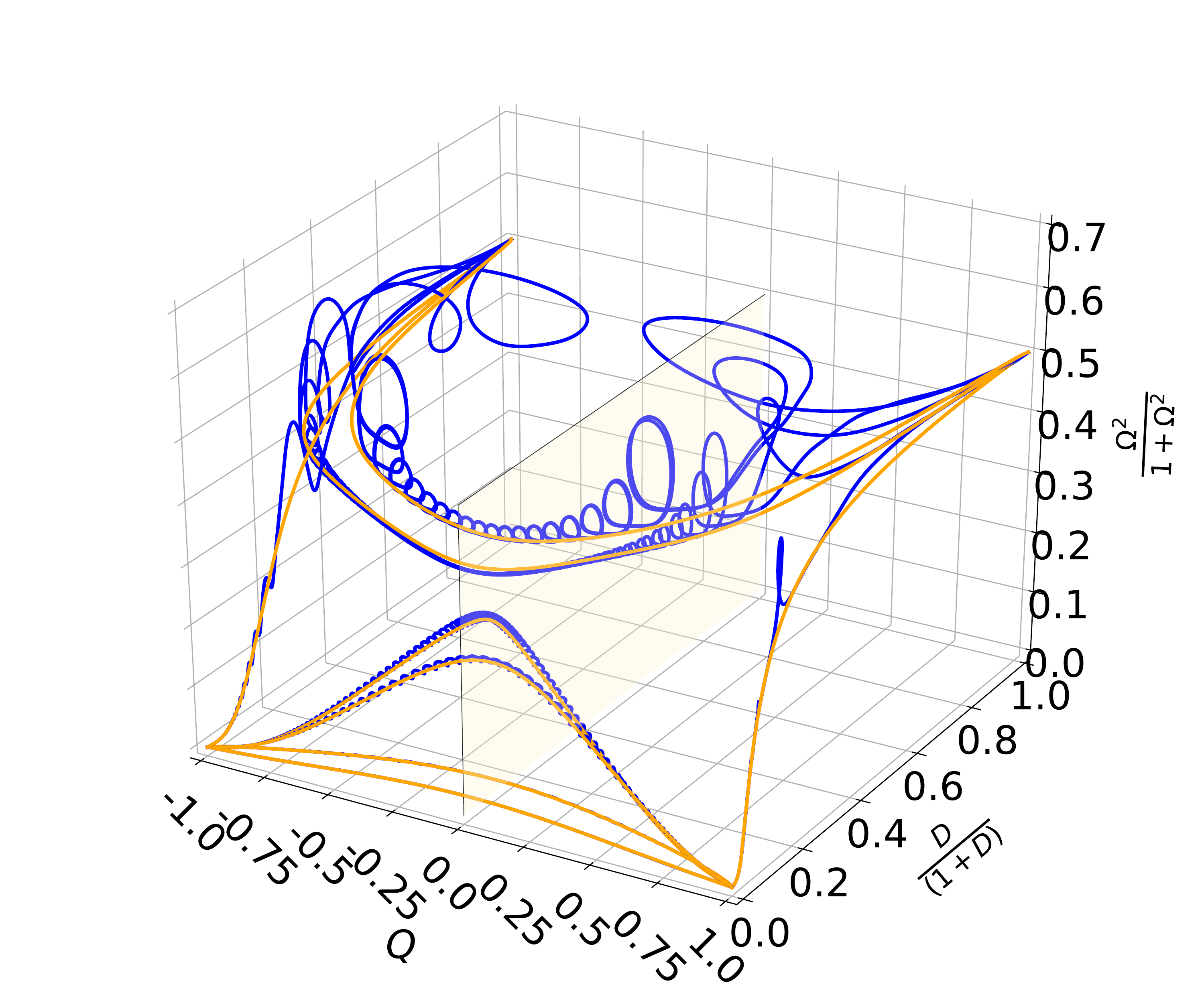}}
    \subfigure[ $\gamma=0.5$.]{\includegraphics[scale=0.25]{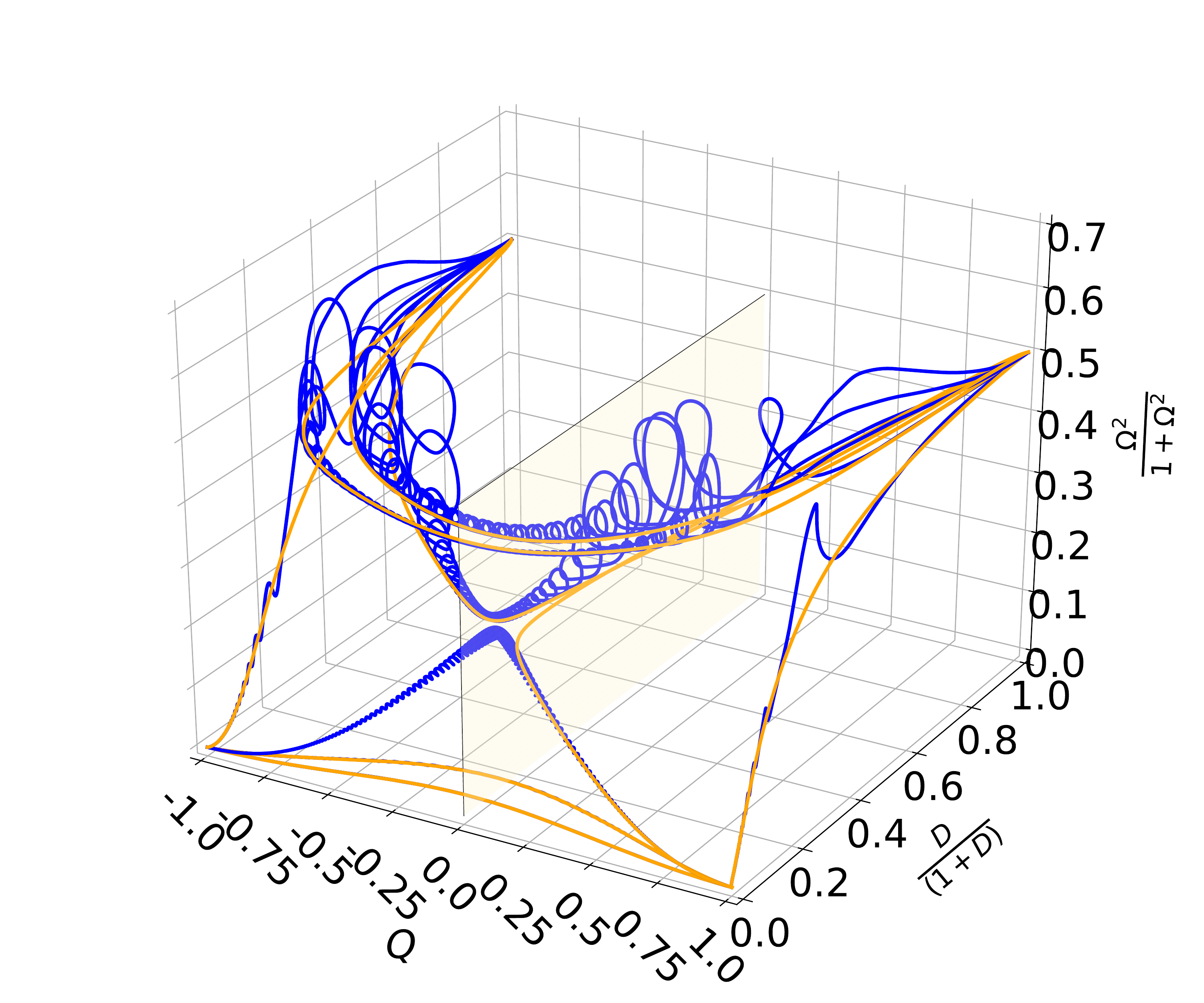}}
     \subfigure[ $\gamma=0.7$.]{\includegraphics[scale=0.25]{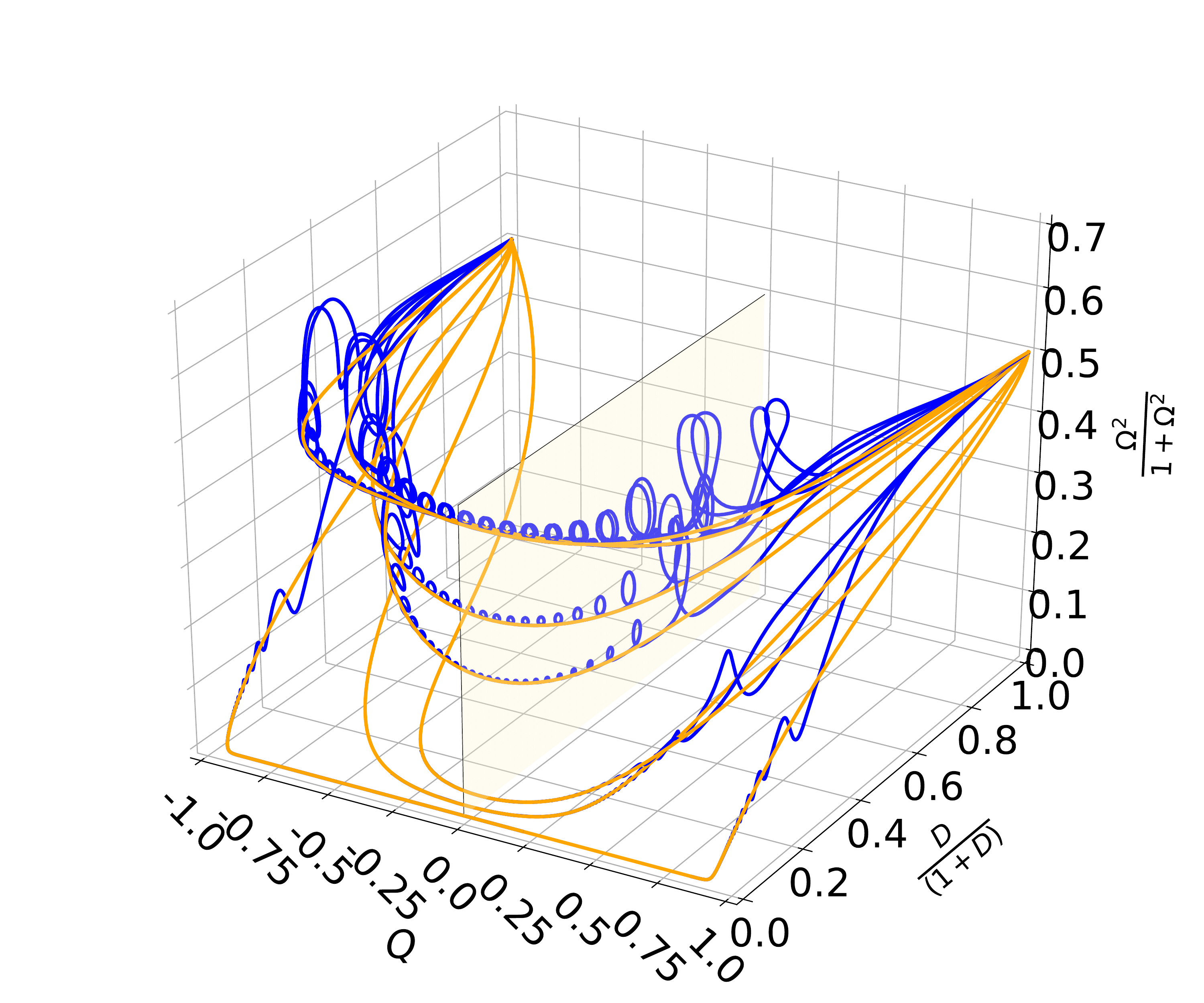}}
    \subfigure[ $\gamma=0.9$.]{\includegraphics[scale=0.25]{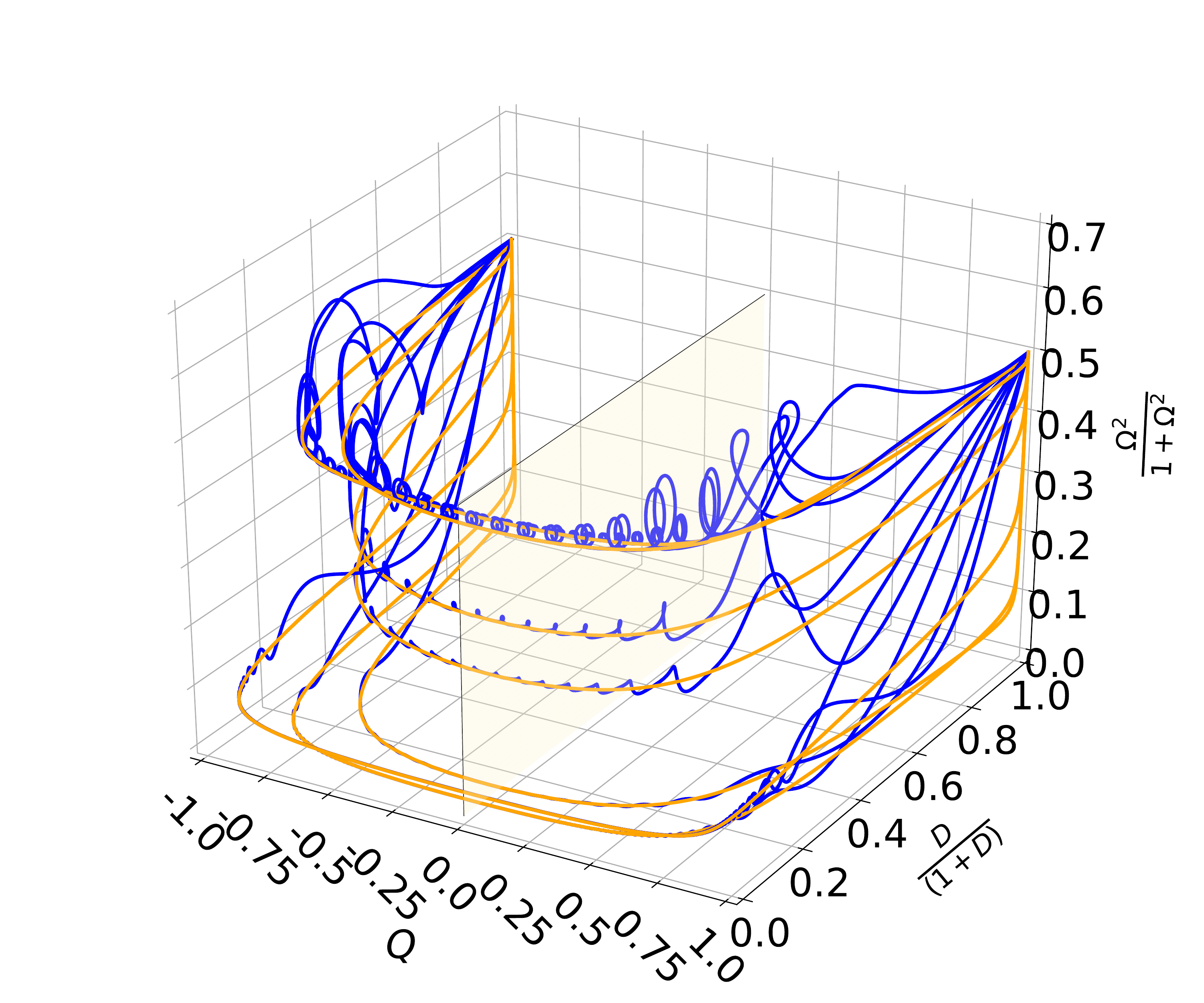}}
    \caption{\label{FIGURE25} Some solutions of the full system \eqref{unperturbed1FLRWClosed} (blue) and time--averaged system \eqref{avrgsystFLRWClosed} (orange) for the closed FLRW metric when $\gamma=0.1$,  $0.3$, $0.4$, $0.5$, $0.7$, and $0.9$, respectively; all of them represented in the space $(Q,D/(1+D),\Omega^{2}/(1+\Omega^{2}))$. We have used for both systems the initial data sets presented in Table \ref{Tab5a}.}
\end{figure*}

\begin{figure*}[h!]
    \centering
    \subfigure[ $\gamma=1.1$.]{\includegraphics[scale=0.24]{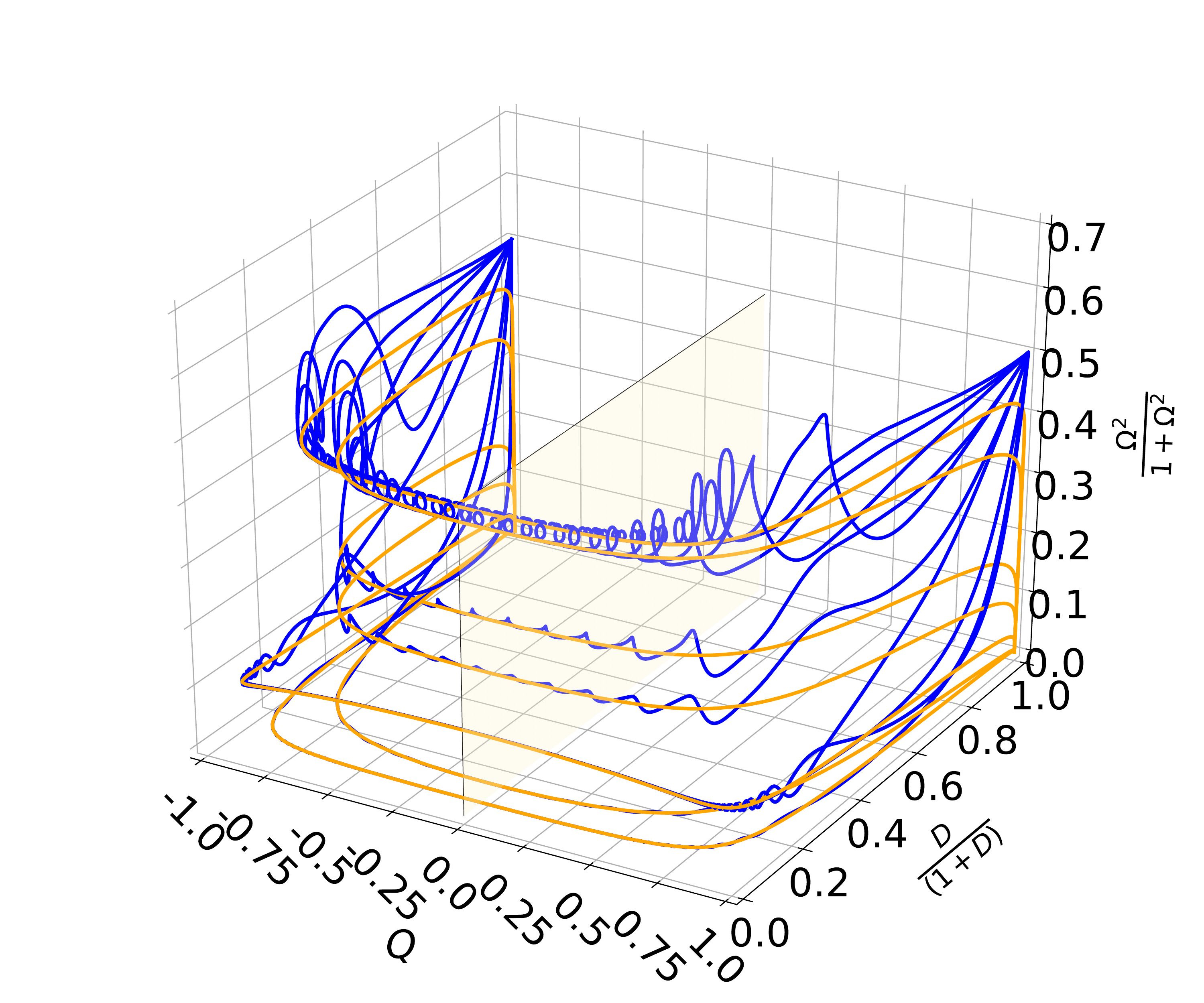}}
   \subfigure[ $\gamma=1.3$.]{\includegraphics[scale=0.24]{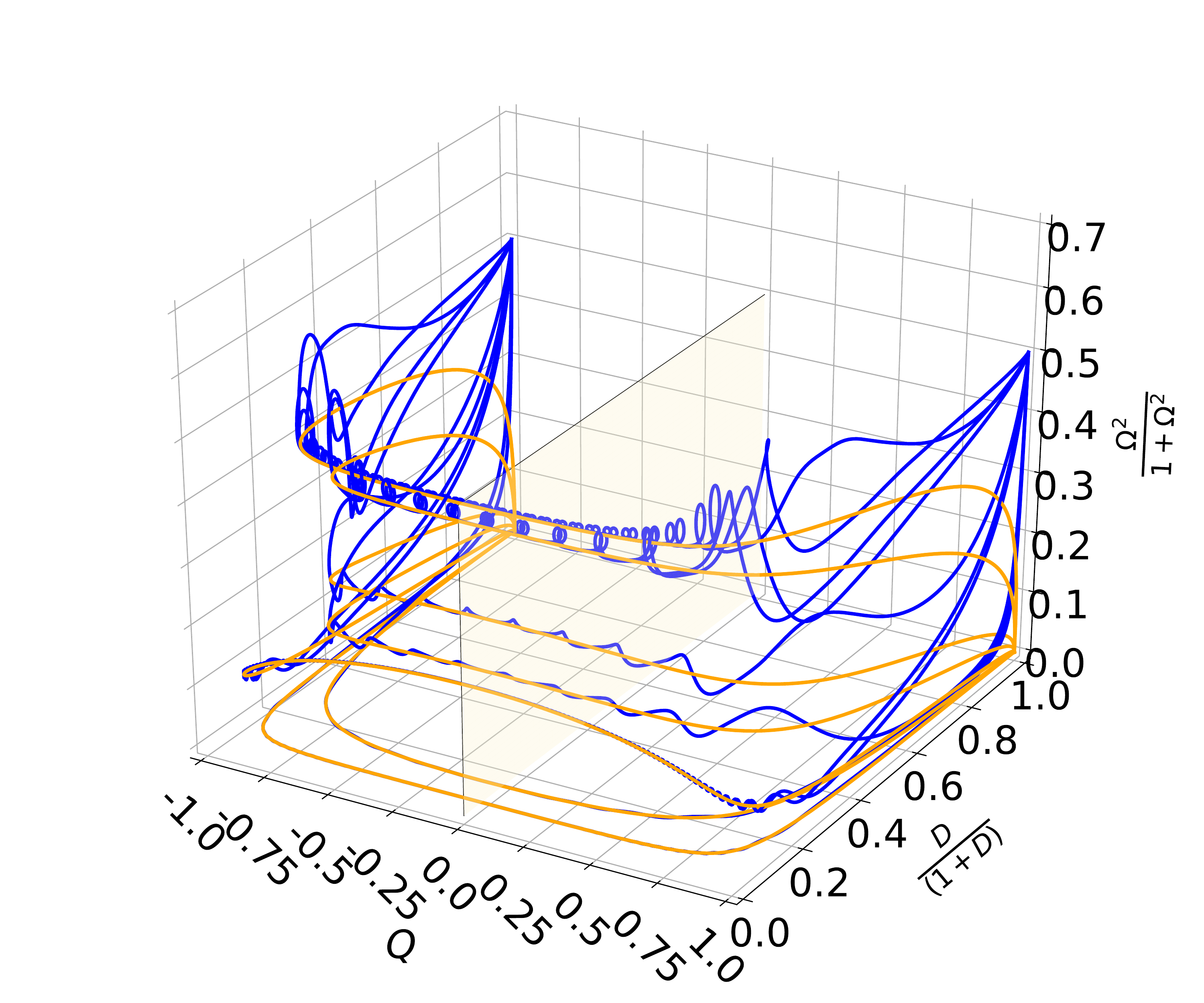}}
    \subfigure[ $\gamma=1.4$.]{\includegraphics[scale=0.24]{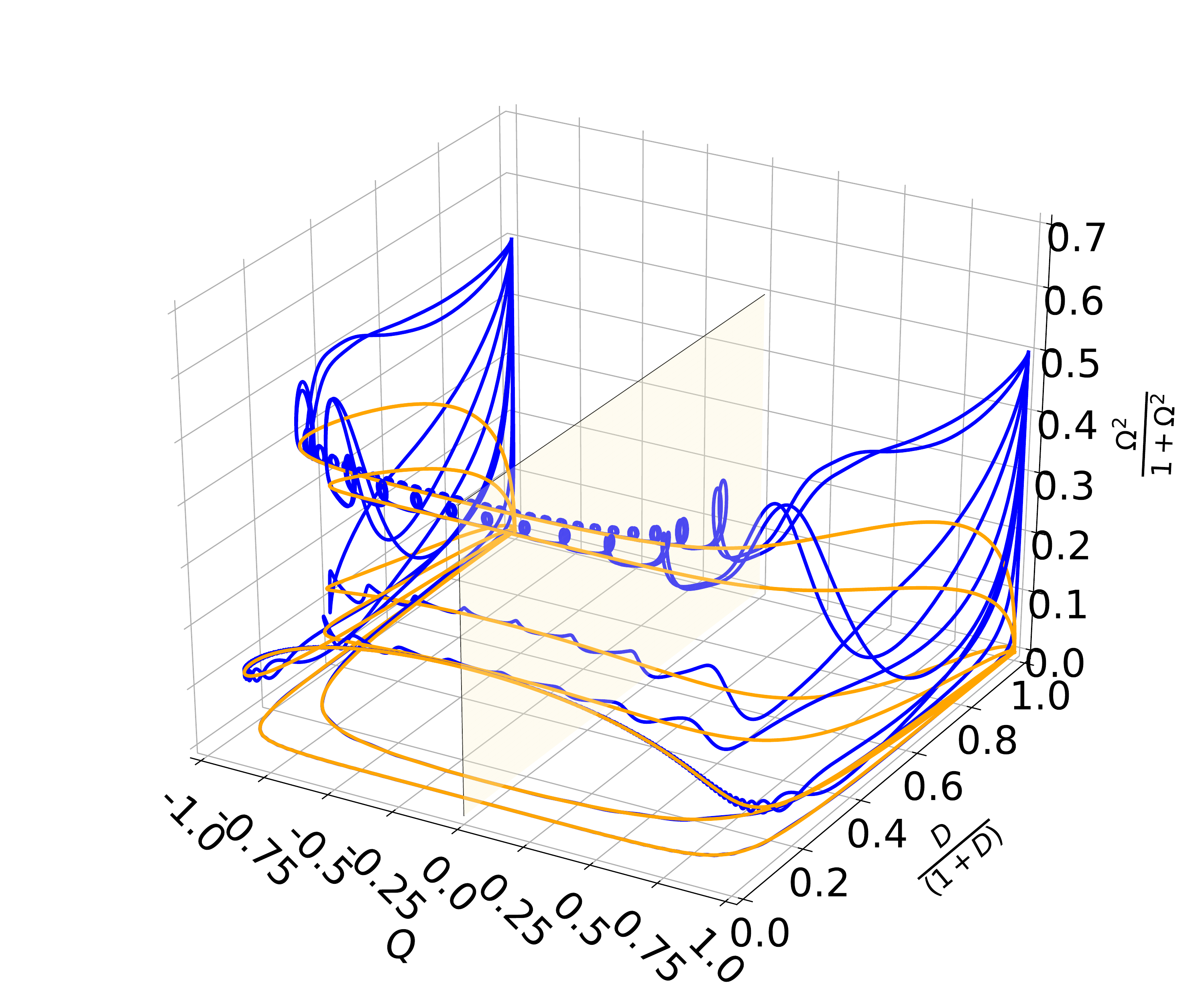}}
    \subfigure[ $\gamma=1.5$.]{\includegraphics[scale=0.24]{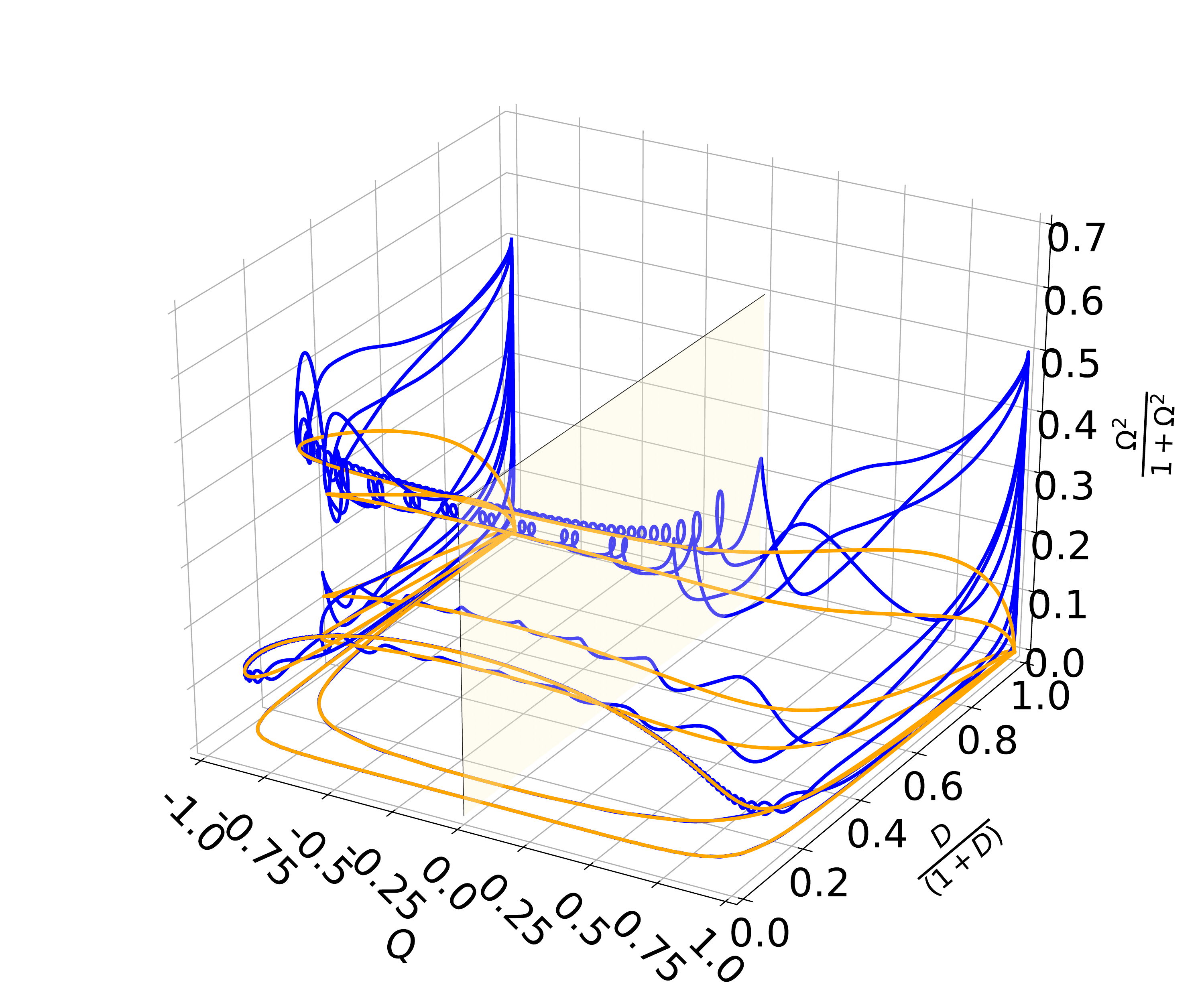}}
         \subfigure[ $\gamma=1.7$.]{\includegraphics[scale=0.24]{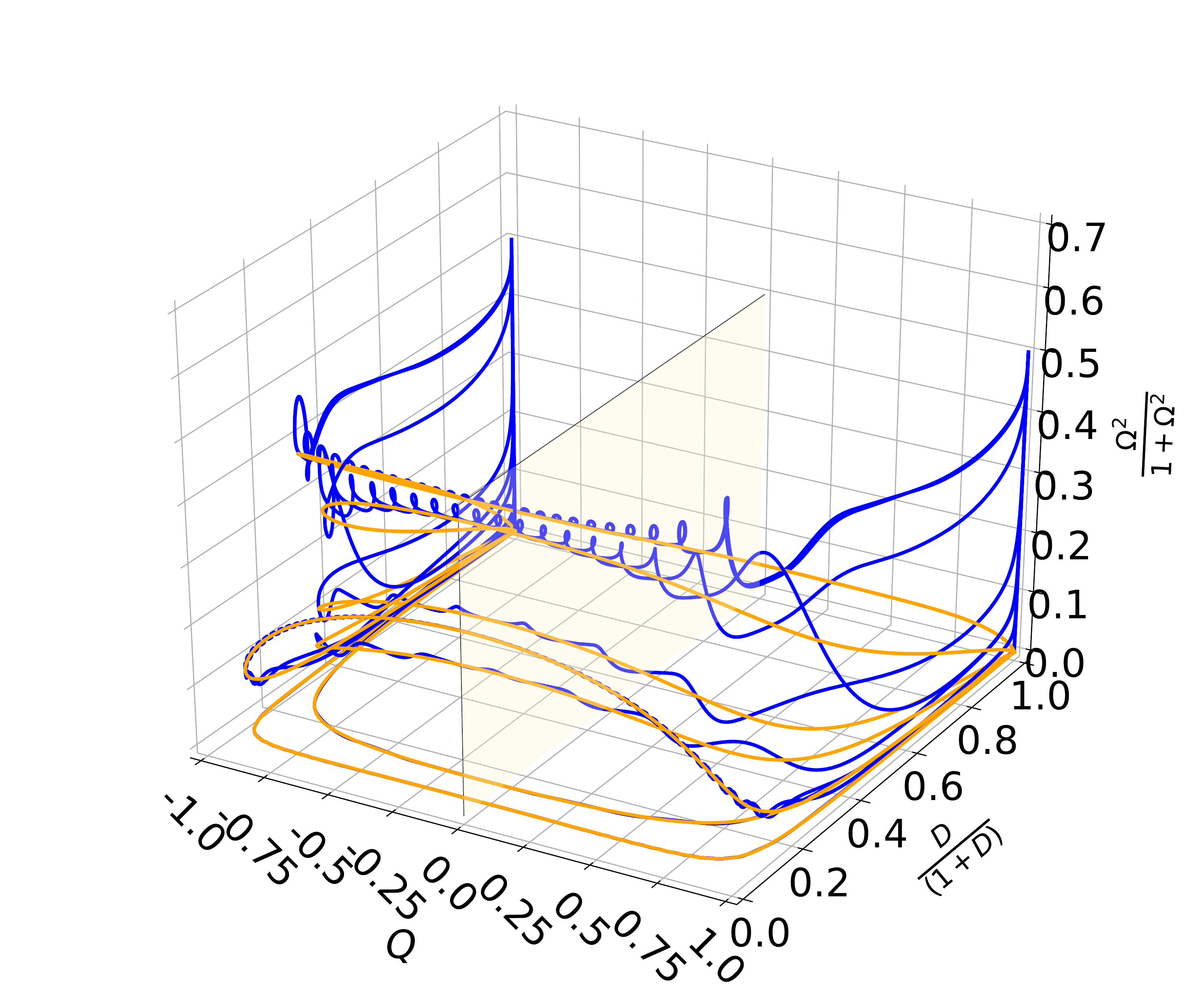}}
     \subfigure[ $\gamma=1.9$.]{\includegraphics[scale=0.24]{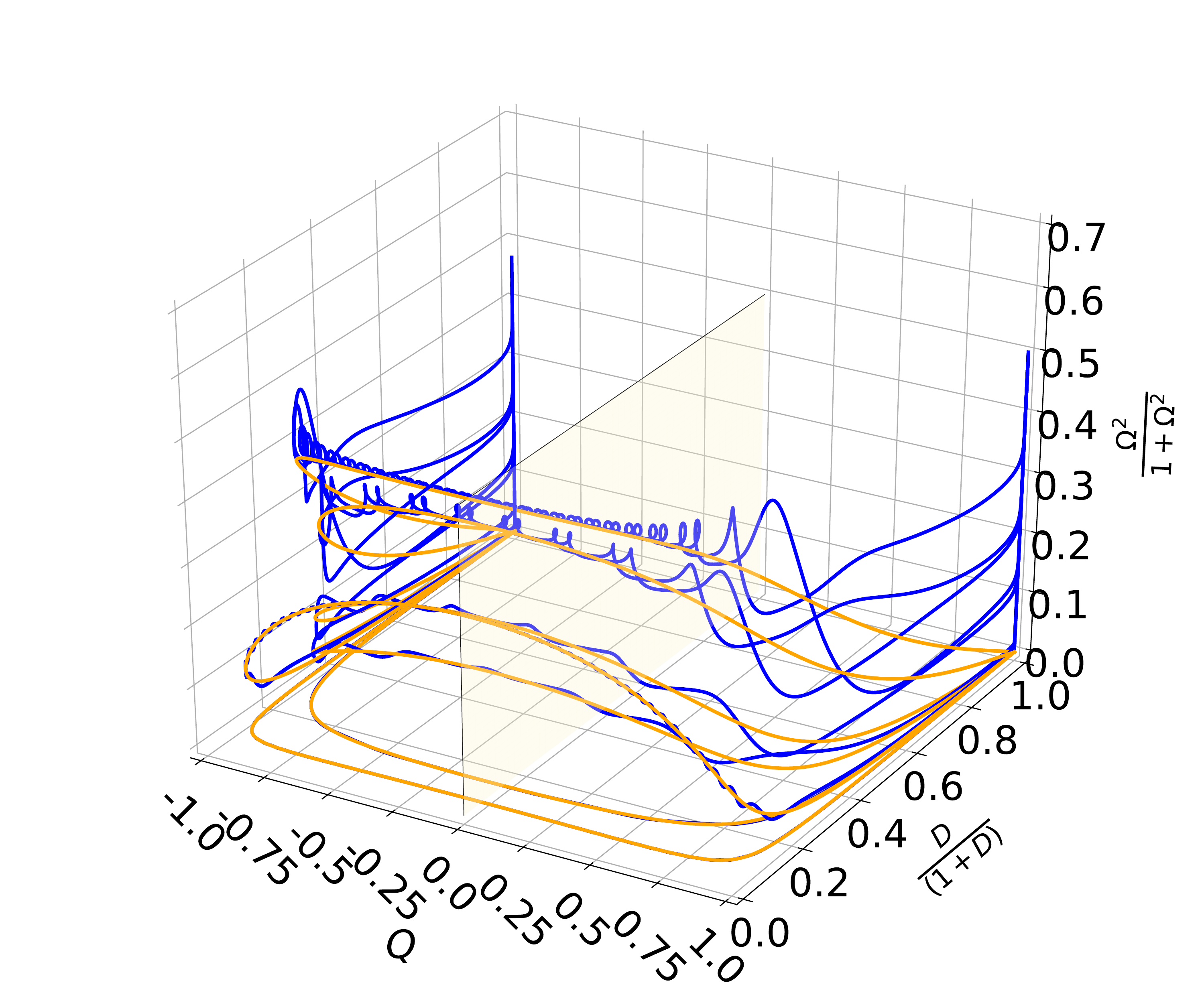}}
     \caption{\label{FIGURE27} Some solutions of the full system \eqref{unperturbed1FLRWClosed} (blue) and time--averaged system \eqref{avrgsystFLRWClosed} (orange) for the closed FLRW metric when $\gamma=1.1$,  $1.3$, $1.4$, $1.5$, $1.7$, and $1.9$, respectively; all of them represented in the space $(Q,D/(1+D),\Omega^{2}/(1+\Omega^{2}))$. We have used for both systems the initial data sets presented in Table \ref{Tab5a}.}
\end{figure*}

For the FLRW metric with positive curvature ($k=+1$) we integrate
\begin{enumerate}
    \item the full system given by \eqref{unperturbed1FLRWClosed}.
  \item The time--averaged system \eqref{avrgsystFLRWClosed}.
\end{enumerate}
For case $\gamma=1$,  for which the differential equation for $\Omega$ in time--averaged system \eqref{avrgsystFLRWClosed} becomes trivial, we integrate  time--averaged system \eqref{eqaverg4D}.
\newline
Independently of the value of $\gamma$, we use as initial conditions  ten data set presented in Table \ref{Tab5a}, where data sets $I$, $II$, and $VII$ are the symmetrical counterpart with respect to $Q$ of data sets $i$, $ii$, and $vii$.
\begin{table}
\caption{\label{Tab5a} Here we show ten initial data sets for the simulation of full system \eqref{unperturbed1FLRWClosed} and time--averaged system \eqref{avrgsystFLRWClosed} for $\gamma\neq 1$ and of system \eqref{eqaverg4D} for $\gamma=1$, for the FLRW metric with positive curvature ($k=+1$). All the conditions are chosen in order to  fulfill the inequalities $0\leq\Omega\leq 1$ and $-1\leq Q\leq 1$.}
\footnotesize\setlength{\tabcolsep}{9pt}
    \begin{tabular}{lcccccc}\hline
Sol.  & \multicolumn{1}{c}{$D(0)$} & \multicolumn{1}{c}{$\Omega^2(0)$} & \multicolumn{1}{c}{$Q(0)$} &  \multicolumn{1}{c}{$\varphi(0)$}  & \multicolumn{1}{c}{$t(0)$}  \\\hline
        i & $0.1$ & $0.9$ & $0.65$ & $0$ & $0$ \\
        I & $0.1$ & $0.9$ & $-0.65$ & $0$ & $0$ \\
        ii & $0.1$ & $0.1$ & $0.95$ & $0$ & $0$ \\
        II & $0.1$ & $0.1$ & $-0.95$ & $0$ & $0$ \\
        iii & $0.1$ & $0.4$ & $0$ & $0$ & $0$ \\
        iv & $0.02$ & $0.02$ & $0.25$ & $0$ & $0$ \\
        v & $0.1$ & $0.02$ & $0.25$ & $0$ & $0$ \\
        vi & $0.1$ & $0.25$ & $0.15$ & $0$ & $0$ \\
        vii & $0.1$ & $0.76$ & $0.2$ & $0$ & $0$ \\
        VII & $0.1$ & $0.76$ & $-0.2$ & $0$ & $0$ \\ \hline
    \end{tabular}
\end{table}
\noindent
In Figures \ref{Figure16}, \ref{Figure17}, \ref{Figure18}, and \ref{Figure19} we present projections of some solutions of  full system \eqref{unperturbed1FLRWClosed} and time--averaged system \eqref{avrgsystFLRWClosed} for $\gamma\neq 1$ and of system \eqref{eqaverg4D} for $\gamma=1$ in the $(Q, D/(1+D), \Omega^{2}/(1+\Omega^{2}))$ space  with their respective projection when $D=0$. For both systems the same initial data sets from Table \ref{Tab5a} were considered. Figures \ref{fig:ClosedFLRWCC3D} and \ref{fig:ClosedFLRWCC2D} show solutions for a fluid corresponding to CC ($\gamma=0$). Figures \ref{fig:ClosedFLRWDust3D} and \ref{fig:ClosedFLRWDust2D} show solutions for a fluid corresponding to dust ($\gamma=1$). Figures \ref{fig:ClosedFLRWRad3D} and \ref{fig:ClosedFLRWRad2D} show solutions for a fluid corresponding to radiation ($\gamma=\frac{4}{3}$). Figures \ref{fig:ClosedFLRWStiff3D} and \ref{fig:ClosedFLRWStiff2D} show solutions for a fluid corresponding to stiff fluid ($\gamma=2$). Figures \ref{Figure16}  and \ref{FIGURE25} show how solutions of full system (blue lines) follow the track of solutions of  averaged system (orange lines) for the whole $D$-range. Figures \ref{Figure17}, \ref{Figure18}, \ref{Figure19}, and \ref{FIGURE27}  are evidence that main theorem presented in Section \ref{SECT:II} is fulfilled for FLRW metric with positive curvature ($k=+1$) only when $D$ is bounded. That is,  solutions of full system (blue lines) follow the track of solutions of  averaged system (orange lines) for the time interval $t D =\mathcal{O}(1)$. However, when $D$ becomes infinite ($T\rightarrow 1$) and for $\gamma\geq 1$   solutions of   full system (blue lines) depart  from   solutions of   averaged system (orange lines) as $D$ becomes large. This is clear because when $D$ becomes large, the approximation obtained under the assumption of small  $D$ fails.

\FloatBarrier
\bibliographystyle{alpha}

\end{document}